\documentclass[linenumbers]{aa}  

    \usepackage{graphicx}
    \usepackage{txfonts}
    \usepackage[colorlinks = true, allcolors=blue]{hyperref}
    \usepackage{stfloats}
    \usepackage{booktabs}
    \usepackage{orcidlink}
    \usepackage{sidecap}
    \usepackage{pdflscape}
    \newcommand\arcdeg{\mbox{$^\circ$}}%
    \newcommand{\kms}{km s$^{-1}$}%
    \newcommand{\Msun}{\hbox{M$_\odot$}}
    \defcitealias{Risbud2025AA...694A.258R}{Paper I}
    \usepackage{amsfonts}
    \usepackage[normalem]{ulem}
    \usepackage{stfloats}   

\begin{document}

\title{Tidal Tails of Nearby Open Clusters}
\subtitle{II. A Review of Simulated Properties and the Reliability of Observational Catalogues}

\titlerunning{Properties and Reliability of Tidal Tails}
\authorrunning{Jadhav et al.}

\author{Vikrant V. Jadhav\orcidlink{0000-0002-8672-3300}\inst{1,2},
    Dhanraj Risbud\orcidlink{0000-0001-9174-2883}\inst{1},
	Pavel Kroupa\orcidlink{0000-0002-7301-3377}\inst{1,2},
	Wenjie Wu\orcidlink{0009-0007-9593-1041}\inst{1}
}
\institute{
	Helmholtz-Institut für Strahlen- und Kernphysik, Universität Bonn, Nussallee 14-16, D-53115 Bonn, Germany\\
	\email{vjadhav@astro-uni-bonn.de}, \email{pkroupa@uni-bonn.de}
	\and
	Astronomical Institute, Faculty of Mathematics and Physics, Charles University, V Holešovičkách 2, CZ-180 00 Praha 8, Czech Republic
}

\date{Received January 13, 2023; accepted June 27, 2023}


\abstract
{
    Recent studies using \textit{Gaia} data have reported tidal tail detections for tens to hundreds of open clusters. However, a comprehensive assessment of the reliability and completeness of these detections is lacking.

}
{
    This work aims to summarise the expected properties of tidal tails based on $N$-body simulations, review the reliability of tidal tail detections in the literature, and grade them according to a set of diagnostic tests. We also provide an overview of the general characteristics of tidal tails available in the literature.

}
{
    We used a grid of 68--20000 \Msun\ simulated clusters and analysed the formation and evolution of the tidal tails.
    We compiled 122 catalogues from the recent literature, encompassing 58 unique clusters within 500~pc of the Sun. We employed various tests based on photometric, morphological and dynamical signatures and comparisons with simulated clusters to grade the tidal tails as gold, silver, and bronze. One of the primary tests was to measure apparent torsion in the Galactocentric XY plane.
}
{
	Based on the simulations, we analysed the complex morphology of the tidal tails and their properties (such as their size, span, stellar types, number density, and mass function) at various cluster masses and ages. During the first 100--200 Myr of evolution, the tails typically form a characteristic \reflectbox{S} shape, with an amplitude that scales with cluster mass. The tail span increases at a rate of $\approx$4 times the initial velocity dispersion, and the {near-tail} {(within 100 pc of the cluster)} is predominantly populated by recent escapees.
    In evaluating {122} published tidal tail catalogues, we found that {15} gold-quality catalogues and {55} silver-quality catalogues passed the majority of the tests. The remaining 51 catalogues were graded as bronze; care should be taken before using these catalogues for further analysis. The age, metallicity, binary fraction, and mass function of stars in the tails were generally consistent with those of their parent clusters.
}
{   
    The simulations presented here provide first-order approximations of the structure and evolution of the tidal tails.
    The gold and silver-grade catalogues (69 catalogues of 40 clusters) represent reliable samples for detailed analyses of tidal tails. Future spectroscopic and astrometric data from large-scale surveys will be essential for further validation and for leveraging tidal tails as tracers of cluster dissolution and the Galactic potential.
}


\maketitle
%

\section{Introduction} \label{sec:introduction}

As star clusters evolve and move within the Milky Way, they evaporate stars due to internal dynamics, energy equipartition, and Galactic interactions \citep{Kroupa2001MNRAS.321..699K, Lada2003ARA&A..41...57L, Baumgardt2003MNRAS.340..227B, Oh2016A&A...590A.107O}. This leads to the formation of tidal tails that extend ahead of and behind the cluster along its Galactic orbit \citep{Just2009MNRAS.392..969J, Kupper2010MNRAS.401..105K}.

Several methods have been developed to identify tidal tails in open clusters.
\citet{2019AA...621L...2Roser_Hyades}, \citet{2019AA...627A...4Roser_Praesepe} and \citet{Risbud2025AA...694A.258R} used the convergent point method. 
\citet{2019AA...624L..11Furnkranz} used the density-based algorithm \textsc{dbscan}.
\citet{2019AA...621L...3Meingast} used overdensities in the 3D space velocities to identify Hyades tails.
\citet{2019ApJ...877...12Tang}, \citet{2020ApJ...889...99Zhang}, \citet{2021ApJ...912..162Pang} and \citet{2022ApJ...931..156Pang} used unsupervised Self-Organizing-Map technique \textsc{StarGO} to 5D data.
{\citet{Oh2020MNRAS.498.1920O} modelled the internal kinematics of Hyades astrometry to reanalyse the \citet{2019AA...621L...2Roser_Hyades} and \citet{2019AA...621L...3Meingast} catalogues.
\citet{2021AA...645A..84Meingast} and \citet{Sharma2025}} used proper motions and 3D clustering using \textsc{dbscan}.
\citet{Jerabkova2021AA...647A.137J} used the compact convergent point method and comparison with $N$-body simulations. We analysed their M1 and M5 model based catalogues independently.
\citet{Bhattacharya2022MNRAS.517.3525B} used \textsc{ml-moc}, which is based on the k-nearest neighbour algorithm and the Gaussian mixture model, and identified extended structures around 46 clusters (19 as tidal tails, 26 as corona but no clear tails).
\citet{2022MNRAS.514.3579Boffin} used combination of the convergent point method and \textsc{dbscan}.
\citet{2023AA...679A.105Vaher} used back-propagation of 6D coordinates of field stars and clusters to identify cluster escapees.
{\citet{Olivares2023AA...675A..28O} used Gaussian mixture models of 6D astrometric phase space to detect members of clusters, tails, and moving groups.}
\citet{2024arXiv240618767Kos} used comparisons with probabilistic cluster dissolution models.

However, all these methods are limited by the data availability and quality. \textit{Gaia} has provided 5D information for $>1$ billion stars \citep{2016A&A...595A...1_Gaia_Instrument, Gaia2023A&A...674A...1G}; however, the 6D information is limited due to a lack of accurate radial velocities. 
{The parallax-based distances can be derived for spherical objects for up to a few kpc \citep{Olivares2025A&A...693A..12O, Liu2025AJ....169..116L}. However, the tidal tails are not spherically symmetric and have complex morphology (see Section~\ref{sec:morphology}); thus, simple prior distributions cannot be used for converting \textit{Gaia} parallaxes to distances. $N$-body based priors might be an option (similar to the approach by \citealt{Jerabkova2021AA...647A.137J}); however, that introduces its own biases.
This work aims to test the validity of such methods; hence, we selected the most model-independent method of converting parallaxes to distances.} The accuracy of {individual} \textit{Gaia} parallaxes is reliable {(with precision of 10 pc or better)} only within $\lessapprox$500 pc from the Sun, restricting 6D analyses primarily to the solar neighbourhood \citep{Piecka2021arXiv210707230P}.

True tidal tails are coeval and chemically homogeneous with their parent cluster. However, without high-resolution spectroscopic data, metallicity cannot be widely used as a discriminant (e.g. \citealt{Piatti2025arXiv250524522P}). The age estimation for individual stars remains complex. Techniques such as gyrochronology \citep{Sha2024ApJ...977..103S}, and asteroseismology \citep{Soderblom2010ARA&A..48..581S} can be informative but are not feasible for large samples. Recently, \citet{Xu2025arXiv250417744X} used 3D information to select a golden sample of tidal tails in 8 clusters among the 389 open clusters in \citet{Tarricq2022A&A...659A..59T}. Nevertheless, a detailed analysis of other published tail catalogues is essential to reliably use them for studying stellar evaporation and probing the Galactic environment.

{Here,} we first present the general properties of tidal tails based on $N$-body simulations of 68--20000~\Msun\ clusters in the solar neighbourhood (Section~\ref{sec:properties_of_tails}). We then assess the reliability of published tidal tail catalogues (Section~\ref{sec:reliability}) using photometric, morphological, and dynamical {features} as diagnostic tools.

\section{Data and Methods} \label{sec:data}

\subsection{\textit{Gaia} DR3 data}

We collected available catalogues of open cluster tidal tails from recent studies {which focused on the detection of extended structure around nearby open clusters (listed in Table~\ref{tab:literature_table}).
Table~\ref{tab:literature_table} also gives a brief description about references, their data quality, fraction of stars with 6D information, and the median span of their respective tidal tail catalogues.}
There are {582} different catalogues for 499 clusters among these references, which were classified as tidal tails or extended structures. We limited our analysis to clusters with distances closer than 500 pc {exhibiting reliable individual 3D astrometry}. 
Thus, the final sample is {122} catalogues of 58 unique clusters.

The cluster properties (age, distance, position, and velocity) {were adopted from \citet{Hunt2024A&A...686A..42H} except for the cluster LP 2429 (for which \citealt{2022ApJ...931..156Pang} was used).}
We cross-matched all the catalogues with \textit{Gaia} DR3 to obtain the astrometry and with the catalogue of \citet{2021AJ....161..147Bailer-Jones} to obtain geometric distances. The geometric distances were preferred instead of inverting parallax due to the unavailability of the prior distribution for a cluster with undefined tidal tails and to avoid negative distances.
The recommended cuts were applied to the catalogues for selecting high-quality cluster members {as given in Table~\ref{tab:literature_table}.}

\subsection{Simulation data}

We created a grid of clusters with masses in the range of 68 to 20000 \Msun\ to analyse the tidal tail properties and comparison with the simulations. The $N$-body simulations were performed using \textsc{PeTar} \citep{2020MNRAS.497..536Wang}. The clusters were initialised using \textsc{McLuster} \citep{Kupper2011MNRAS.417.2300K} with primordial mass segregation \citep{Baumgardt2008ApJ...685..247B}, no primordial binaries, solar metallicity, and the canonical initial mass function (IMF; \citealt{2001MNRAS.322..231KroupaIMF}).
The stars were evolved till 200--5000 Myr based on their mass (i.e. dissolution) and observational requirements using stellar evolution via \textsc{bse} \citep{Hurley2000MNRAS.315..543H, Hurley2002MNRAS.329..897H, 2013ascl.soft03014H, Banerjee2020A&A...639A..41B} in the Galactic potential. 
{In this study, we adopted the Galactic potential \texttt{MWPotential2014}, which comprises of three components: 
(i) The bulge with mass of 5 G\Msun\ as a spherical power-law density profile with exponent of $-1.8$ and cut-off radius of 1.9 kpc. 
(ii) The disk with mass of 68 G\Msun\ with the potential as proposed by \citet{Miyamoto1975PASJ...27..533M}. 
(iii) The dark matter halo model proposed by \citet{Navarro1996ApJ...462..563N} with scale radius of 16 kpc.
We used the default solar position ($R_{GC}=8$ kpc) and velocity (220 \kms) parameters. The relative contributions of the aforementioned three components at the Solar position are 0.05, 0.60, and 0.35, respectively.
(see \citealt{Bovy2015ApJS..216...29B} for more details)}. 

The initial conditions are given in Table~\ref{tab:sim_setup}. 
{To explore the `true' properties of tidal tails, }
the simulated data (without observational noise) were directly used to present the analysis of tidal tails in Section~\ref{sec:properties_of_tails}. {Synthetic noise was added for comparison with observations in Section~\ref{sec:reliability}. See Section~\ref{sec:selecting_matching_simulation} for the details.}
When necessary, the clusters were rotated around the Galactic centre to make the Galactic azimuth $0\arcdeg$ for easier comparison (i.e., $y_{galactocentric}=0$).

\begin{table*}[]
    \centering
    \caption{Initial conditions of the $N$-body simulations.}
    \begin{tabular}{p{3.5cm}p{5.5cm}cc}
    \toprule
        Model name & Initial mass [\Msun] & $xyz_{galactocentric}$ [pc] & $V_xV_yV_{z\ galactocentric}$ [km s$^{-1}$]\\ \hline
        M68, M82, ..., M20000 &    68,    82,   100,   121,   146,   177,   215,   261,   316,   383,   464,   562,   681,   825,  1000,  1211,  1467,  1778,  2154,  2610,  3162,  3831,  4641,  5623,  6812, 20000& -8000, 0, 0 & 0, -220, 0\\
        M20000\_e & 20000 & -7067.59, 2899.3, 99.92 & -110.96, -193.72, 5.33\\     \bottomrule

    \end{tabular}
    \tablefoot{Models M68—M20000 are a grid of clusters on circular orbits, while M20000\_e is on an eccentric (Praesepe-like) orbit. The initial number density within the half-mass radius was kept the same at 306 stars pc$^{-3}$ for all simulations.}
    \label{tab:sim_setup}
\end{table*}

\subsection{Transformations and derived parameters}

We used \textsc{astropy} to perform predefined coordinate transformations between the \texttt{icrs}, \texttt{galactic}, and \texttt{galactocentric} frames. The Solar position in Galactocentric coordinates was fixed to: \texttt{galcen\_distance} = 8122 pc, \texttt{z\_sun = 20.8 pc} and \texttt{galcen\_v\_sun} = (12.9, 245.6, 7.78) km s$^{-1}$ \citep[and references therein]{astropy:2022}.
The absolute magnitudes were calculated using individual stellar distances (based on \citealt{2021AJ....161..147Bailer-Jones}) instead of cluster distance due to the extended nature of the tidal tails and relatively small errors in the parallax (due to the selection criteria by the original authors).
The cluster's orbit was derived from the present-day position and velocity using \textsc{galpy} assuming the \texttt{mwpotential2014} Milky Way potential \citep{Bovy2015ApJS..216...29B}.
{The orbits were integrated using $<0.01$ Myr time-steps for up to $\pm$80 Myr.}
The local tangential projection was used to calculate the angular distance from the cluster centre ($r_{sky}$), radial ($\mu_R$) and tangential ($\mu_T$) components of the proper motions (similar to \citealt{Jadhav2024A&A...687A..89J}).

Measuring the distance from the cluster for tidal tails is not trivial due to their complex morphology (see Section~\ref{sec:morphology}). A simple linear distance from the cluster centre fails to capture the true structure once the tails extend over several kiloparsecs, as tail stars trace a near-circular locus along the cluster's Galactic orbit. Hence, we defined a new metric, $distance\_along\_orbit$, which quantifies the distance between the cluster centre and the point on the numerically integrated Galactic orbit (computed using \textsc{galpy}) that is closest to the given source, measured along the orbit itself.
Once the distance between the cluster and particle reaches half of the Galactic orbit, the classification of the particle into leading or trailing tail is non-trivial (and requires dynamical information such as velocities and the growth rate of the tails). Hence, we limit the measurements to $\pm45\%$ of the Galactic orbit.
Fortunately, the total span of our longest tail is smaller than the Galactic orbit (see Figure~\ref{fig:distance_along_orbit_appendix}), so this issue does not affect the present study.

The corresponding $distance\_from\_orbit$ is defined as the distance of the source from the cluster's Galactic orbit. Note that this is a directionless separation in the 3D space.
Figure~\ref{fig:distance_along_orbit_appendix} shows examples of these measurements for models M20000 and M20000\_e. 
The panels (c) and (f) showing $distance\_from\_orbit$ vs $distance\_along\_orbit$ distributions demonstrate the complex morphology of the tails.
The specific angular momentum ($sL_z$) and the total specific energy ($sE_{total}$) of the $N$-body particles were calculated using their velocities and the Milky Way potential.
High-velocity ejections (likely due to close dynamical interactions or large supernova kicks) from the cluster can have significantly different Galactic orbits and result in deviant $distance\_from\_orbit$. Such sources were ignored while analysing the general tidal tail population.
{The last row in Figure~\ref{fig:distance_along_orbit_appendix} shows the effect of adding \textit{Gaia}-like noise to the data, which leads to a reduction in the number of fainter and distant stars, and a significant increase in the astrometric noise with distance.}

We defined the Galactic Azimuth ($\phi_{galactocentric}$) as the angle created by the Sun-Galactic centre-star on the Galactic plane. The radial velocity in the Galactic plane ($v_{r, galactocentric}$) is defined as the velocity from the vantage point of the Galactic centre. For cluster analysis, we measured the relative azimuth ($\Delta\phi_{galactocentric}$) and relative radial velocity ($\Delta v_{r, galactocentric}$) by subtracting the cluster mean parameters.

The cluster centre was identified from the centre of density, similar to the methodology of \citet{Harfst2007NewA...12..357H}. The tidal radius ($R_{tidal}$) was calculated based on the iterative calculations of Eq.~\ref{eq:r_tidal} until the values converged,
\begin{equation} \label{eq:r_tidal}
    R_{tidal} = R_{Galaxy}\left(\frac{M_{cluster}}{3\times M_{Galaxy}}\right)^{1/3}.
\end{equation}
Here, $M_{cluster}$ is the total cluster mass within $R_{tidal}$, $R_{Galaxy}$ is the Galactocentric distance of the cluster, and $M_{Galaxy}$ is the mass of the Galaxy.
The stars inside the tidal radius were considered as cluster members for the $N$-body models yielding $M_{cluster}$.

We estimated the half-mass radius ($R_{1/2}$) and half-mass population size ($N_{1/2}$) based on all the stars within the tidal radius.
The two-body relaxation time ($T_{relax}$) and the number of $T_{relax}$ passed ($\nu_{relax}$) were calculated as follows \citep[Eq. 2-63]{Spitzer1987degc.book.....S}:
\begin{equation}
    \begin{split}
        T_{half\_mass\_relaxation} &= \frac{1}{\text{ ln}(0.4 N_{cluster})}\sqrt{0.138 \frac{N_{cluster} R_{1/2}^{3}}{G\overline{M}_{cluster}} }\\
        \nu_{relax} &= Age_{cluster}/T_{half\_mass\_relaxation}
    \end{split}
\end{equation}
where $\overline{M}_{cluster}$ is the mean mass of the particles in the cluster, $N_{cluster}$ is the total number of particles in the cluster within $R_{tidal}$, and 0.4 is the Coulomb factor. No cuts were applied to the simulated clusters to remove faint stars, which are typically absent from observed clusters.

\subsection{Unresolved binary fraction}

Unresolved binaries appear brighter and redder than the primary star \citep{Jadhav2021AJ....162..264J}. Thus, the colour-magnitude diagram (CMD) can be used to identify such binaries. An accurate absolute CMD is necessary for reliable measurements. As the previous research applied stringent quality cuts based on parallax errors, we can reliably use the \textit{Gaia} based \texttt{r\_med\_geo} distances to measure the distance modulus. The extinction and metallicity values were taken from \citet{Bossini2019A&A...623A.108B}. As the binary analysis depends on isochrone fitting, we limit our binarity analysis to the clusters present in their catalogue.

Due to known issues with the \textit{Gaia} BP filter at fainter magnitudes, we chose the G vs (G - G$_{RP}$) CMD for these calculations \citep{Riello2021A&A...649A...3R, Jadhav2021MNRAS.503..236J}. We compared the observed CMDs with parsec isochrones \citep{Bressan2012MNRAS.427..127B}; however, the isochrones are never a perfect fit to the cluster main sequence \citep{Rottensteiner2024A&A...690A..16R}. Hence, we used Gaussian process regression (\textsc{robustgp}; \citealt{Li2020ApJ...901...49L, Li2021A&C....3600483L}) to identify the empirical main sequence ridge line. Magnitude cuts were given to the data to remove stars near and above the main sequence turn-off.

The G--mass relation from the isochrone was propagated to the ridge line. The ridge line was then used to calculate G$_{RP, ridge\ line}$. The mass--magnitude relations were then used to create ridge lines for different mass ratios, q $\in$ [0, 0.1, 0.2, ... 1.0]. We defined a parameter $cmd\_distance$, which is the distance from the region defined by the ridge lines of $q=0$ and $q=1$ in the CMD ($cmd\_distance$ is zero within this region). Sources can fall outside this region due to errors or stellar evolution. To be conservative, we interpolated the mass, q, luminosity, and temperature using \textsc{scipy}'s \citep{Virtanen2020NatMe..17..261V} smooth piecewise cubic interpolator ($CloughTocher2DInterpolator$; \citealt{ALFELD1984169}) for sources with $cmd\_distance$ = 0. A nearest-neighbour interpolator ($NearestNDInterpolator$) was used to interpolate for sources with $cmd\_distance$ $>0$. The quadratic summation of photometric errors, parallax errors, and the $cmd\_distance$ was used to measure a total CMD error, and it was used to calculate the errors in the q and mass values (see Figure~\ref{fig:binary_fraction_demo_appendix} for details).

To measure the binary fraction (BF), we limit the selection to sources within $cmd\_distance$ $<0.1$, and the CMD region where all systems for a given primary mass were within previous selections. The BF ($BF_{q>0.5}$) was then defined as the number of sources with $q>0.5$ divided by the total sources within the selection.

\section{Properties of tidal tails from simulations} \label{sec:properties_of_tails}

\subsection{Morphology of tidal tails} \label{sec:morphology}

\begin{figure*}
    \centering
    \includegraphics[scale=0.55]{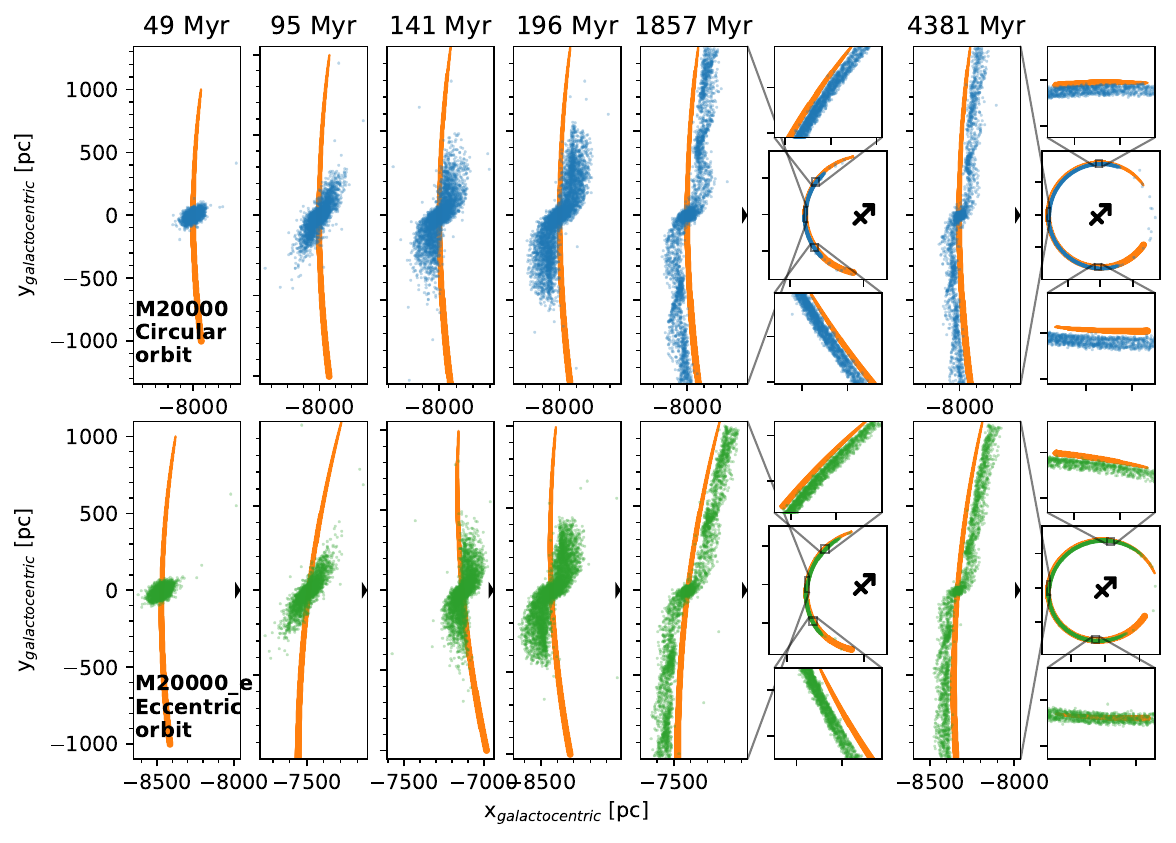}
    \includegraphics[scale=0.55]{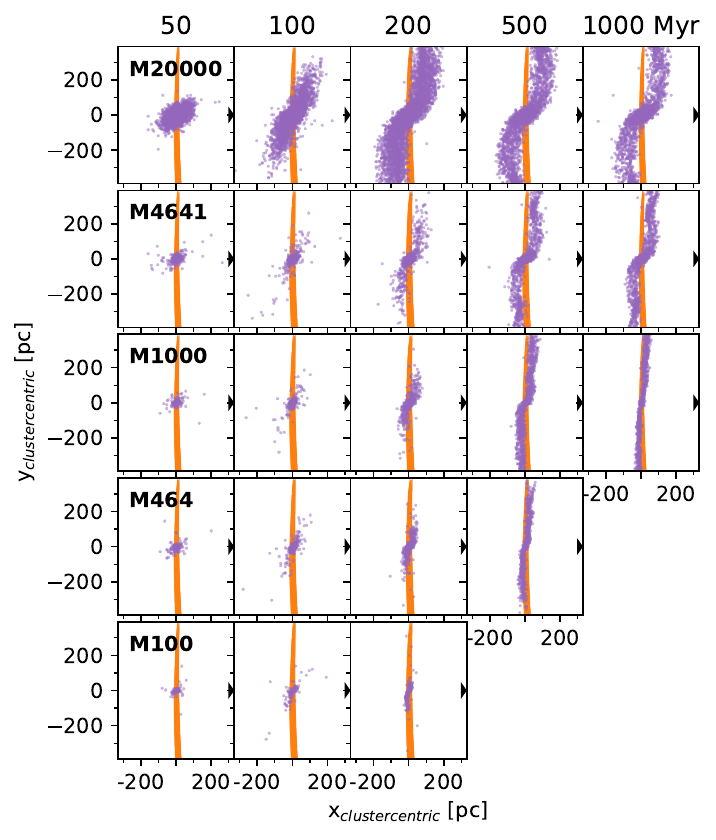}
    \caption{Evolution of tidal tail morphology.
    The top left (blue) models: M20000 (on circular orbit). 
    The bottom left (green) models: M20000\_e (on eccentric orbit).
    The right (purple) models: Grid of M20000, M4641, M1000, M464, and M100 clusters on circular orbits.
    The different columns show the clusters at the age written above each panel.
    For the 1857 and 4381 Myr models, the whole simulation (middle right), a zoomed-in section of the cluster (middle left), and arbitrary zoomed-in sections along the leading (top right) and trailing (bottom right) tail are shown to compare the position of the tidal tail and the orbit at large distances.
    The cluster orbit (orange) and the Galactic centre position (black) are shown for reference. All clusters are rotated and placed on the X-axis for easier comparison.}
    \label{fig:morphology}
\end{figure*}

Figure~\ref{fig:morphology} shows the evolution of tail morphology. The left plots show the tail structure for clusters on circular (M20000) and eccentric (M20000\_e) orbits at various ages. The right plots show the tail morphology for various initial masses and ages.
Movies showing the full morphological evolution of the M1000, M20000, and M20000\_e are available as online supplementary material. (see Figure~\ref{fig:movie_screenshot_appendix} for a screenshot).

\begin{figure*}
\centering
    \includegraphics[width=0.8\linewidth]{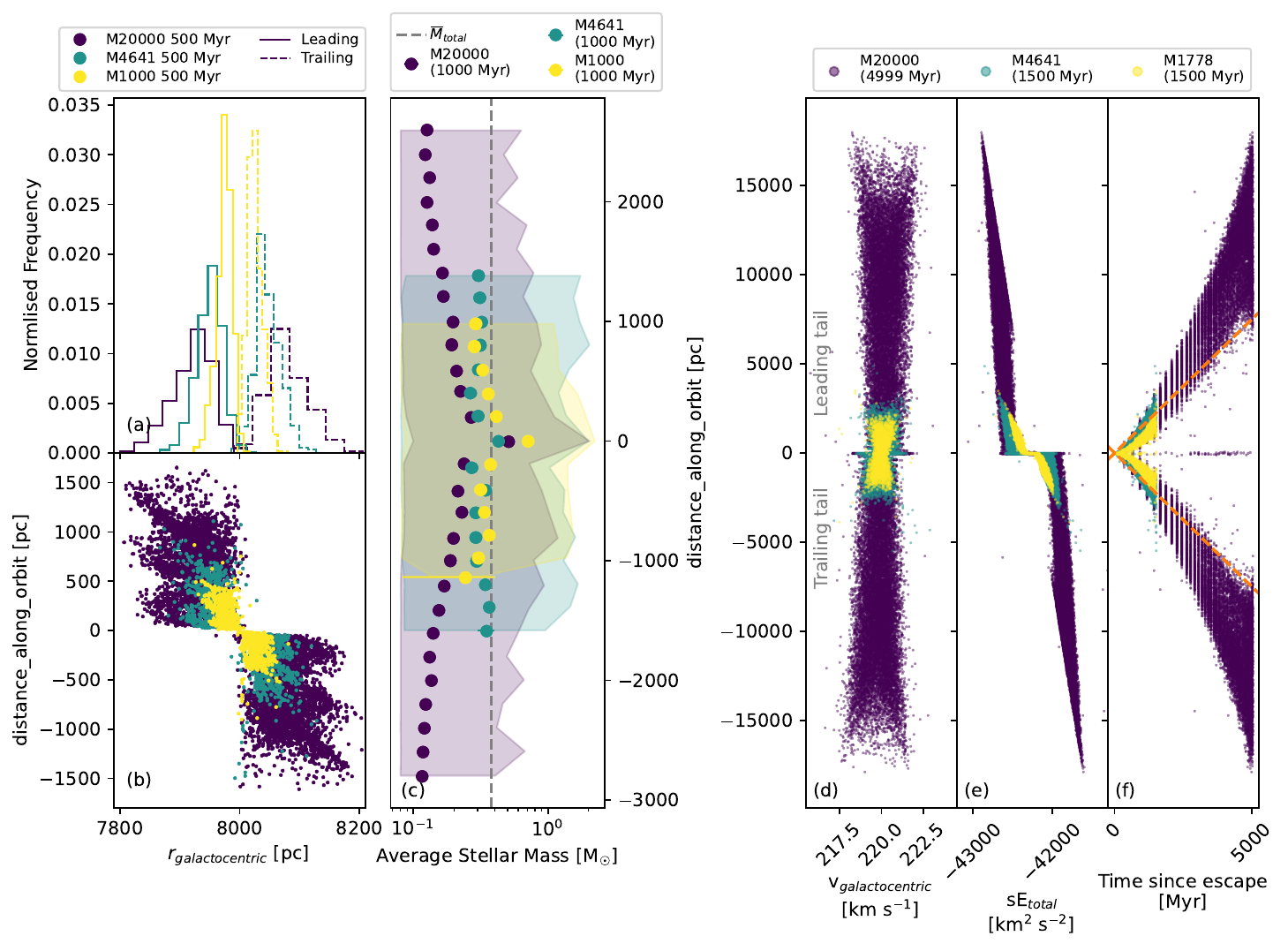}
    \includegraphics[width=0.8\linewidth]{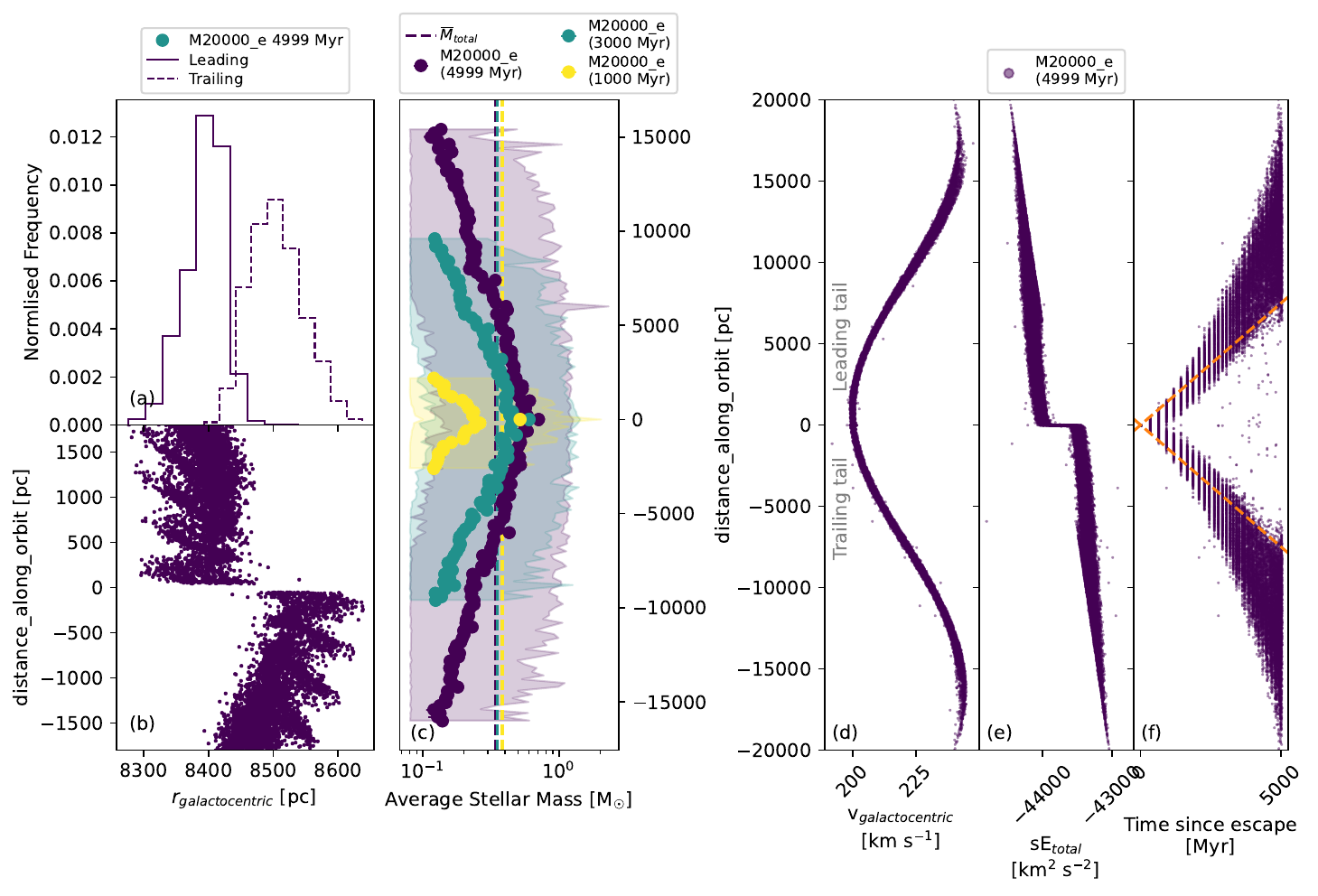}
    \caption{
    {Plots regarding the amplitude of the \reflectbox{S} shape (a--b), average mass along the tail (c), and kinematics of the tail (d--f).}
    (a) $r_{galactocentric}$ histogram for the M20000, M4641 and M1000 models at an age of 500 Myr. The leading (solid lines) and trailing (dashed lines) are shown separately.
    (b) The distribution of $distance\_along\_orbit$ with $r_{galactocentric}$ for the same models in (a). The central 2$R_{tidal}$ region is omitted while plotting the distributions in (a) and (b).
    (c) Average stellar mass as a function of \textit{distance\_along\_orbit} {in 200 pc bins} for the models M20000, M4641 and M1000 (at an age of 1000 Myr). {The shaded regions indicate 5--95 percentile values of the stellar masses. The dashed lines show the average mass of all stars in each simulation.}
    (d--f) The variation of $distance\_along\_orbit$ with $v_{galactocentric}$ (d) total specific energy (e) and the time since escape (f) for M20000, M4641, and M1778 for their oldest snapshot. The orange dashed lines in panel (f) represent stars moving with a speed of 1.5 pc Myr$^{-1}$ (= 1.47 km s$^{-1}$) away from the cluster. The top row shows the results for M20000, M4641, and M1778.
    The bottom row shows the results for the M20000\_e model.
    }
    \label{fig:averge_mass}
\end{figure*}

The clusters develop extended tidal tails after $\approx$50 Myr. For a circular cluster orbit, the leading tail lies inside the cluster's orbit (slightly lower energy/more bound orbit) while the trailing tail lies outside (slightly higher energy orbit/loosely bound; see Figure~\ref{fig:averge_mass} e). These individual orbits are effectively the same; thus, the tail particles create a locus along these two orbits.

In the case of an eccentric cluster orbit, each escaped particle follows a unique orbit; thus, their locus does not lie along any particular Galactic orbit. For low eccentricity orbits (e.g., for most open clusters), the loci of escaped particles (i.e., the tail) lie close to the cluster's orbits. For highly eccentric orbits (e.g., some globular clusters), the early escapees can have significantly different orbits, and thus their positions are independent of the present-day cluster orbit. However, we can detect the tails up to only a kpc in observations of open clusters; hence, these detected tails should lie near the cluster's orbit (inwards for the leading tail and outwards for the trailing tail).

The changes in the tail morphology are similar for all the clusters tested (68--20000 \Msun) and are based on the cluster age and not on $\nu_{relax}$. This is because of the similarities in the escape velocities, which are the dominant factor in shaping the tidal tails.

\subsection{The \reflectbox{S} shape} \label{sec:s_shape}

The simulated tidal tails show the typical \reflectbox{S} shape near the cluster centre (also seen in observations of open clusters: \citealt{2019AA...621L...2Roser_Hyades} and globular clusters: \citealt{Odenkirchen2003AJ....126.2385O}).
The \reflectbox{S} shape of the tails develops during the first 100--200 Myr, as the total tail span reaches $\geq500$ pc. The shape is a result of the different Galactic orbits of the escapees.
Figure~\ref{fig:averge_mass} (a) and (b) show the distribution of $r_{galactocentric}$ for the tail region. It shows that the `amplitude' of the \reflectbox{S} shape (width of the \reflectbox{S} shape in Figure~\ref{fig:morphology}) increases with the cluster mass. 
This is due to the higher velocities required to escape the more massive cluster, which leads to more distinct orbits for the escapees in more massive clusters and vice versa. 
As the cluster loses mass, the amplitude of the \reflectbox{S} shape decreases and eventually becomes close to zero when the escape velocity from the cluster becomes close to zero (as seen in the M1000 model at 1000 Myr in Figure~\ref{fig:morphology}).

For eccentric orbits, $r_{galactocentric}$ cannot be used as a proxy for $distance\_from\_orbit$. And $distance\_from\_orbit$ is a directionless quantity; thus, measuring the amplitude of the \reflectbox{S} shape in the XY plane is not easy.
However, by visually inspecting the XY distributions, similar behaviour (larger amplitude for larger cluster mass, decreasing amplitude near cluster dissolution, and the presence of the \reflectbox{S} shape) is seen in the eccentric orbits (see in Figure~\ref{fig:averge_mass} a \& b of the M20000\_e model). The only difference is that the \reflectbox{S} shape is tilted according to the local Galactic orbit (see the M20000\_e model at 2076 Myr in Figure~\ref{fig:morphology}).

\subsection{Average stellar mass across the tidal tail}

Figure~\ref{fig:averge_mass} (c) shows the average stellar mass decreases with $|distance\_along\_orbit|$ for the M20000 model indicating that the lower mass stars are preferentially evaporated and the losses decrease the average mass at larger \textit{distance\_along\_orbit} (as expected from theory; \citealt{Kroupa2008LNP...760..181K}). 
{The lower-mass clusters evaporate more high-mass stars compared to higher-mass clusters, indicative of the stronger mass segregation in higher-mass clusters.}
The same trend is not clearly seen for lower mass clusters. The lower escape velocity and poorer statistics lead to masking the trend. Similar behaviour is seen in the mass function (MF) evolution discussed in Section~\ref{sec:mf} (the present-day MF is also referred to as the MF throughout this text).
This effect might be difficult to detect in observations due to significant incompleteness and poor populations of {the near-tails}. \footnote{We refer to the tails within tidal radius to 100 pc as the near-tails and tails outside the 100 pc region as far-tails because we typically have information about only the near-tails of open clusters.} 

A similar behaviour is seen for clusters in eccentric orbits (Figure~\ref{fig:averge_mass} c of the bottom row).
{The panel also shows that the average mass in the younger tails, at a given $distance\_along\_orbit$, is smaller than the average mass in older tails, due to the preferential ejection of lower mass stars in the beginning.}

\subsection{Galactic dynamics of the tail}

Figure~\ref{fig:averge_mass} (d) and (e) show the variation of Galactocentric velocity ($v_{galactocentric}$) and the total specific energy for the individual particles of various clusters at their oldest snapshot. The $v_{galactocentric}$ (and the specific kinetic energy) of the leading tail is higher than the trailing tail, while the total specific energy shows an inverse trend (i.e., the leading tail is more bound to the Galaxy). The different circular orbits preferred by the two tails cause these differences.

Figure~\ref{fig:averge_mass} (f) shows the time since the escape. The {near-tail} is mostly new escapees, while the first escapees occupy the farthest regions of the tail. The orange dashed line shows the distribution if all particles were lost at 1.5 pc Myr$^{-1}$. As the early escapees had larger escape velocities (due to the higher cluster mass), their distances from the cluster have increased proportionally.

An eccentric orbit allows for the exchange of the kinetic and potential energy, hence the $v_{galactocentric}$ distribution shows larger deviations (see Figure~\ref{fig:averge_mass} d of the bottom row). Otherwise, the eccentric clusters also show the same behaviour of larger escape velocities in the early evaporations, and the {near-tail} is mostly populated by recent losses.

\begin{figure}
    \centering
    \includegraphics[width=0.99\linewidth]{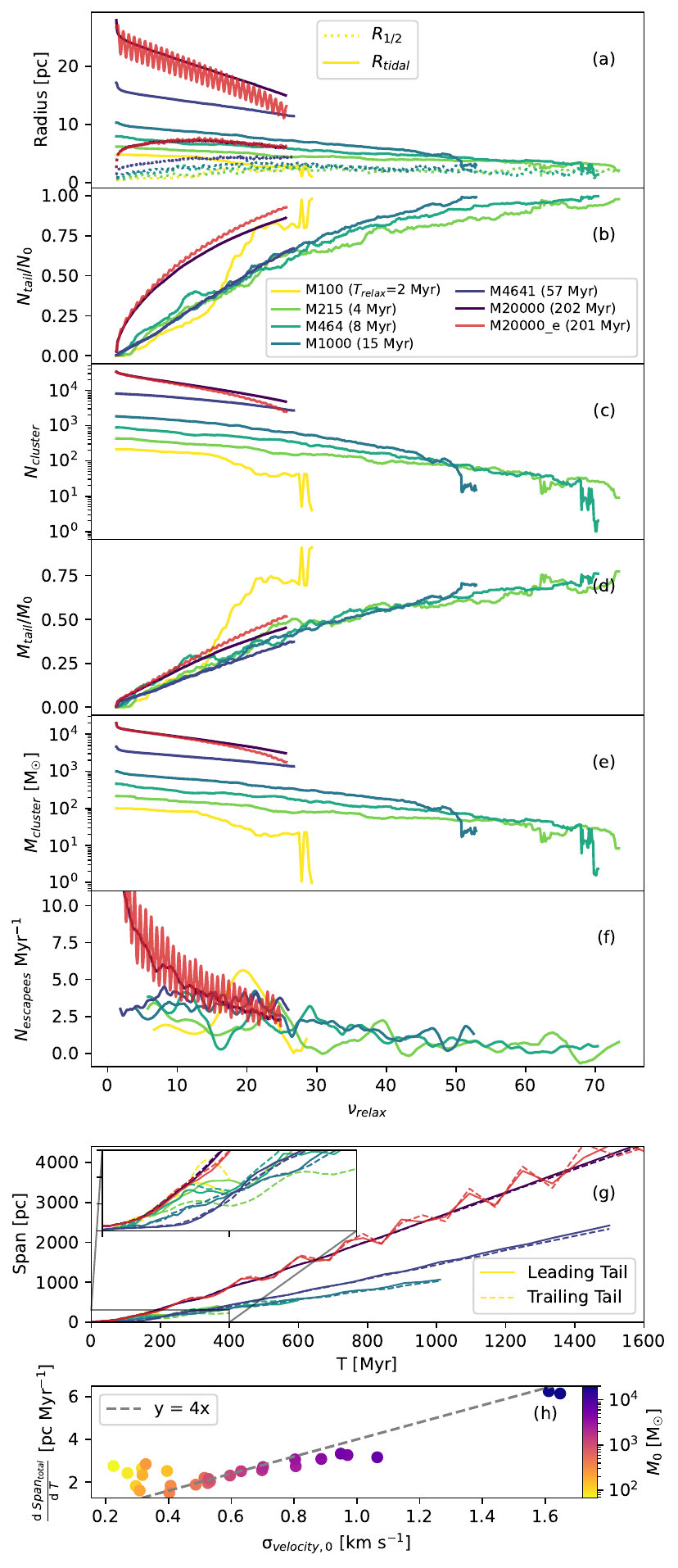}
    \caption{Evolution of tail and cluster parameters for various models. 
    (a) The tidal radius (solid curve) and half mass radius (dotted curve).
    (b) Number of stars in the tail normalised by the total number of stars at the beginning.
    (c) Number of stars in the cluster. 
    (d) Mass of the tail normalised by the total mass at the beginning.
    (e) Mass of the cluster.
    (f) Number of averaged escapees per Myr. 
    (g) Span of the leading (solid curves) and trailing (dashed curves) tail.
    (h) Variation of the rate of increase in the total tail span ($span_{total}$) with the initial velocity dispersion ($\sigma_{velocity, 0}$) of the clusters, coloured according to the initial mass ($M_0$). See Figure~\ref{fig:number_escapee_span_appendix} for comparison with the analytical formulae.}
    \label{fig:number_escapee_span}
\end{figure}

\subsection{Evolution of the cluster radii}

Figure~\ref{fig:number_escapee_span} (a) shows the evolution of the tidal radius and the half-mass radius. The tidal radii depend on the cluster mass and slowly decrease as the cluster loses mass due to stellar evaporation and evolution. The half-mass radius increases in the beginning but stays relatively constant thereafter. The cluster can be considered dissolved when both radii reach $\approx0$ pc. Note that the calculations of $R_{tidal}$ and $R_{1/2}$ are affected by low statistics when the cluster is near dissolution. As expected, the tidal radii oscillate significantly due to changes in $r_{galactocentric}$ of the M20000\_e cluster.

\subsection{Tail stellar count and mass across time}

Figure~\ref{fig:number_escapee_span} (b) and (c) show the evolution of the number of stars in the tail and the cluster (also see Figure~\ref{fig:number_escapee_span_appendix} a and b).
During early evolution, the population of the tidal tail increases approximately linearly with time. The clusters lose half their members in $\approx15T_{relax}$ and eventually dissolve after $\approx30T_{relax}$. 
Empirically, this can be described as follows:
\begin{equation} \label{eq:n_tail}
\begin{split}
    N_{cluser} (T) &\approx N_{initial} \left[1-\text{tanh}\left(\frac{T}{kT_{relax}}\right)\right], \\
    N_{tail} (T) &\approx N_{initial} \text{ tanh}\left(\frac{T}{kT_{relax}}\right),
\end{split}
\end{equation}
where $k$ is roughly of 20--50. The $k$ also accounts for the reduction in the total number due to stellar evolution. The formula was chosen because of its simplicity (linearity near $T=0$ and asymptotically approaches an upper bound).
These empirical formulae were based on simulations with $R_{galactocentric}\approx8$ kpc. More analysis of clusters at different $R_{galactocentric}$ is required to establish if this relation is universally applicable.

The evolution of tail mass shows similar changes on similar timescales to the number of particles in the tail (Figure~\ref{fig:number_escapee_span} d \& e). 
However, the quick mass loss due to stellar evolution in the beginning and the preferential ejection of lower mass stars induces the sharp decrease in the cluster mass during the early ($\nu_{relax}\leq1$) evolution. \citet{Lamers2005A&A...441..117L} provided an analytical formula for the evolution of cluster mass, which can be modified to estimate the mass of the tails. For a solar metallicity cluster on a circular orbit and ignoring the stellar evolution (as most of the tail is composed of low mass stars) in the tails, the mass of the tail can be estimated as follows (see Figure~\ref{fig:number_escapee_span_appendix} c and d):

\begin{equation} \label{eq:m_tail_lamers}
    \begin{split}
        \mu_{ev} (T) &\coloneqq \frac{M_{cluster}}{M_{initial}}= 1-10^{(\text{log}T - 7)^{0.255} - 1.805} \text{ for } T>12.5 \text{ Myr}\\
        \frac{M_{cluster} (T)}{M_{initial}} &\approx \left[\{\mu_{ev}(T)\}^\gamma - \frac{\gamma T}{T_0}\frac{\Msun}{M_{initial}}\right]^{1/\gamma}\\
        \frac{M_{tail} (T)}{M_{initial}} &\lesssim \mu_{ev} - \left[\{\mu_{ev}(T)\}^\gamma - \frac{\gamma T}{T_0}\frac{\Msun}{M_{initial}}\right]^{1/\gamma}
    \end{split}
\end{equation}
where $\mu_{ev}$ is the fractional mass of the cluster at time $T$ (in yr) due to only stellar evolution losses, $\gamma = 0.62$, and $T_0$ is a constant depending on the tidal field and the orbital ellipticity (with values of 0.1--30 Myr). Note that the Eq.~\ref{eq:m_tail_lamers} based estimations result in unphysical values near the birth and dissolution of the cluster.

The M20000\_e model shows a larger population in the tail due to the smaller $R_{tidal}$ during the Galactic orbit. The individual parameter values of all parameters oscillate near the M20000 values due to the orbital motion, but the cluster mass loss proceeds faster. As most clusters follow a non-circular orbit around the Milky Way, deriving generalised analytical formulae based on the initial position, velocity, mass, and Galactic potential is not possible. We recommend that the formulae given above can only be used to get the first-order approximations.

\subsection{Number of escapees with time}

{\citet{2008MNRAS.387.1248Kupper} analytically derived $N_{escapee}$ and found it to depend only weakly on the cluster population ($\propto \log[N_{tidal}(T)]$).}
Alternatively, based on the Eq.~\ref{eq:n_tail}, the number of escapees can be described as:

\begin{equation}\label{eq:escapees}
\begin{split}
    N_{escapee} (T) &\approx \frac{N_{initial}}{kT_{relax}} \text{ sech}^2\left(\frac{T}{kT_{relax}}\right)
\end{split}
\end{equation}

The moment a star escapes from a cluster is difficult to determine, especially in observations. Even in simulations, a star well outside the tidal radius could be simply following an eccentric orbit around the cluster. For simplicity, we counted the difference in the number of stars in the tidal radius for sequential snapshots to measure the number of escapees in a cluster. {Since only a few stars escape per Myr—which is the interval between successive simulation snapshots—and given the uncertainties in estimating the tidal radius, the number of escapees can fluctuate, sometimes appearing negative, particularly for clusters on eccentric orbits. To smooth these variations, we applied a rolling average over a 10–50 Myr timescale when generating the plots.}
Figure~\ref{fig:number_escapee_span} (f) shows the averaged number of escapees and Figure~\ref{fig:number_escapee_span_appendix} (e) shows the comparison with Eq.~\ref{eq:escapees}.
Due to the scaling relations between $N_{initial}$ and $T_{relax}$, the escape rate is higher for higher mass clusters. However, the escape rate is similar (2--5 stars/Myr) for all tested clusters after the early evolution ($\nu_{relax}\geq5$). 
The escape rate decreases with time, eventually reaching zero when the cluster dissolves.
Following the changes in $R_{tidal}$ (and $R_{galactocentric}$), the number of escapees oscillates for the eccentric model.

\begin{figure*}
	\centering
	\includegraphics[width=0.99\linewidth]{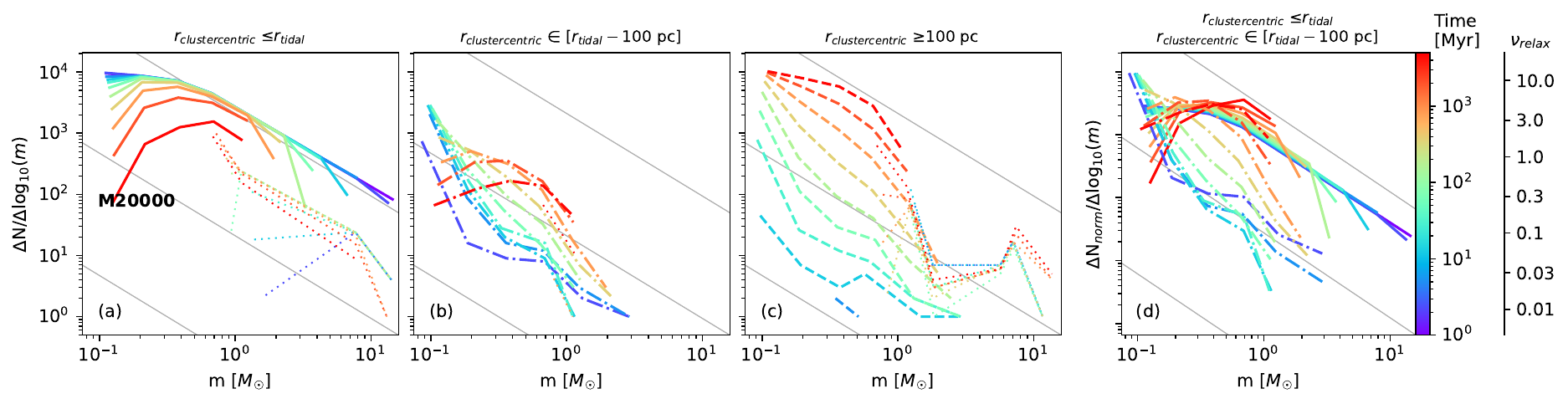}
	\includegraphics[width=0.99\linewidth]{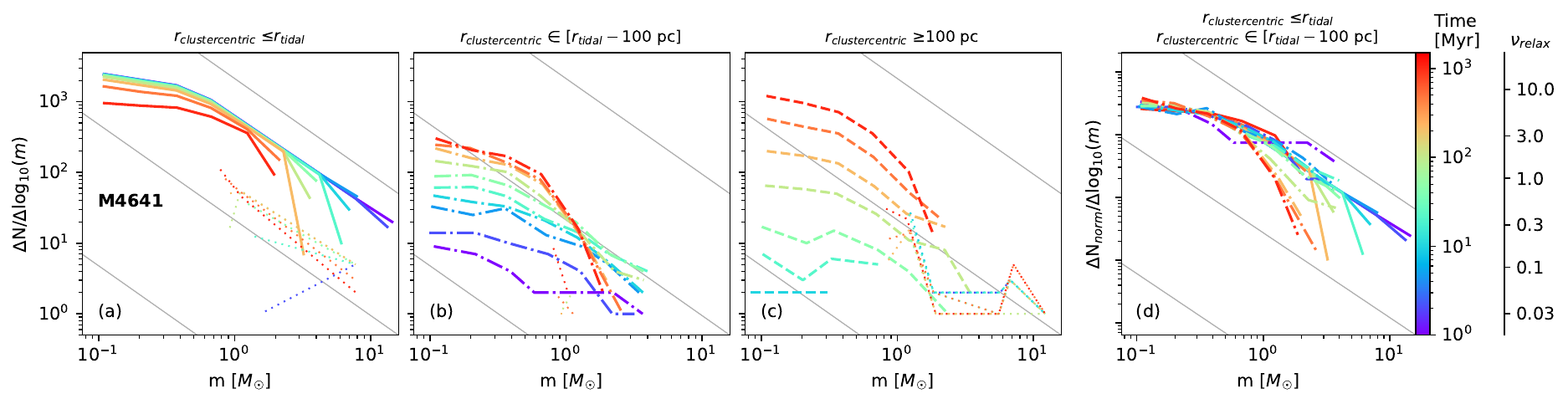}
	\includegraphics[width=0.99\linewidth]{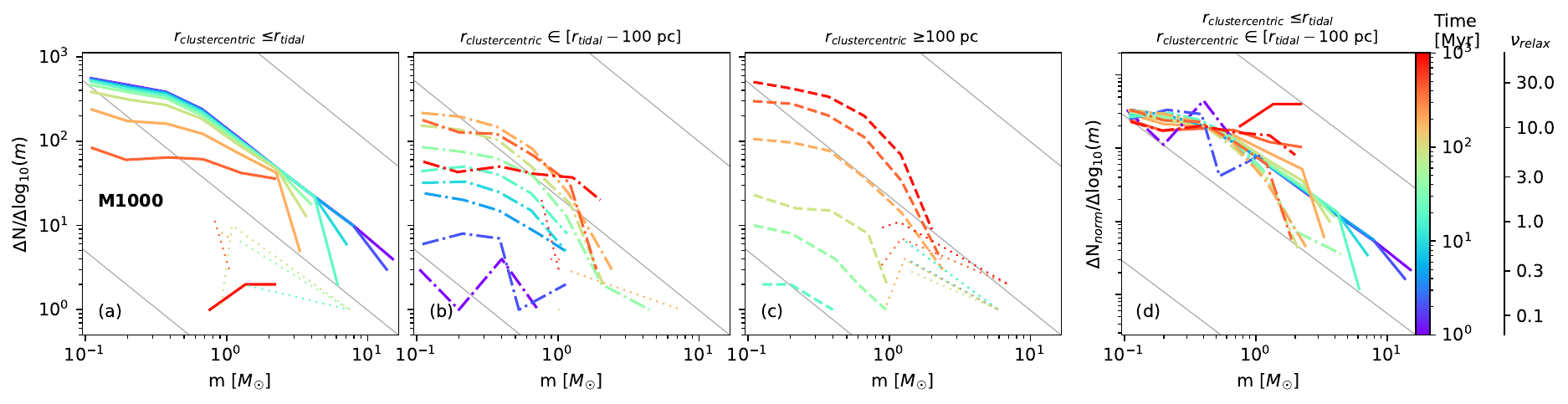}
	\caption{Evolution of the MF for the cluster, {near-tail and far-tail} for the M20000 (top row), M4641 (middle row) and M1000 (bottom row) models.
        (a) Evolution of the MF of the cluster. The colours represent the cluster age (blue: younger, red: older). The solid lines show the MFs of the non-degenerate (ignoring white dwarfs, neutron stars, and black holes) population.
	(b) Evolution of the MF within the observationally detectable {near-tail} region. The non-degenerate MF is shown using dash-dotted lines.
	(c) Evolution of the MF in the {far-tails} region. The non-degenerate MF is shown using dashed lines.
	(d) Normalised (scaled by the total number of stars) MF of the cluster (solid lines) and {near-tail} (dash-dotted lines).
        {The dotted lines show the MFs of degenerate stars in (a)--(c) panels.}
        Grey lines with the \citet{Salpeter1955ApJ...121..161S} IMF slope ($\Gamma=-1.35$) are shown as for reference.
	}
	\label{fig:tail_mf}
\end{figure*}

\subsection{Evolution of the length/span of tidal tails}

Figure~\ref{fig:number_escapee_span} (g) shows the increase in the span of tidal tails. 
There are many high-velocity outlier escapees in the simulations that need to be removed while measuring the length of the tails. Additionally, the end of the tails is a diffuse structure without any sharp cut-off.
Hence, we defined the span of the leading tail as the 99th percentile value of $distance\_along\_orbit$ (eliminating the outliers at the tip of the tail) and the span of the trailing tail as the negative of the 1st percentile value. The total span is defined as the range of 1st--99th percentile values of the $distance\_along\_orbit$.

The massive clusters have longer tails for a given age. This is driven by the higher escape velocities of the tail stars for the higher mass cluster. 
The total span of the tails is determined by the initial escapees, which had the highest escape velocities. Their kinetic energy remains unchanged (for circular orbits) after leaving the cluster, hence the rate of change of the total span is constant after the early evolution. The typical rate of increase in the total span was found to be $\approx4$ times the initial velocity dispersion (Figure~\ref{fig:number_escapee_span} h). Correspondingly, the leading and trailing tails grow at half the rate. This is equivalent to an escape velocity of roughly twice the velocity dispersion.

The trailing tails (for circular orbits) have a slightly shorter span than the leading tail; however, the differences are within the measurement errors. Detailed analysis of the number density along the tails is required to determine if the effect is real.

For eccentric orbits, measuring the tail span is not trivial, as the escapees do not always follow the current orbit. However, the $distance\_along\_orbit$ values measured using the standard method show a similar increase in the spans with local oscillations around the circular model's values.

\subsection{Mass function of tidal tails and the cluster} \label{sec:mf}

The IMF of the model clusters was taken from \citet{2001MNRAS.322..231KroupaIMF} as a broken power law. Figure~\ref{fig:tail_mf} shows the evolution of the MF for the M20000 cluster, the {near-tail and far-tail region}. The cluster MF resembles the IMF for the first 20--50 Myr of evolution. During the dynamical evolution, low-mass stars are preferentially evaporated, making the cluster MF top-heavy. And the top-heavy-ness increases as the cluster evolves. Consequentially, the tail MF becomes bottom-heavy with time. The earliest tails are the most bottom-heavy, however, as the cluster evolves ($\nu_{relax} \geq1$), the tail MF flattens significantly while still staying bottom-heavy compared to the cluster. {However, the differences in MF of low mass clusters ($M_0\lessapprox1000$ \Msun) and their tails might be undetectable due to lesser mass segregation, stochasticity, and poor statistics (see Figure~\ref{fig:tail_mf}).}

\subsection{Density of tidal tails across time}

\begin{figure*}
	\centering
	\includegraphics[width=0.7\linewidth]{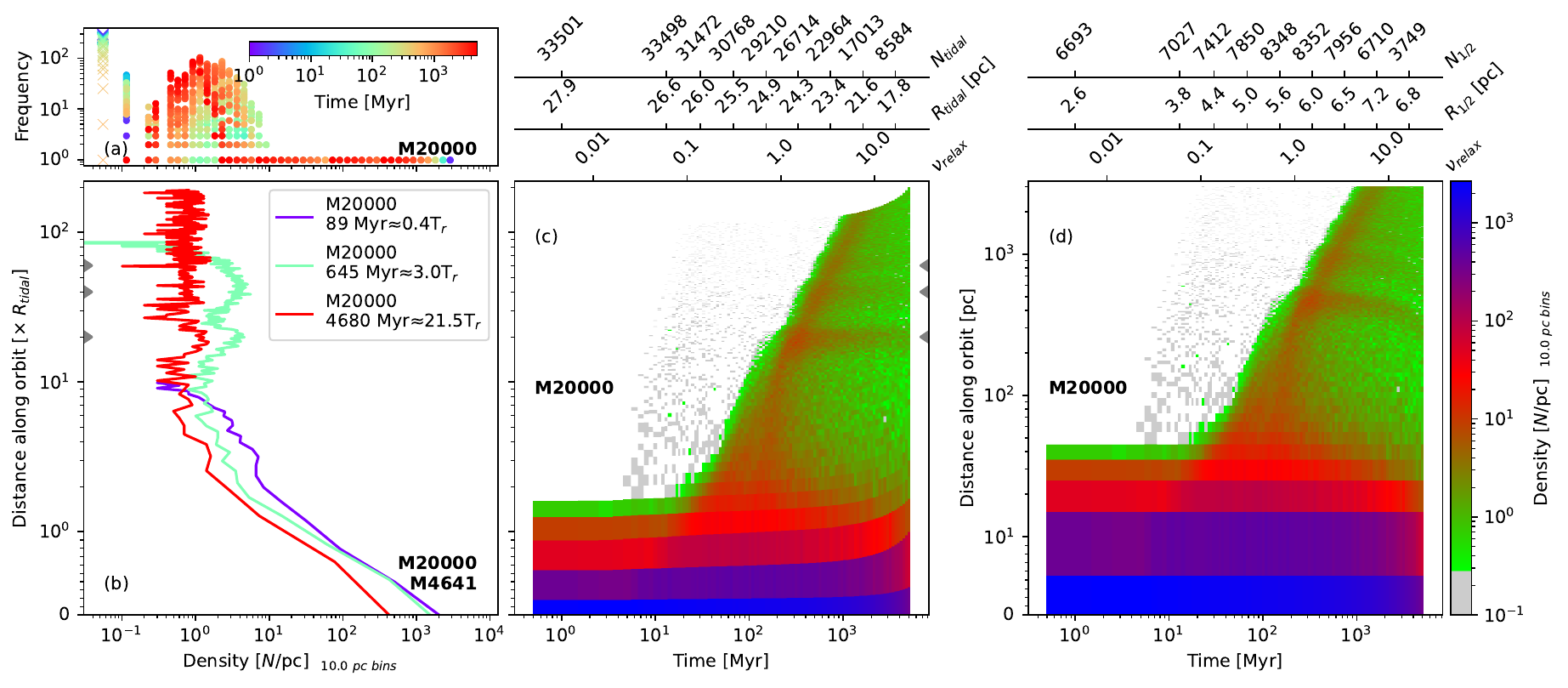}
	\includegraphics[width=0.7\linewidth]{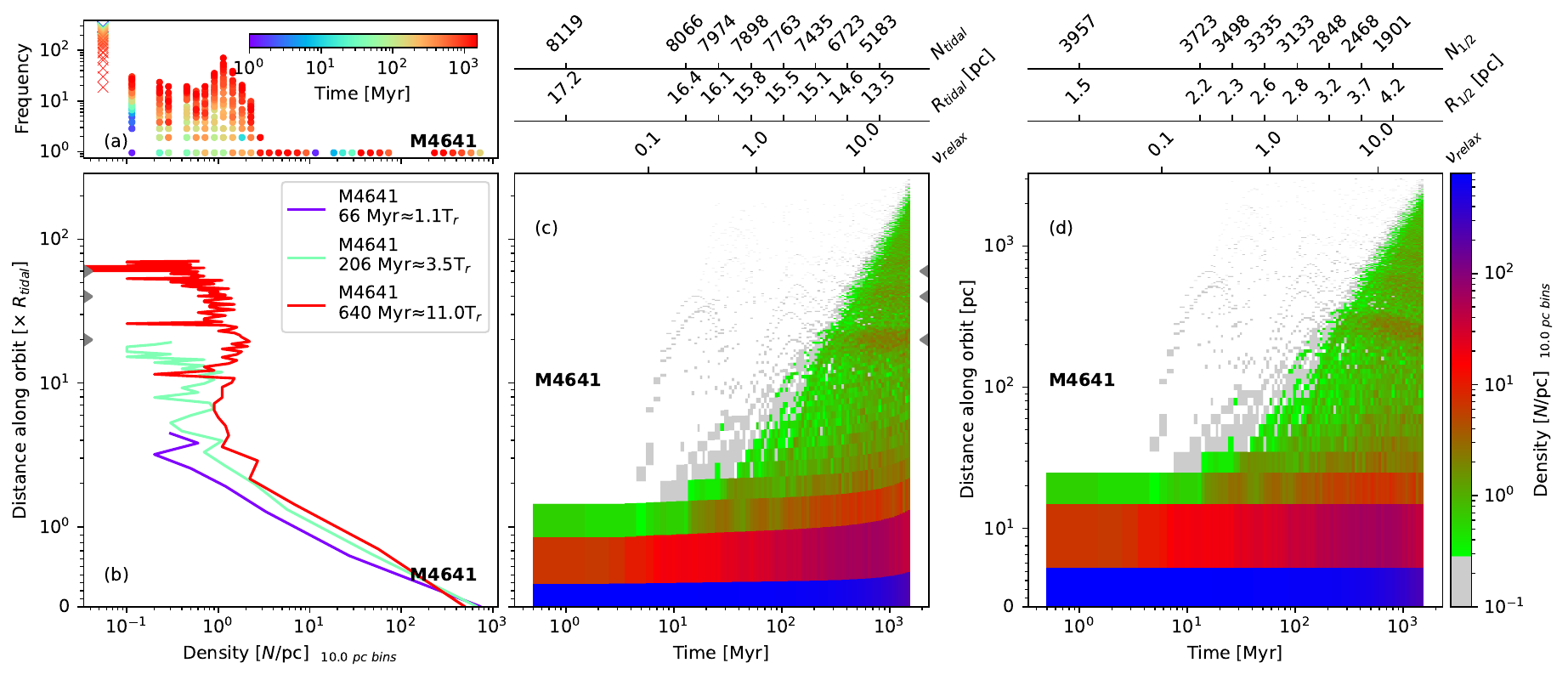}
	\includegraphics[width=0.7\linewidth]{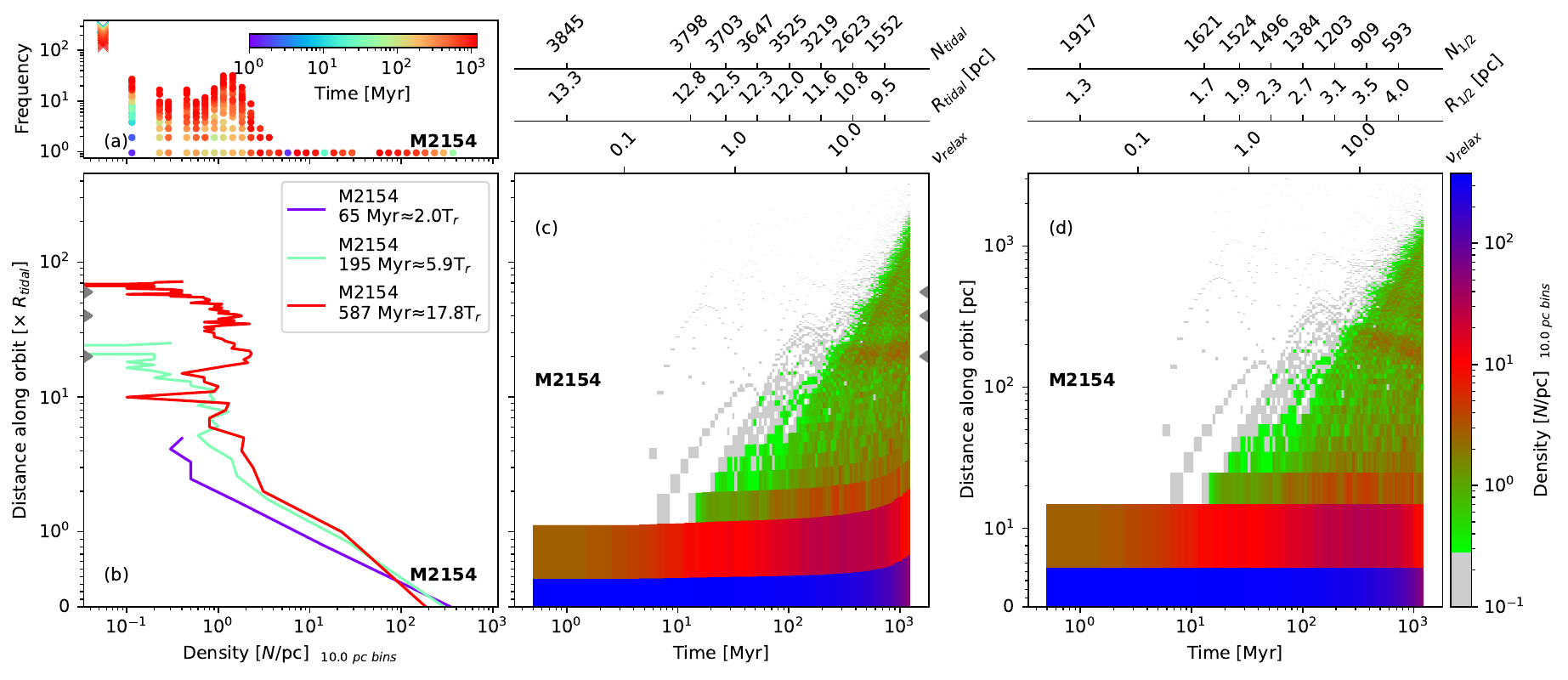}
	\includegraphics[width=0.7\linewidth]{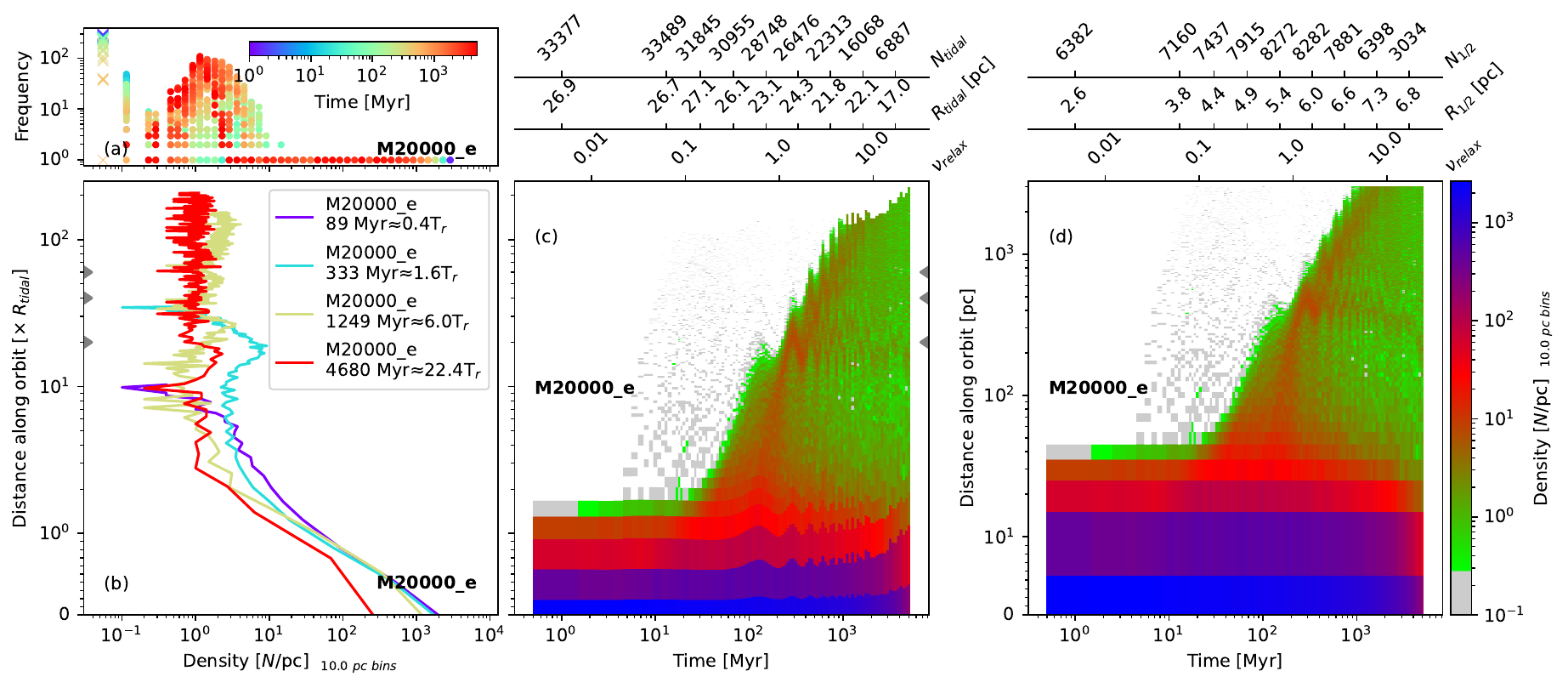}
	\caption{
    {Linear number density (N/pc) within the cluster tails as calculated by binning $distance\_along\_orbit$ every 10 pc.}
    (a) {Frequency of the number density in all bins coloured according to the cluster age. The frequency of empty bins is shown by crosses at the left end of the plot.}
    (b) The number density for M20000 (solid lines) and M4641 (dotted lines) at various ages. 
    (c) Evolution of the tail density in M20000 as a function of $distance\_along\_orbit$ (as a multiple of $R_{tidal}$). 
    (d) Density variation in M20000 as a function of the $distance\_along\_orbit$ (in pc).
    The grey colour denotes the region with $<3$ particles per bin.
    The $\nu_{relax}$, $R_{1/2}$, $R_{tidal}$, $N_{1/2}$, and $N_{tidal}$  are indicated for reference. The 20, 40, and 60 $R_{tidal}$ are marked by grey triangles as visual guides near overdensities.
    The first, second, third, and fourth rows show the results for M20000, M4641, M2154, and M20000\_e models, respectively.}
    \label{fig:density}
\end{figure*}

We define the linear number density as the number of stars per pc as measured along the orbit (using $distance\_along\_orbit$). A 1D density determination was chosen over a 2D/3D density determination due to the complex morphology of the tails.
Figure~\ref{fig:density} (b) shows the number density in M20000 and M4641 along the orbit at various ages. As seen from Figure~\ref{fig:number_escapee_span} (f), the number of escapees and thus the tail density becomes similar for the two models after $\approx5T_{relax}$.
Figure~\ref{fig:density} (c) shows the same 1D number density with $distance\_along\_orbit$ as a multiple of $R_{tidal}$ for all simulated ages. 
The cluster has the highest density while the tail has a density of $\approx 0.1-10$ star pc$^{-1}$ (Figure~\ref{fig:density} (a)). The panel also shows the multiple overdensities near $20$, $40$ and $60R_{tidal}$ (also noted by \citealt{2008MNRAS.387.1248Kupper}). 
Panel (c) shows that the overdensities are not at the same physical distance in pc, and they come closer as $R_{tidal}$ decreases. 
As the density becomes smaller than 1 star pc$^{-1}$, measuring it and retaining the density fluctuations becomes challenging due to poor statistics. More detailed analysis is required to understand the changes in the tidal tail density.
    
\subsection{Epicyclic overdensities}

\citet{2008MNRAS.387.1248Kupper} (also see \citealt{Capuzzo2005AJ....129.1906C, Just2009MNRAS.392..969J, Kupper2010MNRAS.401..105K, Kupper2012MNRAS.420.2700K}) noted that there are overdensities in the tidal tails (at $\approx40 R_{tidal}$ distance) corresponding to ``places where escaping stars slow down in their epicyclic motion away from the star cluster''. The M20000 simulation shows that multiple such overdensities exist at roughly $20$, $40$ and $60R_{tidal}$ (Figure~\ref{fig:density} c and Figure~\ref{fig:distance_along_orbit_appendix} c). The density within these regions is maximum near $\nu_{relax}\approx1-5$ and decreases as the cluster evolves.
Figure~\ref{fig:density} (for M2145 and M4641) shows that smaller clusters also have such overdensities and at similar relative distances. Physically, the overdensities come closer to the cluster during ageing due to their smaller tidal radii at later stages. The number density within these overdensities is smaller than M20000 but is within the same order of magnitude.

The location of Epicyclic overdensities is not at a constant $xR_{tidal}$ for a cluster on an eccentric orbit (M20000\_e in Figure~\ref{fig:density} and Figure~\ref{fig:distance_along_orbit_appendix} i). However, their location oscillates around similar values as for the circular orbit model (see \citealt{Kupper2012MNRAS.420.2700K} for a detailed theoretical examination).

\subsection{Stellar evolutionary stages in the tail}

\begin{figure*}
    \centering
    \includegraphics[width=0.7\textwidth]{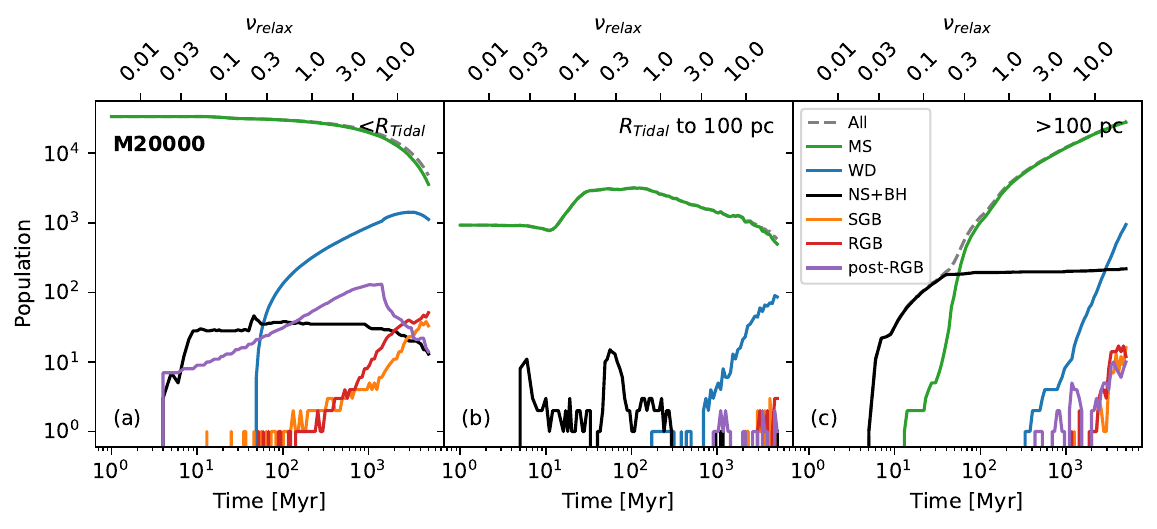}
    \includegraphics[width=0.7\textwidth]{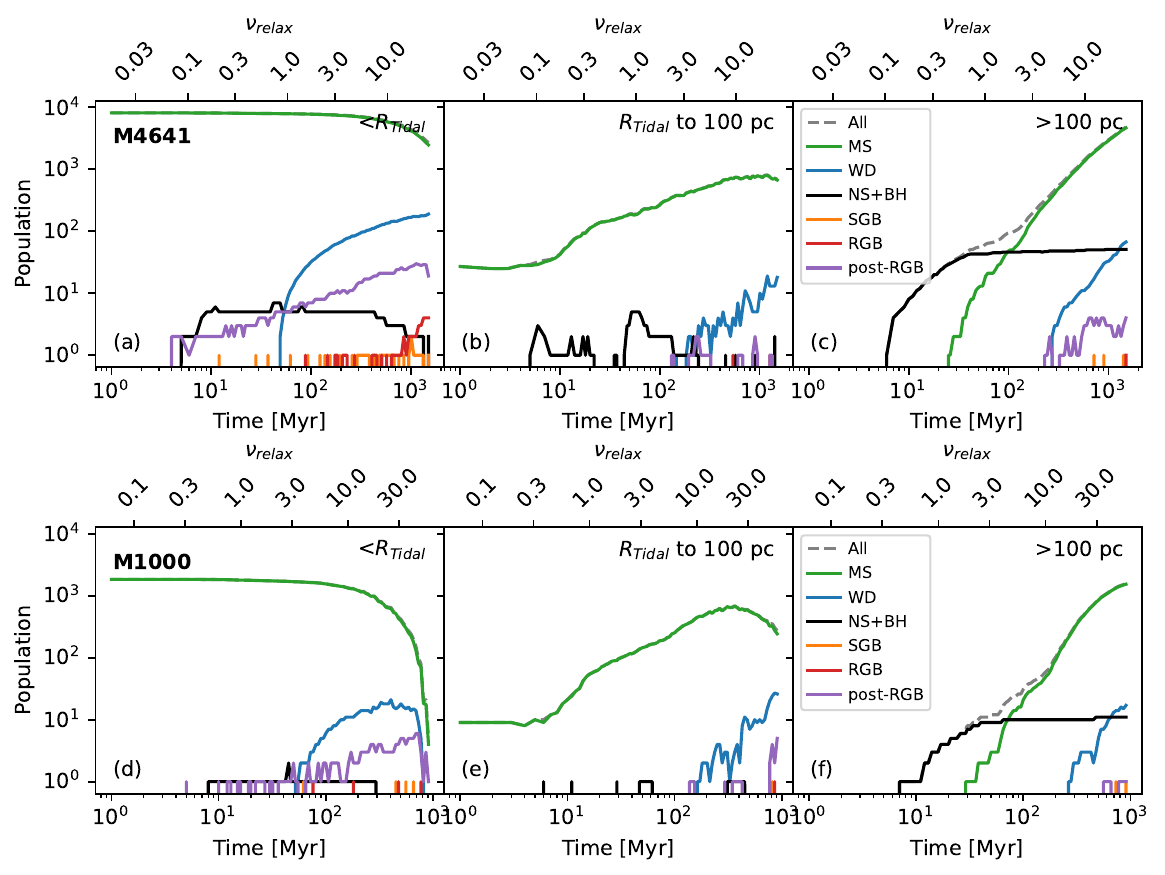}
    \caption{Stellar type population changes in the cluster (a), {near-tail} (b), and the far-tail (c) for the M20000 model. Population evolution is shown for all stars, main sequence (MS), white dwarf (WD), neutron stars (NS) + black holes (BH), subgiant (SGB), red giant (RGB), and post-RGB population. The $\nu_{relax}$ is shown for reference. 
    The results for M20000, M4641, and M1000 are shown in the top, middle, and bottom panels, respectively.}
    \label{fig:bse_type}
\end{figure*}

Figure~\ref{fig:bse_type} shows the evolution of stellar types within the cluster and its tidal tails. The {near-tail} is primarily populated by main-sequence stars, along with a smaller fraction of stellar remnants. These remnants tend to remain bound to the cluster until dynamical evolution reduces their relative mass, allowing them to escape.

The {near-tail} notably lacks subgiants and red giants, as these stars represent the most massive objects in the simulation at the time of the evolution---the even more massive main sequence stars having already evolved. In contrast, the {far-tail} can host more giants, largely due to its larger overall stellar population. Additionally, early giant progenitors, which were not among the most massive stars at the time, could more easily escape during the cluster's initial evolution.

White dwarfs become the dominant stellar remnant population within a cluster after the first 50 Myr. However, the earliest-forming white dwarfs are relatively massive and therefore remain gravitationally bound to the cluster. The white dwarf progenitors---being among the most massive stars at the time of transition---are typically centrally concentrated due to mass segregation. As a result, the number of white dwarfs in the tidal tails is significantly lower than in the cluster itself.

This scenario changes as the cluster approaches dissolution, at which point the entire stellar population---including white dwarfs---gradually evaporates into the tidal tails, leading to an increase in tail white dwarfs. However, since white dwarfs cool and become faint over time \citep{Fontaine2001PASP..113..409F}, the number of observable white dwarfs is subject to strong detection biases and must be estimated with care. If the tail white dwarfs were formed in the cluster and subsequently evaporated, their cooling age is expected to increase with distance from the cluster, with an upper limit set by the escape velocity (of the order of 1--2 pc Myr$^{-1}$).

\subsection{Criteria for selecting tidal tail members}

The majority of the above analysis assumed a circular orbit around the Milky Way. However, most open clusters have non-circular orbits and also have oscillations above and below the Galactic disc mid-plane. Thus, specific angular momentum ($sL_z$) and specific total energy ($sE_{total}$) are the best parameters to select members of the cluster and the tidal tails. A conservative 5$\sigma$ cut based on the initial values works throughout the evolution because the standard deviation of the $sL_z$ and $sE_{total}$ distributions does not increase significantly with time. This also removes any high-velocity ejections from the simulations, which lead to problems while visualising and analysing the general tail population.

An alternative method, which does not depend on the Galactic potential and knowing the velocities, is using the combination of position-dependent $distance\_along\_orbit$ and $distance\_from\_orbit$ as the selection criteria. This approach can be useful when comparing with observations where 6D astrometry is not known.

\begin{equation}
    \begin{split}
        \text{With velocities:}&\quad sL_z \in 5\sigma\ \textsc{and} \  sE_{total}\in 5\sigma \\
        \text{Without velocities:}&\quad  distance\_from\_orbit<400\text{ pc} \ \textsc{and} \\ 
        &\quad distance\_along\_orbit \neq \textit{NaN}
    \end{split}
\end{equation}

\subsection{Additional caveats}

Additional physical (e.g. gas expulsion process, initial stellar density, primordial BF, binary properties, different initial mass segregation levels, Galactocentric distance, clumpy Galactic potential, interaction with spiral arms and molecular clouds, gravitational models) and computational (e.g. floating point precision, choice of the code) phenomena can further change the properties of the tidal tails. The first 10 Myr of evolution can be drastically different based on these assumptions; however, the rest of the simulated behaviour should be similar in any setup, given a reliable $N$-body code.
The eccentricity of the orbit has a significant impact on the properties of the cluster and the tail; however, calculating this effect analytically or empirically is not trivial {\citep{2008MNRAS.387.1248Kupper, Kupper2012MNRAS.420.2700K}}. Dedicated $N$-body simulations (based on present-day properties) are required to properly understand the history of a particular star cluster.
The analysis presented in Section~\ref{sec:properties_of_tails} did not include any observational errors. The effect of these uncertainties needs to be accounted for while comparing with real data.
However, the analysis presented here should hold as a first-order approximation for the given initial masses (for clusters in the solar neighbourhood). 

\begin{figure*}
	\centering
    \includegraphics[width=0.95\linewidth]{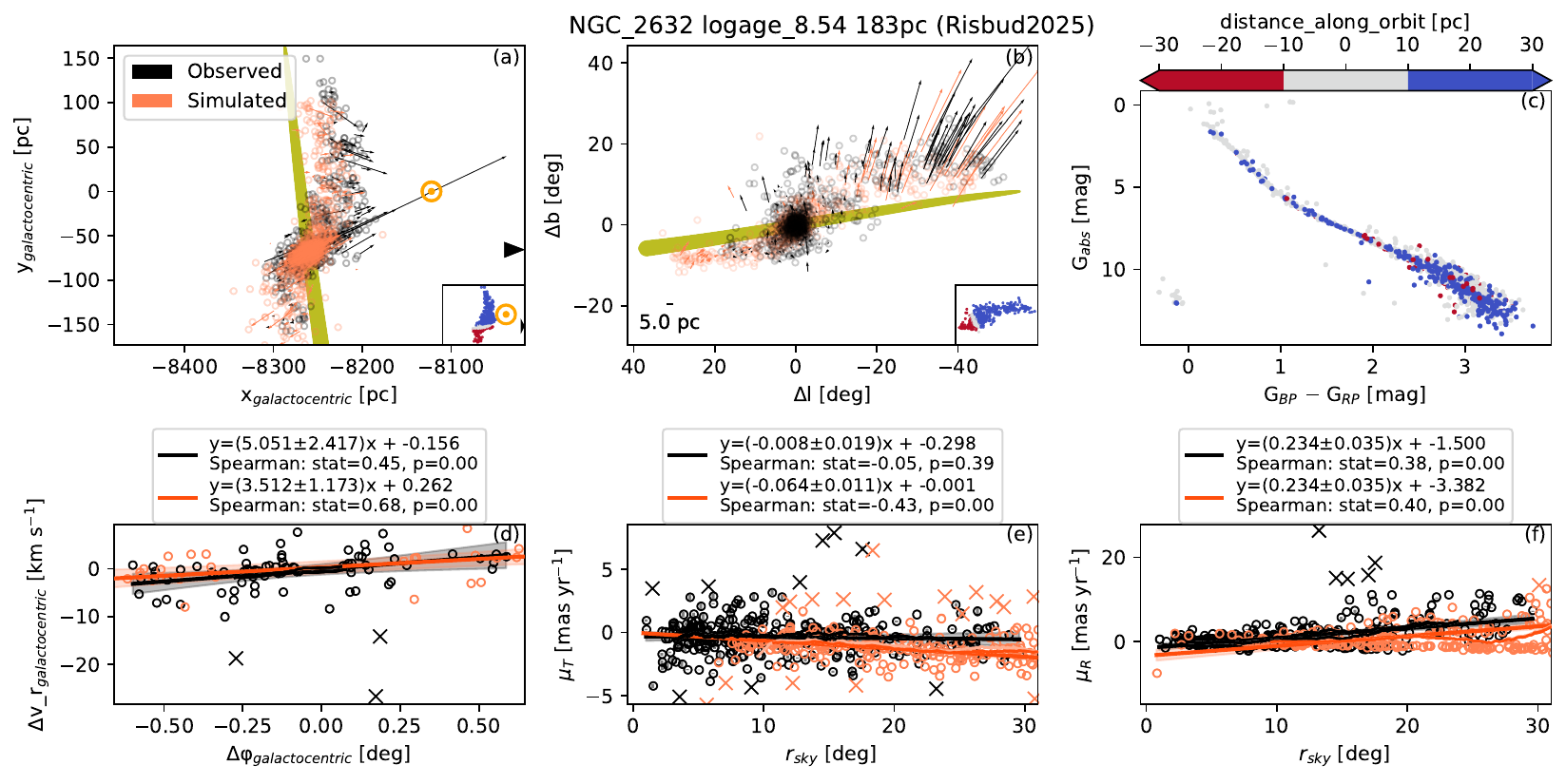}
			\includegraphics[width=0.95\linewidth]{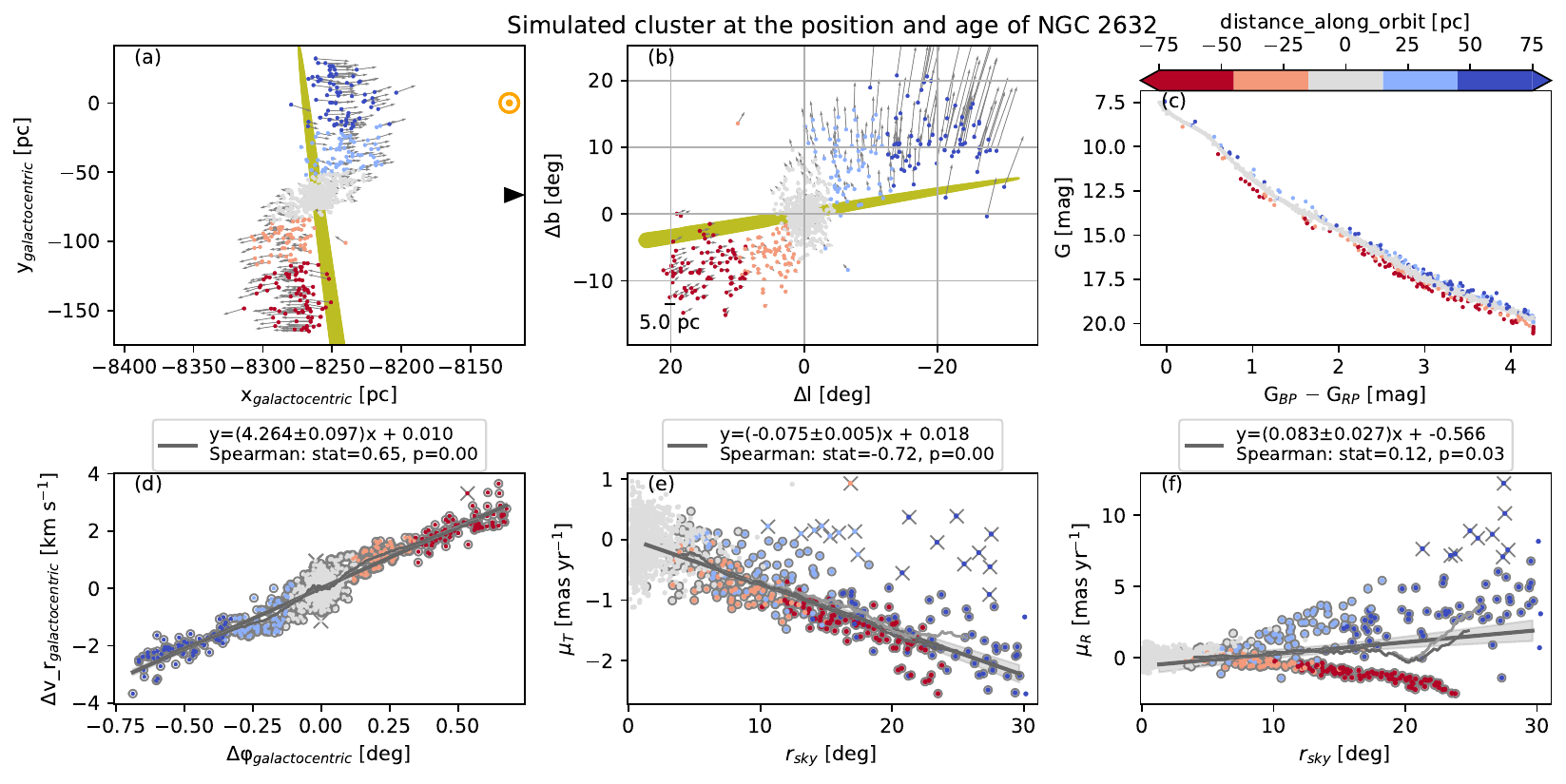}
	\caption{Diagnostic plots for a silver (NGC 2632) cluster in the top part and a noise-free simulated cluster (at the position and age of NGC 2632) in the bottom part. (a) Spatial distribution in the Galactocentric plane. The observed cluster (black) and the simulated (with \textit{Gaia}-like noise) cluster (coral) members are shown with arrows indicating the cluster-centric velocities. The Galactic orbit is shown by the olive arrow. (b) Distribution in tangential Galactic coordinates. (c) Absolute CMD coloured according to distance along the orbit. The same colour scheme is used for the small plots in the corners of (a) \& (b). (d) Variation of $V_{r, galactocentric}$ with $\phi_{galactocentric}$. (e) $R$-$\mu_T$ distribution. (f) $R$-$\mu_R$ distribution. The central 20 pc region is omitted in (d)--(f) plots, and the outlier-rejected regression and Spearman correlation test results are given in the legend. The rolling average is plotted as a fainter line in (d)--(f).
    For the simulated results:
    (i) the simulated CMD uses apparent magnitudes;
    (ii) all points are coloured according to the $distance\_along\_orbit$.
    }
	\label{fig:combo_real}
\end{figure*}

\begin{table*}
	\centering
	\caption{Reliability of tidal tails in literature catalogues.}
    \small
	\begin{tabular}{lp{140mm}}
\toprule
Reference & Clusters \\
\midrule
\textit{Gold sample} (16) &  \\
\citet{Jerabkova2021AA...647A.137J} & Melotte\_25 (M1), Melotte\_25 (M5) \\
\citet{2024arXiv240618767Kos} & Melotte\_22, NGC\_1039, Platais\_3, UPK\_612 \\
\citet{2019AA...621L...3Meingast} & Melotte\_25 \\
\citet{2021AA...645A..84Meingast} & Blanco\_1, Theia\_517 \\
\citet{Oh2020MNRAS.498.1920O} & Melotte\_25 \\
\citet{Risbud2025AA...694A.258R} & Alessi\_96, Blanco\_1, Melotte\_25, Stock\_10 \\
\citet{2019AA...621L...2Roser_Hyades} & Melotte\_25 \\
\citet{2020ApJ...889...99Zhang} & Blanco\_1 \\
\hline \textit{Silver sample} (55) &  \\
\citet{Bhattacharya2022MNRAS.517.3525B} & Melotte\_20, NGC\_1662, NGC\_1901, Stock\_12, Teutsch\_35,
 Theia\_517 \\
\citet{2022MNRAS.514.3579Boffin} & NGC\_752 \\
\citet{2019AA...624L..11Furnkranz} & Melotte\_111 \\
\citet{2024arXiv240618767Kos} & ASCC\_99, Alessi\_9, Alessi\_96, Blanco\_1, Collinder\_350,
 Herschel\_1, IC\_4756, Mamajek\_4, Melotte\_111, Melotte\_25,
 NGC\_2422, NGC\_3532, NGC\_6475, NGC\_6633, Renou\_23,
 Ruprecht\_147, Stock\_12, Stock\_2, Teutsch\_35, Theia\_517,
 UBC\_32 \\
\citet{2021AA...645A..84Meingast} & Melotte\_20, Melotte\_22, NGC\_2516 \\
\citet{Olivares2023AA...675A..28O} & Melotte\_111 \\
\citet{2022ApJ...931..156Pang} & Alessi\_3, Collinder\_350, LP\_2429, Theia\_517 \\
\citet{Risbud2025AA...694A.258R} & ASCC\_101, Alessi\_9, Melotte\_111, Melotte\_22, NGC\_2632,
 NGC\_6633, Platais\_10, Roslund\_6, Ruprecht\_147, Stock\_2,
 Theia\_517 \\
\citet{2019AA...627A...4Roser_Praesepe} & NGC\_2632 \\
\citet{2019ApJ...877...12Tang} & Melotte\_111 \\
\citet{2023AA...679A.105Vaher} & IC\_2391, IC\_2602, Melotte\_20, Melotte\_22, NGC\_2451A \\
\hline \textit{Bronze sample} (51) &  \\
\citet{Bhattacharya2022MNRAS.517.3525B} & ASCC\_101, Alessi\_24, Alessi\_5, BH\_164, BH\_99, Herschel\_1,
 IC\_2602, IC\_4665, IC\_4756, Melotte\_22, NGC\_1039, NGC\_2451B,
 NGC\_3228, NGC\_6475, Roslund\_6, Ruprecht\_98, Stock\_1,
 Trumpler\_10 \\
\citet{2024arXiv240618767Kos} & ASCC\_101, Alessi\_3, Gulliver\_20, NGC\_1662, NGC\_1901, NGC\_2516,
 NGC\_2632, NGC\_752, Ruprecht\_98, Stock\_1, UPK\_20, UPK\_305,
 UPK\_350 \\
\citet{2021AA...645A..84Meingast} & IC\_2391, IC\_2602, NGC\_2451A, NGC\_2547, Platais\_9 \\
\citet{2021ApJ...912..162Pang} & NGC\_2516, NGC\_2547, NGC\_6633, Ruprecht\_147 \\
\citet{2022ApJ...931..156Pang} & IC\_4756, Mamajek\_4 \\
\citet{Risbud2025AA...694A.258R} & Alessi\_3, Collinder\_350, RSG\_4, UPK\_545 \\
\citet{2023AA...679A.105Vaher} & Blanco\_1, NGC\_2516, NGC\_2547, Platais\_9, Theia\_517 \\
\bottomrule
\end{tabular}

	\label{tab:classification}
\end{table*}

\subsection{Key features of tidal tails}

We used synthetic observations of the M2154 model placed around the Sun (3 spherical grids at 200, 300, and 500 pc) to observe the expected behaviour of {near-tails}. The tail populations showed the following characteristics:
\begin{itemize}
	\item The tidal tails are elongated along the cluster's Galactic orbit.
	\item For low eccentricity orbits, the leading tail was always found to be inside the Galactic orbit, while the trailing tail goes outside the orbit.
	\item The distribution of $\Delta\phi_{galactocentric}$-$\Delta v_{r, galactocentric}$ values always showed a linear relation due to the inward motion of the leading tail and the outward motion of the trailing tail. This will be referred to as torsion in the XY plane. This behaviour is independent of cluster position in the sky and is only dependent on the accuracy of Galactocentric position and velocity measurements.
	\item An equivalent effect can be seen in the sky plane for some clusters (typically near $b\approx0\arcdeg\text{ or }180 \arcdeg$). This will be referred to as torsion in the sky plane.
	\item If the tail(s) lie along the line of sight, then their sky distribution is much smaller in extent, leading to poorer performance for any test in the sky plane. For example, if one of the tails was along the line of sight, then the $\mu_T$ of that tail might be amplified and the $R$-$\mu_T$ plot becomes non-linear.
	\item The number of stars with observed radial velocities is the limiting factor for the velocity-based tests. Similarly, the parallax errors are the limiting factor for analysis based on the Galactocentric coordinates.
	\item The absolute CMD shows that the tail and cluster have the same cluster turn-off (a sign of a coeval formation).
\end{itemize}

\section{Reliability of the observed tidal tails} \label{sec:reliability}

\subsection{Tests for reliable tidal tails}

Based on the literature and simulations, we devised the following flags for identifying tidal tails:
\begin{enumerate}
	\item \texttt{f\_no\_plx\_issue}: The tidal tail extension cannot be attributed to \textit{Gaia} parallax errors (see \citealt{Xu2025arXiv250417744X} for an attempt at quantitatively measuring this effect).
	\item \texttt{f\_xy\_extension}: The tidal tails should be an extended structure in the Galactocentric XY coordinates.
	\item \texttt{f\_xy\_shape}: The XY distribution should match with simulated clusters. This tests if the extension is along the Galactic orbit and if the leading tail should be closer to the Galactic centre compared to the trailing tail \citep{Dinnbier2022ApJ...925..214D}.
	\item \texttt{f\_xy\_torsion}: The Galactic azimuth ($\Delta \phi_{galactocentric}$) vs radial Galactocentric velocity ($\Delta v_{r, galactocentric}$) distribution should show a positive correlation and match with the simulation. To avoid noise due to the large number of cluster members and their velocity dispersion, only the tidal tail members were fitted.
	\item \texttt{f\_sky\_extension} Extended structure in the sky plane.
	\item \texttt{f\_sky\_torsion}: The torsion in the sky place due to different velocities of the two tidal tails: angular radius from the cluster centre ($r_{sky}$) vs tangential proper motion ($\mu_T$) should show some correlation. The on-sky torsion is significantly affected by cluster position; hence, we compared the slopes of the observed data with simulated data. To avoid inaccuracies due to the spherical to tangential plane transformations, the stars $>30$\arcdeg\ away from the cluster centre were ignored. Similar to \texttt{f\_xy\_torsion}, only tail members were fitted.
	\item \texttt{f\_cmd}: The CMD of the tails and cluster should be similar.
	\item \texttt{f\_expansion}: The leading and trailing tails should show outward motion from the cluster, which is an indication of increasing proper motion ($\mu_R$) with distance from the cluster centre ($R$). Similar to \texttt{f\_sky\_torsion}, only tidal tail members within 30\arcdeg\ were used.
\end{enumerate}

\begin{figure*}
	\centering
	\includegraphics[width=0.95\linewidth]{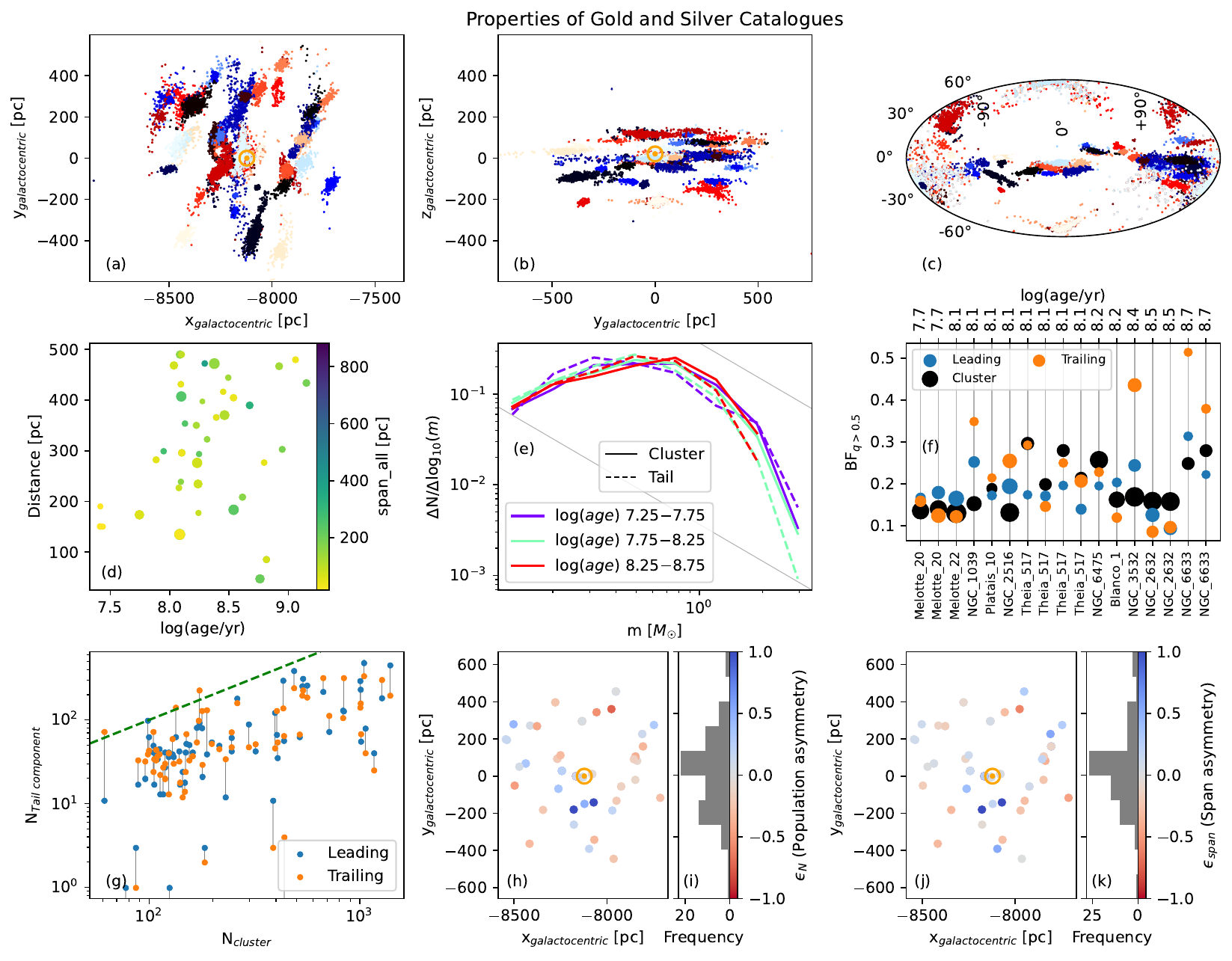}
	\caption{
    Properties of observed tidal tails in gold and silver catalogues. 
    (a) Spatial distribution of tidal tails and clusters (coloured arbitrarily) in the Galactic XY plane.
    (b) Spatial distribution of tidal tails and clusters in the Galactic YZ plane.
    (c) Spatial distribution of tidal tails and clusters in the Galactic coordinate plane.
    (d) The distribution of cluster age and distance from the Sun, coloured according to the total span and sized according to $N_{cluster}$. 
    (e) MF of all stars in the clusters (solid lines) and tail (dashed lines), binned according to the cluster age. {The grey lines indicate Salpeter IMF slope.}
    (f) Variation of BF for clusters and the tidal tails (ordered according to the cluster age). Only the clusters with $<$40\% errors in all three BFs are shown. The marker size corresponds to the population size.
    (g) Distribution of the detected tail population with the cluster population. The leading (blue) and trailing (orange) tails are shown connected by a solid grey line. The dashed green line represents $N_{tail}=N_{cluster}$. 
    (h) Clusters coloured according to normalised asymmetry in the tail populations ($\epsilon_N$).
    (i) Histogram of $\epsilon_N$.
    (j) Clusters coloured according to normalised asymmetry in the tail spans ($\epsilon_{span}$).
    (k) Histogram of $\epsilon_{span}$.
    Similar plots for the bronze catalogues are given in Figure~\ref{fig:general_properties_appendix}.
    }
	\label{fig:general_properties}
\end{figure*}

\subsection{Selecting a simulation for comparison} \label{sec:selecting_matching_simulation}

Based on the analysis of the simulated clusters, the tidal tails evolve uniquely based on the cluster mass, age, and orbit 
(note that the near-tail morphology is less sensitive to the cluster age)
. Ideally, to compare an observed cluster with simulations, one would need to match all three features. Unfortunately, due to the numerical and stochastic nature of stellar dynamics and uncertainties concerning the Galactic potential, predicting the initial conditions is not trivial.
For simplicity, we used a model with a circular orbit (from Table~\ref{tab:sim_setup}) which closely matched the present-day age and the number of particles within the cluster.

For comparisons with the observed \textit{Gaia} data, synthetic observations of the simulated cluster were performed to create \textit{Gaia}-like photometric and astrometric parameters.
The simulated clusters were placed at a given position and velocity using translations and rotations. The cluster was first rotated around the Galactic centre to match the required Galactic azimuth. And then it was translated (in position and velocity) to match the present-day astrometry of the observed clusters. A small rotation around the cluster centre was performed so that the angle between the ($x_{galactocentric},y_{galactocentric}$) and ($v_{x, galactocentric},v_{y,galactocentric}$) was the same as the real cluster. This accounts for the tilting seen in the local tail morphology of eccentric orbits (Figure~\ref{fig:morphology}).

The \textit{Gaia} magnitudes were calculated by assuming the stars follow the mass-magnitude relation of a parsec isochrone of corresponding age \citep{Bressan2012MNRAS.427..127B}. For simplicity, all stars were assumed to be main-sequence stars. As the magnitudes were only used for observational error estimations and the cut-off at the faint end, this assumption does not impact the overall study.
Noise was added to the observables based on median errors vs Gmag distributions of \textit{Gaia} DR3: parallax, pmra and pmdec from \citet{Gaia2023A&A...674A...1G}, radial velocity from \citet{Katz2023A&A...674A...5K}, and photometric errors in G, BP and RP from \citet{Riello2021A&A...649A...3R}.

We applied the following cuts to the simulated data to obtain $N_{sim,\ noisy}$ (number of stars in a simulated cluster with \textit{Gaia}-like noise):
\begin{equation}
    \begin{split}
        \texttt{Plx}-3\cdot\texttt{s\_Plx} < parallax<\texttt{Plx}+3\cdot\texttt{s\_Plx}, \\
        r_{sky} < (\texttt{rt}+\texttt{rJ})/2, \\
        phot\_g\_mean\_mag < 20 \text{ mag}
    \end{split}
\end{equation}
where \texttt{Plx} (cluster parallax), \texttt{s\_Plx} (standard deviation of cluster members), \texttt{rt} (tidal radius) and \texttt{rJ} (Jacobi radius) are taken from \citet{Hunt2024A&A...686A..42H}; $parallax$, $r_{sky}$ and $phot\_g\_mean\_mag$ are from the synthetic observations. We verified that the cluster members lie within the same cuts in the \citet{Hunt2024A&A...686A..42H} catalogue.
The parallax-based cut-off was chosen over a simple $r_{3D}$ (3D Cartesian radius) because most clusters are elongated in the Cartesian space due to parallax uncertainties.

The size of the tidal population ($N_{cluster}$) in the simulated cluster was used to select approximately nearby snapshots based on the cluster age and number of particles within the cluster. $N_{sim,\ noisy}$ was calculated for these nearby snapshots.
The snapshot with the least distance in the Age--N phase space ($[\Delta Age/e\_Age]^2+[\Delta N/e\_N]^2$) was chosen as the matching simulation.
This ensured that the statistical properties of the simulated cluster (and the tail) are similar to the observed cluster.

\subsection{Testing observed clusters}

For simplicity and due to imprecise tidal radius measurements, we defined the leading tail, cluster, and trailing tail as sources with $distance\_along\_orbit$ larger than $10$ pc, within $-10$ to $10$ pc, and less than $-10$ pc, respectively. 
Many observational catalogues have sources with large $distance\_from\_orbit$ and near-zero $distance\_along\_orbit$. As such sources are likely not tail members, ignoring such sources during analysis does not affect the overall results.

We created diagnostic plots for each cluster similar to Figure~\ref{fig:combo_real}. The figure shows the diagnostic plots for the observed cluster and a simulated cluster with \textit{Gaia}-like noise. The \texttt{f\_no\_plx\_issue}, \texttt{f\_xy\_shape} were judged visually and manually. \texttt{f\_xy\_extension} and \texttt{f\_sky\_extension} were determined by the presence of stars outside the tidal radius (accounting for the parallax errors). \texttt{f\_xy\_torsion} and \texttt{f\_sky\_torsion} were determined by comparing the slopes of respective distributions to the matching simulation within errors. The bottom part of Figure~\ref{fig:combo_real} shows the diagnostic plots for a simulated cluster without observational noise. The diagnostic plots for all 122 cluster catalogues are available in Appendix~\ref{sec:appendix_supplimentary}.

In practice, we found that all clusters and tails had compatible CMDs, and the expansion of the tails had a very poor signal. Hence, these two flags (\texttt{f\_cmd} and \texttt{f\_expansion}) were not used in the final assessment.
We summed the 6 pertinent flags (with equal weightage), creating a new flag (\texttt{f\_all}), and graded the clusters as follows.
Gold: The tidal tails passing all tests with \texttt{f\_all} = 6.
Silver: Reliable tidal tails with \texttt{f\_all} $\in$ [4,5].
Bronze: Unreliable tidal tails with \texttt{f\_all} $\in$ [1,2,3].

There are 16 catalogues with gold tidal tails (9 unique clusters), 55 catalogues with silver tidal tails (36 unique clusters), and 51 catalogues with bronze tidal tails (40 unique clusters). Overall, there are 40 unique clusters in the Gold+Silver sample.
Note that the grading is dependent on the modelling of the cluster, visual inspection, and the detection method. For example, \citet{Jerabkova2021AA...647A.137J} and \citet{2024arXiv240618767Kos} tails are more likely to pass the \texttt{f\_no\_plx\_issue}, \texttt{f\_xy\_shape}, \texttt{f\_xy\_extension}, and \texttt{f\_sky\_extension} flags because they used models (which already pass these tests) for tail recovery. \citet{2023AA...679A.105Vaher} tails are likely to fail the \texttt{f\_xy\_torsion} and \texttt{f\_xy\_torsion} tests because their tail populations and spans are small, and thus their slope comparisons can fail. Hyades (Melotte 25) is likely to fail the on-sky rotation test because of its large sky extent.
Readers are advised to use these flags, accounting for the inherent biases in the different methods. Overall, we recommend that all non-bronze catalogues are reliable enough for analysing the tails.

\section{Discussion} \label{sec:discussion}

\subsection{General statistics of the observed tails}

We measured the basic properties of real clusters and their tidal tails (individual flags, population numbers, spatial span, etc.). The table containing the overall properties of the clusters ({clusters.dat}) and a master source catalogue ({sources.dat}) created using all literature catalogues is available at the CDS. Appendix Table~\ref{tab:sources_columns_appendix} gives the column descriptions.

Figure~\ref{fig:general_properties} shows the overall properties of the gold and silver catalogues. Panels (a)-(c) show the spatial distribution of the clusters and tidal tails. The cluster+tail typically lie near the Galactic disc and have a Z range of $\pm$200 pc.
Panel (d) shows the distribution of cluster age and distance from the Sun, coloured according to the span of the overall structure and sized according to the number of sources in the cluster region. The longest tails are seen in Melotte 25, Melotte 111, NGC 2632, and clusters near the 400 pc region. The extensions in the closest clusters are due to better data, while the extensions in the farthest ones are dominated by the parallax-based elongation.
Panel (e) shows the MF of the cluster and tail population separated into 3 age bins. The averaged MFs are similar for the cluster and the tail. However, detailed analysis (accounting for incompleteness and accurate mass calibration) of individual cluster MFs is required to validate the MF evolution seen in the simulations.
Panel (f) shows the BF for all non-bronze clusters with reliable estimates.
Panel (g) shows the comparison of the number of sources within the cluster and the tails. In most clusters, the tidal tails are the smaller population.
Panels (h--k) show the asymmetry in the leading and trailing tails. We defined the population asymmetry ($\epsilon_N$) and span asymmetry ($\epsilon_{span}$) as follows \citep{Pflamm2023A&A...671A..88P}:
\begin{equation}
    \begin{split}
        \epsilon_N &= \frac{N_{leading}-N_{trailing}}{N_{leading}+N_{trailing}}  \\
        \epsilon_{span} &= \frac{span_{leading}-span_{trailing}}{span_{leading}+span_{trailing}} 
    \end{split}
\end{equation}

Figure~\ref{fig:general_properties_appendix} shows that many clusters near $l\approx-90$ fall into the bronze grade. This could be due to the line-of-sight alignment of the tails and cluster.

\subsection{Binary content of the tidal tails} \label{sec:binary_fraction}

Based on theoretical simulations, the BF of tidal tails decreases with cluster evolution, eventually reaching an approximately similar value to the cluster \citep{Wirth2024A&A...691A.143W}.
We estimated the photometric BF for clusters with well-fitted isochrones (clusters present in \citealt{Bossini2019A&A...623A.108B}). Figure~\ref{fig:binary_fraction_demo_appendix} shows a graphical example of the mass ratio estimation and the $BF_{q>0.5}$. The presented values represent the unresolved BF of the whole main sequence.

Figure~\ref{fig:general_properties} (f) shows the BF in non-bronze clusters with acceptable Poisson errors.
The BF in the tidal tails is higher than the cluster for $\approx50\%$ of the catalogues. Within this small sample, the BF appears to increase with the cluster age, indicating higher binary retention in the clusters as they are generally heavier. There is no clear correlation found between the BF and cluster population.
Analysis in a similar mass range and using a more complete sample is required to ascertain any links between the BFs of the cluster and the tidal tail.

\subsection{Asymmetry in the tidal tails}

The asymmetry in the tidal tails has been studied theoretically \citep{Pflamm2023A&A...671A..88P} and observationally \citep{Kroupa2022MNRAS.517.3613K, 2024ApJ...970...94Kroupa} in the context of testing gravitational theories. Figure~\ref{fig:general_properties} (h) and (j) show the asymmetry in the tail populations and tail spans as a function of cluster position. 
Panel (i) shows that the majority of catalogues (38:31) have more populated leading tails. Panel (k) shows that the majority of the clusters (41:28) have longer leading tails.

The more populated leading tails support the prediction by Milgromian dynamics \citep{Kroupa2022MNRAS.517.3613K, 2024ApJ...970...94Kroupa, Pflamm2025A&A...693A.127P}. However, the length of the tidal tail and differing magnitude limits across the length of tidal tails also affect these measurements. 
Similar plots for bronze clusters (Figure~\ref{fig:general_properties_appendix} h--k) show that their $\epsilon_{N}$ and $\epsilon_{span}$ values are highly correlated with the cluster position around the Sun. 
Thus, understanding the reliability of the tails is crucial before analysing the trends seen in the tail populations.

The detection of tidal tails depends on the relative velocities of the stars with the Sun, the observational uncertainties (a function of solar position), and the method of identifying the tails (see \citealt{Risbud2025AA...694A.258R}).
The analysis of all these biases is recommended before comparing these published spans and populations with simulations.

\subsection{Radial velocity vs $\phi_1$}

\citet{Risbud2025AA...694A.258R} noted that many clusters showed a trend of decreasing radial velocity with the tail-aligned longitude ($\phi_1$).
All richly populated clusters in their sample had the same trend.
However, further analysis of simulated clusters showed that there is no universal relation between the radial velocity and $\phi_1$. And the observed trend was simply a coincidence, caused by the small sample. The figure demonstrating the variation of $\phi_1$ with radial velocity for clusters placed spherically around the Sun and at the positions of \citet{Risbud2025AA...694A.258R} clusters is available in the Appendix~\ref{sec:appendix_supplimentary}.

\subsection{Using metallicity and age as a test for membership}

The tidal tails and clusters should have similar metallicity. Using the \textit{Gaia} DR3 \texttt{mh\_gspphot}, we found that the metallicities of the tail were comparable to the cluster. Precise spectroscopic metallicities for a large sample would improve the outlier detection. However, non-standard evolution of stars can lead to differences in metallicity, and thus, metallicity alone cannot be used to reject any outliers.

As individual stellar ages are not easily available, we can only analyse the whole tail and the cluster as an ensemble.
Based on the CMD turn-off, all tidal tails have similar ages to the cluster. However, the CMDs give a lower bound to the age, and thus, they cannot be used to confirm the coeval nature of individual stars. The cluster neighbourhood could also be populated by stars born from the same (or nearby) molecular cloud at a similar time. These stars will also pass any age tests, including the computationally/observationally expensive methods such as gyrochronology and asteroseismology. And thus, the age test is only useful for rejecting tidal tails with obvious and distinct turn-off sequences.

\subsection{Comparison across detection methods}

The catalogues collected from literature used a variety of methods for tidal tail detection. The supervised and unsupervised clustering techniques use overdensities in the 5D or 6D astrometric space. As the simulations show that the dynamics of the tail are different from the cluster and the on-sky parameters are highly dependent on the Solar position, only small portions of the tails can be recovered using such methods. The convergent point method utilises the convergent point to rescale observed astrometry and increases the phase space density. This method leads to the longest tidal tails based purely on observational data. The compact convergent point method and other $N$-body simulation-based comparisons can lead to longer spans; however, these are affected by the inherent assumptions from the $N$-body model and are difficult to verify, as there are no easy tests to validate the tidal tails. Comparisons with probabilistic models are limited by their assumptions and all the biases/problems of the $N$-body-based methods.

Among the analysed catalogues, the different techniques result in the following Gold:Silver:Bronze ratios.
(i) 5D Astrometric density-based techniques
5:16:24.
(ii) The convergent point-based methods
5:13:4. 
(iii) $N$-body model based methods
{2:0:0}.
(iv) Probabilistic model-based method
{4:21:18}.
(v) Backtracing stars
0:5:5 
{(see Table~\ref{tab:literature_table} for the literature and the corresponding techniques)}.

{The convergent point method has produced some of the richest and highest-quality catalogues (longer than 100 pc and up to 50\arcdeg\ in the sky; Table~\ref{tab:literature_table}). \citet{Jerabkova2021AA...647A.137J} introduced the compact convergent point method for which $N$-body models are needed to map the distant non-linear tidal tails.
However, the complex morphology requires a very accurate $N$-body model. The primary aspects that need to be matched with observations are the present velocity vector, orbit, $N_{tidal}$, and age (with decreasing order of importance). While models can improve detection, they also introduce biases and may identify comoving stars that are not genuine tail members. Distinguishing between these contaminants and true tail stars is challenging due to uncertainties in stellar ages and metallicity spreads. 
}

{The comparison with the probabilistic model \citep{2024arXiv240618767Kos} resulted in many bronze catalogues. Likely causes include simplified assumptions about stellar evaporation and neglect of cluster mass as a parameter. The simple assumptions also do not perfectly recreate the \reflectbox{S} shape. A more accurate analytical description of the \reflectbox{S} morphology as a function of cluster mass, age, and orbit would significantly improve performance.}

{5D+ astrometric techniques heavily use radial velocities and can detect tidal tails with spans of $>100$ pc. However, such techniques have only been applied for clusters within 200 pc of the Sun. Extending such work using all available radial velocity data would be useful in increasing and improving the known tidal tail catalogues.}

{5D astrometric density-based techniques can detect tidal tails in clusters out to a few kpc, though often at the cost of reduced tail span, completeness, and purity. Validating such short or distant tails is especially difficult due to low statistics and observational/model uncertainties. These methods also struggle to recover tails extending more than $20\arcdeg$ on the sky unless projection effects are carefully handled.}

{6D astrometric backtracing techniques \citep{2023AA...679A.105Vaher} can recover a fraction of the tail population, particularly stars near the cluster and moving radially outward. However, tail stars do not always exhibit such motion (see e.g. \citealt{Jerabkova2021AA...647A.137J}), limiting the method’s applicability.
The absence or uncertainties in radial velocities and distances further reduce effectiveness. And the limited statistics hinder catalogue validation.
Analysis of the 6D astrometry of Hyades tails (found by other techniques) might provide insights into the reliability and improvements to the backtracing techniques.}

{Overall, convergent point-based methods--especially when assisted by accurate $N$-body or analytical models--remain the most effective for detecting tidal tails. Though such analysis is limited to a few hundred parsecs near the Sun. 
For more distant clusters, 5D density-based techniques are the most promising for increasing the number of tail detections.
However, assessing the completeness and purity of existing catalogues remains a major challenge. Asymmetric tail morphologies, combined with direction-dependent uncertainties (particularly in distance and radial velocity), introduce further biases. 
Testing the performance of various techniques on realistic synthetic data is essential to characterise recovery rates, incompleteness, and purity.}

\section{Summary and Conclusions} \label{sec:conclusions}

$N$-body simulations provide baseline expectations for the properties of tidal tails. The key findings from these simulations are as follows:
The morphology of tidal tails is notably complex. The tails of a cluster on a circular orbit create circular loci, while the tails of clusters on eccentric orbits create non-circular loci. 
The {near-tail} creates a \reflectbox{S} shape whose amplitude depends on the cluster mass. 
Escapees in the leading tails tend to follow more tightly bound orbits, while those in the trailing tails are generally more loosely bound.
The tails also exhibit over-densities near 20, 40, and 60$R_{tidal}$, though these features diminish in prominence for lower-mass clusters that cannot continuously supply sufficient escapees. While the average stellar mass in the tails decreases with increasing distance from the cluster, detecting this trend observationally remains challenging due to incompleteness, stochastic variation, and poor statistics.

The {near-tail}s are populated by recent escapees, while the tips host the earliest escapees, which also move with higher relative speeds, similar to the higher velocity dispersion of the cluster. The population increase in the tail can be approximately described by a tanh($T$) function, while the total span of the tails increases linearly with time (at a rate of $\approx4\sigma_{velocity,0}$). The number of escapees depends on the cluster mass in the early evolution ($\nu_{relax}<5$). For dynamically evolved clusters, however, both escape rate and tail density exhibit only weak mass dependence.

The MF of the {near-tail} and the cluster are different in the early evolution, while they both become flatter and top-heavy as the dynamical evolution leads to the evaporation of lower mass stars to the {far-tail} region. The {near-tail} is primarily composed of main-sequence stars, with a notable deficit of giants---consistent with the fact that, at a given simulation time, giants are among the most massive and thus more centrally retained stars. These statements remain generally valid for clusters on eccentric orbits. Their properties oscillate around the values estimated for the clusters of similar mass on equivalent circular orbits.

{Lower and higher mass clusters have similar 1D number density, location of epicyclic overdensity (as a multiple of $R_{tidal}$), and a similar rate of escapees for the majority of the evolution (except near the birth and dissolution of the cluster). However, the morphology of the tails, the width of the \reflectbox{S} shape, the mass function difference between the cluster and near-tail, and the escape velocity are cluster mass-dependent. Therefore, while tidal tails of low-mass simulated clusters may be a suitable replacement for comparing properties such as number density and escapee counts, they may not accurately represent the overall morphological features seen in more massive clusters.}

These conclusions are based on one grid of simulations. Additional physical processes and numerical factors may affect these results. Nonetheless, the presented analysis serves as a valuable first-order approximation of tidal tail behaviour.

Guided by synthetic observations of these simulations, we developed a set of tests to evaluate the validity of tidal tail catalogues in the recent literature.
We examined the quality of 122 tidal tail catalogues (corresponding to 58 unique clusters) published in the recent literature. We used the morphological, photometric, and dynamical properties of the clusters to assess the comoving and coeval nature of the clusters and the dynamical signatures (such as torsion in the Galactic and sky plane). We also compared real data with synthetic observations to establish benchmarks for expected tidal tail behaviour.

Based on the six pertinent tests, the catalogues were graded into gold (15), silver (54), and bronze (51) categories. We recommend that only gold and silver catalogues be used for detailed scientific analysis, while bronze catalogues should be treated with caution or avoided.

All catalogues confirmed coeval and comoving tail populations—unsurprising, as these were often selection criteria in the original studies. Metallicity distributions in the tails were found to be consistent with their parent clusters.

We measured the unresolved BF ($BF_{q>0.5}$) of the cluster and tidal tails to be 10--40\%. This required the use of empirical main-sequence ridge lines to correct for minor deviations from theoretical isochrones. Overall, the BF of the tails was similar to the clusters, the latter showing a weak trend of increasing BF with increasing age.

The non-bronze catalogues have more stars on average in the leading tidal tails, in agreement with the predictions of Milgromian dynamics. However, further careful scrutiny of the data biases and incompleteness is needed before making firm conclusions based on the observational data.

The issue of tail incompleteness remains a major limitation for drawing advanced conclusions from current catalogues. A comprehensive treatment of selection effects and completeness is essential for accurate interpretation.
Upcoming wide-scale surveys (such as LSST, \textit{Gaia} DR4, 4MOST) are expected to vastly improve the depth and quality of available data. These improvements will enable more refined tidal tail catalogues and advance our understanding of star cluster dissolution and the Galactic potential.

\section{Data Availability}

Tables {clusters.dat} and {sources.dat} are only available in electronic form at the CDS via anonymous ftp to \url{https://cdsarc.u-strasbg.fr/} (130.79.128.5) or via \url{http://cdsweb.u-strasbg.fr/cgi-bin/qcat?J/A+A/}. Appendix Table~\ref{tab:sources_columns_appendix} gives the column descriptions.

\begin{acknowledgements}
	We thank the anonymous referee for constructive comments. 
    We thank S. Bhattacharya and H. M. Boffin for providing their catalogues.
	VJ thanks the Alexander von Humboldt Foundation for their support.
	PK acknowledges support through the DAAD Eastern European Exchange Programme.
	This work has made use of data from the European Space Agency (ESA) mission {\it Gaia} (\url{https://www.cosmos.esa.int/gaia}), processed by the {\it Gaia} Data Processing and Analysis Consortium (DPAC, \url{https://www.cosmos.esa.int/web/gaia/dpac/consortium}). Funding for the DPAC has been provided by national institutions, in particular, the institutions participating in the {\it Gaia} Multilateral Agreement. 
    The work used HPC systems (Marvin and dynamix) for data generation and the following tools for the analysis: 
    \textsc{Astropy} \citep{astropy:2022}; 
    \textsc{ClusterTools} \citep{Webb2023JOSS....8.4483W}; 
    \textsc{Galpy} \citep{Bovy2015ApJS..216...29B}; 
    \textsc{NumPy} \citep{2020Natur.585..357Harris}; 
    \textsc{SciPy} \citep{Virtanen2020NatMe..17..261V}; 
    \textsc{topcat} \citep{2005ASPC..347...29TOPCAT}.

\end{acknowledgements}
\bibliographystyle{aa} 
\bibliography{references}

\begin{appendix}

		\section{Supplementary table and figures}
\begin{table*}[b]
    \centering
    \caption{Summary of the various catalogues used in this work.}
    \footnotesize
        \begin{tabular}{llcc cccc}
        \toprule
Reference	&	Techniques			&	$N_{clusters}$	&	$N_{sources}$	 & 	$f_{6D}$	&	$q_{parallax\_over\_error}$	&	$q_{radial\_velocity\_error}$	&	Median span	\\						
	&	[Recommended cuts]			&		&		 & 		&	[1,50,99]	&	[1,50,99]  [\kms]	&		\\ \hline						
\citet{Jerabkova2021AA...647A.137J}	&	$N$-body comparison,	&	2	 & 	1971	 & 	0.51	 & 	15,425,1463 	 & 	0.1,1.1,22	 & 	 881 pc (57°)	\\		&	\multicolumn{6}{l}{	Compact convergent point	}	\\
\citet{2022MNRAS.514.3579Boffin}	&	Convergent point			&	1	 & 	598	 & 	0.46	 & 	3,47,167 	 & 	0.1,2.3,14	 & 	 197 pc (12°)	\\		&	\multicolumn{6}{l}{[	$\texttt{phot\_g\_mean\_mag} \leq 20$ mag	]}	\\
\citet{Risbud2025AA...694A.258R}	&	Convergent point			&	19	 & 	10762	 & 	0.45	 & 	22,123,1207 	 & 	0.1,2.2,13	 & 	 176 pc (16°)	\\						
\citet{2019AA...621L...2Roser_Hyades}	&	Convergent point			&	1	 & 	979	 & 	0.65	 & 	85,654,1941 	 & 	0.1,1.1,26	 & 	 213 pc (56°)	\\		&	\multicolumn{6}{l}{[	$\texttt{Member} \neq \texttt{other} \ \ \textsc{and}\ \  \texttt{Comment} \neq \texttt{Contam.}$	]}	\\
\citet{2019AA...627A...4Roser_Praesepe}	&	Convergent point			&	1	 & 	1393	 & 	0.39	 & 	18,93,450 	 & 	0.2,1.6,12	 & 	 207 pc (27°)	\\						
\citet{2024arXiv240618767Kos}	&	Probabilist model comparison			&	38	 & 	14891	 & 	0.54	 & 	14,103,870 	 & 	0.2,2.9,14	 & 	 155 pc (6°)	\\		&	\multicolumn{6}{l}{[	$\texttt{pbint} > 0.9$	]}	\\
\citet{2019AA...621L...3Meingast}	&	5D+ Astrometric density			&	1	 & 	238	 & 	0.98	 & 	92,994,2016 	 & 	0.1,0.2,1.5	 & 	 180 pc (55°)	\\						
\citet{Oh2020MNRAS.498.1920O}	&	5D+ kinematic modelling			&	1	 & 	1002	 & 	0.62	 & 	70,623,1844 	 & 	0.1,1.4,22	 & 	 219 pc (56°)	\\		&	\multicolumn{6}{l}{[	$\texttt{Pmem} > 0.5$	]}	\\
\citet{Olivares2023AA...675A..28O}	&	5D+ Astrometric density			&	1	 & 	302	 & 	0.52	 & 	7,236,865 	 & 	0.2,1.3,11	 & 	 130 pc (33°)	\\						
\citet{2019ApJ...877...12Tang}	&	5D Astrometric density			&	1	 & 	196	 & 	0.62	 & 	65,341,846 	 & 	0.2,1.2,9	 & 	 73 pc (24°)	\\						
\citet{2019AA...624L..11Furnkranz}	&	5D Astrometric density			&	1	 & 	213	 & 	0.63	 & 	46,329,793 	 & 	0.2,1.3,10	 & 	 64 pc (23°)	\\						
\citet{2020ApJ...889...99Zhang}	&	5D Astrometric density			&	1	 & 	644	 & 	0.29	 & 	16,52,315 	 & 	0.4,2.5,13	 & 	 83 pc (9°)	\\						
\citet{2021ApJ...912..162Pang}	&	5D Astrometric density			&	4	 & 	3596	 & 	0.34	 & 	10,53,258 	 & 	0.1,2.9,17	 & 	 53 pc (2°)	\\						
\citet{2021AA...645A..84Meingast}	&	5D Astrometric density			&	10	 & 	7882	 & 	0.36	 & 	16,118,653 	 & 	0.3,2.8,16	 & 	 119 pc (11°)	\\		&	\multicolumn{6}{l}{[	$\texttt{fc} < 0.5$	]}	\\
\citet{2022ApJ...931..156Pang}	&	5D Astrometric density			&	6	 & 	1620	 & 	0.46	 & 	10,68,358 	 & 	0.1,2.9,14	 & 	 52 pc (2°)	\\		&	\multicolumn{6}{l}{[	$\texttt{Flag}==t$	]}	\\
\citet{Bhattacharya2022MNRAS.517.3525B}	&	5D Astrometric density			&	24	 & 	9910	 & 	0.37	 & 	7,75,479 	 & 	0.3,3,15	 & 	 46 pc (2°)	\\						
\citet{2023AA...679A.105Vaher}	&	Backtracing 6D astrometry			&	10	 & 	1784	 & 	1	 & 	31,228,670 	 & 	0.3,2.5,13	 & 	 20 pc (2°)	\\		&	\multicolumn{6}{l}{[	$\texttt{pmemb} \geq0.1 \ \ \textsc{and}\ \  \texttt{pfug} \geq 0.1$	]}	\\

        \bottomrule
    \end{tabular}
    \tablefoot{The columns are as follows: $N_{clusters}$: number of clusters in the paper; 
    $N_{sources}$: total number of stars in the paper; 
    $f_{6D}$: fraction of stars with 6D information;
    $q_{parallax\_over\_error}$: 1st, 50th and 99th quantiles of $parallax\_over\_error$; 
    $q_{radial\_velocity\_error}$: 1st, 50th and 99th quantiles of ${radial\_velocity\_error}$; 
    median span of all tails in the paper in the Cartesian and sky planes. All the numbers are after giving the 500 pc selection cut and the recommended quality cuts. `5D+' in the technique means that the radial velocity was heavily used, but was not necessary.}
    \label{tab:literature_table}
\end{table*}

\begin{figure*}[b]
    \centering
    \includegraphics[width=0.95\linewidth]{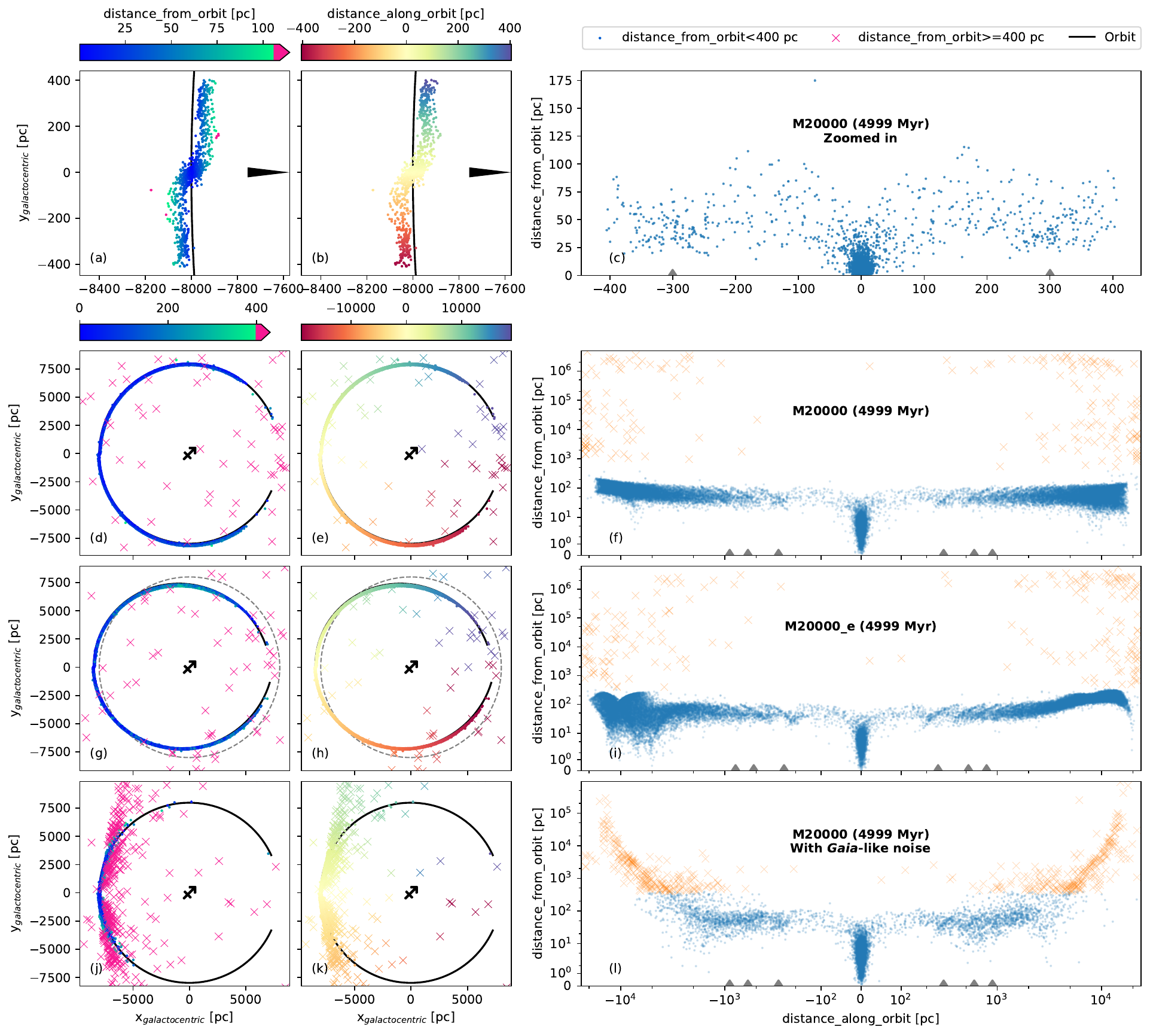}
    \caption{Demonstration of the $distance\_from\_orbit$ (panels a, d, g, and j) and $distance\_along\_orbit$ (panels b, e, h, and k) measurements for M20000 (first, second row), M20000\_e (third row), and a noisy-M20000 (fourth row) at 4999 Myr. 
The stars well outside the average locus ($distance\_from\_orbit \geq 400$ pc) are shown by crosses.
{First and second columns are coloured according to $distance\_from\_orbit$ and $distance\_along\_orbit$, respectively.}
The 8 kpc circle is shown by a grey dashed curve in panels (g) and (h) for reference (the orbit itself is a circle of radius 8 kpc for the model M20000).
The axes in panels (f), (i) \& (l) use Asinh scaling for better visualisation, and the $\pm$20, $\pm$40, and $\pm$60$R_{tidal}$ are marked by the grey triangles along the x axis.} 
    \label{fig:distance_along_orbit_appendix} 
\end{figure*} 

\begin{table*}[]
			\centering
			\caption{Description of columns in {clusters.dat} and {sources.dat} tables available at CDS.}

			\footnotesize

			\begin{tabular}{ll}
				\toprule
				\multicolumn{2}{c}{Description of {clusters.dat} catalogue.}                                           \\
\midrule
cluster & Cluster name \\
reference & Reference of the source catalogue \\
cluster\_id & Unique cluster ID \\
RAdeg, DEdeg & Right ascension and Declination (ICRS) at Ep=2016 (Hunt and Reffert 2024) \\
dist50 & Distance from the Sun (Hunt and Reffert 2024) \\
logAge50 & Logarithm of cluster age (Hunt and Reffert 2024) \\
pmRA, pmDE & Proper motion in R.A. (pmRA*cosDE) and Dec. direction (Hunt and Reffert 2024) \\
RV & Distance (Hunt and Reffert 2024) \\
f\_no\_plx\_issue & Flag indicating absence of elogation due to parallax \\
f\_xy\_extension & Flag indicating extended structure in Galactocentric XY plane \\
f\_xy\_shape & Flag for expected morphology \\
f\_xy\_torsion & Flag for torsion signature in Galactocentric XY plane \\
f\_sky\_extension & Flag for extended structure in sky plane \\
f\_sky\_torsion & Flag for torsion signature in sky plane \\
f\_all & Summation of the 6 pertinent flags \\
grade & Grade of the catalogue (G=gold, S=silver, B=bronze) \\
n\_all & No. of all stars in the catalogue \\
n\_leading & No. of stars in leading tail (LT) \\
n\_cluster & No. of stars in the cluster \\
n\_trailing & No. of stars in trailing tail (TT) \\
n\_reference\_all & No. of all reference stars for binary fraction calculation. Only
     sources with mass. \\
n\_reference\_leading & No. of reference stars in LT \\
n\_reference\_cluster & No. of reference stars in cluster \\
n\_reference\_trailing & No. of reference stars in TT \\
n\_binary\_all & No. of all binary stars (q>0.5). Limited to sources with mass
     estimates. \\
n\_binary\_leading & No. of binary stars in LT \\
n\_binary\_cluster & No. of binary stars in cluster \\
n\_binary\_trailing & No. of binary stars in TT \\
BF\_all, e\_BF\_all & Total binary fraction and its error \\
BF\_leading, e\_BF\_leading & Binary fraction in LT and its error \\
BF\_cluster, e\_BF\_cluster & Binary fraction in cluster and its error \\
BF\_trailing, e\_BF\_trailing & Binary fraction in TT and its error \\
span\_all & Span of all stars along the cluster orbit \\
span\_leading & Span of the LT along the cluster orbit \\
span\_trailing & Span of the TT along the cluster orbit \\
				\midrule
				\multicolumn{2}{c}{Description of {sources.dat} catalogue.}                                            \\
				\midrule
GaiaDR3 & Gaia DR3 source\_id \\
cluster & Cluster name \\
reference & Reference of the source catalogue \\
cluster\_id & Unique Cluster ID \\
grade & Grade of the catalogue (G=gold, S=silver, B=bronze) \\
class & Membership (C=cluster, L=leading tail, T=trailing tail) \\
RAdeg, DEdeg & Right ascension and Declination (ICRS) at Ep=2016 \\
r\_med\_geo & Geometric distance (Bailer-Jones 2021) \\
pmRA, pmDE & Proper motion in R.A. (pmRA*cosDE) and Dec. direction (Gaia) \\
RV & Radial velocity (Gaia) \\
Gmag, BPmag, RPmag & Gaia DR3 G, BP and RP band magnitudes (Gaia) \\
BP\_RP & Gaia DR3 BP-RP colour (Gaia) \\
GMag & Absolute G band magnitude (Gaia) \\
r\_3d & Cartesian distance from the cluster centre \\
v\_3d & Cartesian speed relative to the cluster centre \\
r\_sky & Distance from the cluster centre projected on the sky \\
pm\_R, e\_pm\_R & Radial component of the projected proper motion from the cluster
     centre and its error \\
pm\_T, e\_pm\_T & Tangential component of the projected proper motion from the
     cluster centre and its error \\
delta\_v\_r\_GC & Relative radial velocity in the Galactic plane as seen from the
     Galactic centre \\
delta\_phi\_GC & Relative azimuth. Angle created by the Sun-GC-star on the
     Galactic plane \\
distance\_from\_orbit & Distance between the star and the clusters Galactic orbit \\
distance\_along\_orbit & Distance between the centre and the point closest to the star on
     the Galactic orbit \\
cmd\_distance & Distance from the region defined by the RLs of q=0 and q=1 in the
     CMD \\
q, e\_q & Mass ratio and its error \\
mass\_system\_MLR\_ & Mass using mass-mag relation of parsec ZAMS. NOT to be used for
     science \\
mass\_system, e\_mass\_system & Mass of the source assuming an unresolved binary system and its error \\
mass\_A, e\_mass\_A & Mass of primary component of the unresolved binary system and its error \\
mass\_B, e\_mass\_B & Mass of secondary component of the unresolved binary system and its error \\
				\bottomrule
			\end{tabular}
			\label{tab:sources_columns_appendix}

		\end{table*}

\begin{twocolumn}

\begin{figure}[h]
	\centering
	\includegraphics[width=0.99\linewidth]{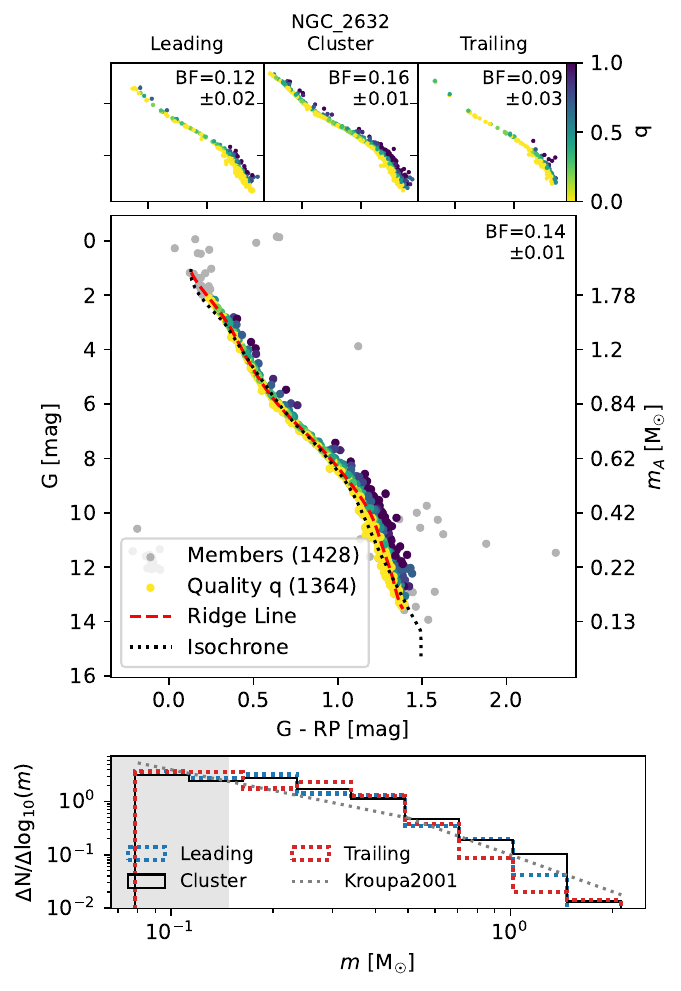}
\caption{Demonstration of BF calculation for NGC 2632. (Top row) Absolute CMDs for leading tail, cluster, and trailing tail. (Middle row) Absolute CMD for all stars coloured according to the mass ratio (q). The isochrone (grey dotted curve) and main sequence ridge line (dashed red curve) are shown for reference. (Bottom row) The MF of the visible members of the cluster and tidal tail. \citet{2001MNRAS.322..231KroupaIMF} IMF is shown for reference. The lower limit of credible mass estimates is indicated by the grey-shaded region.}
	\label{fig:binary_fraction_demo_appendix}
\end{figure} 
	
\begin{figure}
    \centering
    \includegraphics[width=0.99\linewidth]{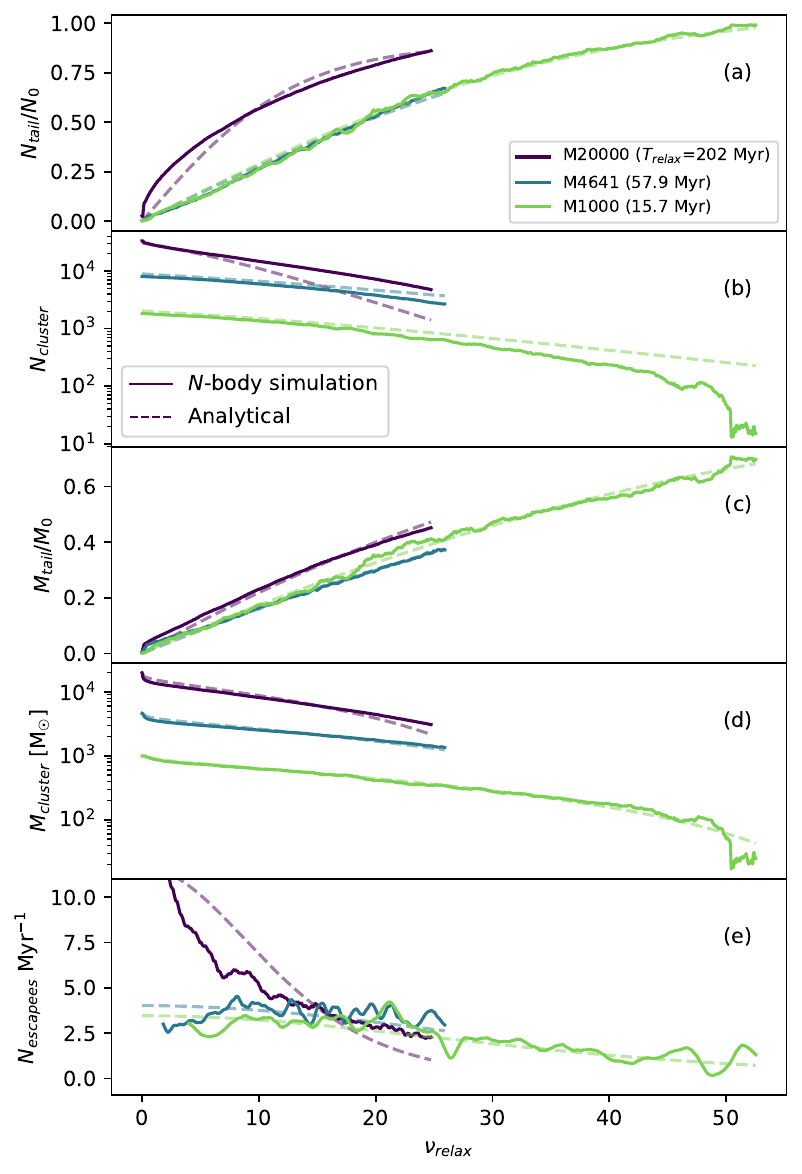}
    \caption{Evolution of tail and cluster parameters for various models and their analytical estimates. 
    (a) Number of stars in the tail normalised by the total number of stars at the beginning (see Eq.~\ref{eq:n_tail}).
    (b) Number of stars in the cluster (Eq.~\ref{eq:n_tail}). 
    (c) Mass of the tail normalised by the total mass at the beginning (see Eq.~\ref{eq:m_tail_lamers}).
    (d) Mass of the cluster (see Eq.~\ref{eq:m_tail_lamers}).
    (e) Number of averaged escapees per Myr (see Eq.~\ref{eq:escapees}). 
    The $N$-body results are shown as solid curves while the analytical estimates based on Eq.~\ref{eq:n_tail}, \ref{eq:m_tail_lamers}, and \ref{eq:escapees} are shown by dashed curves.}
    \label{fig:number_escapee_span_appendix}
\end{figure} 

\end{twocolumn}

\begin{onecolumn}

\begin{figure}
\centering
    \includegraphics[width=0.9\linewidth]{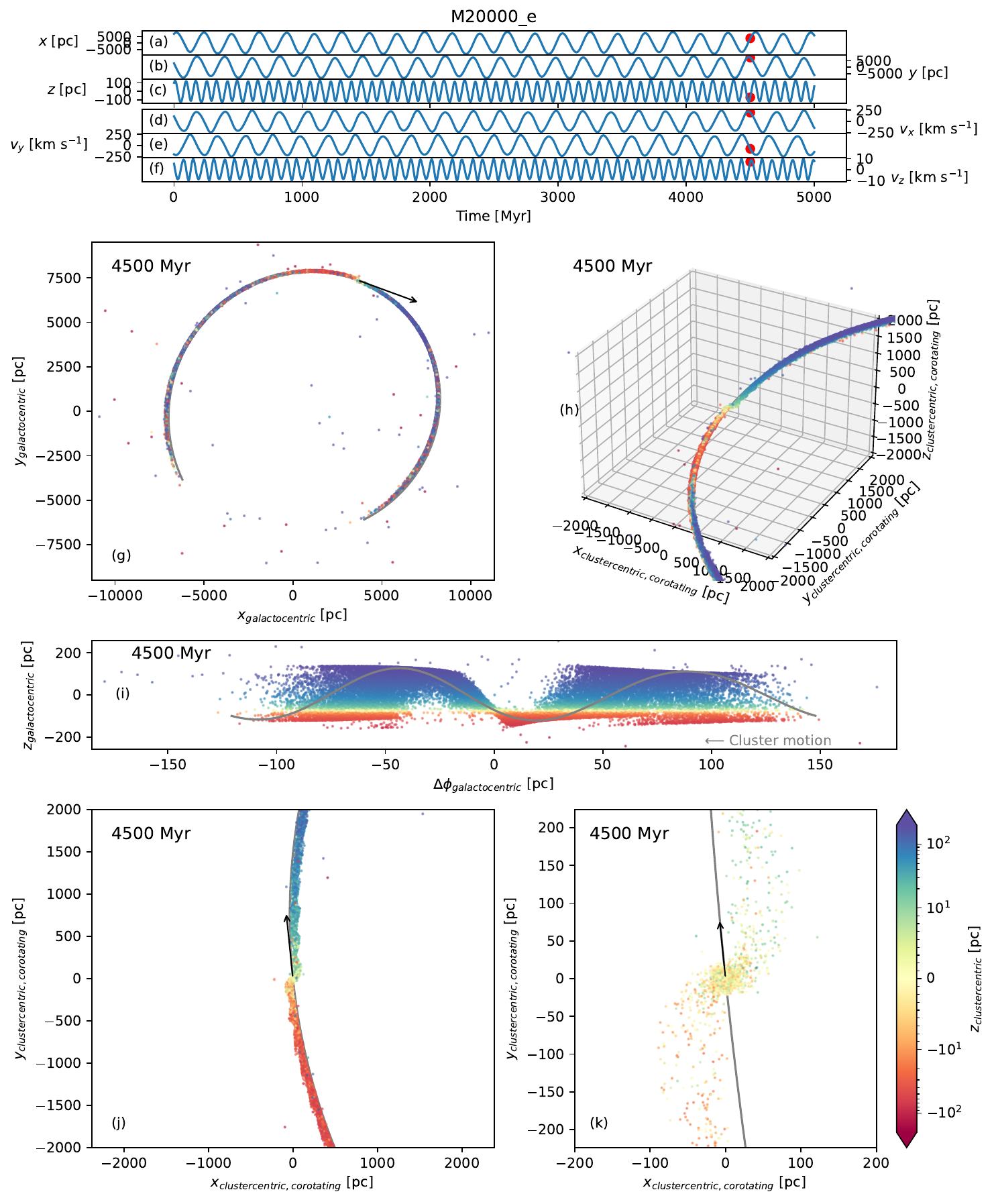}
    \caption{A screenshot of the movies available as online material (for models M1000, M20000, and M20000\_e). 
    The (a) to (f) panels show the total evolution of Galactocentric position ($x$, $y$, $z$) and velocities ($v_x$, $v_y$, $v_z$) with time. The points at a given time of the movie are marked by red dots. 
    (g) The Galactocentric distribution of all particles in the simulations. 
    (h) The 3D distribution of particles in the corotating frame and with the origin at the cluster centre (i.e., the clustercentric corotating frame).
    (i) The distribution of $\Delta\phi_{galactocentric}$ with $z_{galactocentric}$.
    (j) The XY distribution of particles in the clustercentric corotating frame. 
    (k) Same as panel (j) but zoomed in near the cluster centre. 
    The points in panel (g) to (k) are coloured according to their clustercentric z coordinate, and the time of the snapshot is written in the corner. The cluster orbit and velocity are shown as the grey curve and black arrow, respectively.}
    \label{fig:movie_screenshot_appendix}
\end{figure} 

\begin{figure}
	\centering
	\includegraphics[width=0.9\linewidth]{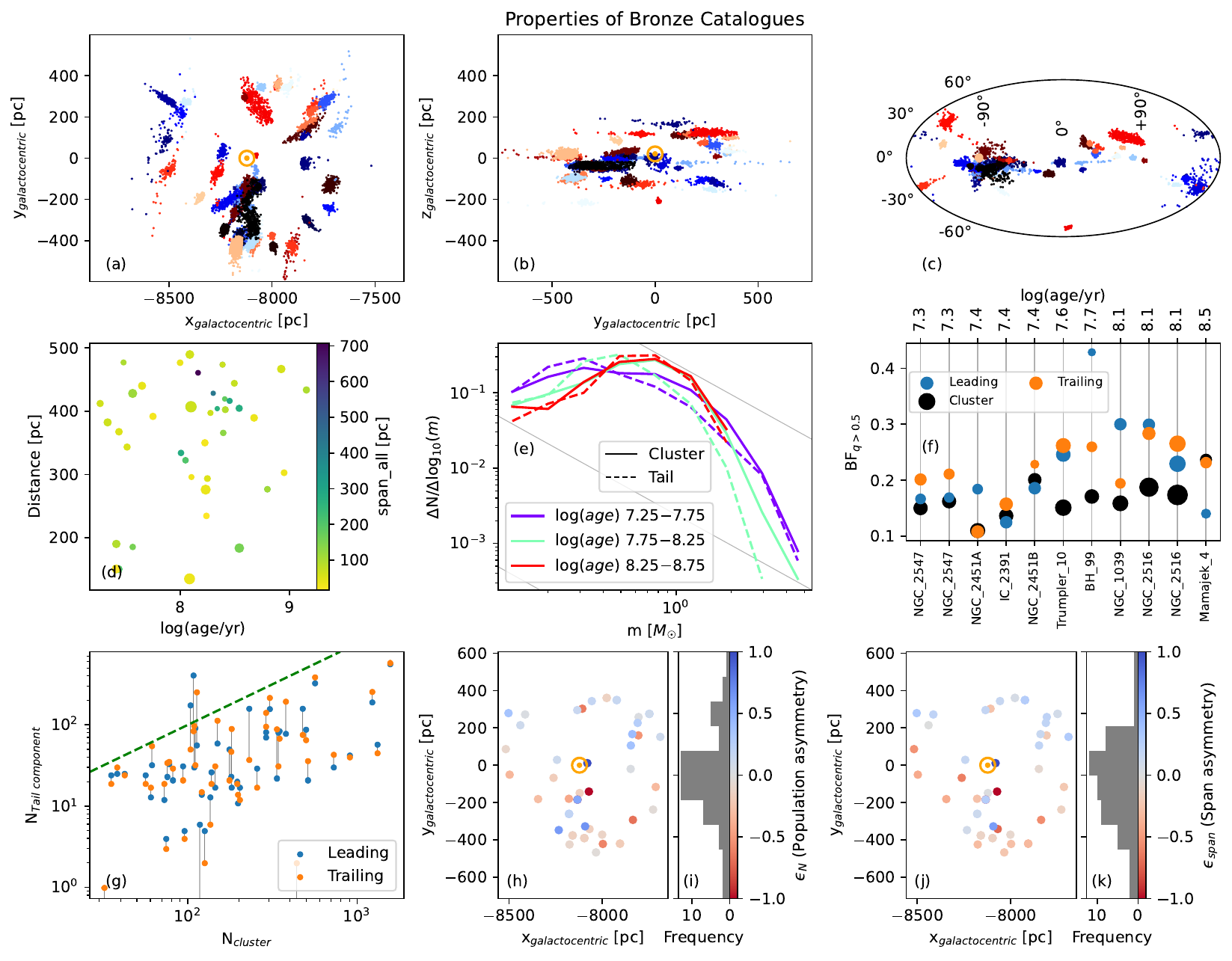}
	\caption{
    Properties of observed tidal tails in bronze catalogues. All details are similar to Figure~\ref{fig:general_properties}.
    }
	\label{fig:general_properties_appendix}
\end{figure} 

\begin{landscape}

\section{Supplementary figures (only available on arXiv)} \label{sec:appendix_supplimentary}

\begin{figure}[ht]
    \centering
    \includegraphics[width=0.8\linewidth]{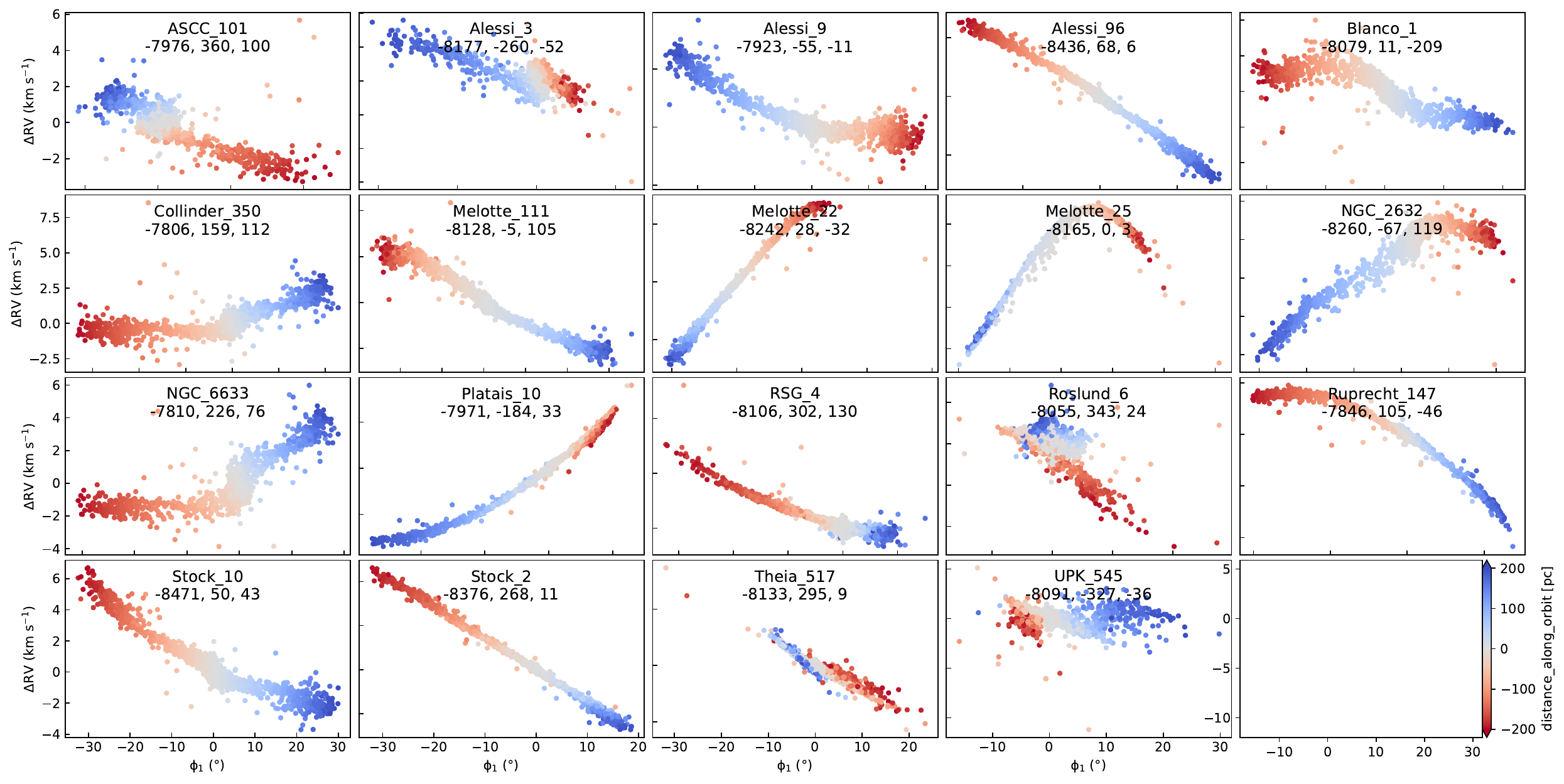}
    \caption{Variation of the radial velocity with $\phi_1$ (tail-aligned longitude) for a simulated cluster (M2154) placed at the positions of Risbud et al. (2025) clusters. The points are coloured according to the $distance\_along\_orbit$ (leading tail = bluer; trailing tail = redder). The numbers at the top of the subplots represent the heliocentric XYZ coordinates (in pc) of the simulated cluster. Note that the x and y limits are different for each subplot.}
    \label{fig:rv_phi_risbud}
\end{figure}

\begin{figure}[ht]
    \centering
    \includegraphics[width=0.9\linewidth]{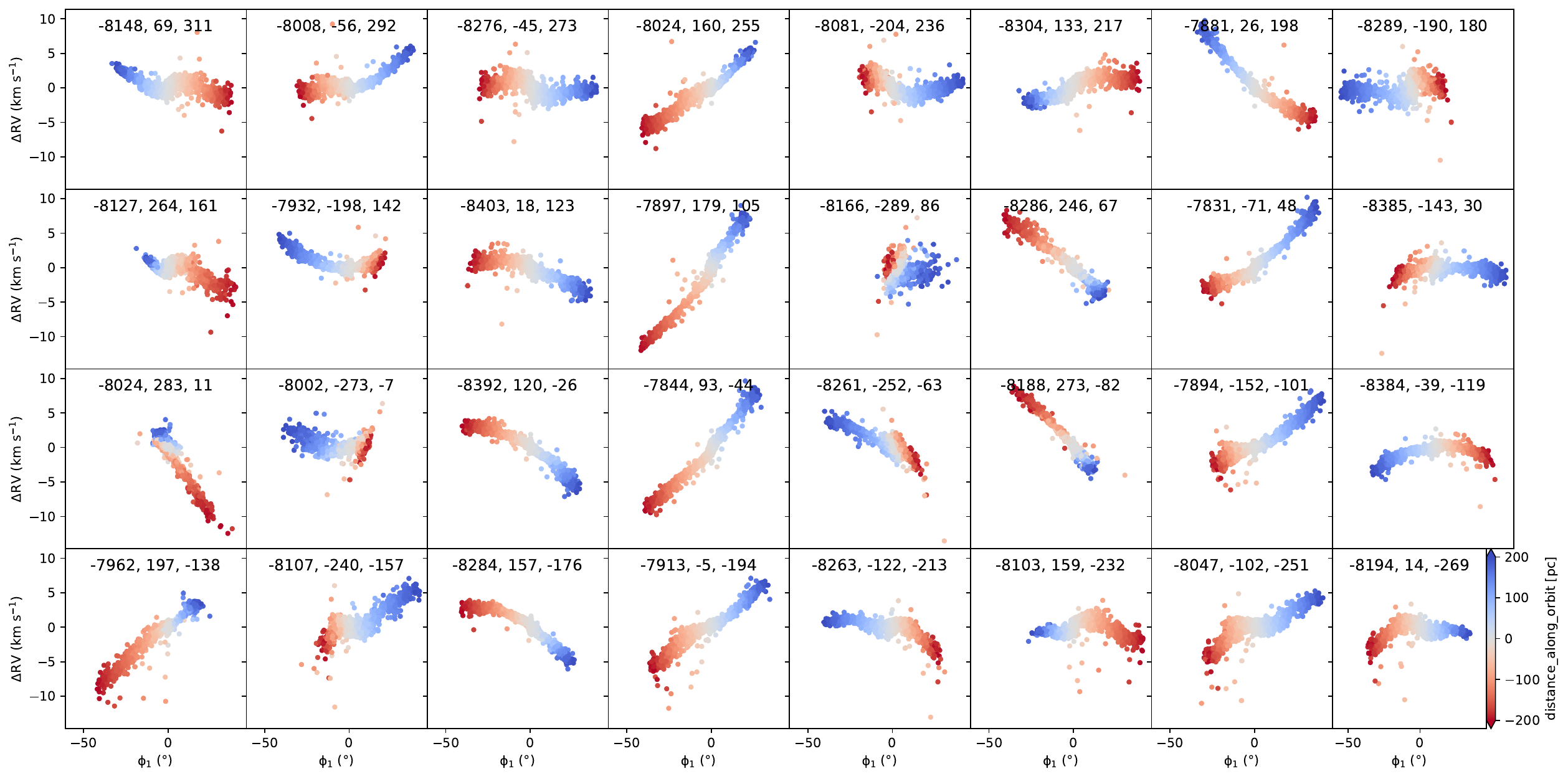}
    \caption{Variation of the radial velocity with $\phi_1$ for clusters placed in a spherical grid around the Sun. All details are similar to Figure~\ref{fig:rv_phi_risbud}. All subplots share the same x and y limits.}
    \label{fig:rv_phi_grid}
\end{figure}

        \begin{figure}
\includegraphics[width=0.5\linewidth]{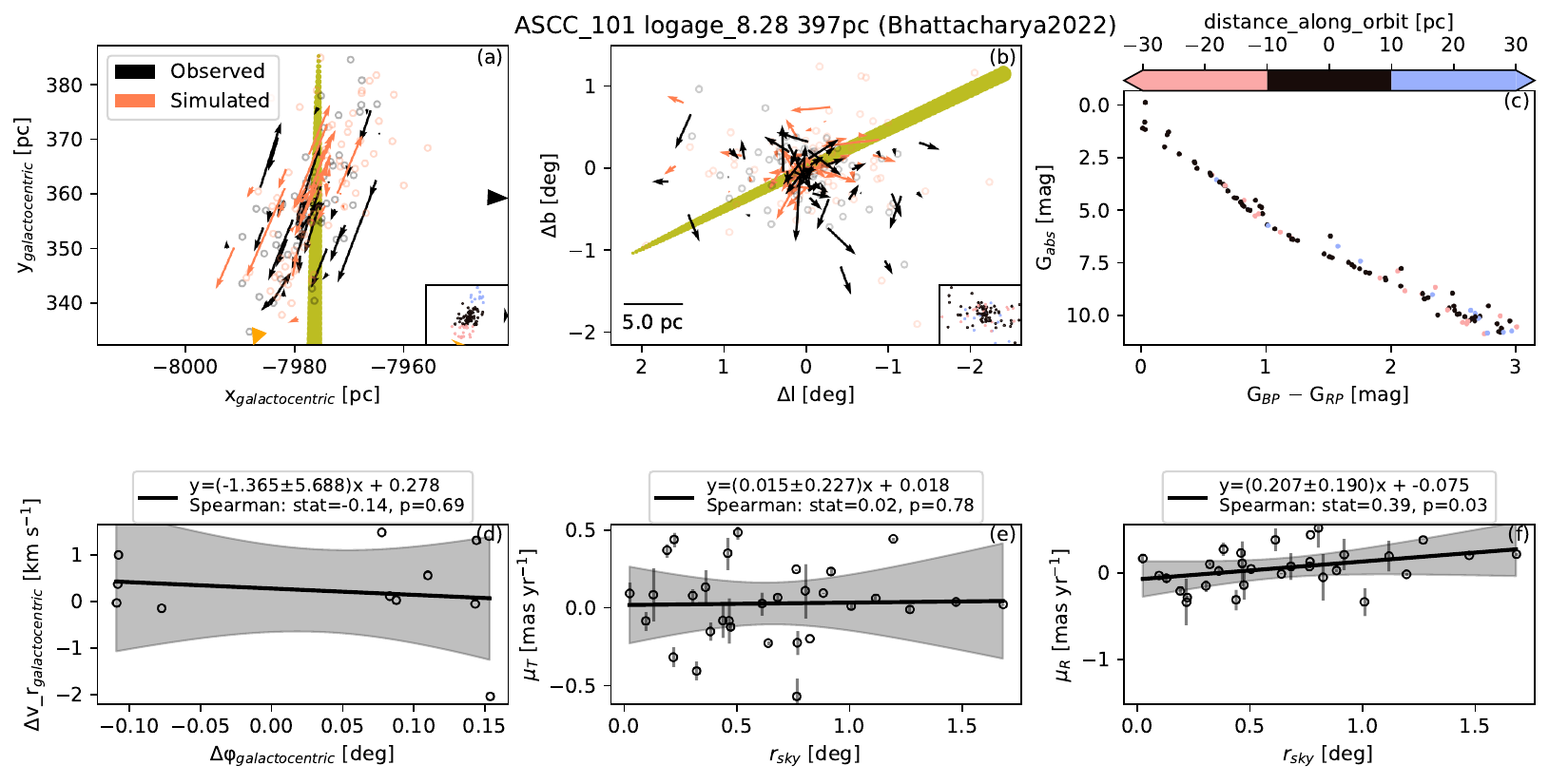}
\includegraphics[width=0.5\linewidth]{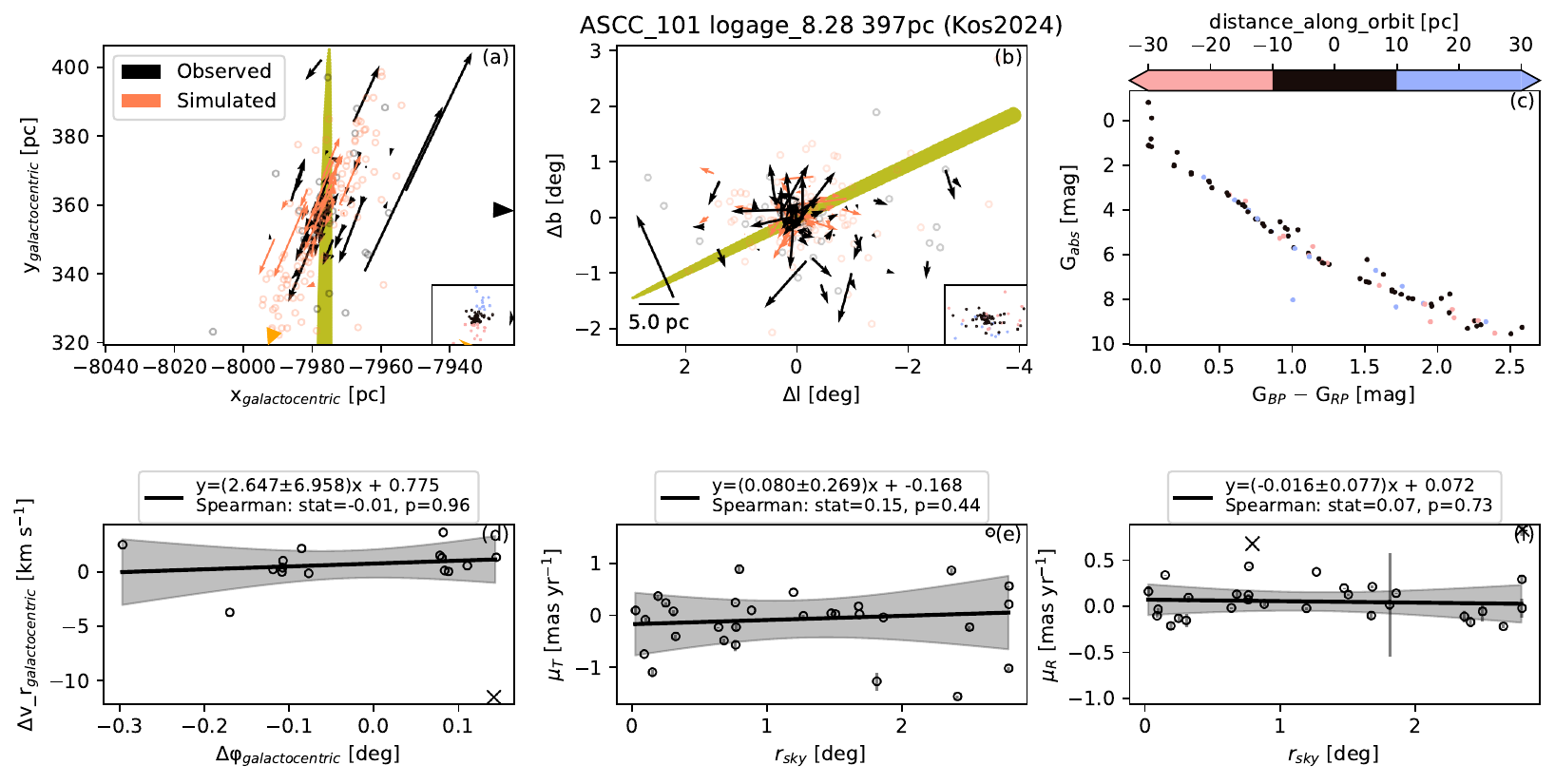}
\includegraphics[width=0.5\linewidth]{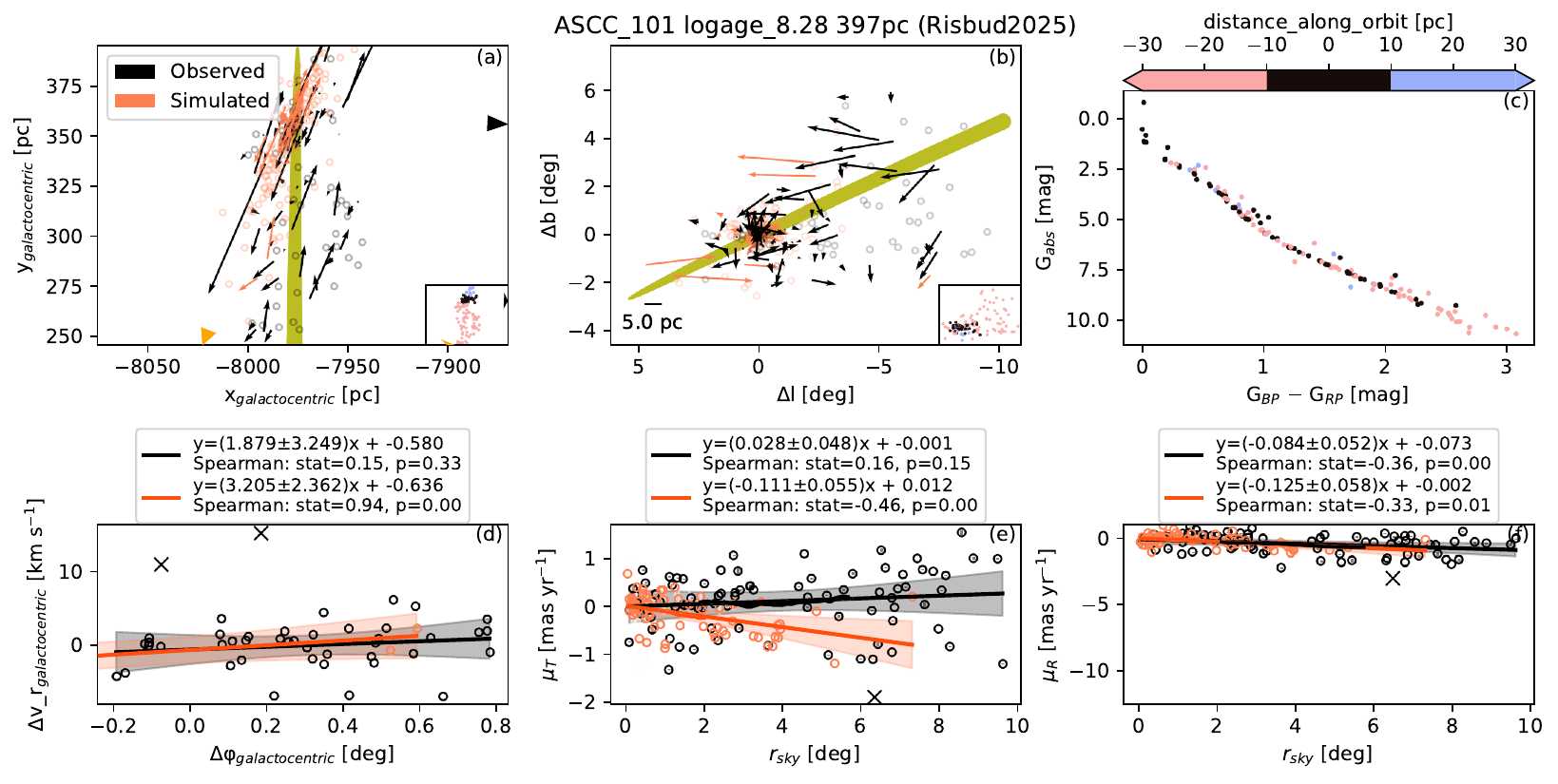}
\includegraphics[width=0.5\linewidth]{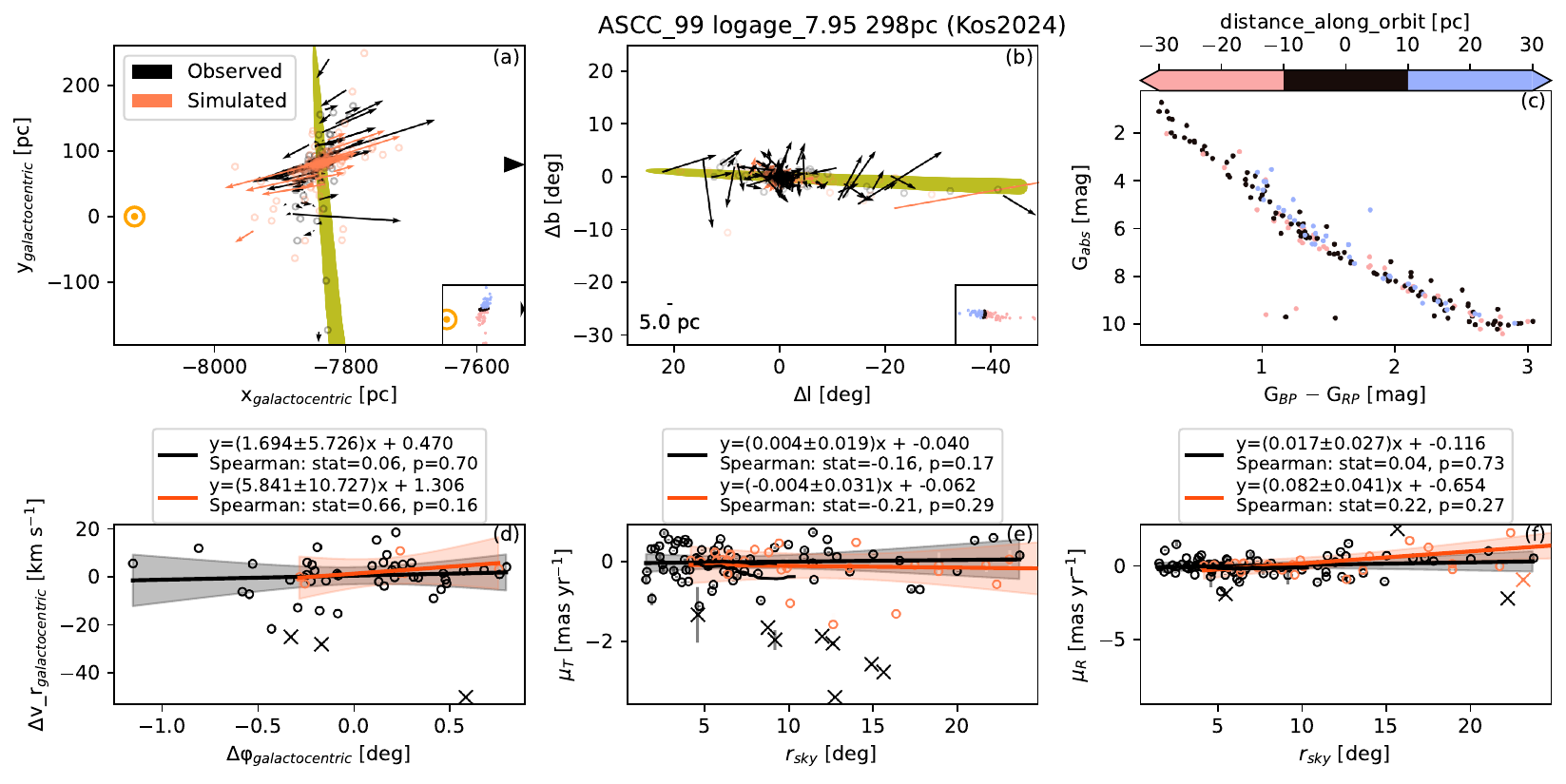}
    \caption{Diagnostic figures for ASCC 101 (Bhattacharya2022), ASCC 101 (Kos2024), ASCC 101 (Risbud2025), ASCC 99 (Kos2024).}
        \label{fig:supplementary.ASCC_99.Kos2024}
        \end{figure}
         
        \begin{figure}
\includegraphics[width=0.5\linewidth]{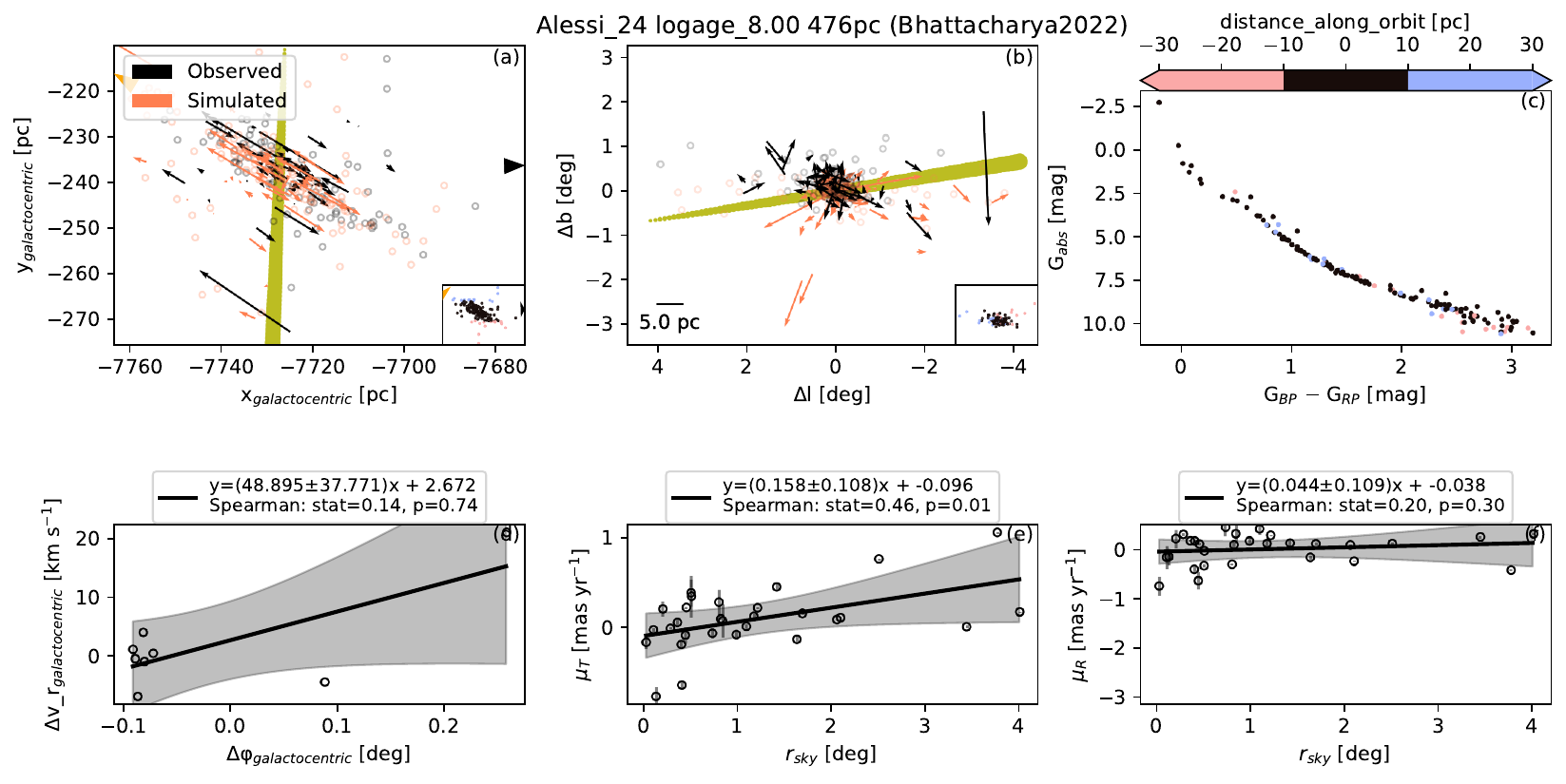}
\includegraphics[width=0.5\linewidth]{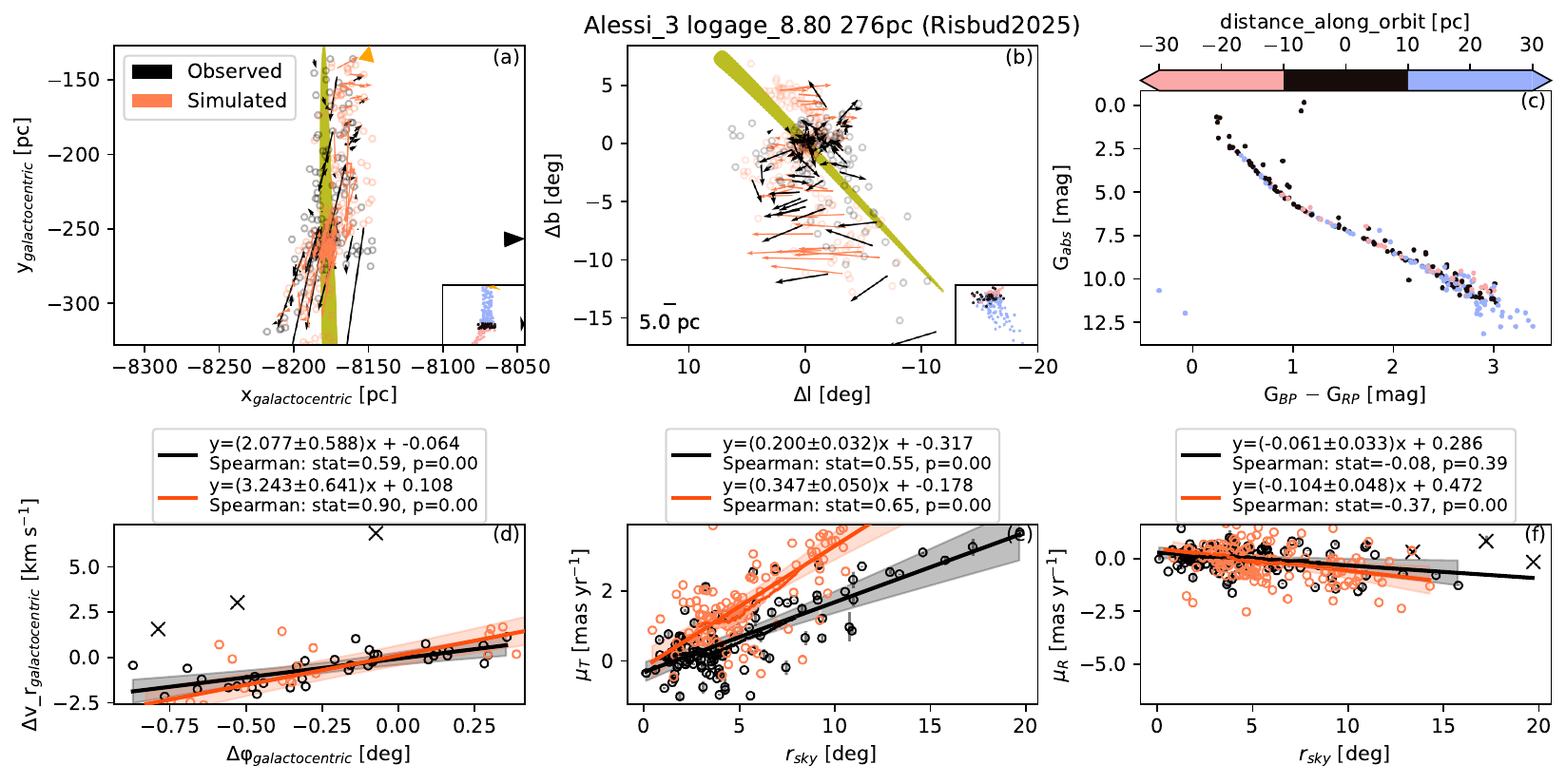}
\includegraphics[width=0.5\linewidth]{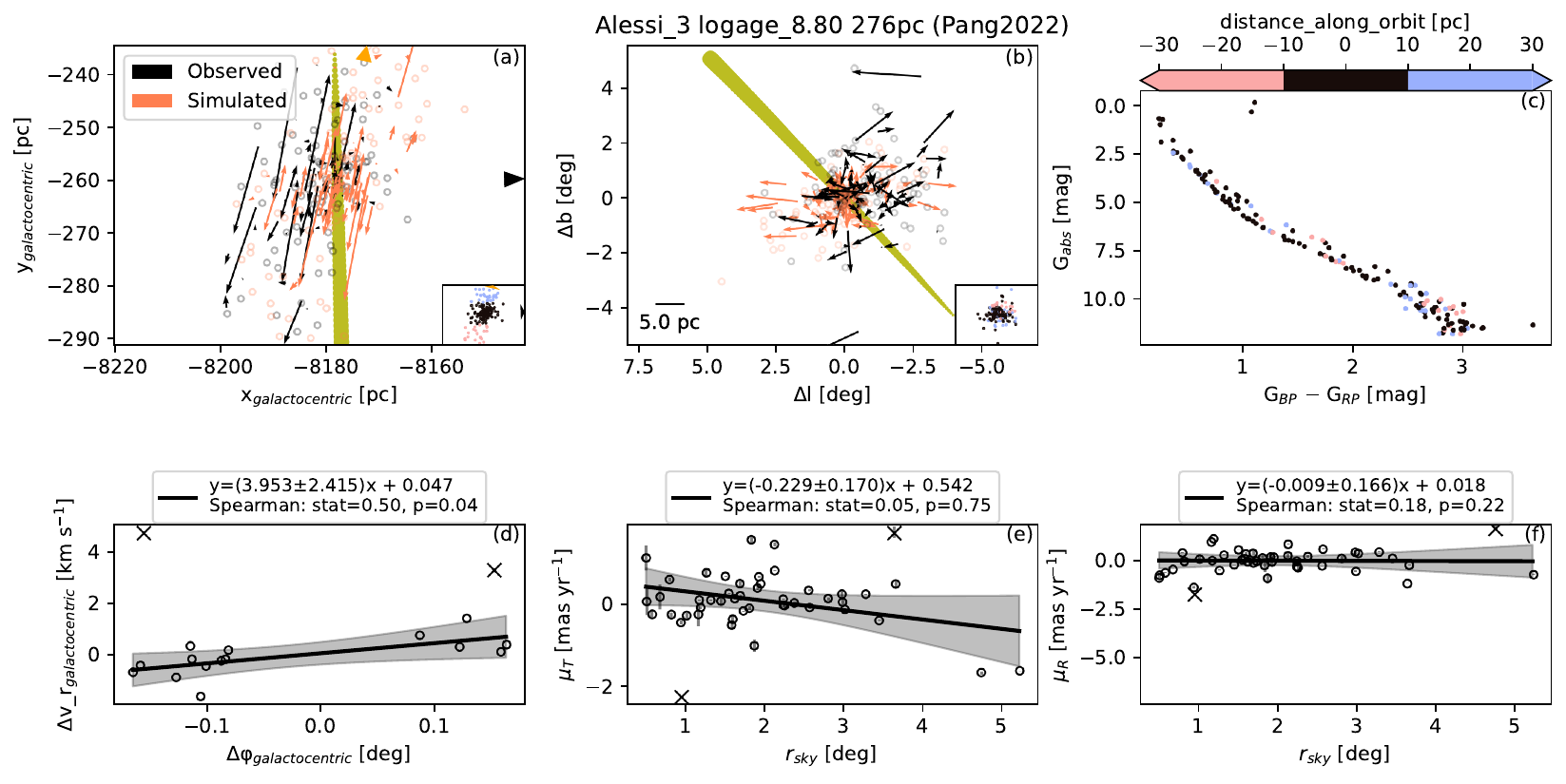}
\includegraphics[width=0.5\linewidth]{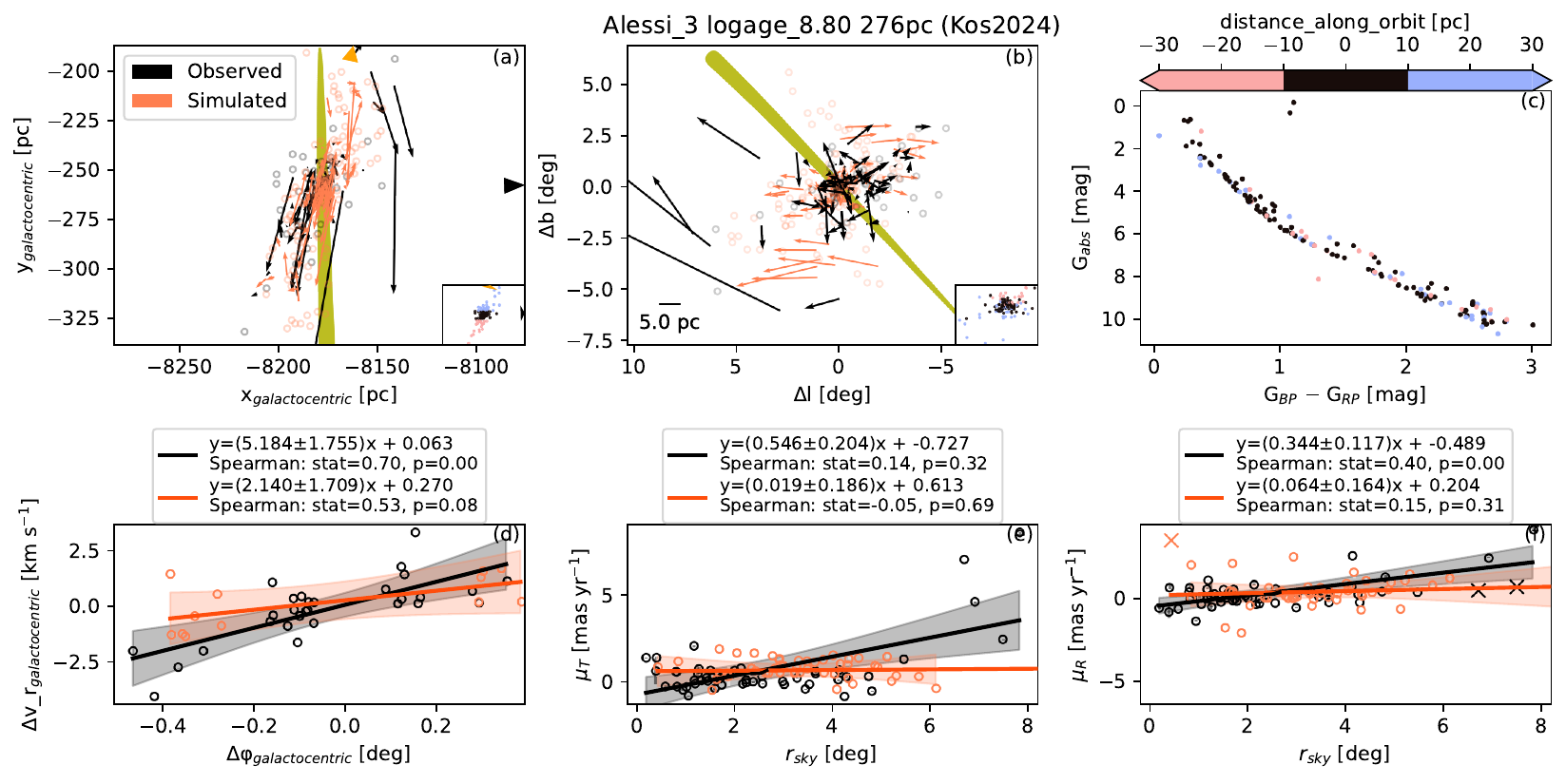}
    \caption{Diagnostic figures for Alessi 24 (Bhattacharya2022), Alessi 3 (Risbud2025), Alessi 3 (Pang2022), Alessi 3 (Kos2024).}
        \label{fig:supplementary.Alessi_3.Kos2024}
        \end{figure}
         
        \begin{figure}
\includegraphics[width=0.5\linewidth]{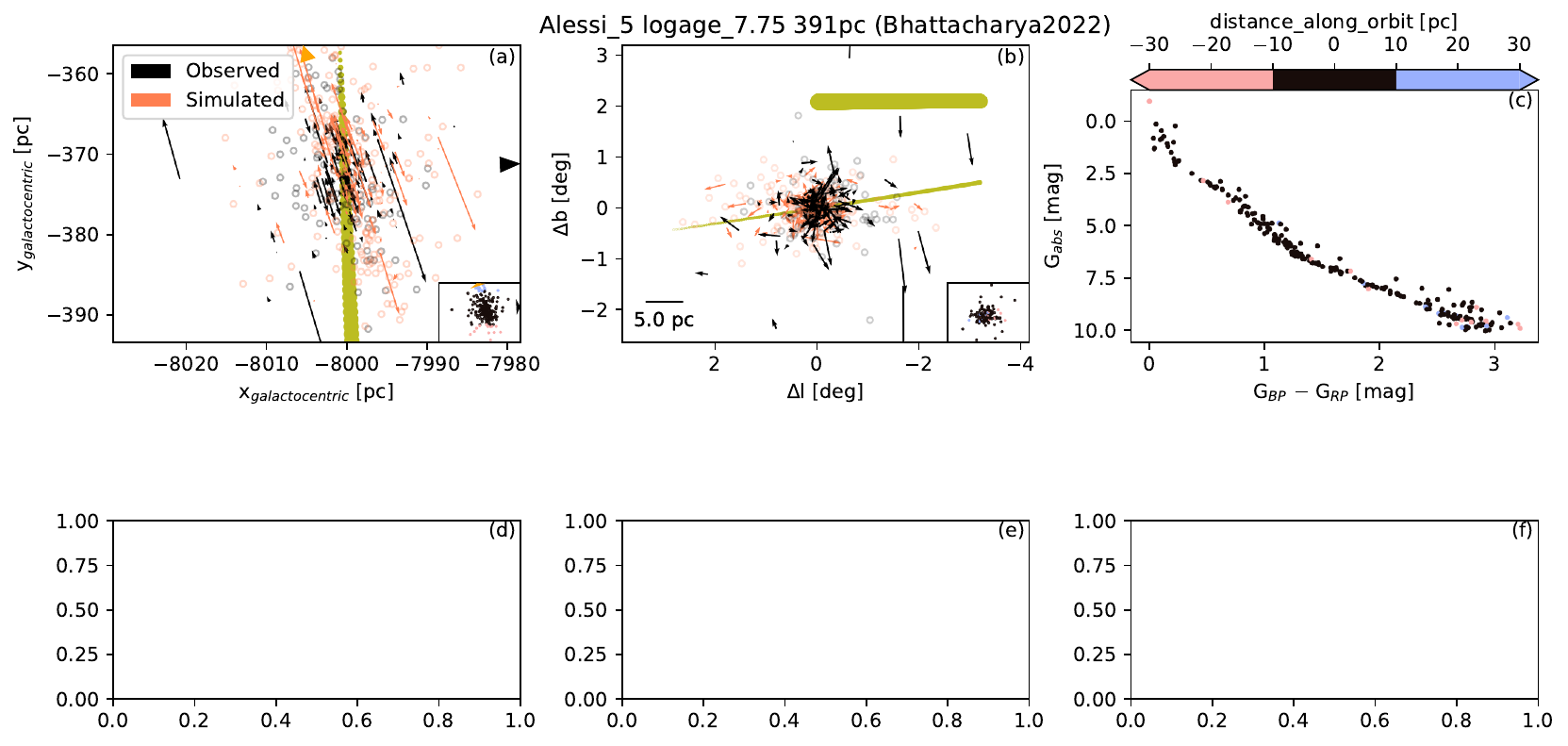}
\includegraphics[width=0.5\linewidth]{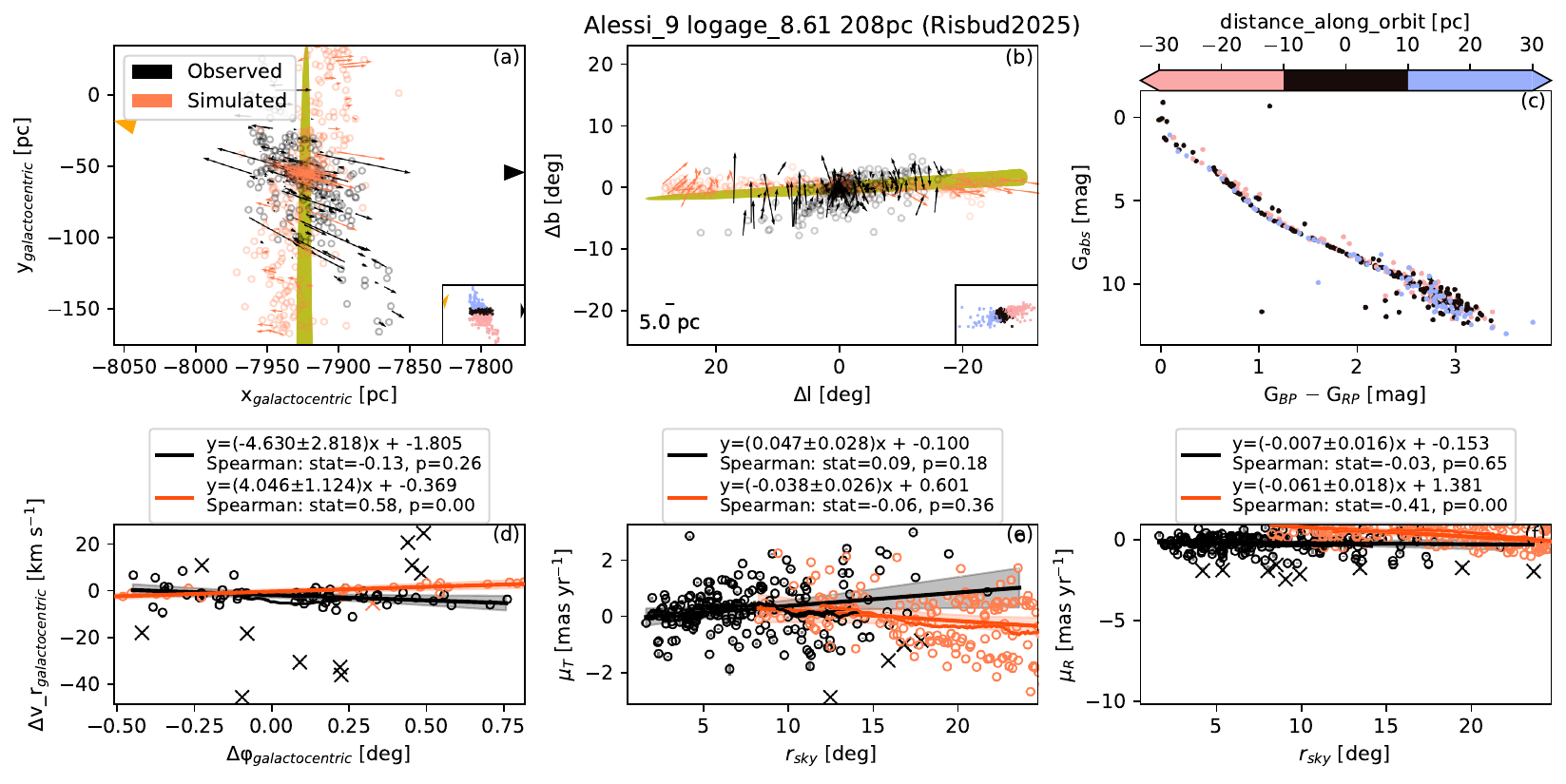}
\includegraphics[width=0.5\linewidth]{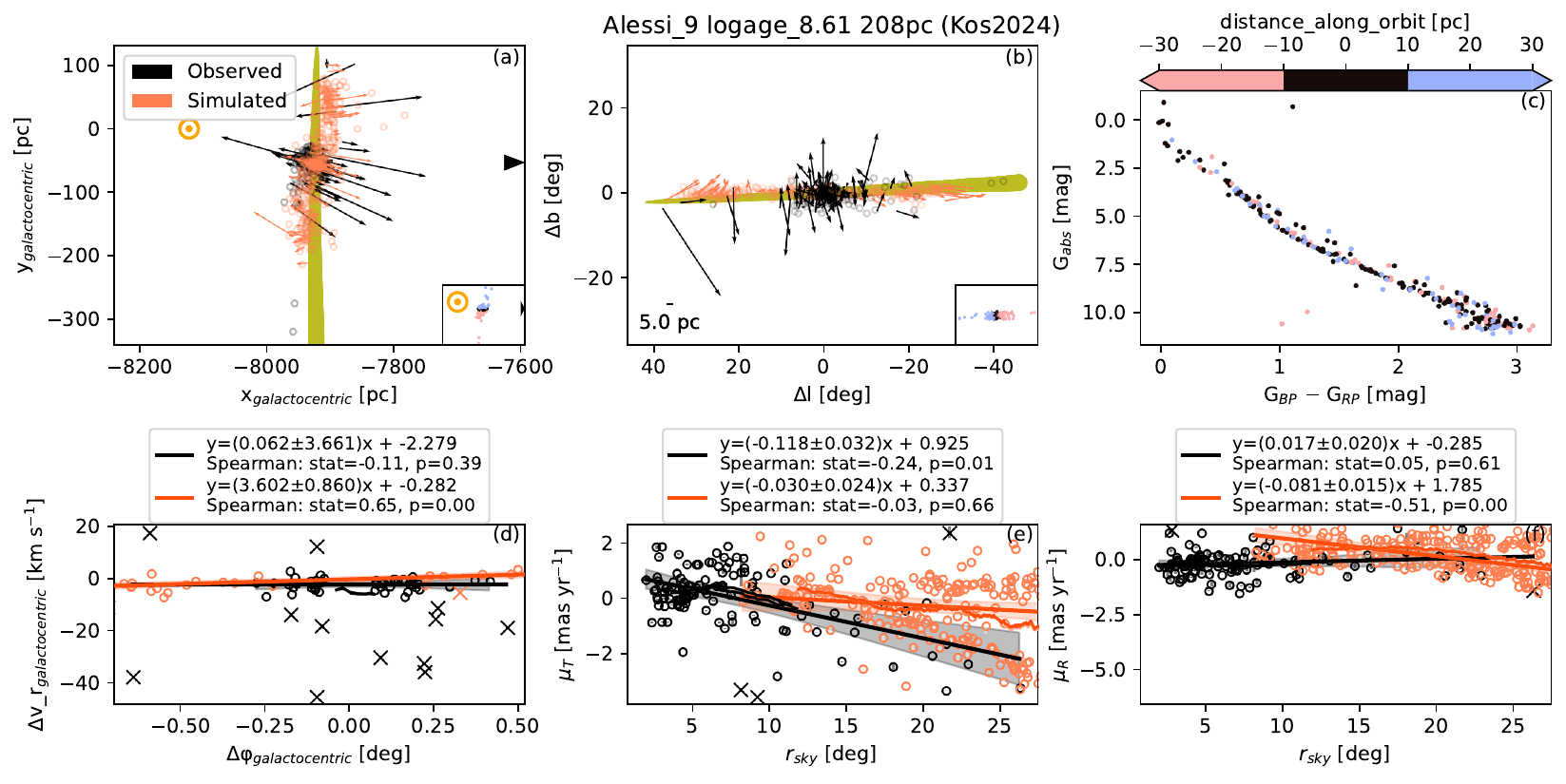}
\includegraphics[width=0.5\linewidth]{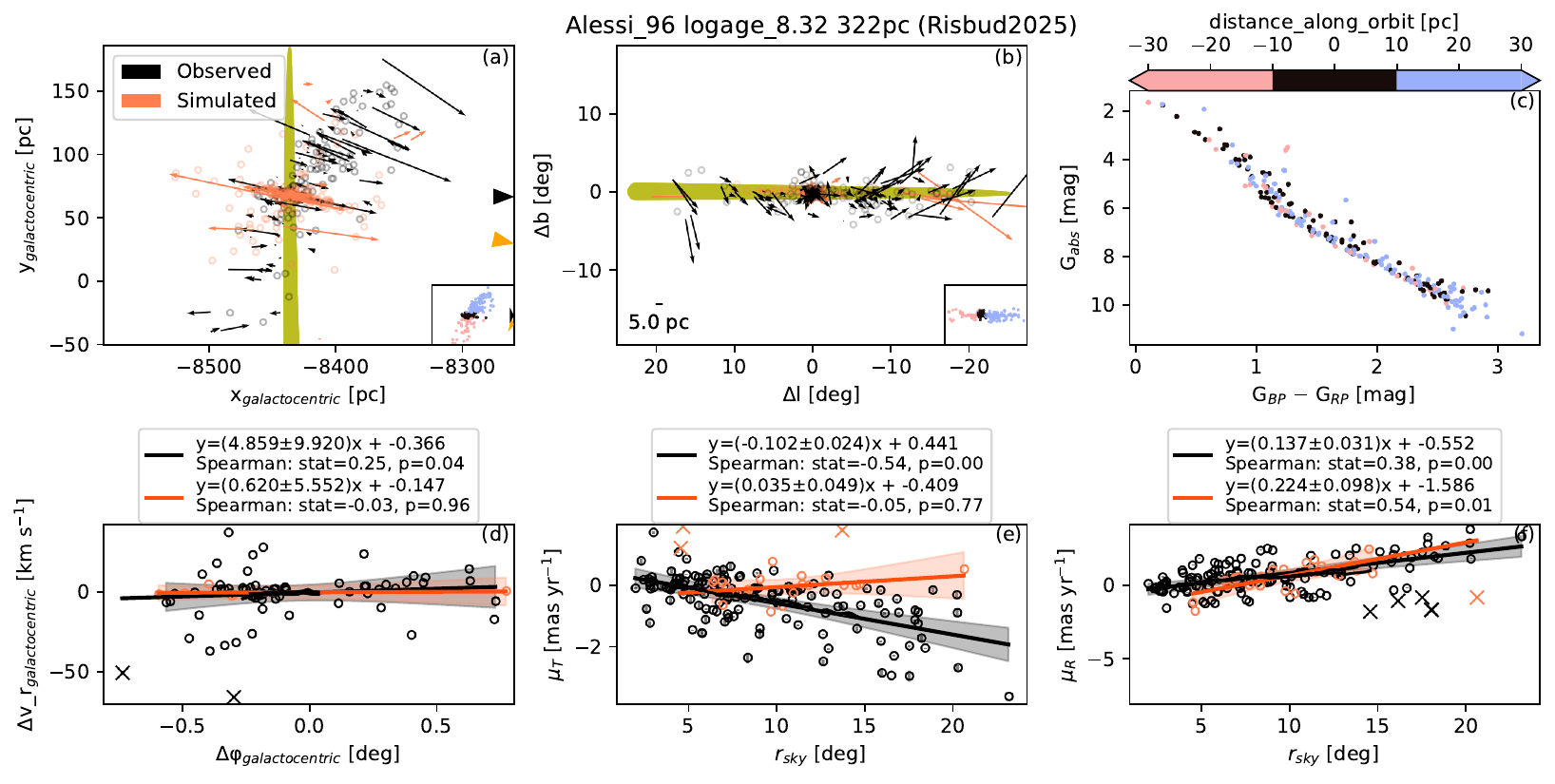}
    \caption{Diagnostic figures for Alessi 5 (Bhattacharya2022), Alessi 9 (Risbud2025), Alessi 9 (Kos2024), Alessi 96 (Risbud2025).}
        \label{fig:supplementary.Alessi_96.Risbud2025}
        \end{figure}
         
        \begin{figure}
\includegraphics[width=0.5\linewidth]{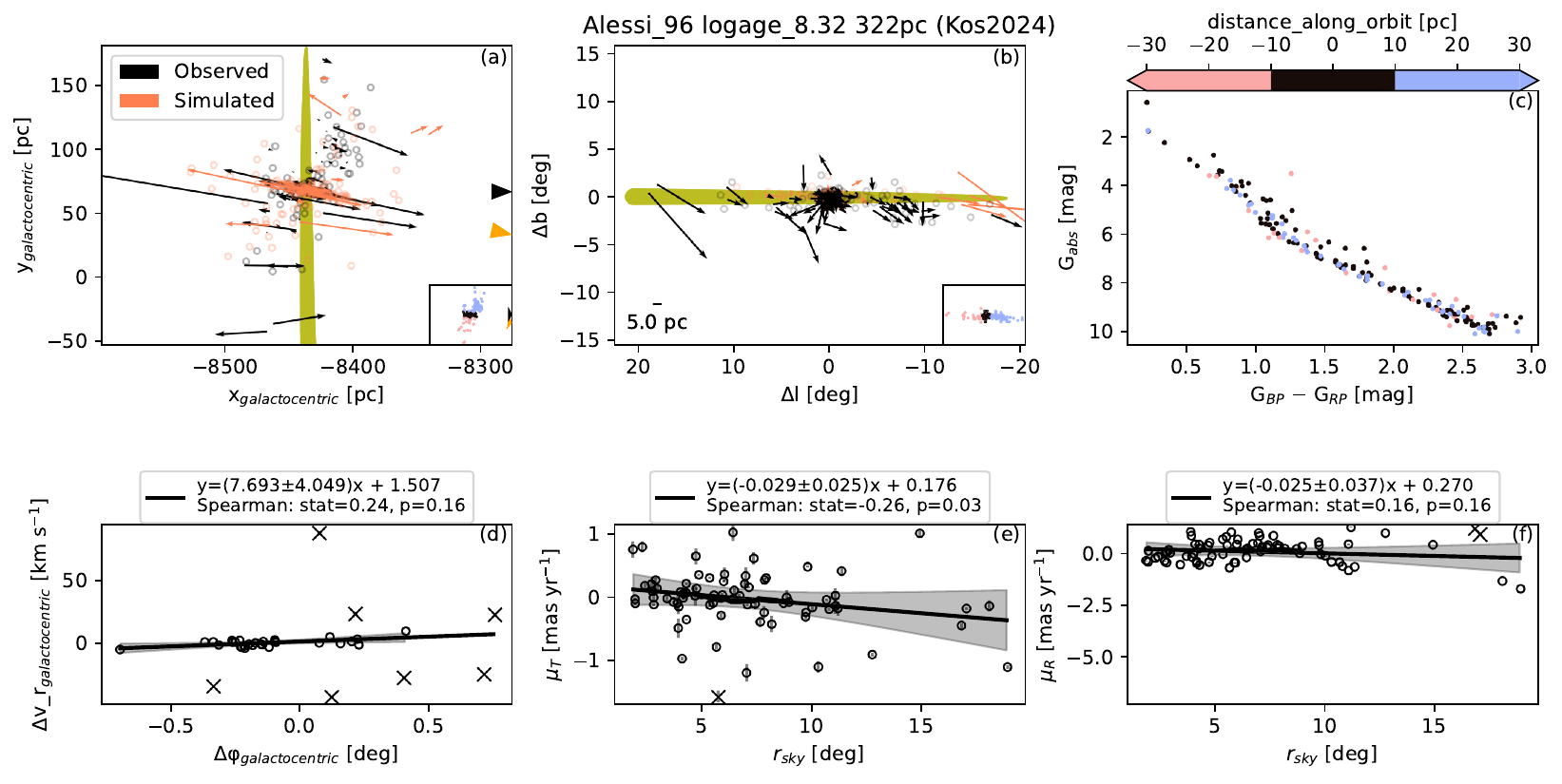}
\includegraphics[width=0.5\linewidth]{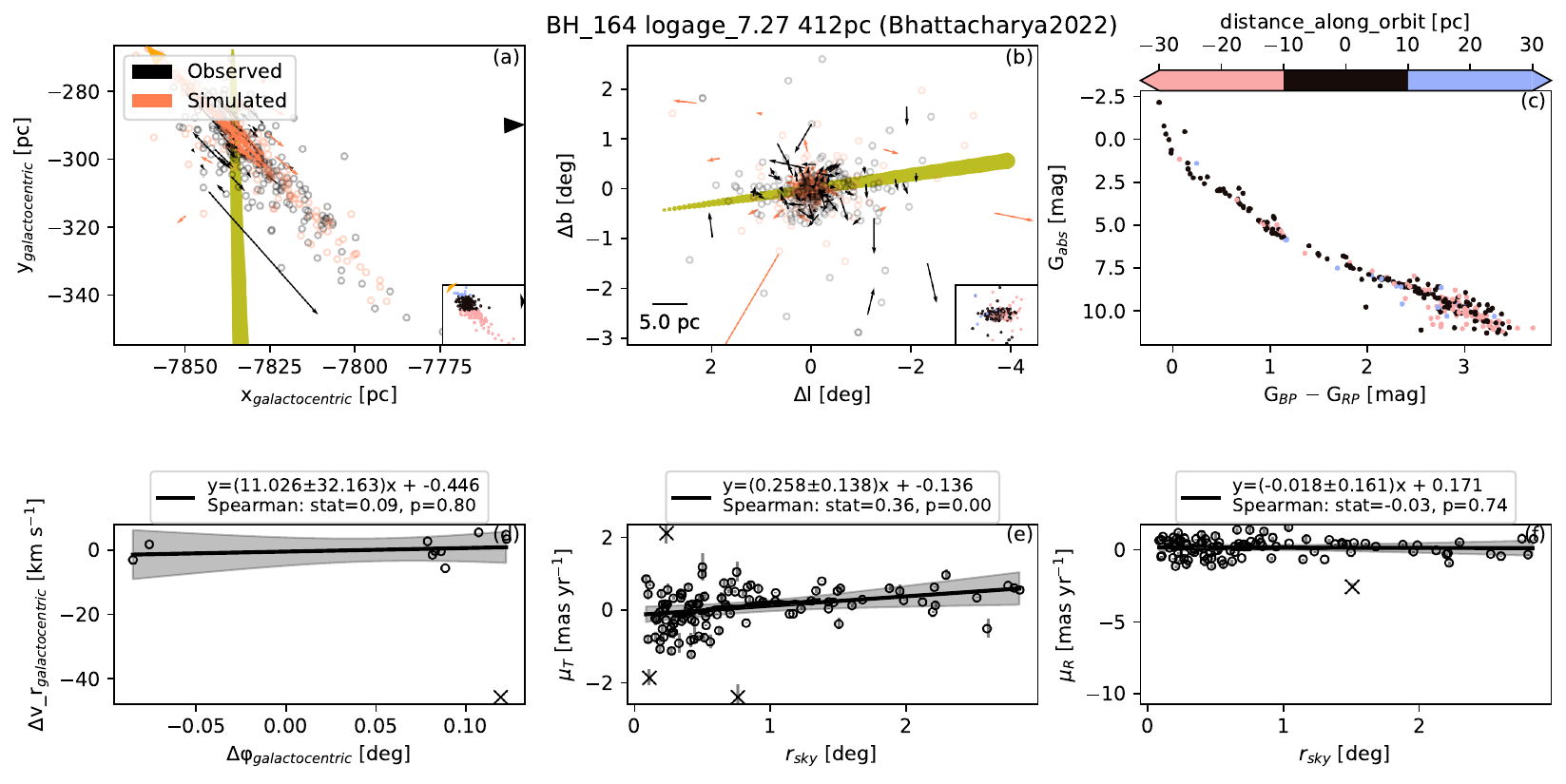}
\includegraphics[width=0.5\linewidth]{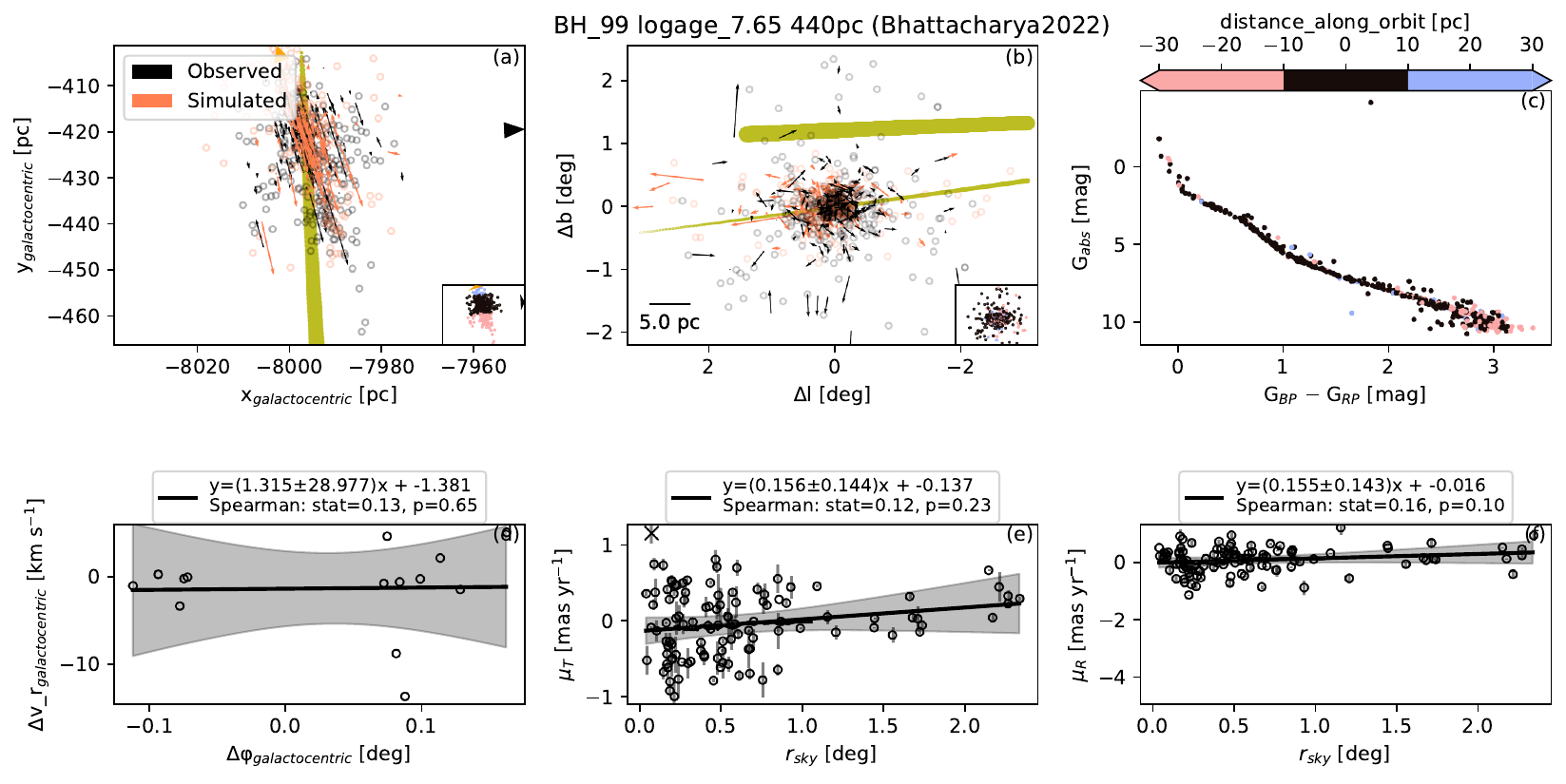}
\includegraphics[width=0.5\linewidth]{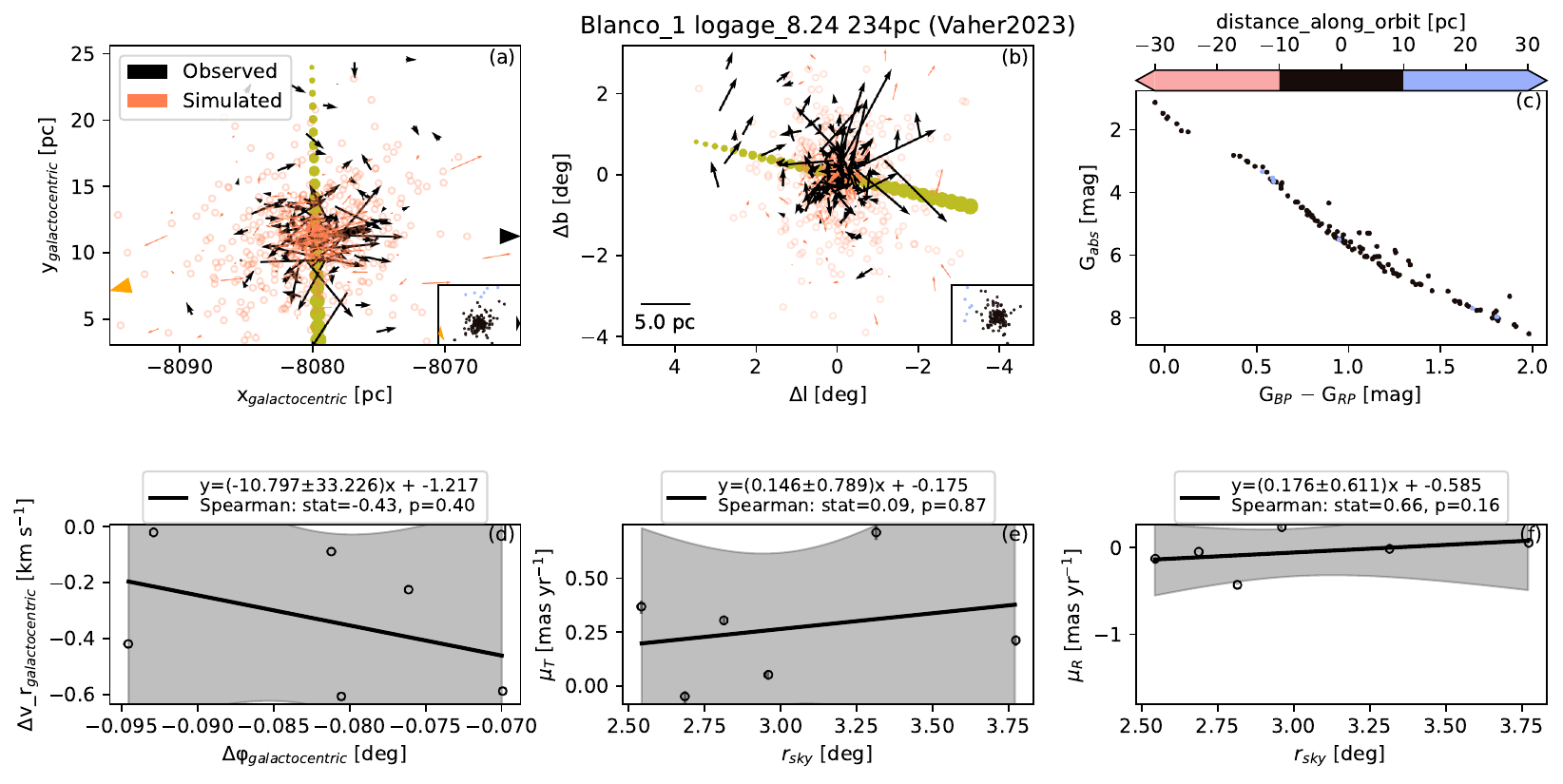}
    \caption{Diagnostic figures for Alessi 96 (Kos2024), BH 164 (Bhattacharya2022), BH 99 (Bhattacharya2022), Blanco 1 (Vaher2023).}
        \label{fig:supplementary.Blanco_1.Vaher2023}
        \end{figure}
         
        \begin{figure}
\includegraphics[width=0.5\linewidth]{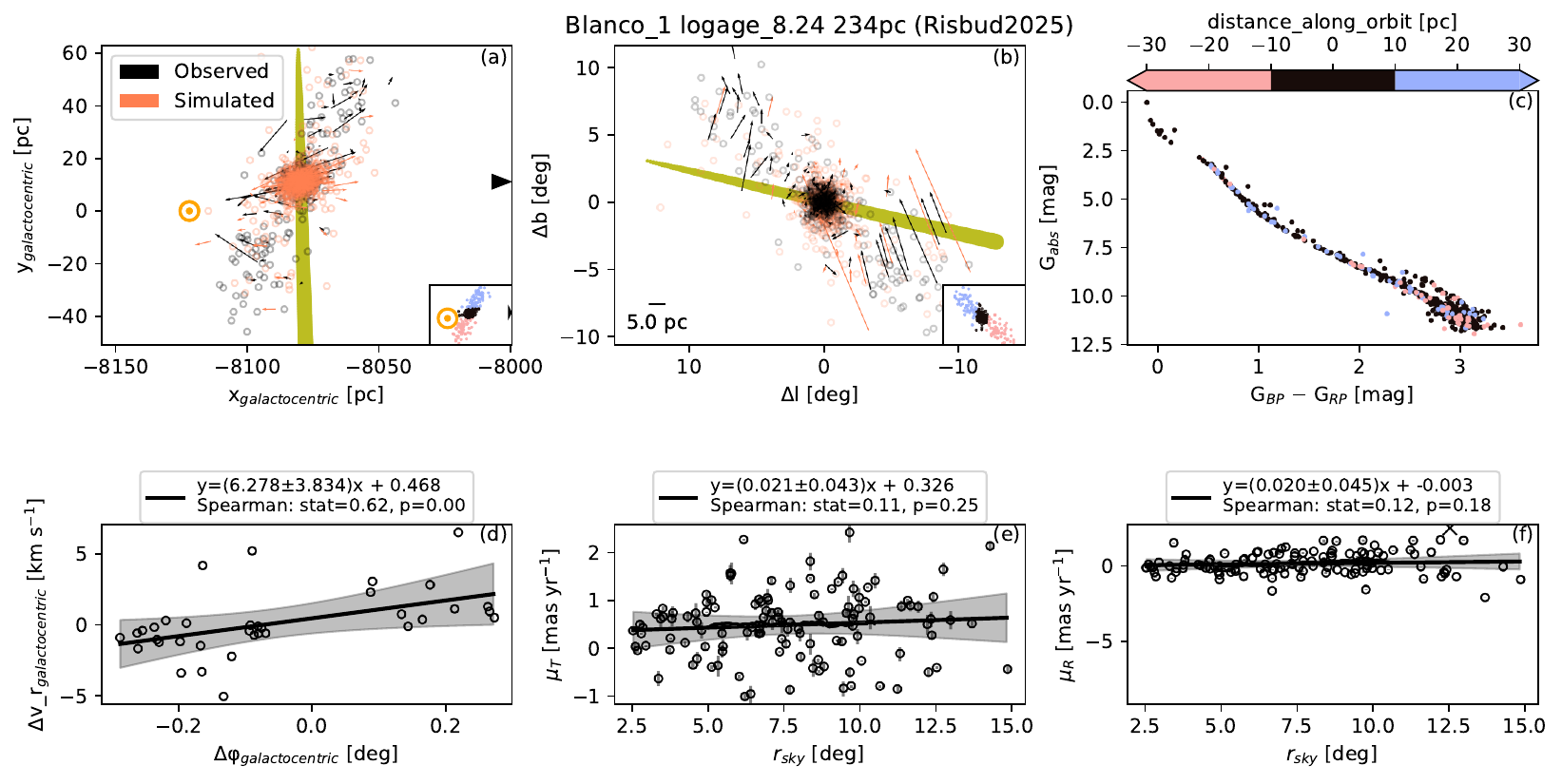}
\includegraphics[width=0.5\linewidth]{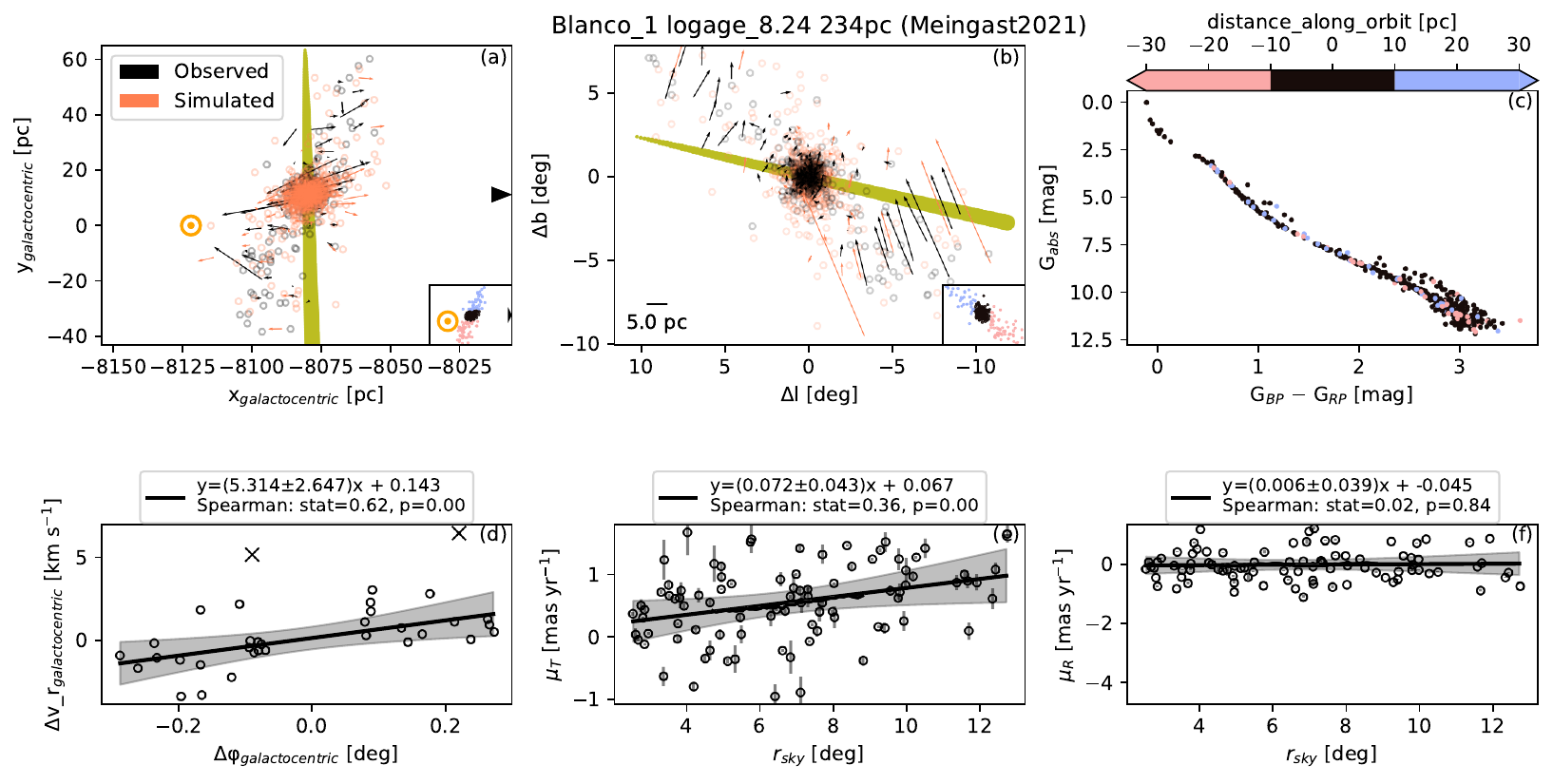}
\includegraphics[width=0.5\linewidth]{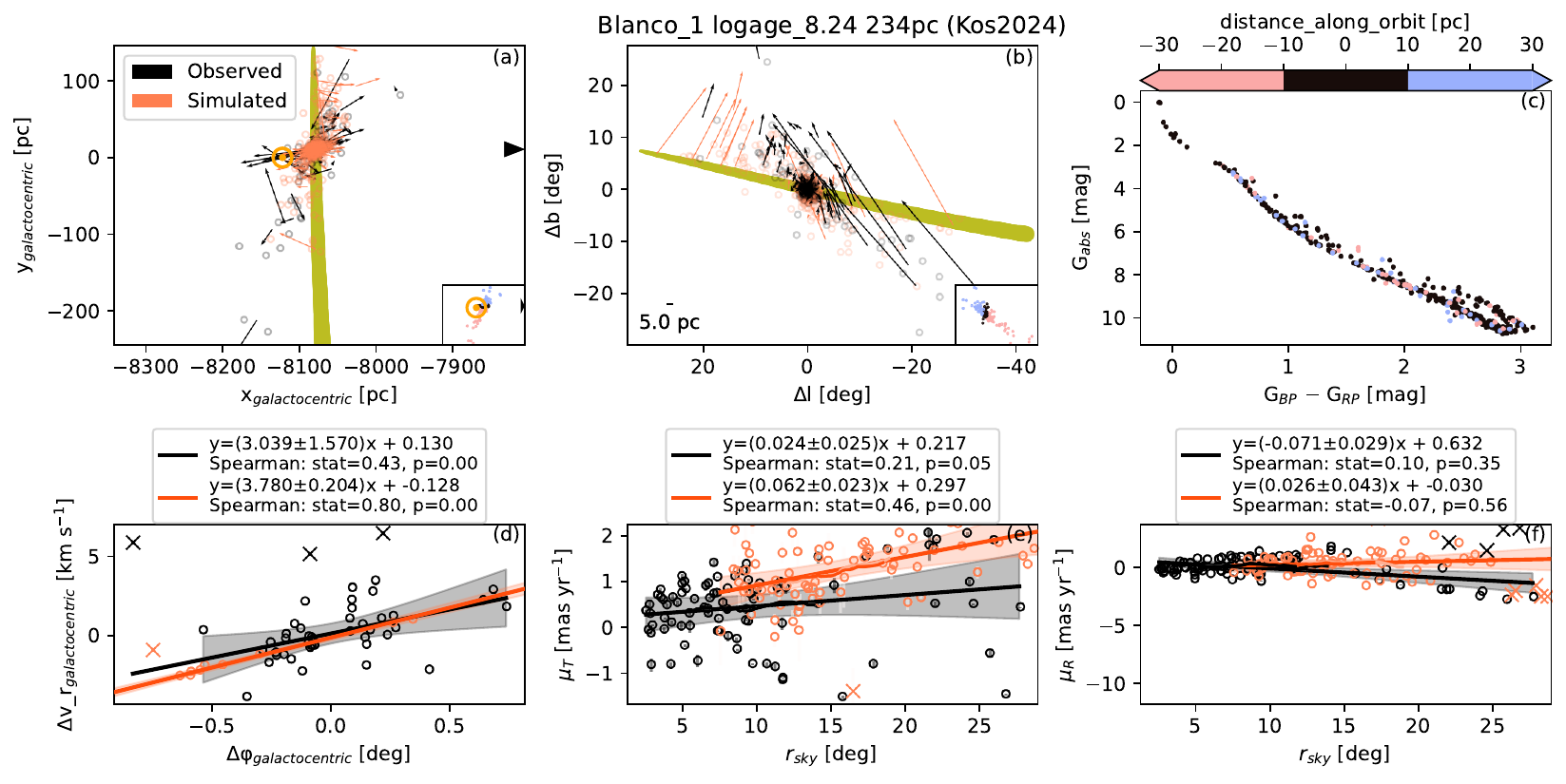}
\includegraphics[width=0.5\linewidth]{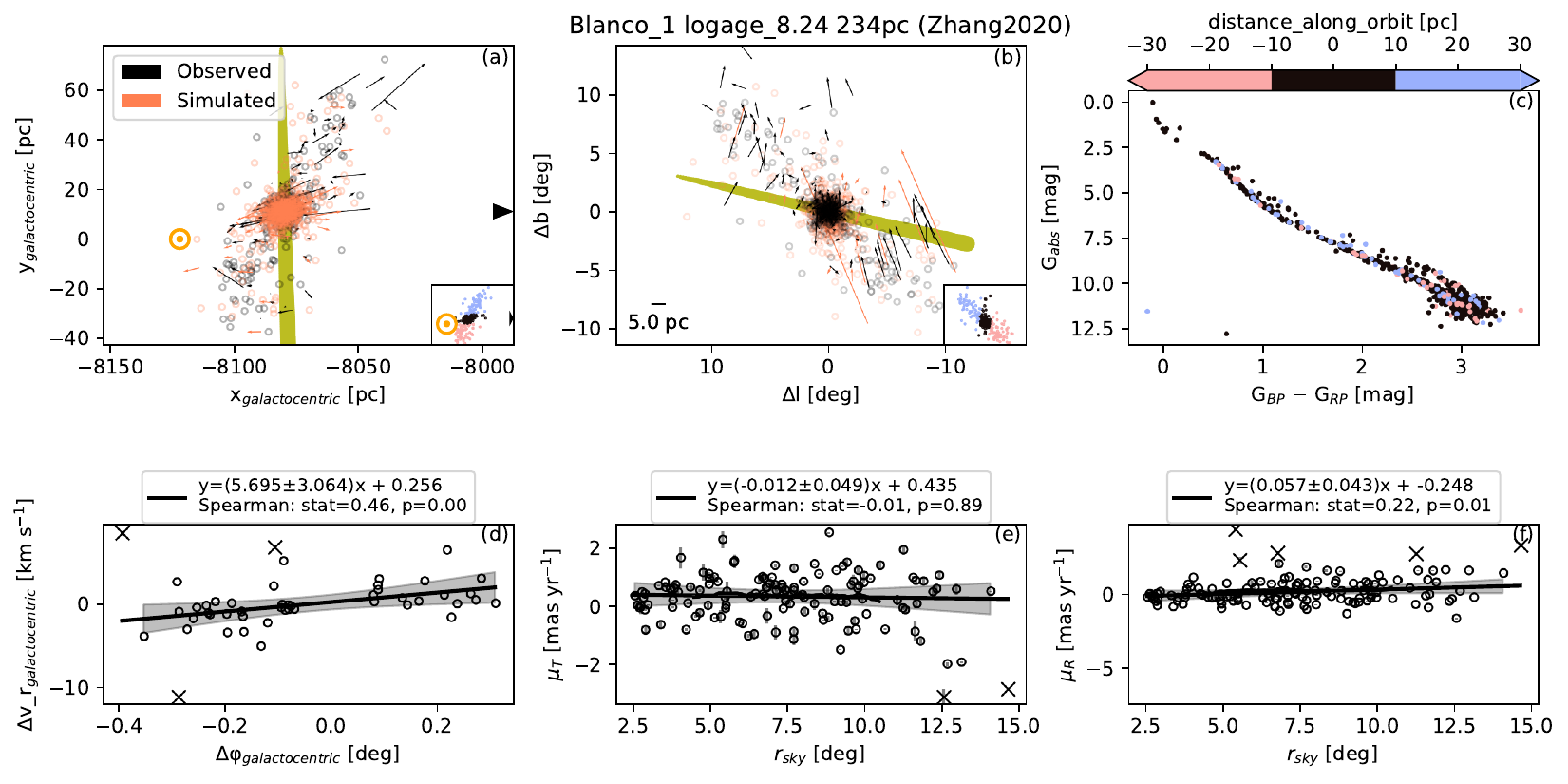}
    \caption{Diagnostic figures for Blanco 1 (Risbud2025), Blanco 1 (Meingast2021), Blanco 1 (Kos2024), Blanco 1 (Zhang2020).}
        \label{fig:supplementary.Blanco_1.Zhang2020}
        \end{figure}
         
        \begin{figure}
\includegraphics[width=0.5\linewidth]{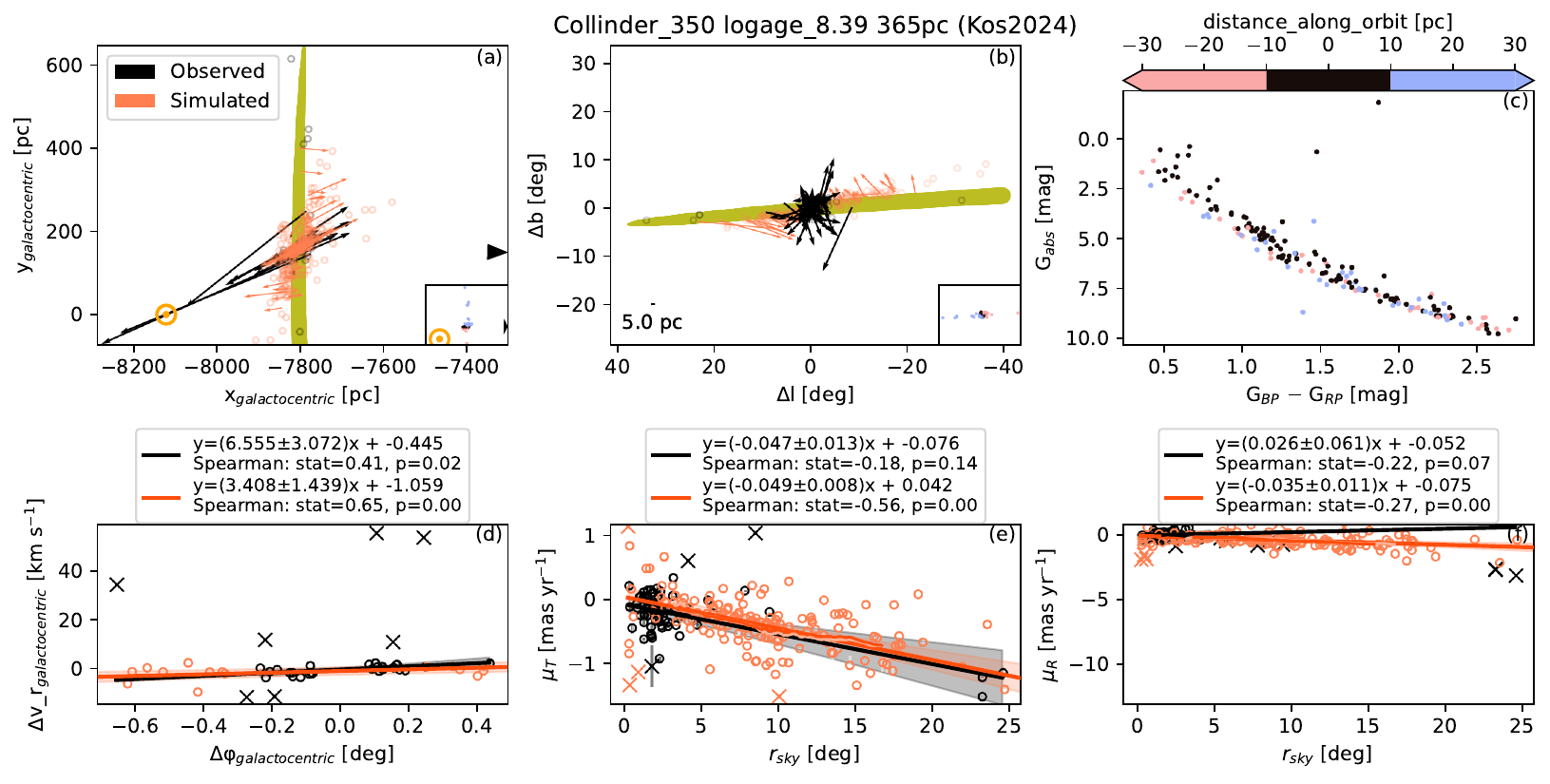}
\includegraphics[width=0.5\linewidth]{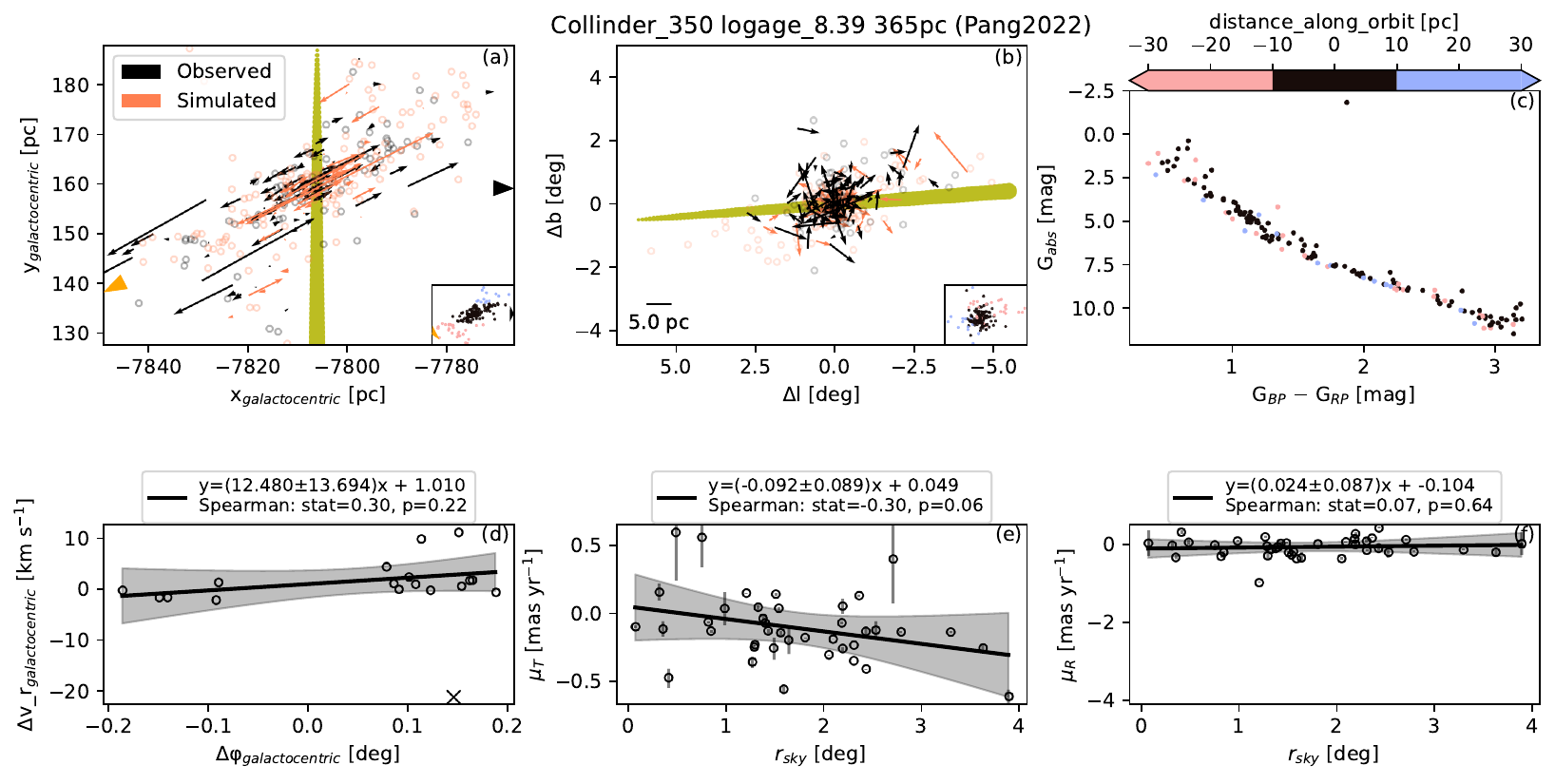}
\includegraphics[width=0.5\linewidth]{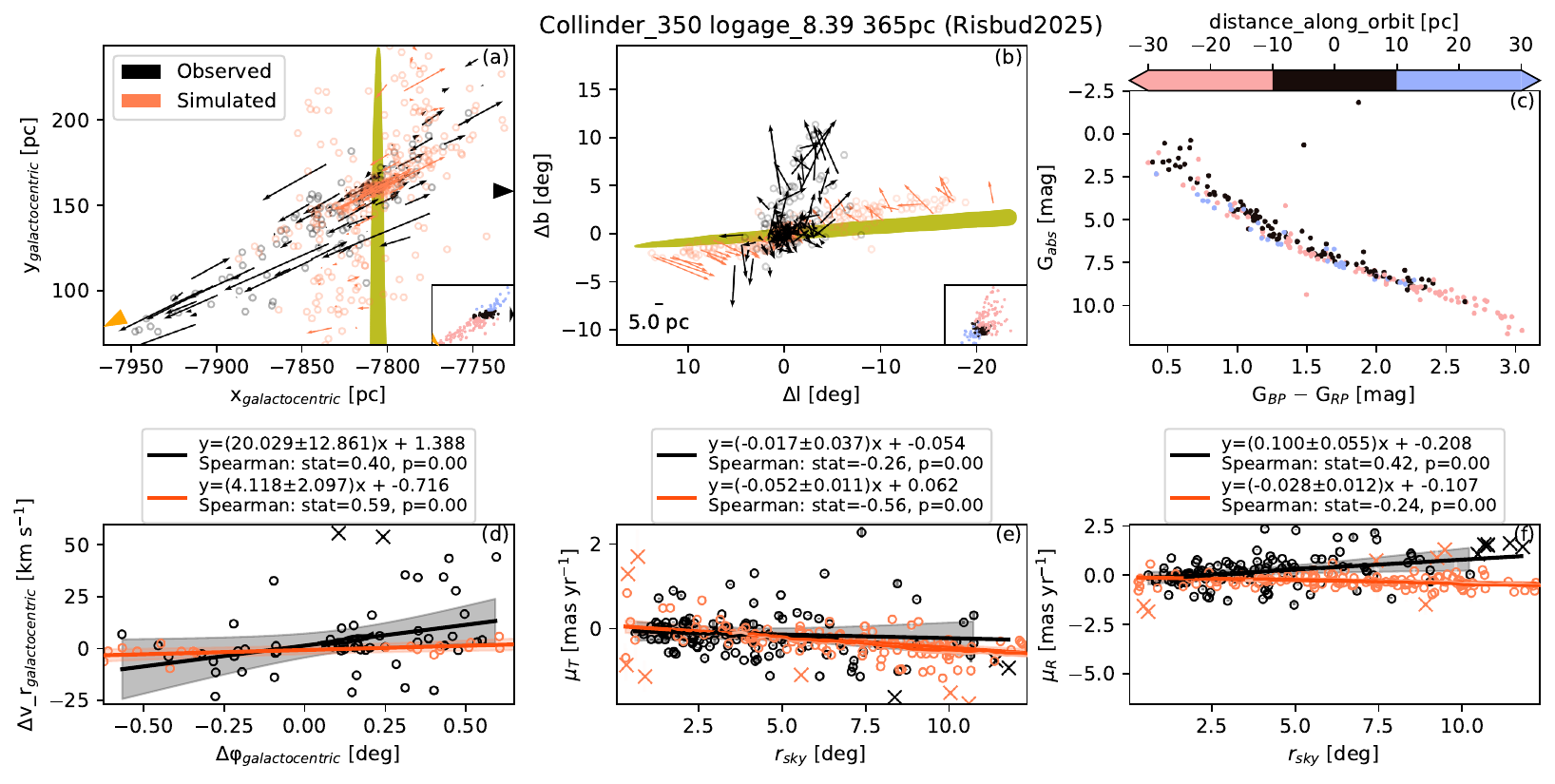}
\includegraphics[width=0.5\linewidth]{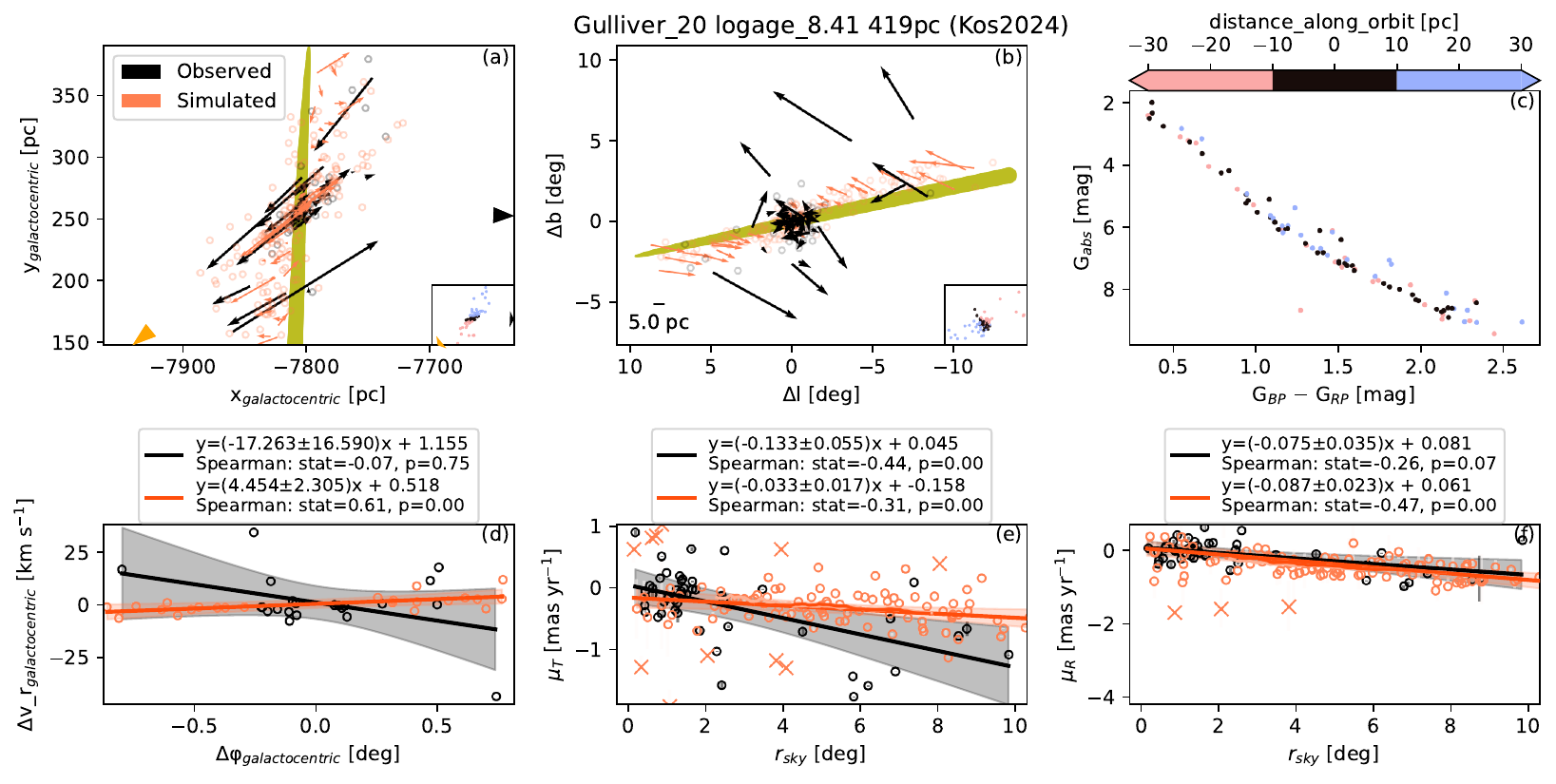}
    \caption{Diagnostic figures for Collinder 350 (Kos2024), Collinder 350 (Pang2022), Collinder 350 (Risbud2025), Gulliver 20 (Kos2024).}
        \label{fig:supplementary.Gulliver_20.Kos2024}
        \end{figure}
         
        \begin{figure}
\includegraphics[width=0.5\linewidth]{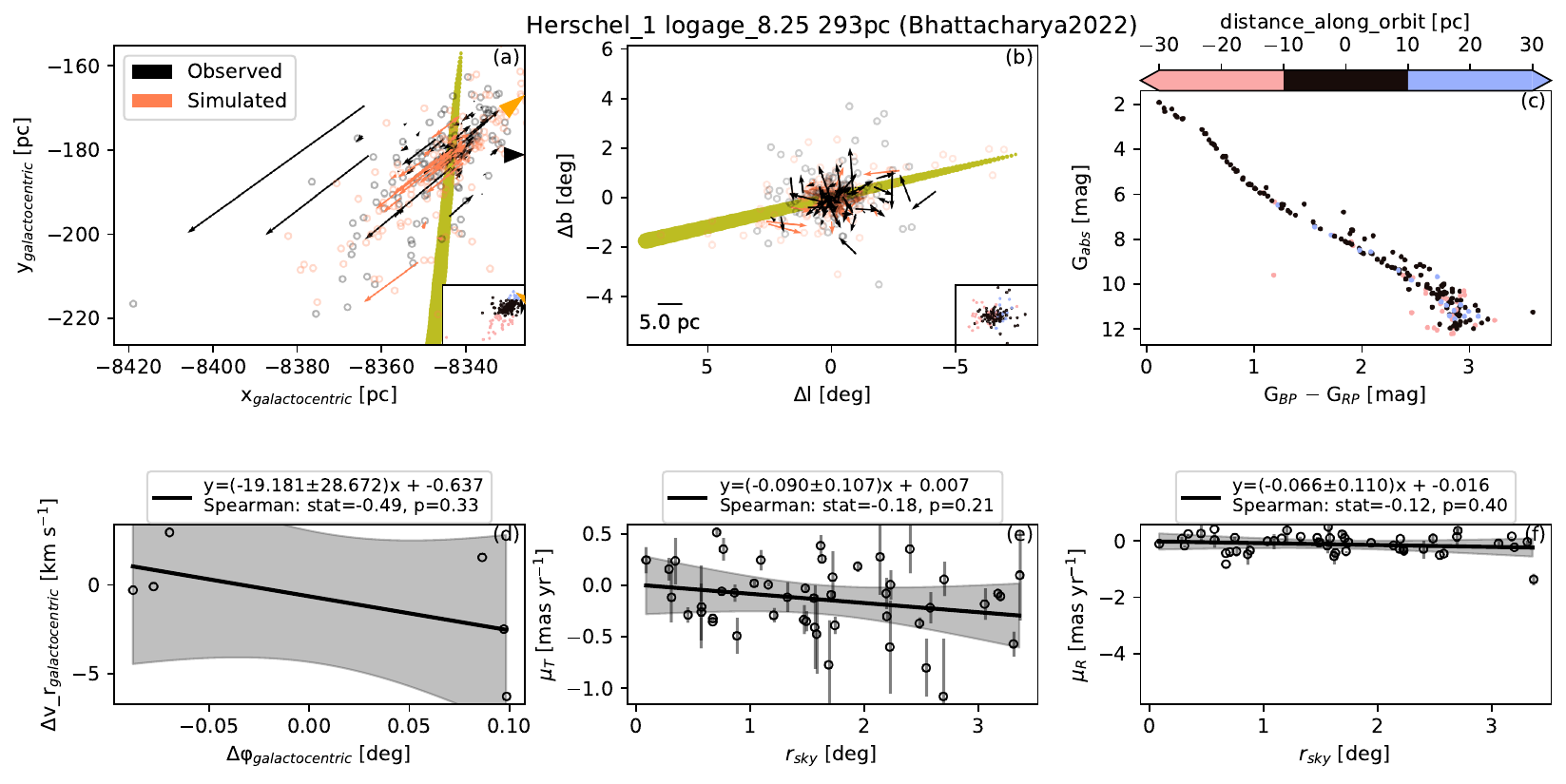}
\includegraphics[width=0.5\linewidth]{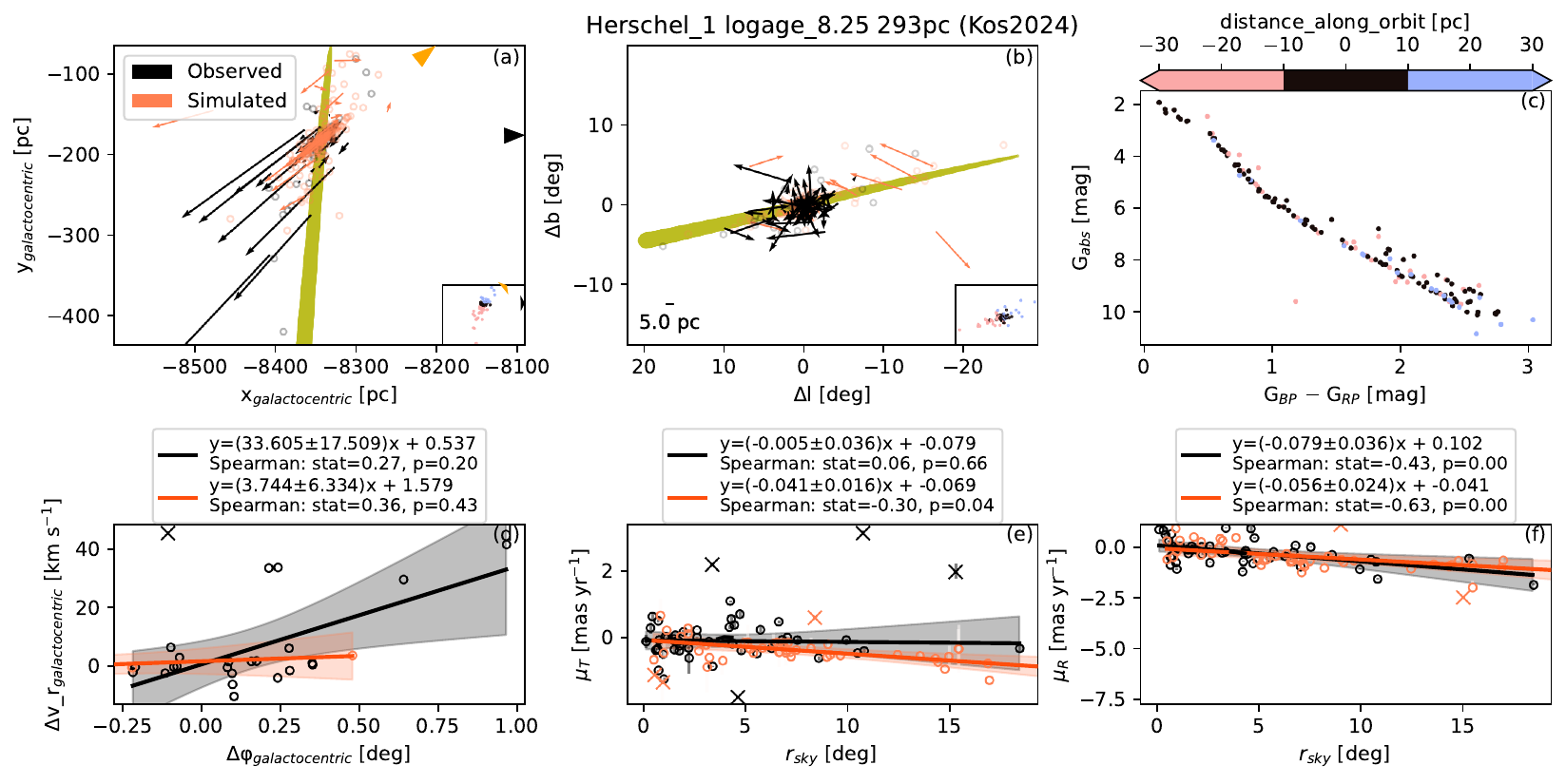}
\includegraphics[width=0.5\linewidth]{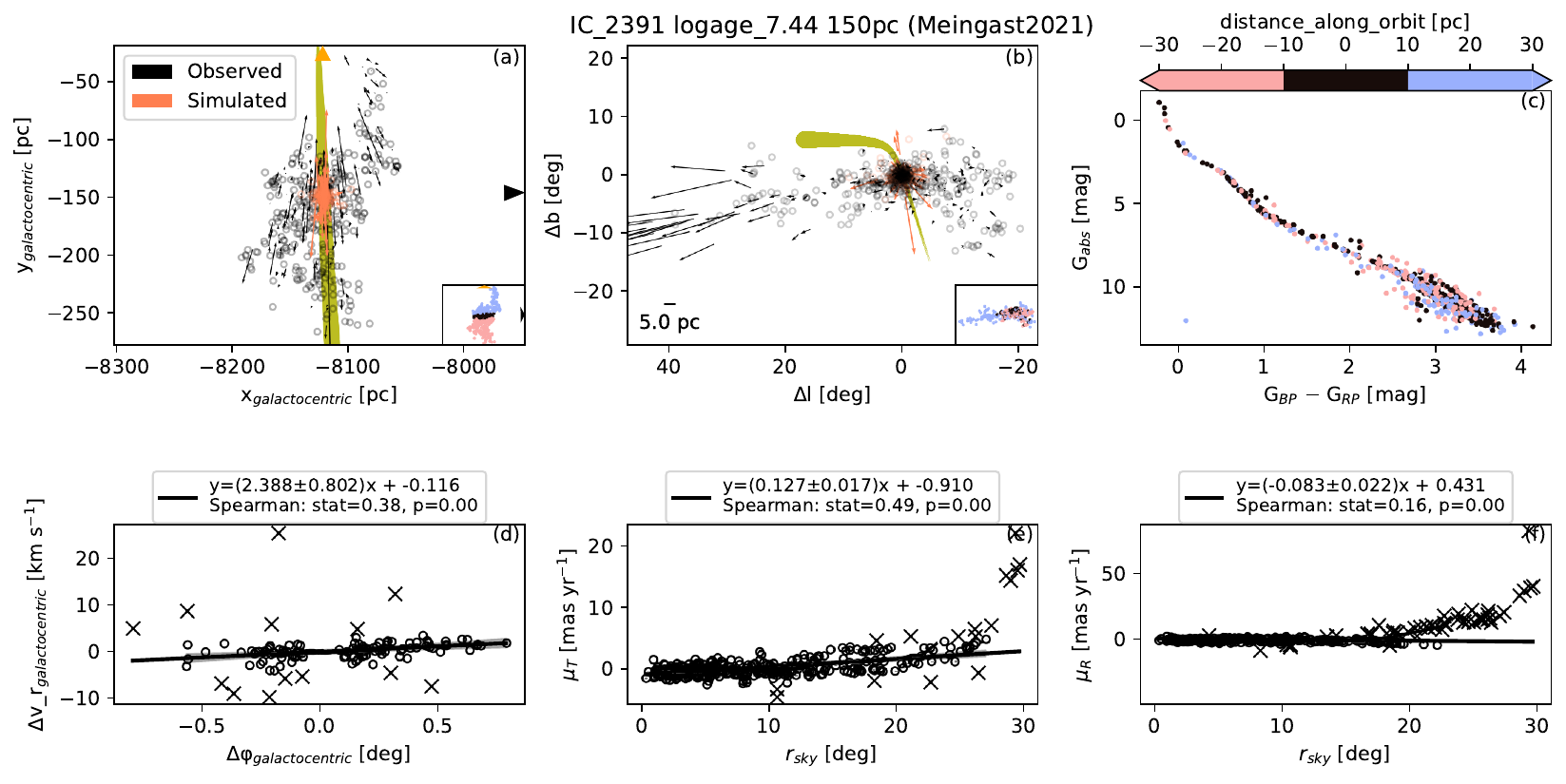}
\includegraphics[width=0.5\linewidth]{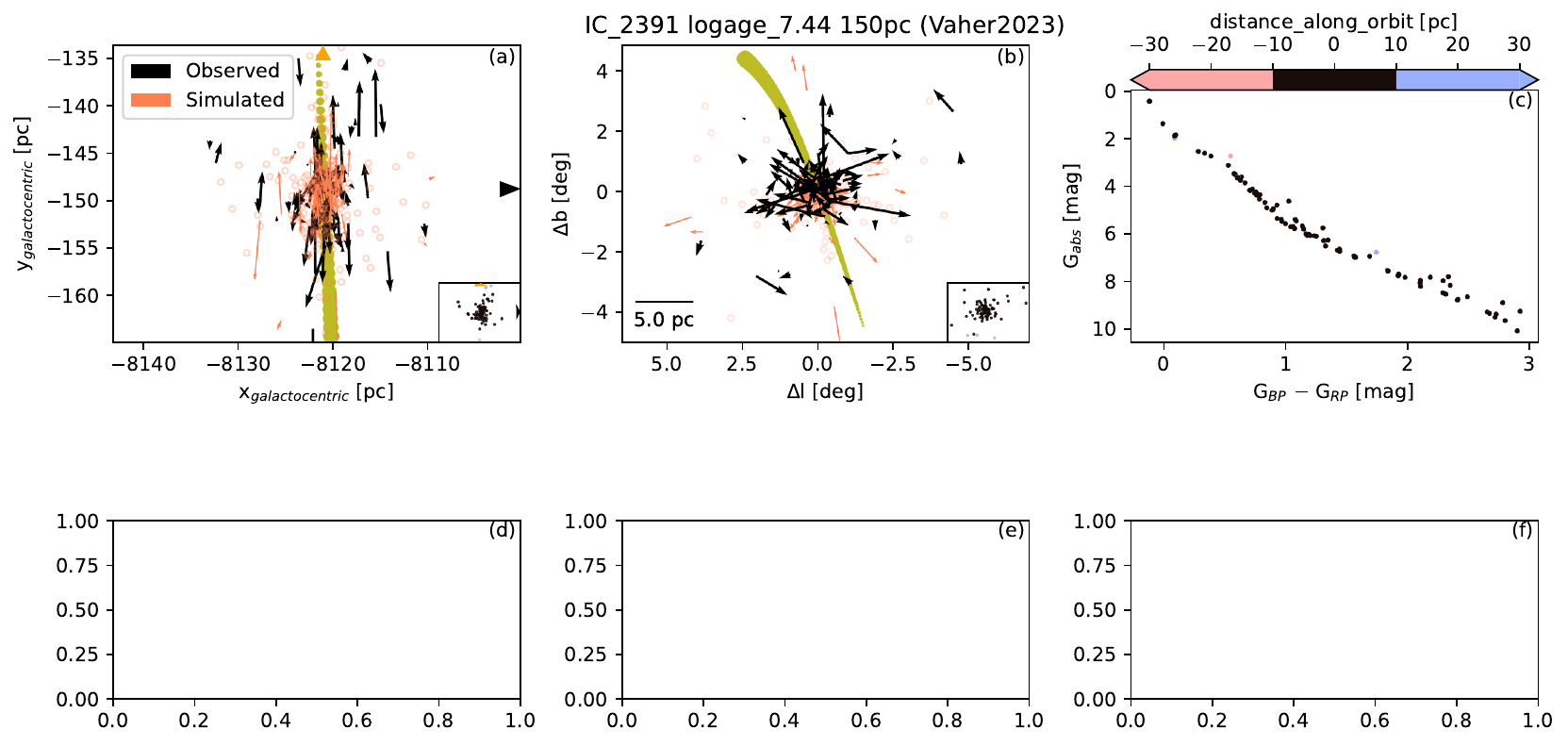}
    \caption{Diagnostic figures for Herschel 1 (Bhattacharya2022), Herschel 1 (Kos2024), IC 2391 (Meingast2021), IC 2391 (Vaher2023).}
        \label{fig:supplementary.IC_2391.Vaher2023}
        \end{figure}
         
        \begin{figure}
\includegraphics[width=0.5\linewidth]{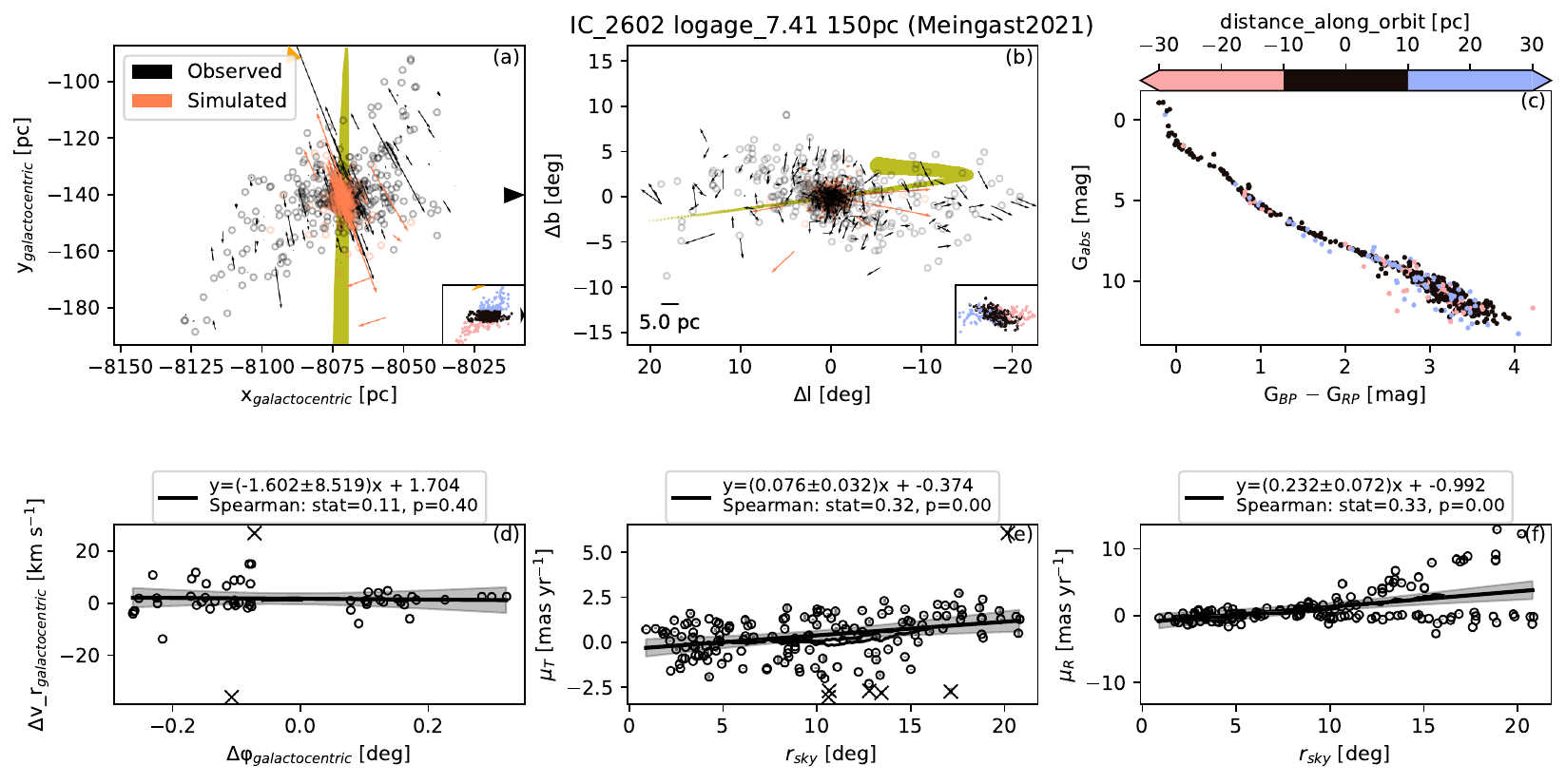}
\includegraphics[width=0.5\linewidth]{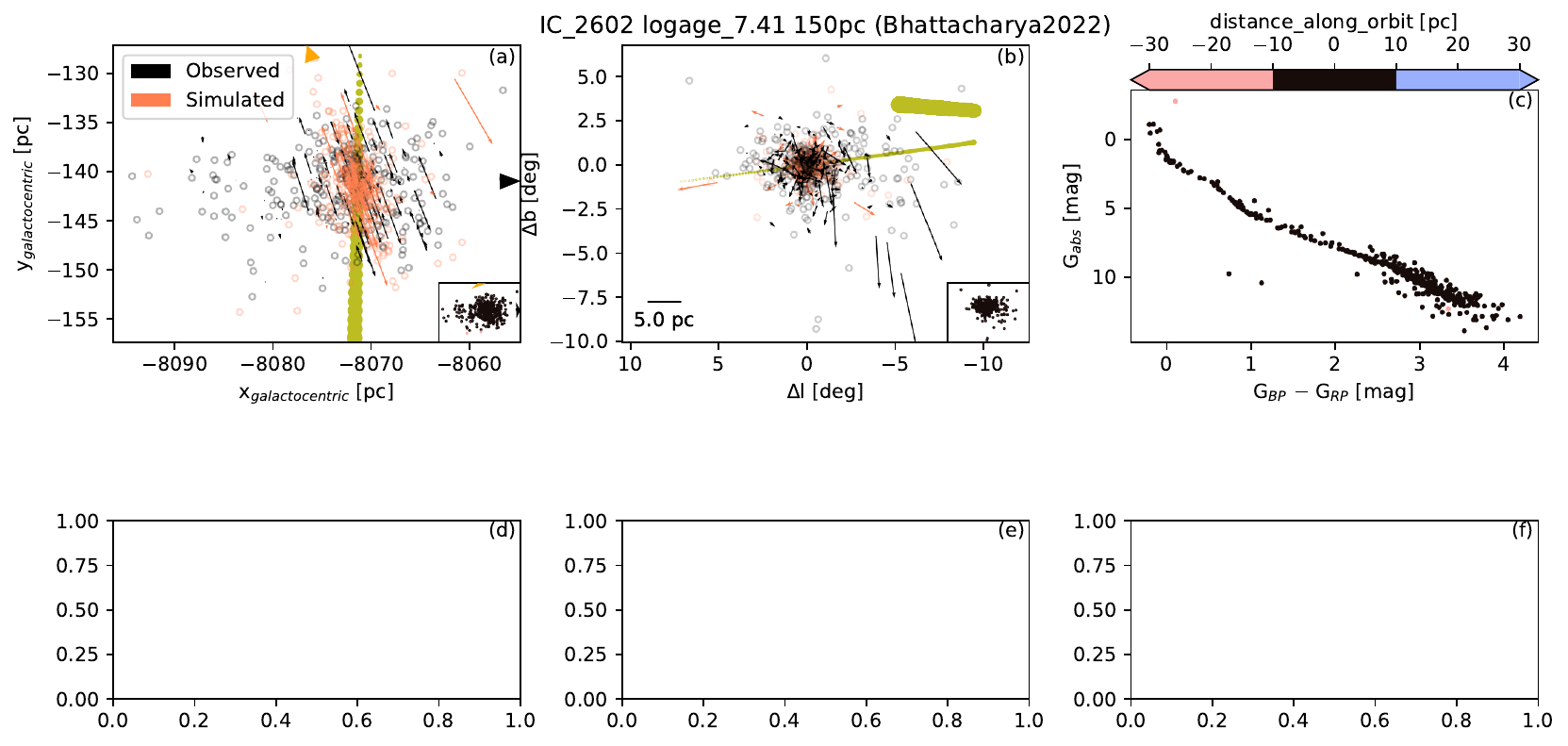}
\includegraphics[width=0.5\linewidth]{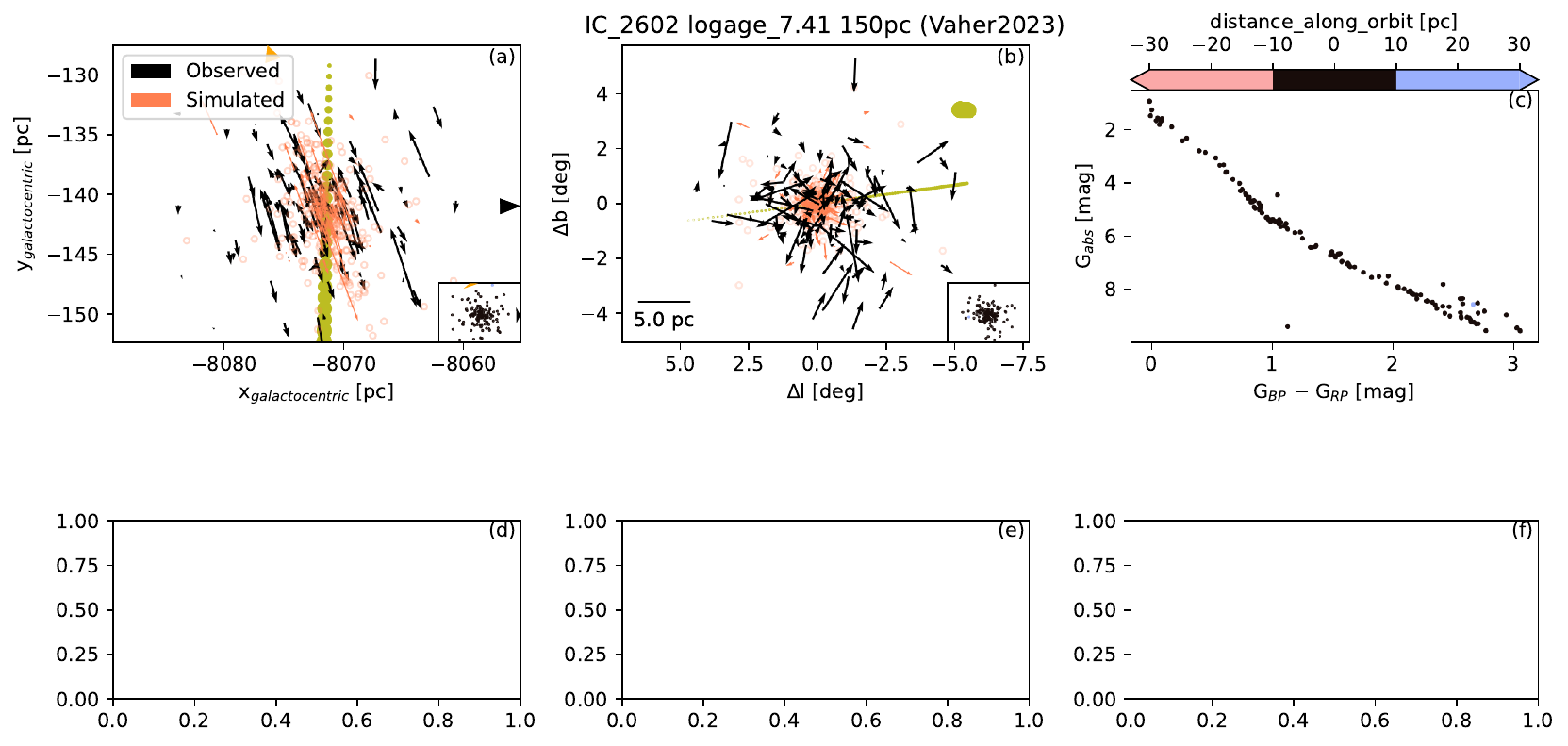}
\includegraphics[width=0.5\linewidth]{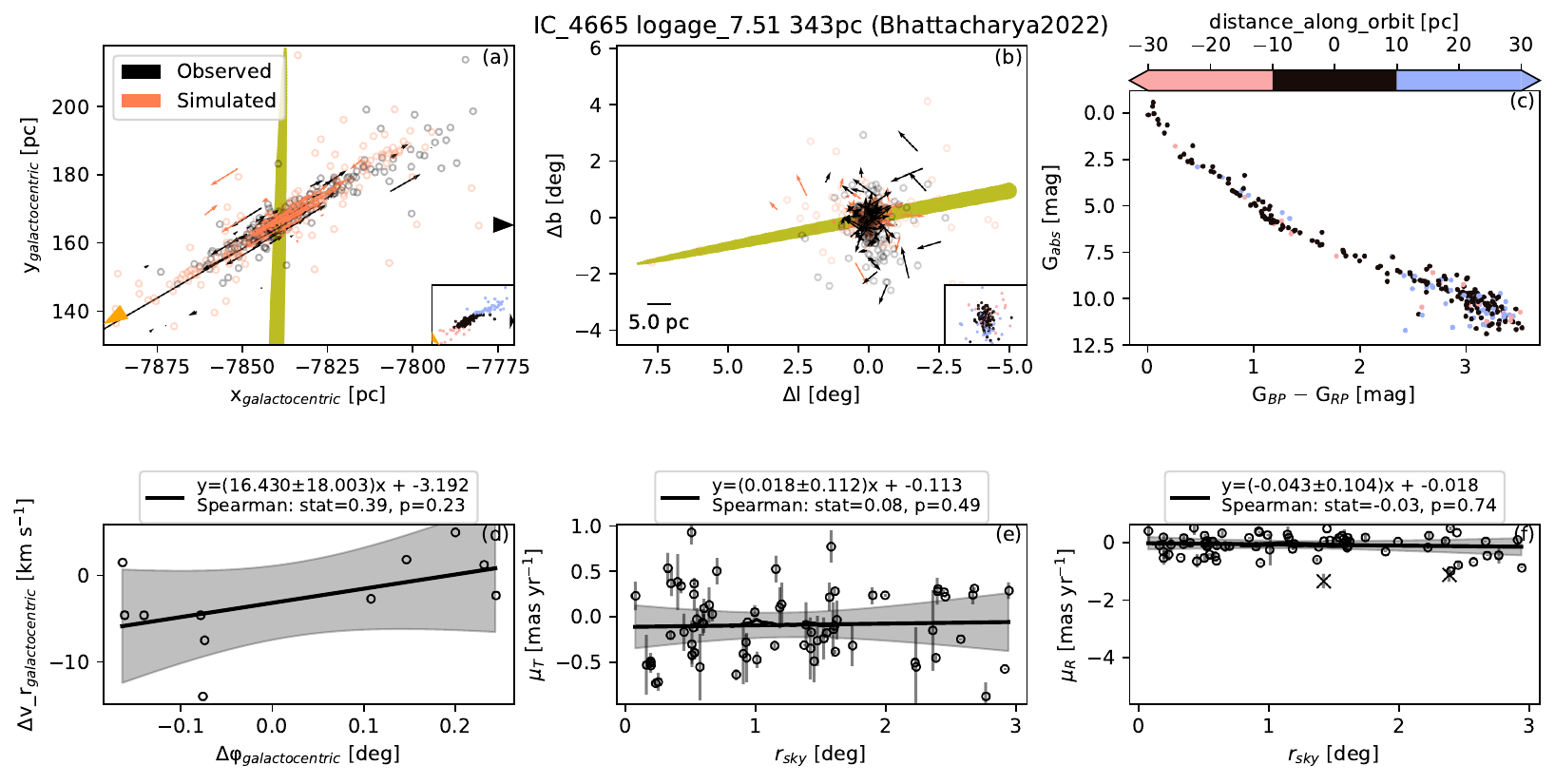}
    \caption{Diagnostic figures for IC 2602 (Meingast2021), IC 2602 (Bhattacharya2022), IC 2602 (Vaher2023), IC 4665 (Bhattacharya2022).}
        \label{fig:supplementary.IC_4665.Bhattacharya2022}
        \end{figure}
         
        \begin{figure}
\includegraphics[width=0.5\linewidth]{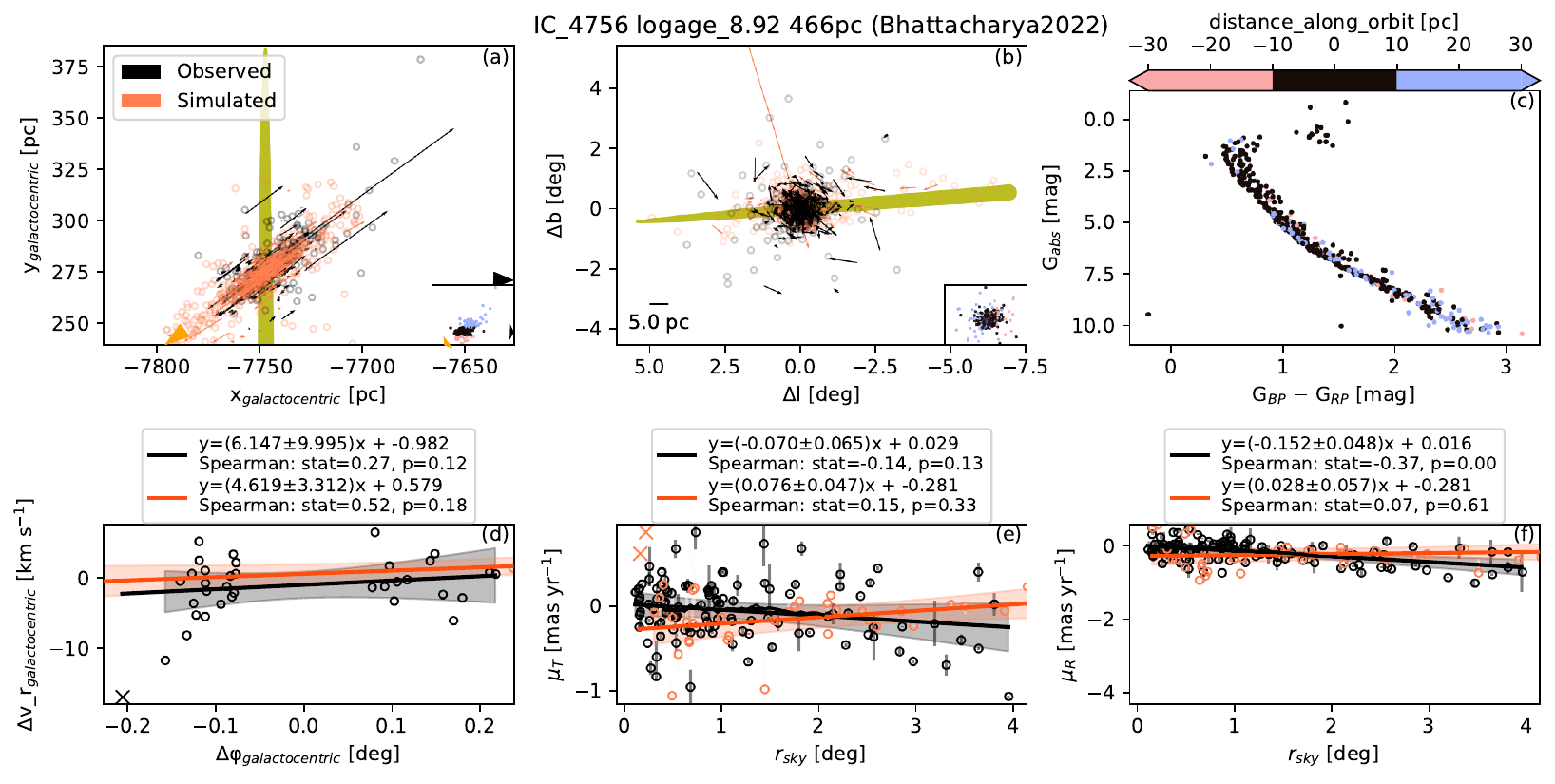}
\includegraphics[width=0.5\linewidth]{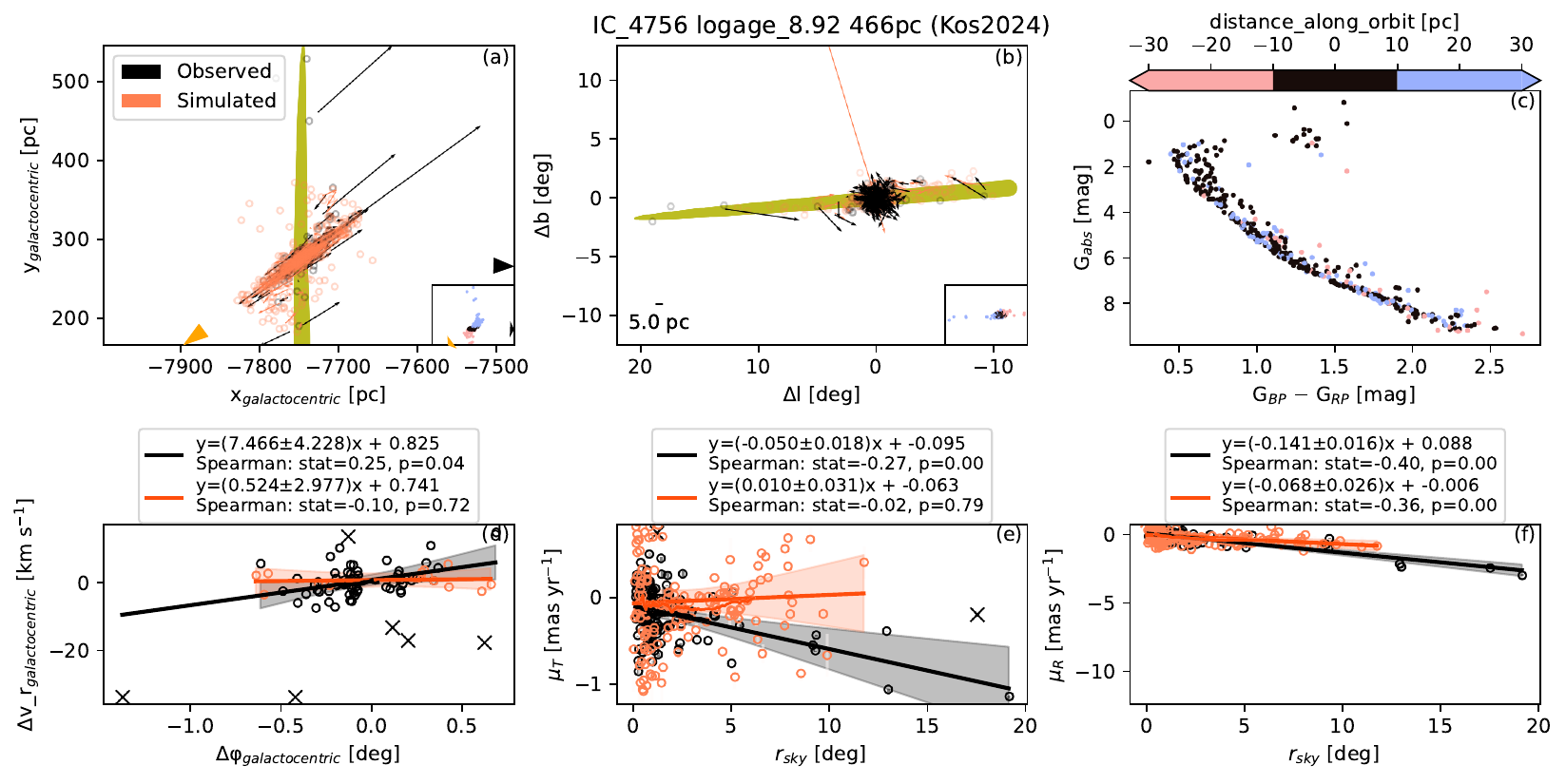}
\includegraphics[width=0.5\linewidth]{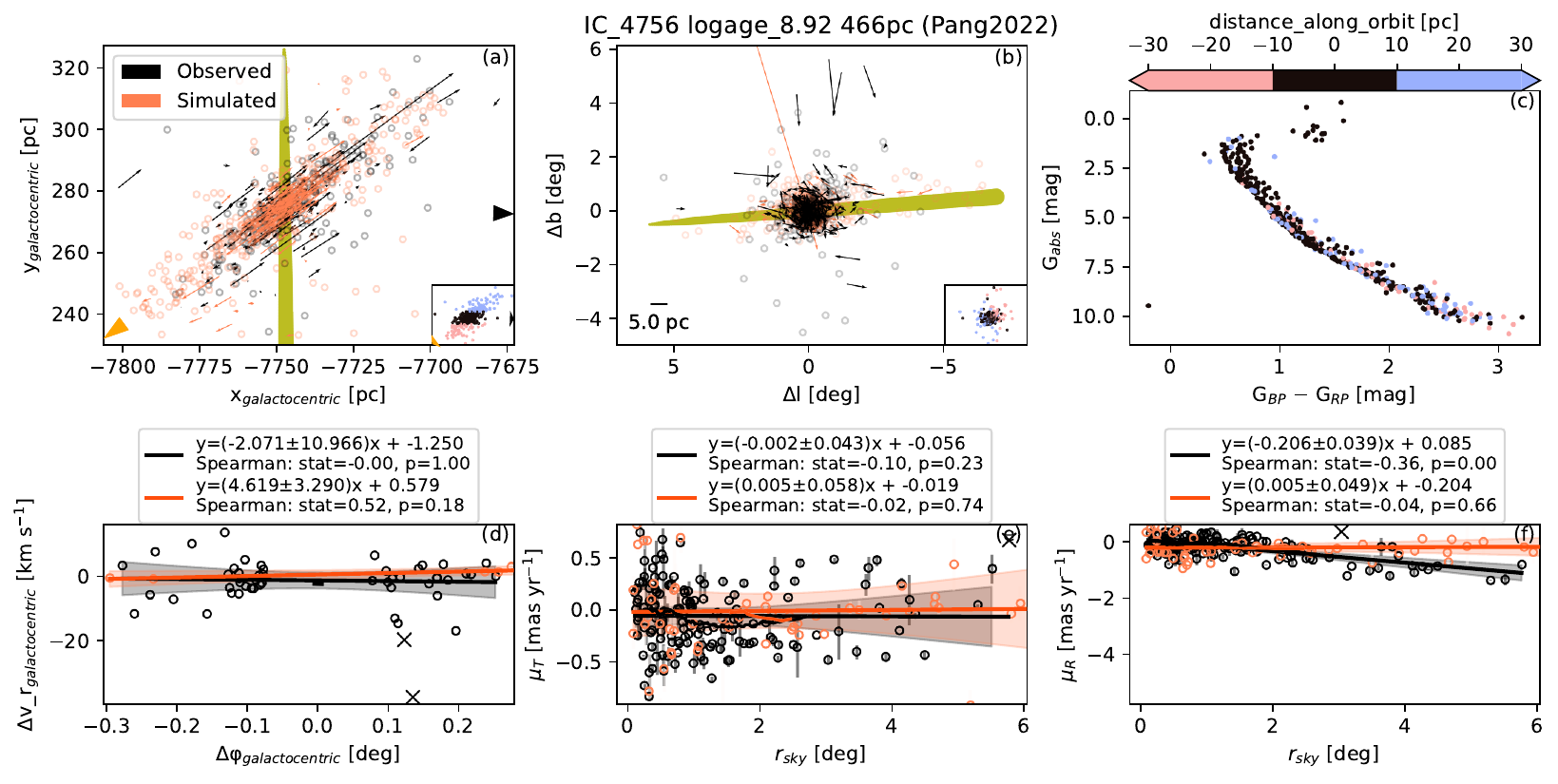}
\includegraphics[width=0.5\linewidth]{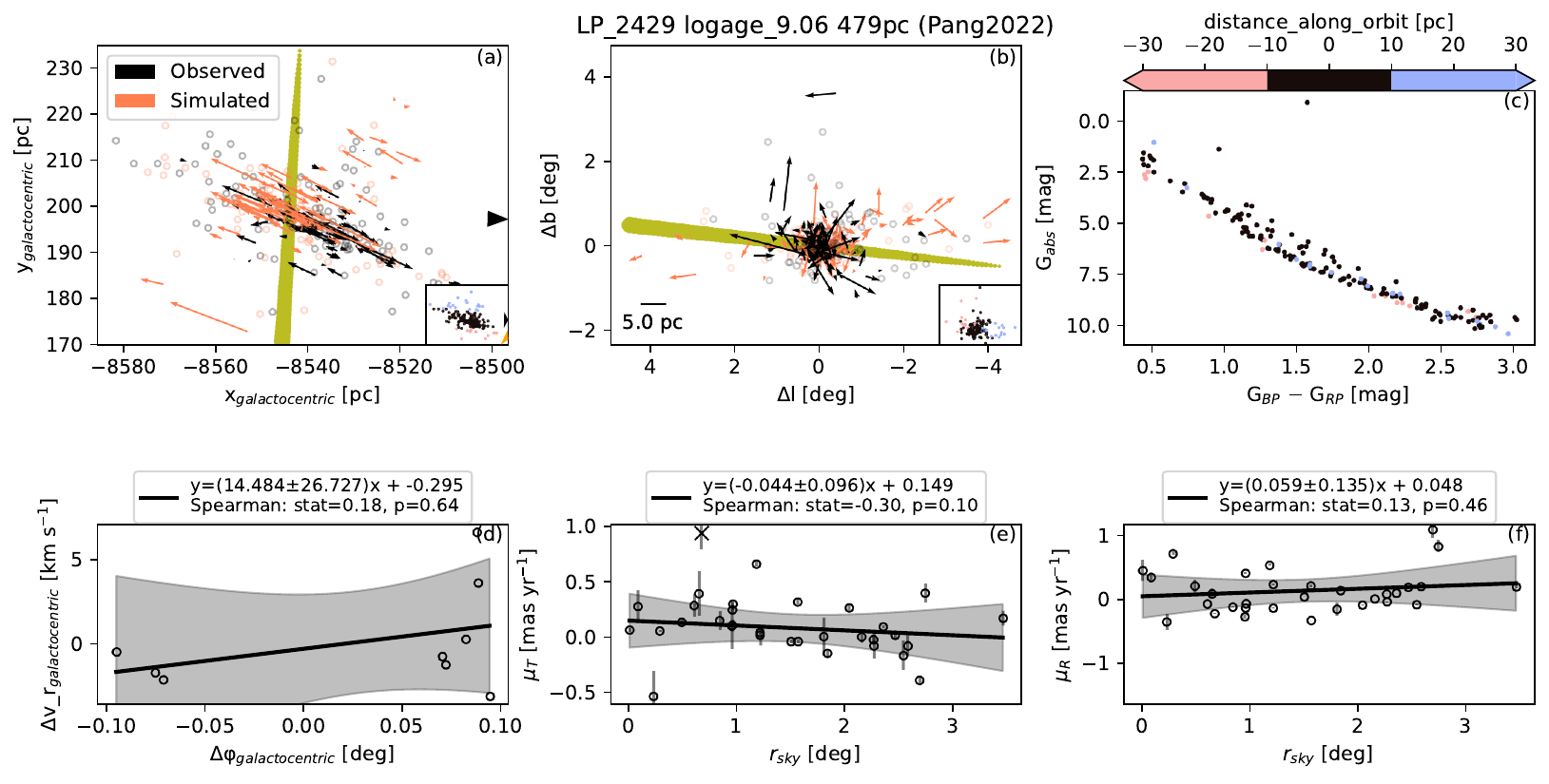}
    \caption{Diagnostic figures for IC 4756 (Bhattacharya2022), IC 4756 (Kos2024), IC 4756 (Pang2022), LP 2429 (Pang2022).}
        \label{fig:supplementary.LP_2429.Pang2022}
        \end{figure}
         
        \begin{figure}
\includegraphics[width=0.5\linewidth]{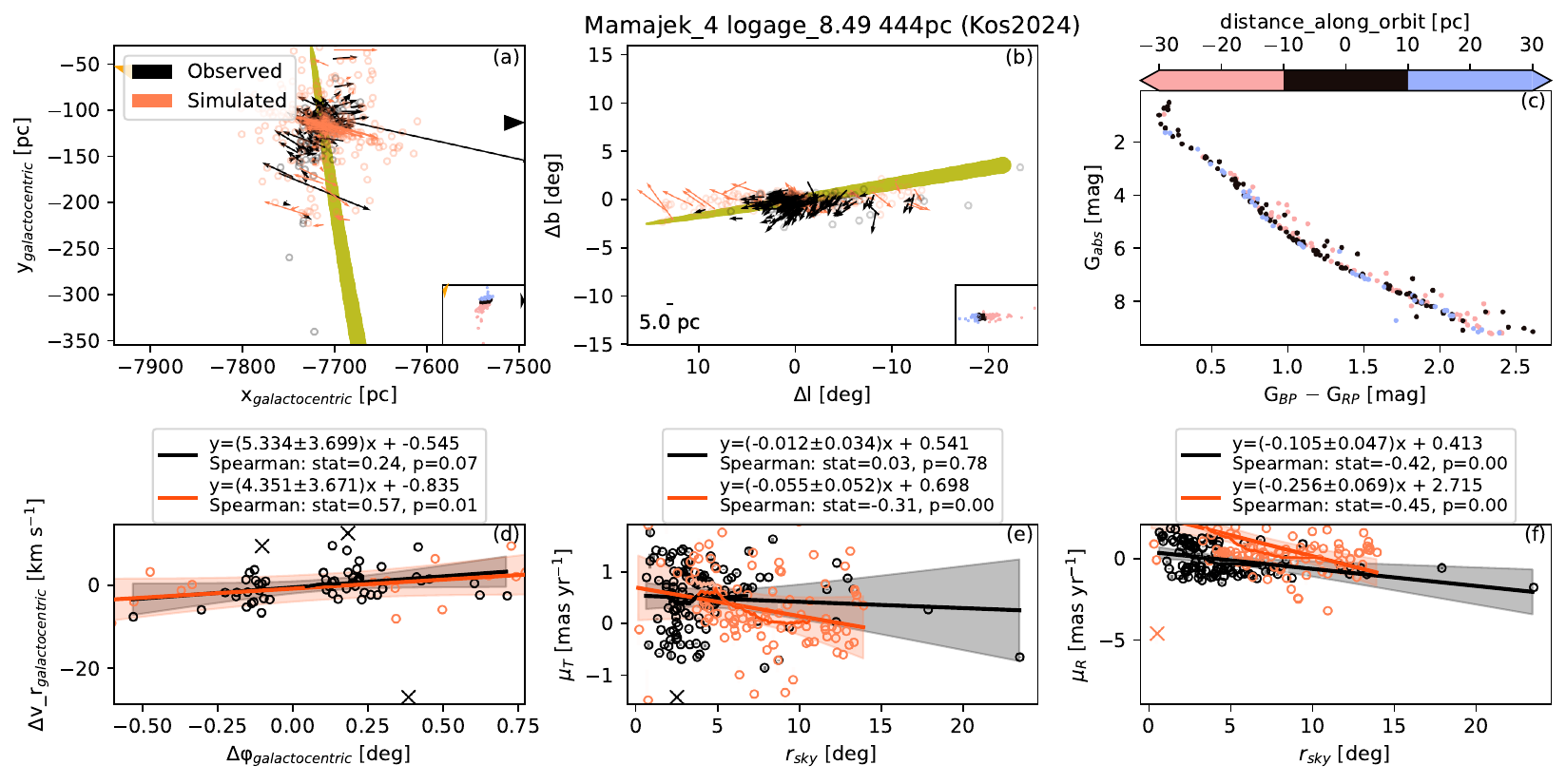}
\includegraphics[width=0.5\linewidth]{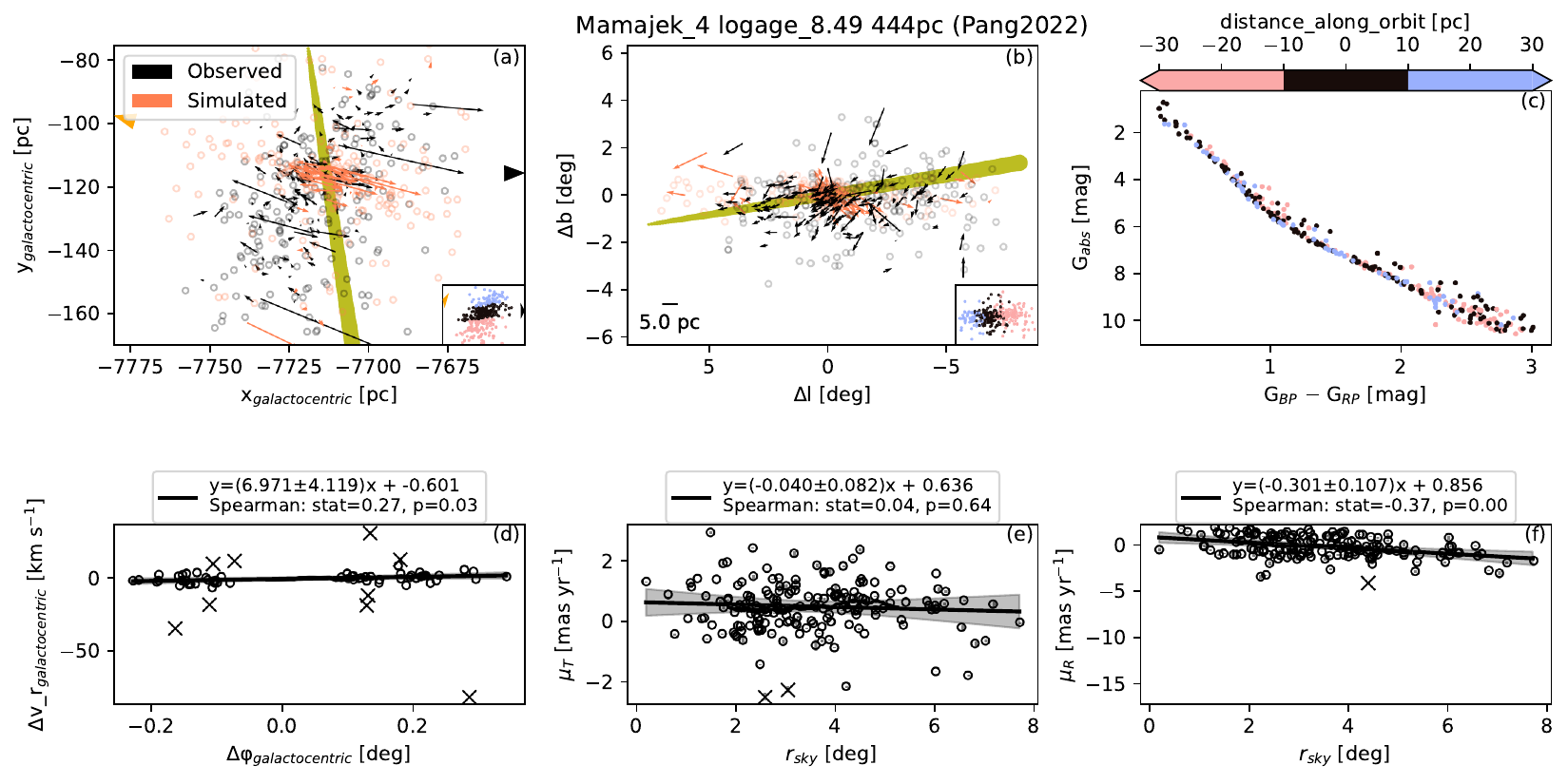}
\includegraphics[width=0.5\linewidth]{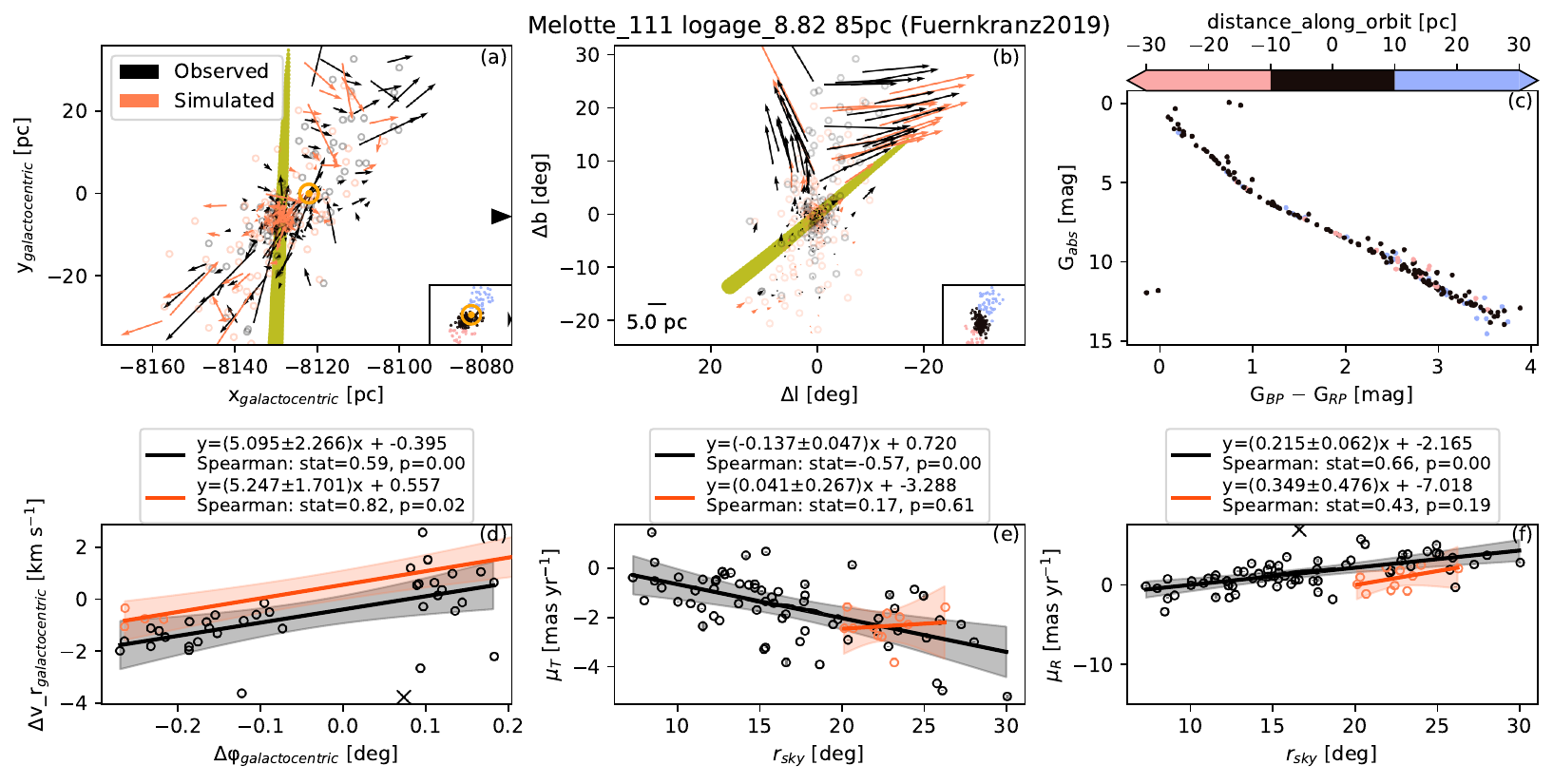}
\includegraphics[width=0.5\linewidth]{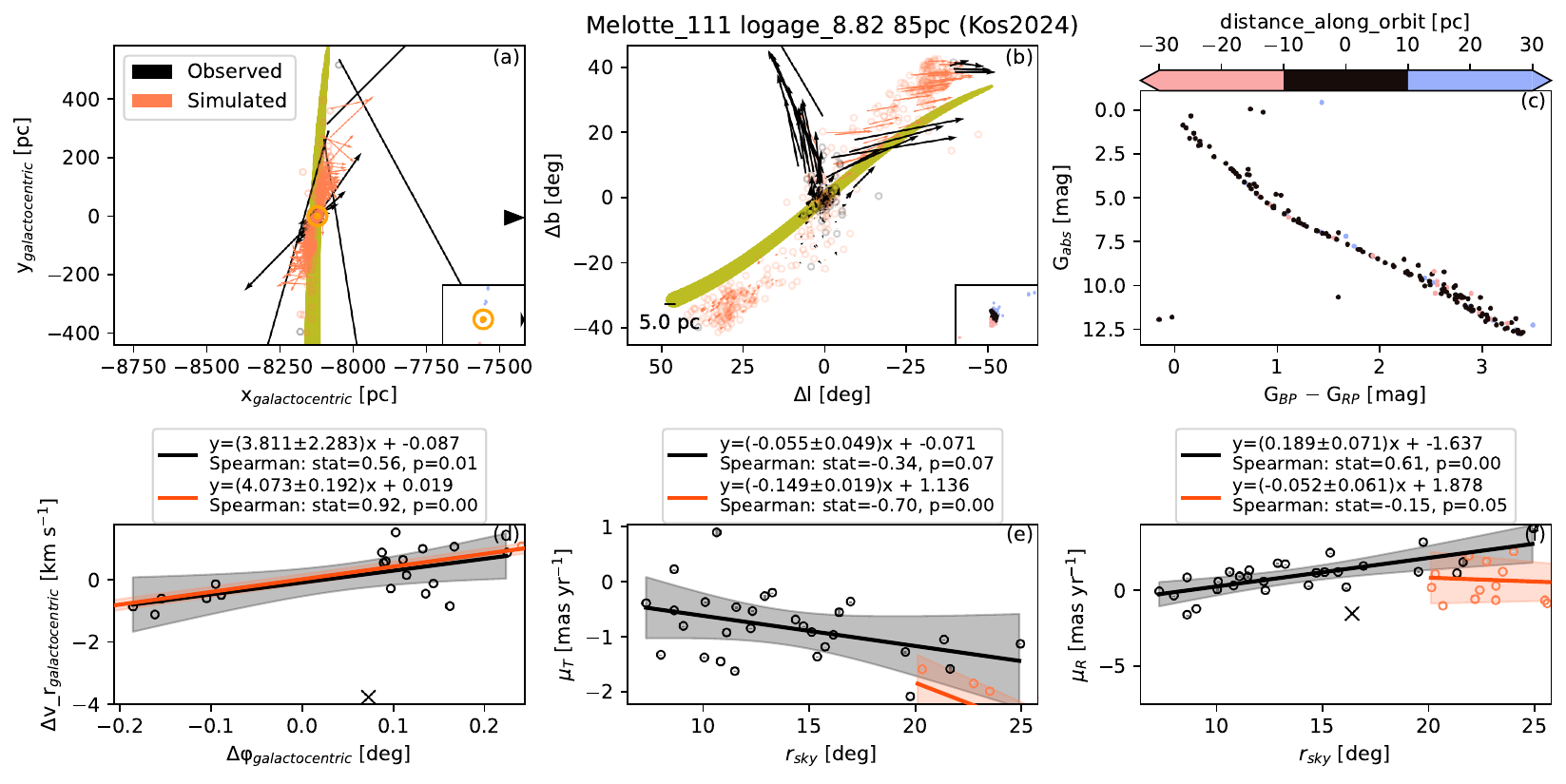}
    \caption{Diagnostic figures for Mamajek 4 (Kos2024), Mamajek 4 (Pang2022), Melotte 111 (Fuernkranz2019), Melotte 111 (Kos2024).}
        \label{fig:supplementary.Melotte_111.Kos2024}
        \end{figure}
         
        \begin{figure}
\includegraphics[width=0.5\linewidth]{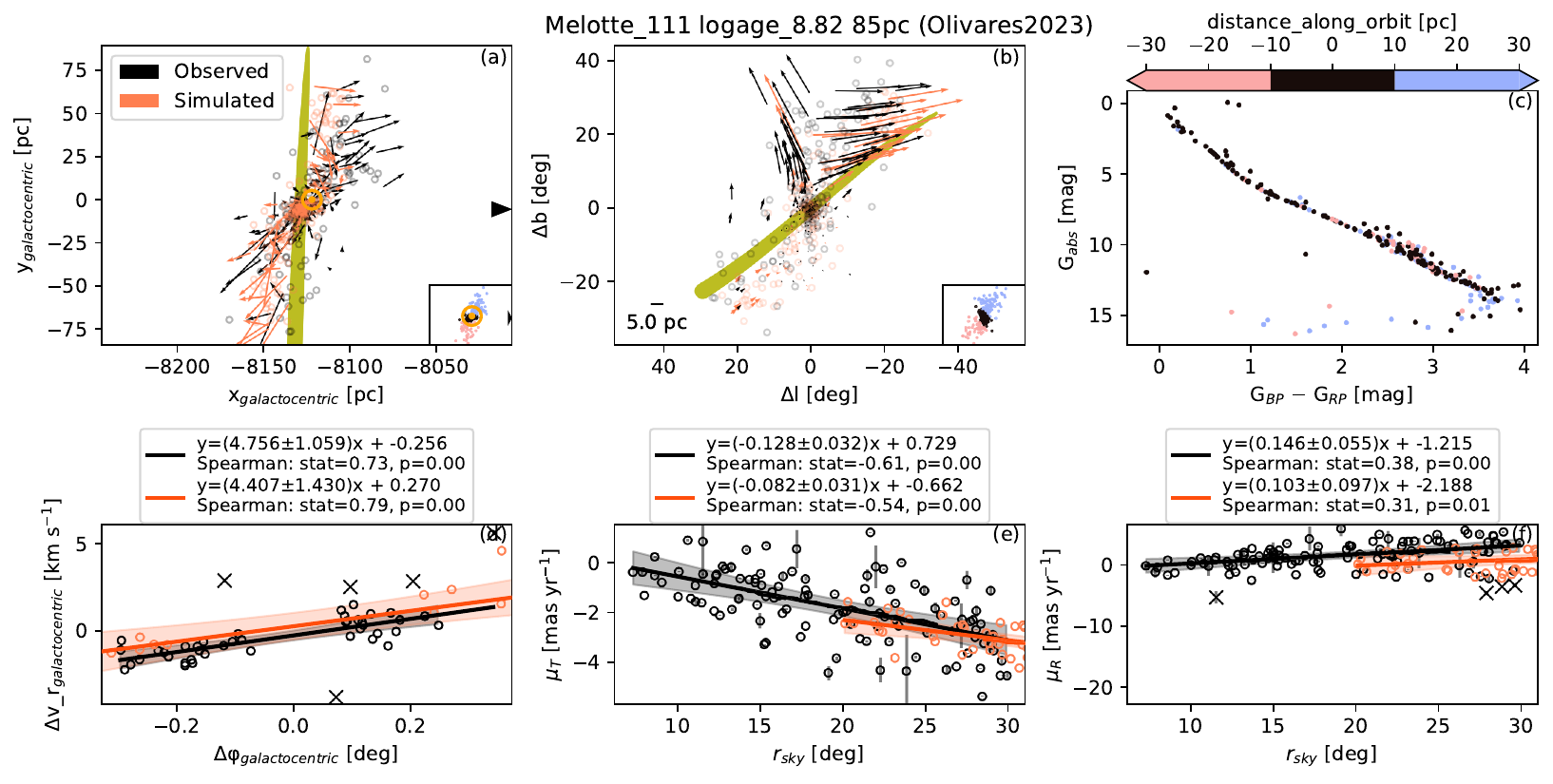}
\includegraphics[width=0.5\linewidth]{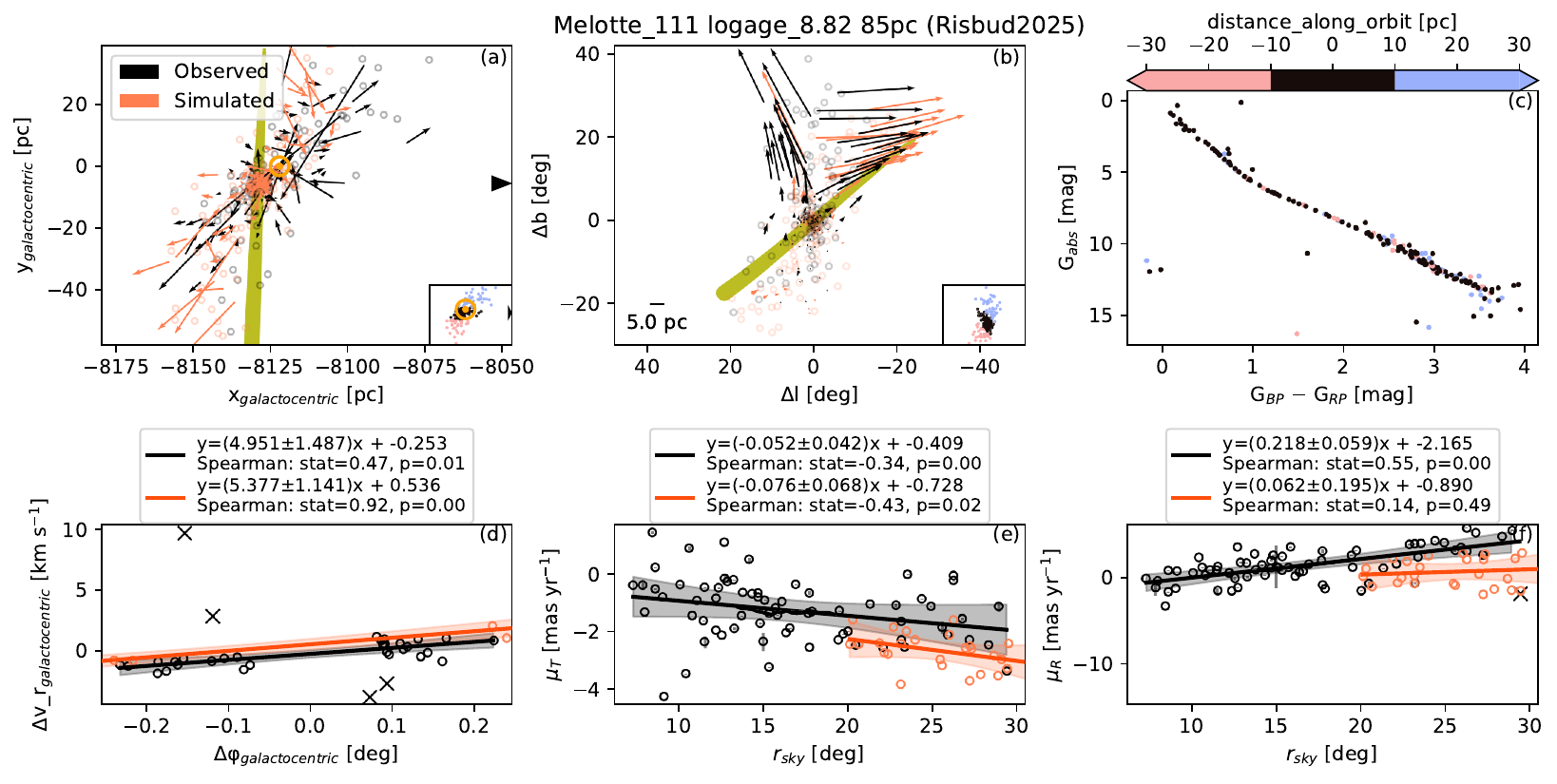}
\includegraphics[width=0.5\linewidth]{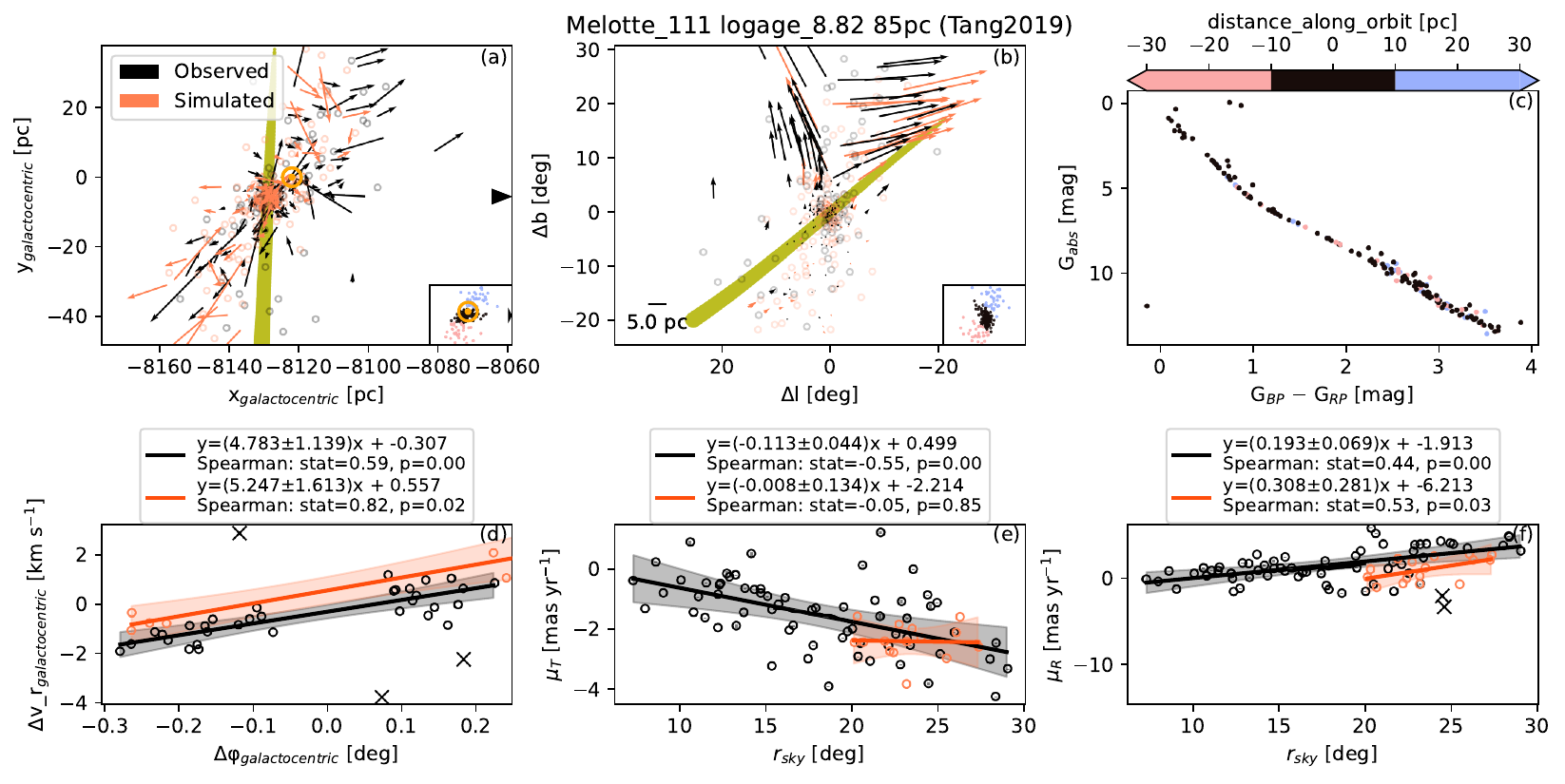}
\includegraphics[width=0.5\linewidth]{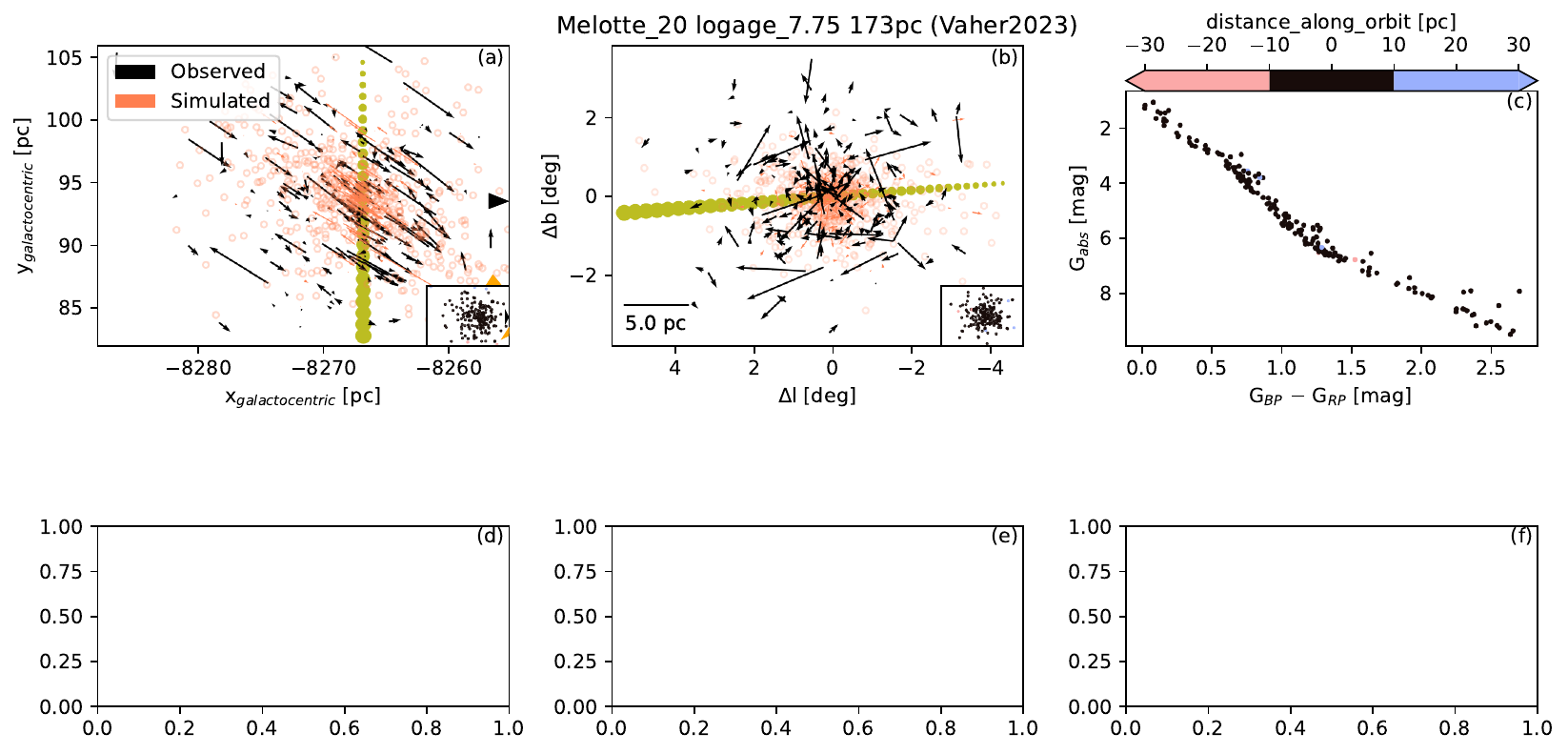}
    \caption{Diagnostic figures for Melotte 111 (Olivares2023), Melotte 111 (Risbud2025), Melotte 111 (Tang2019), Melotte 20 (Vaher2023).}
        \label{fig:supplementary.Melotte_20.Vaher2023}
        \end{figure}
         
        \begin{figure}
\includegraphics[width=0.5\linewidth]{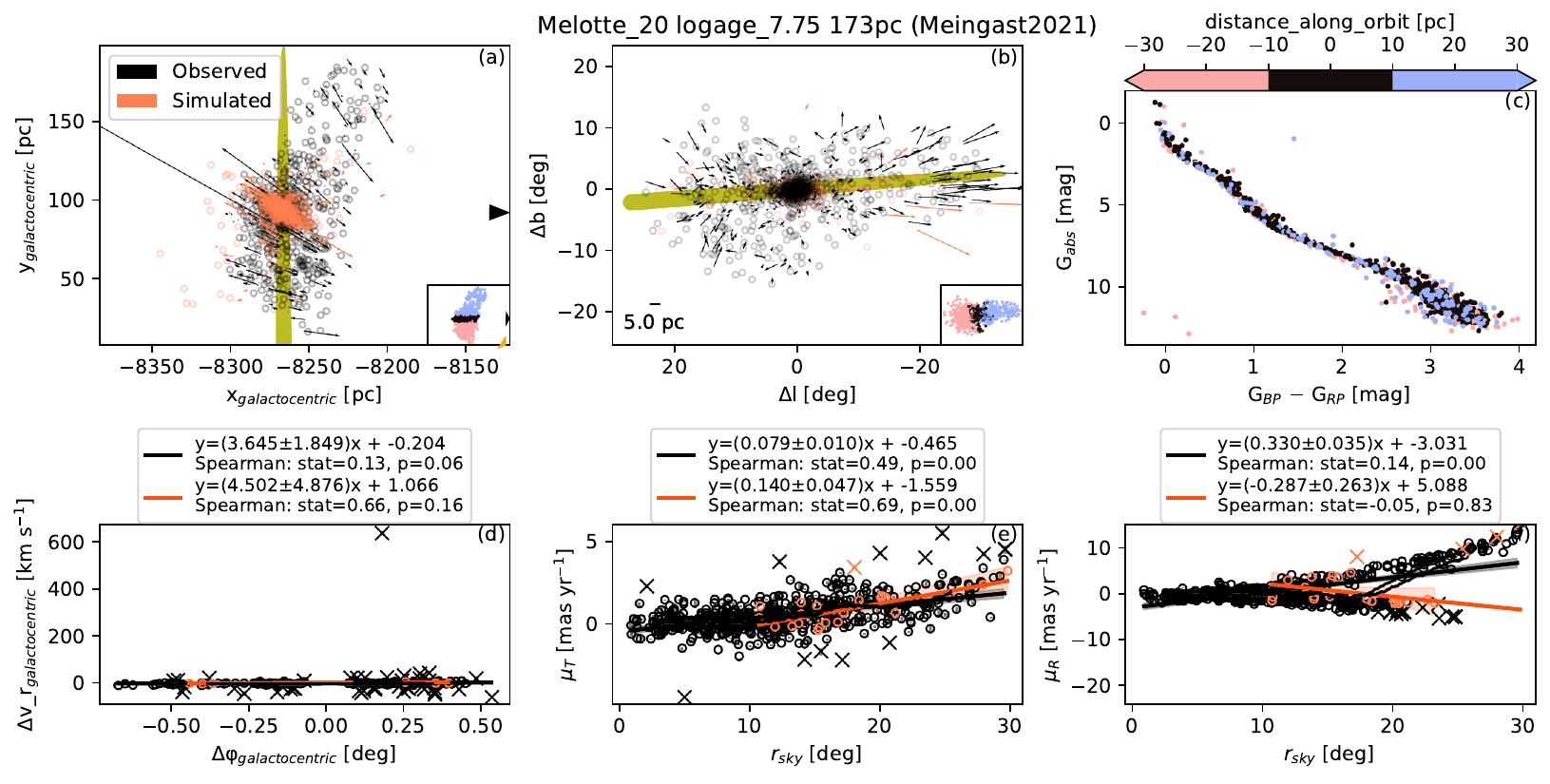}
\includegraphics[width=0.5\linewidth]{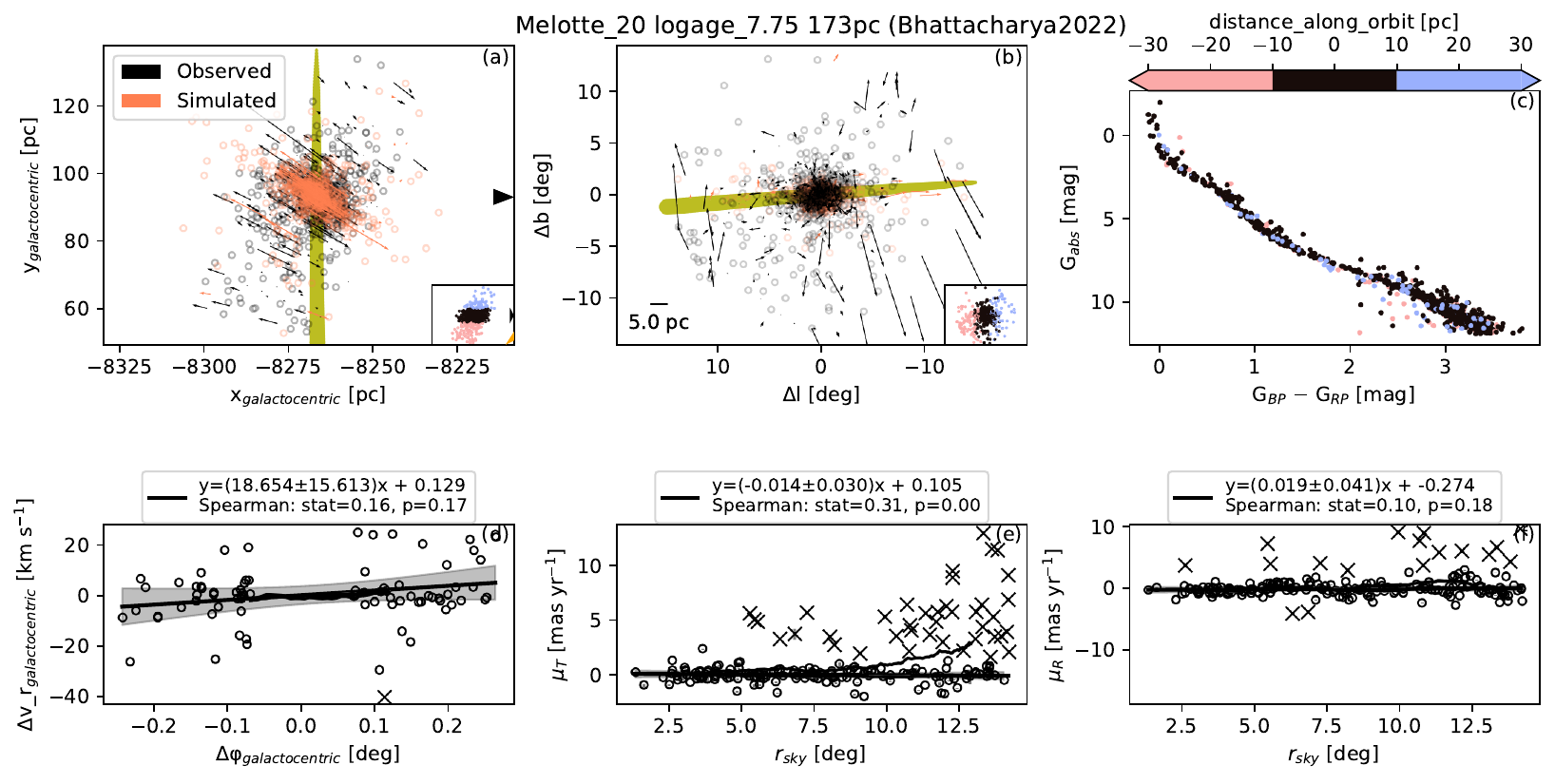}
\includegraphics[width=0.5\linewidth]{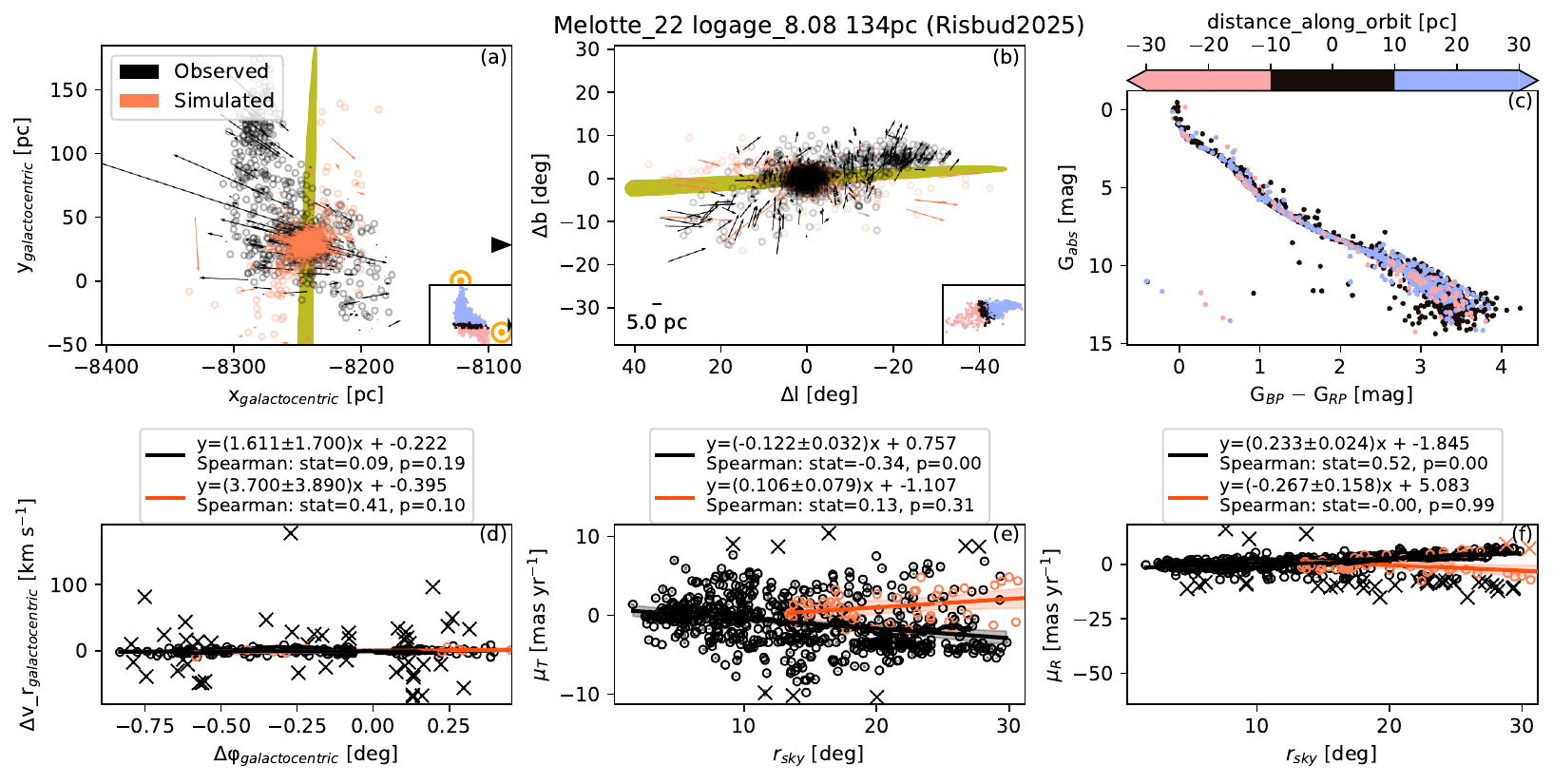}
\includegraphics[width=0.5\linewidth]{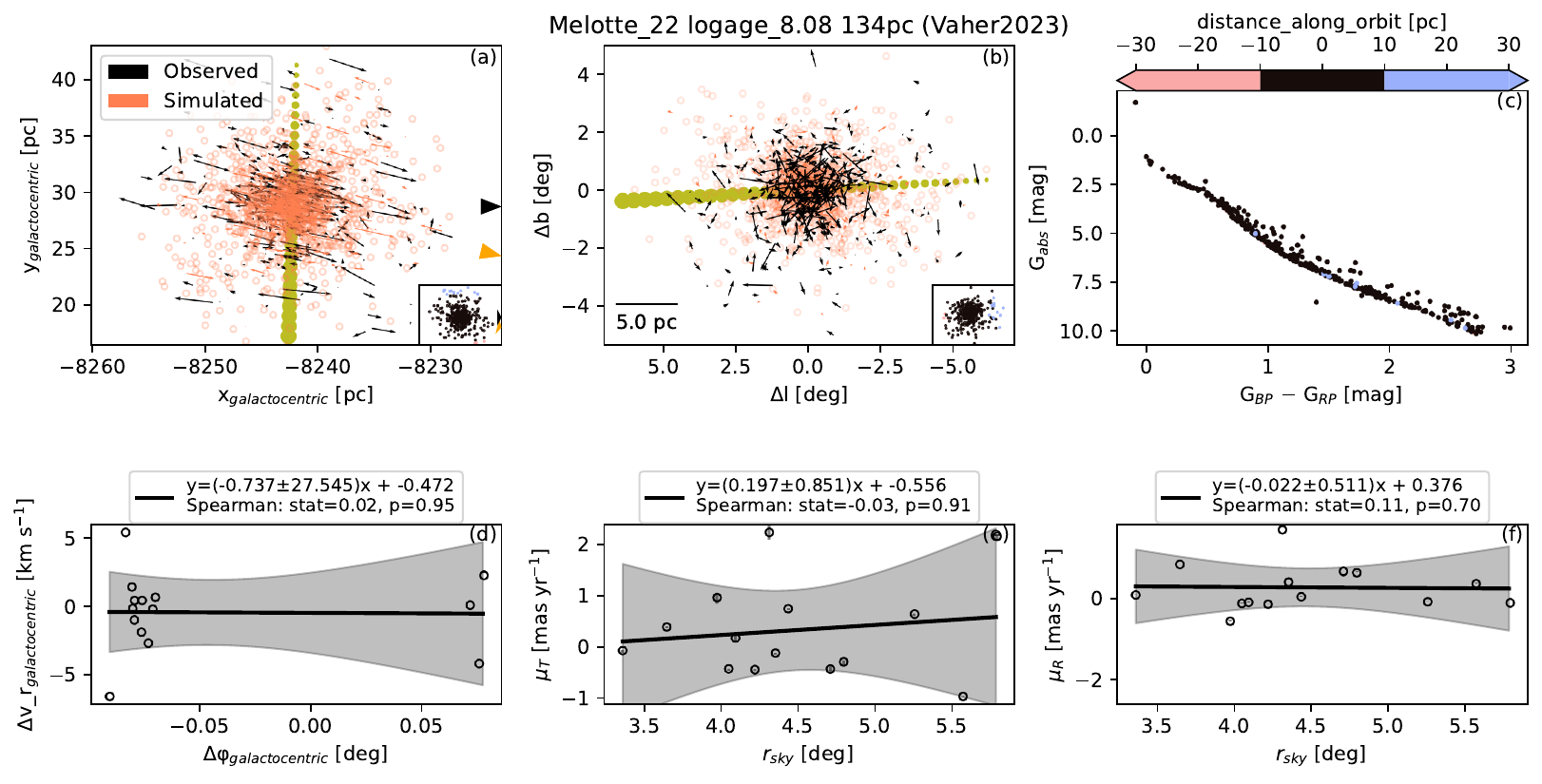}
    \caption{Diagnostic figures for Melotte 20 (Meingast2021), Melotte 20 (Bhattacharya2022), Melotte 22 (Risbud2025), Melotte 22 (Vaher2023).}
        \label{fig:supplementary.Melotte_22.Vaher2023}
        \end{figure}
         
        \begin{figure}
\includegraphics[width=0.5\linewidth]{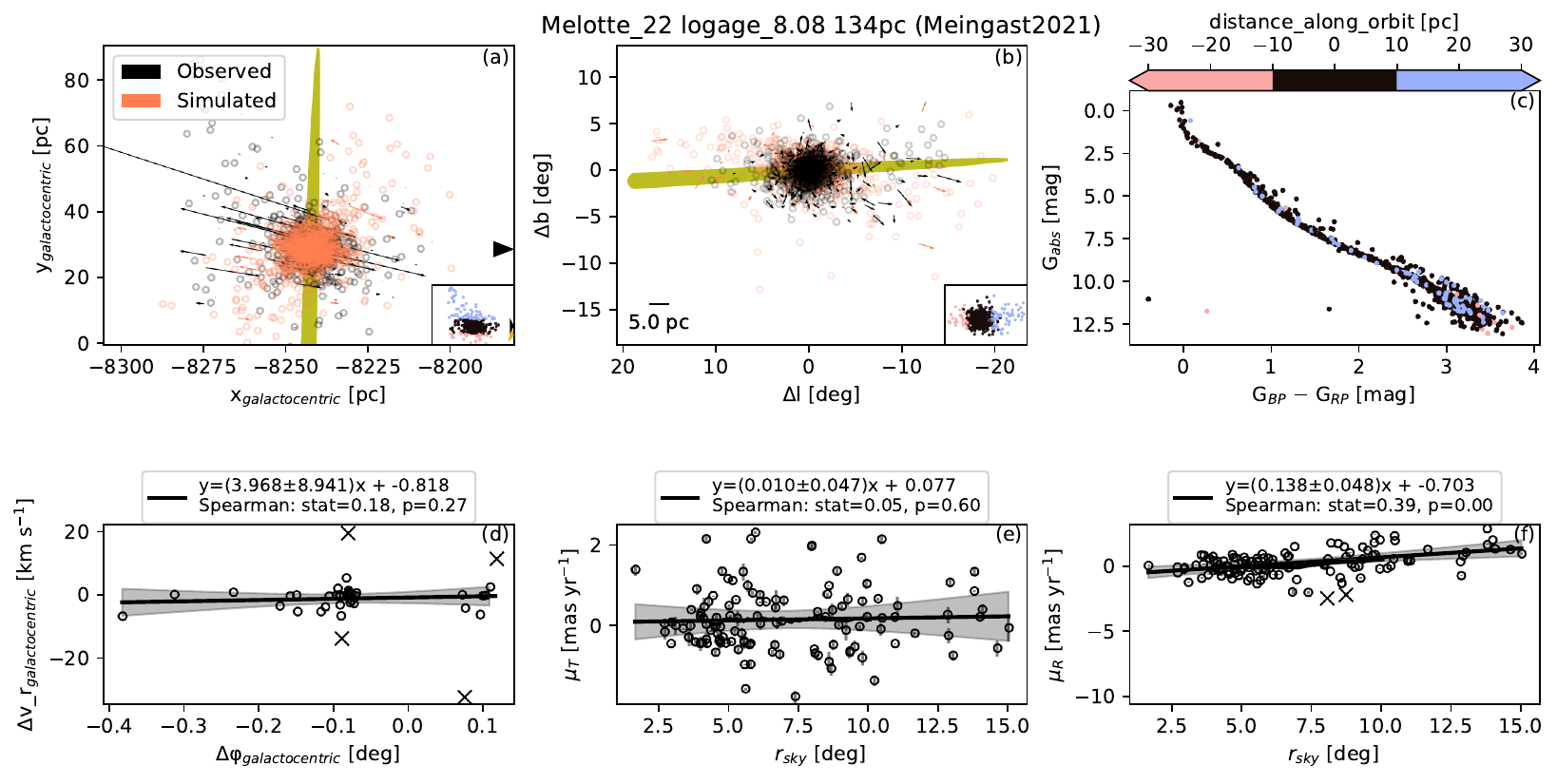}
\includegraphics[width=0.5\linewidth]{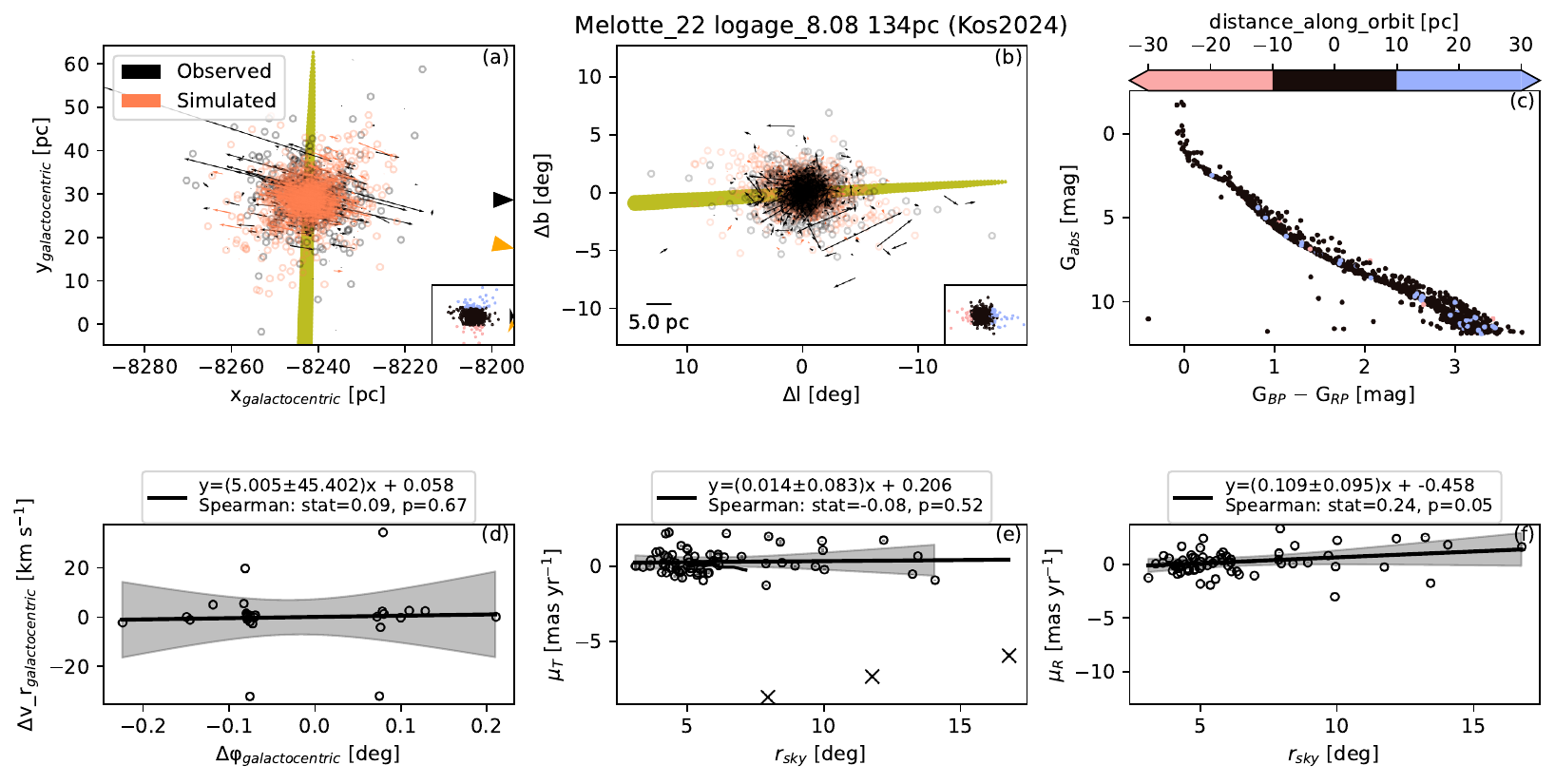}
\includegraphics[width=0.5\linewidth]{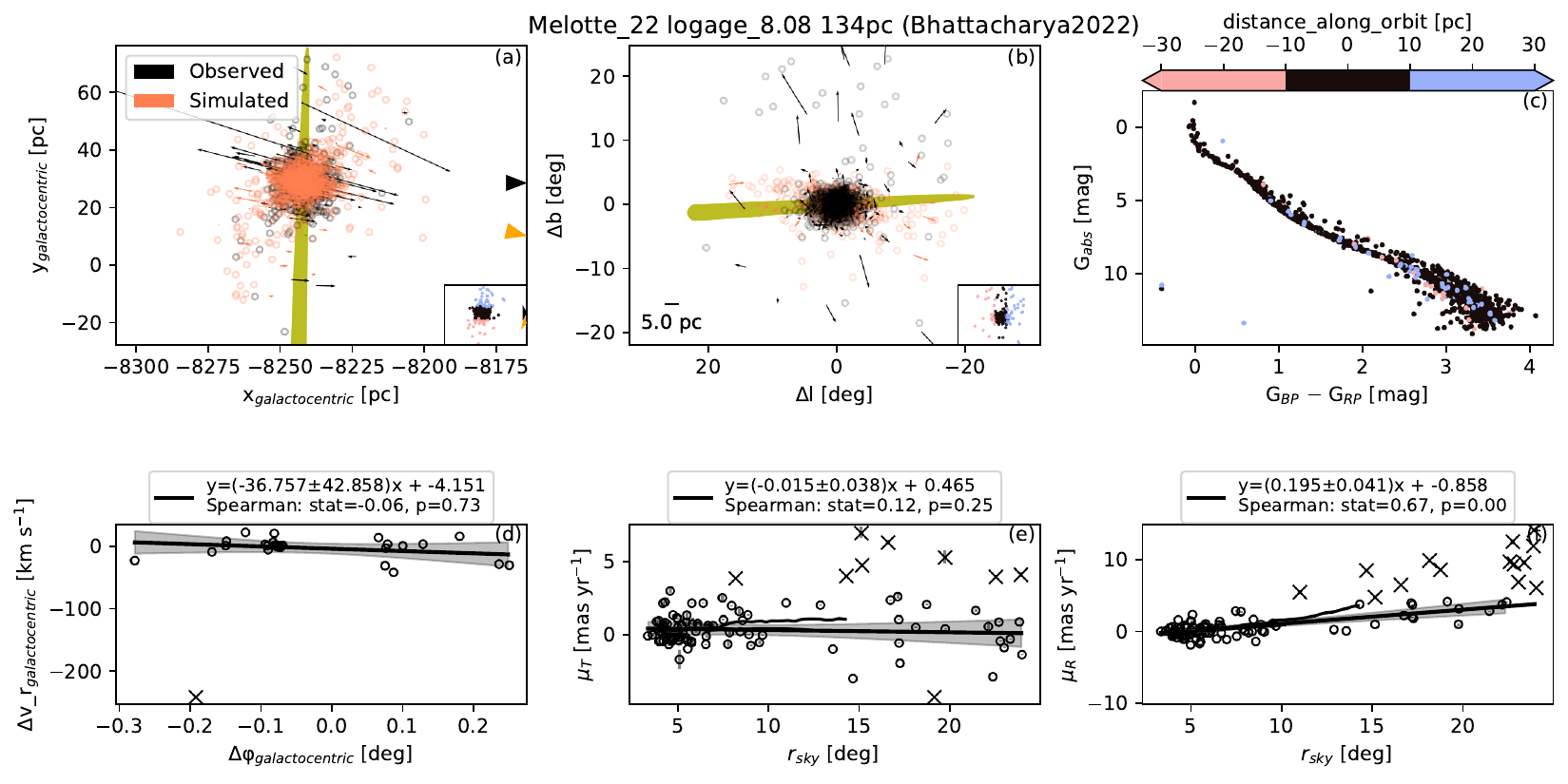}
\includegraphics[width=0.5\linewidth]{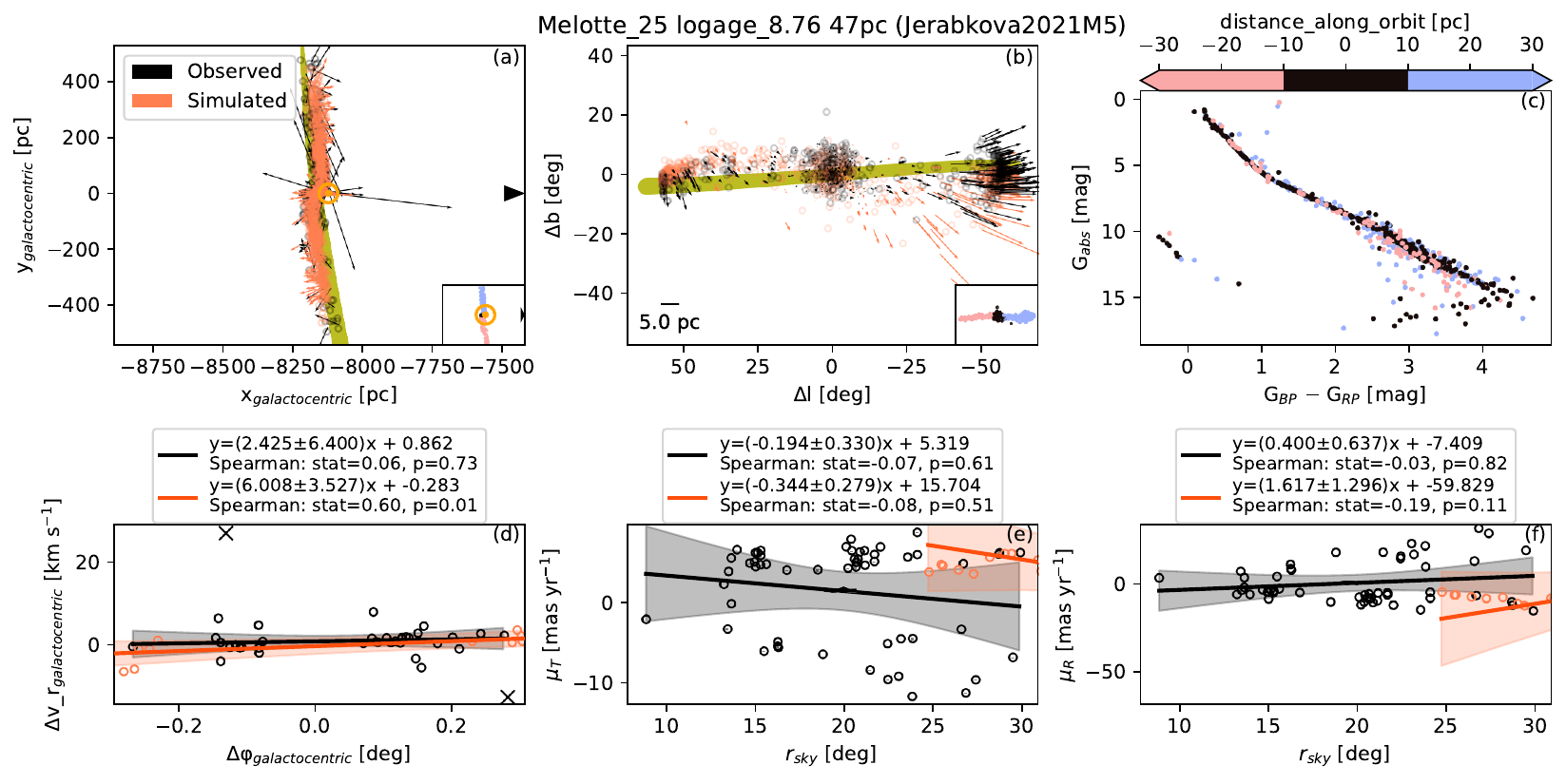}
    \caption{Diagnostic figures for Melotte 22 (Meingast2021), Melotte 22 (Kos2024), Melotte 22 (Bhattacharya2022), Melotte 25 (Jerabkova2021M5).}
        \label{fig:supplementary.Melotte_25.Jerabkova2021M5}
        \end{figure}
         
        \begin{figure}
\includegraphics[width=0.5\linewidth]{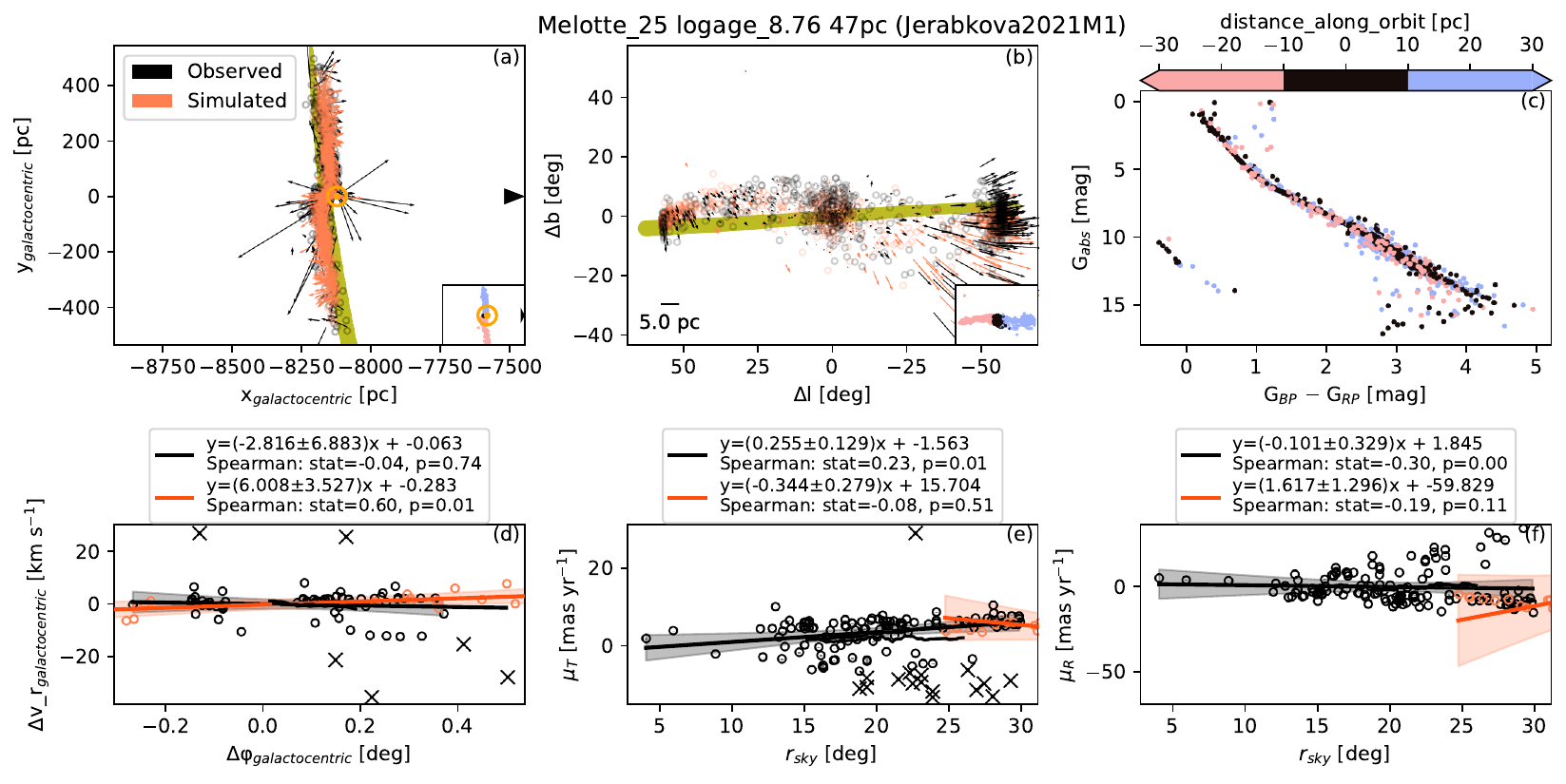}
\includegraphics[width=0.5\linewidth]{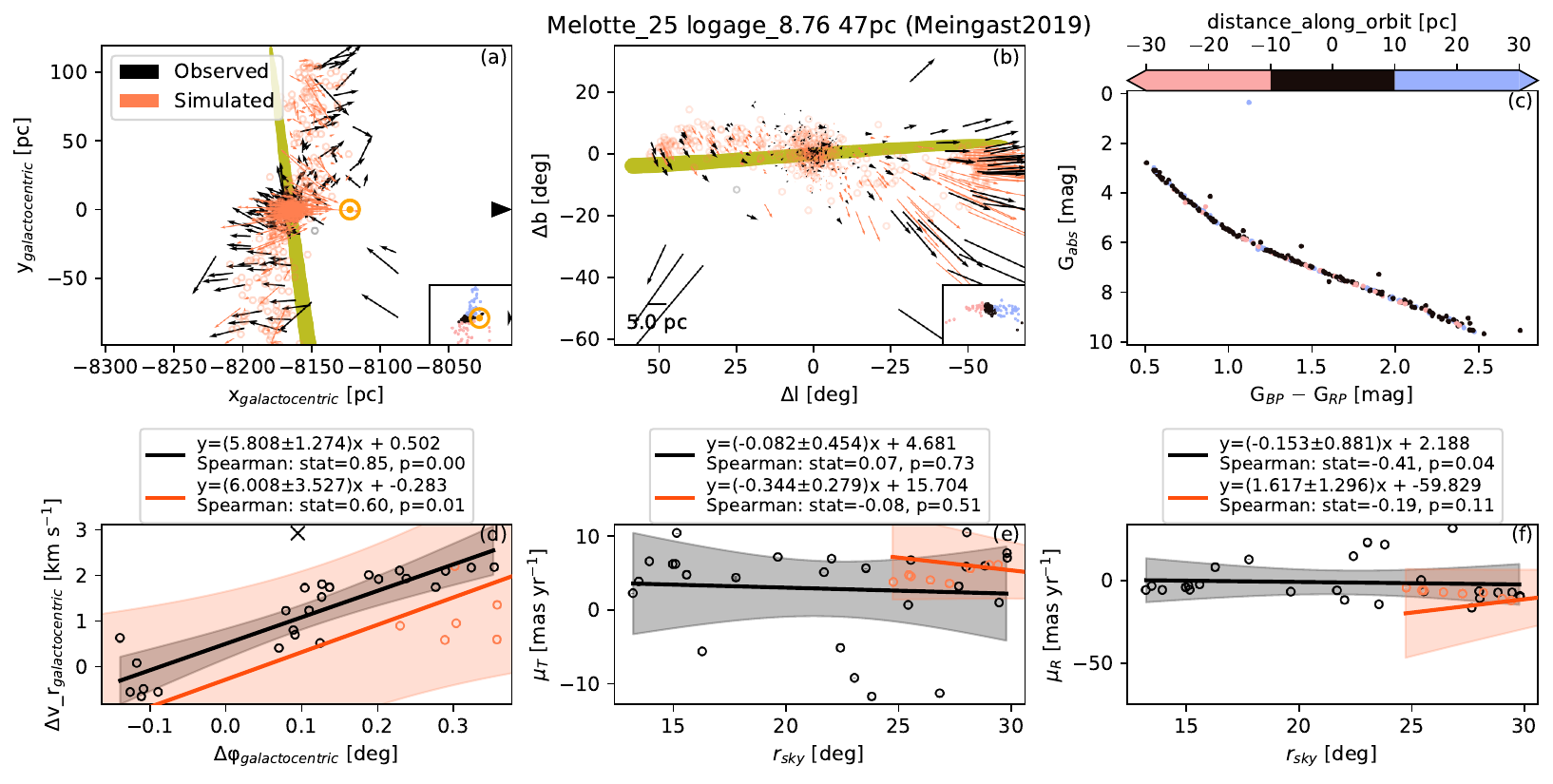}
\includegraphics[width=0.5\linewidth]{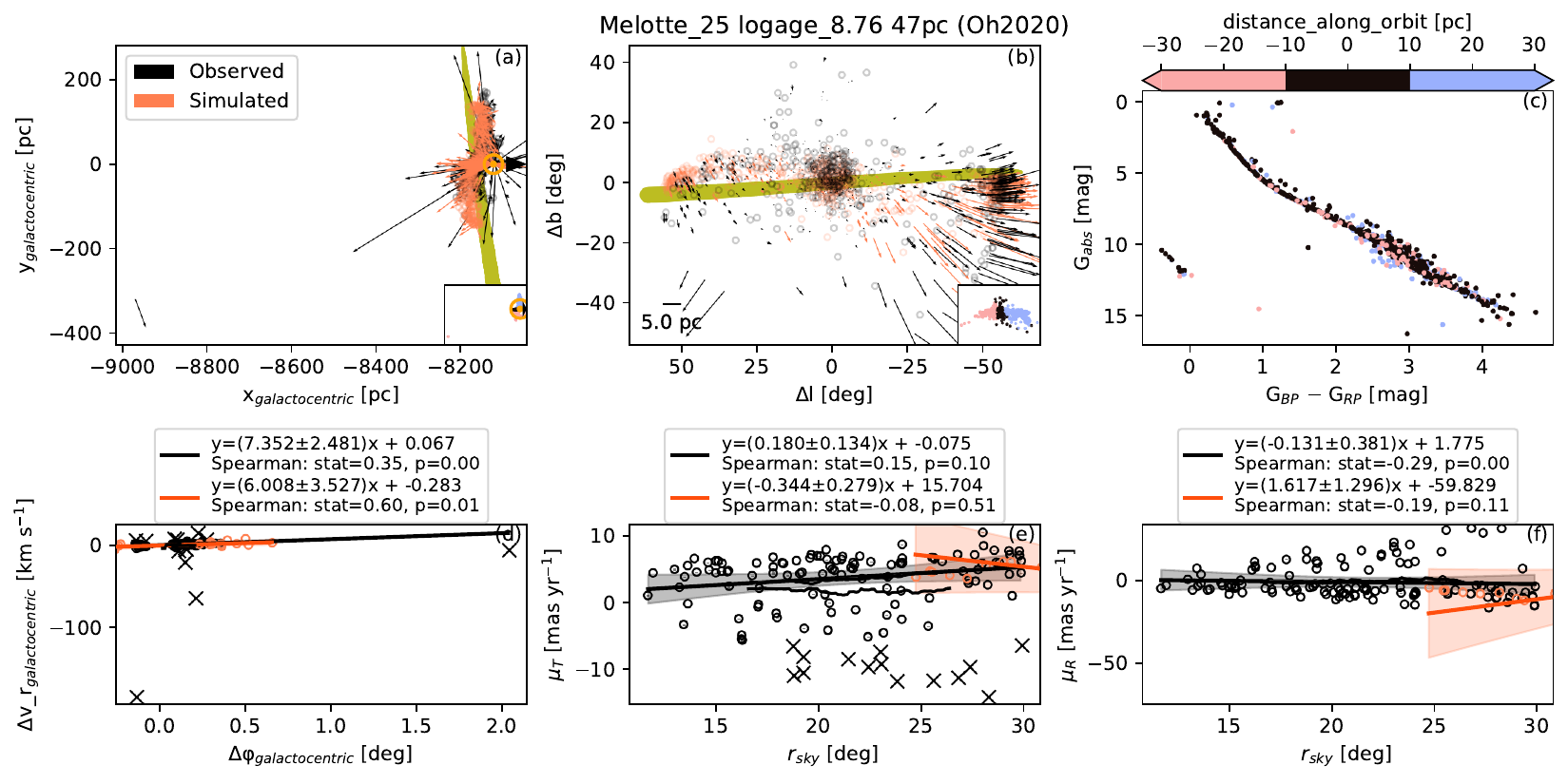}
\includegraphics[width=0.5\linewidth]{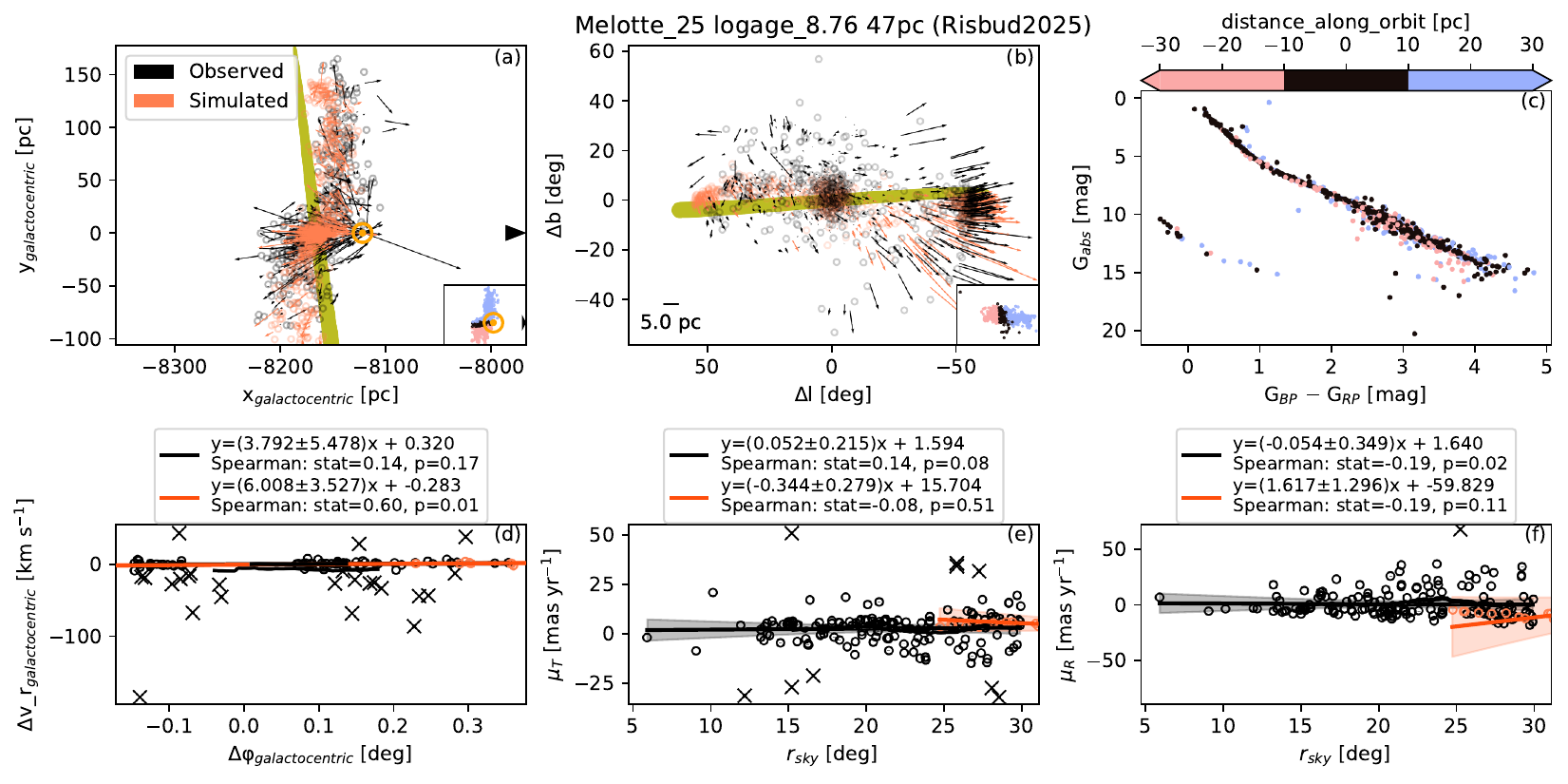}
    \caption{Diagnostic figures for Melotte 25 (Jerabkova2021M1), Melotte 25 (Meingast2019), Melotte 25 (Oh2020), Melotte 25 (Risbud2025).}
        \label{fig:supplementary.Melotte_25.Risbud2025}
        \end{figure}
         
        \begin{figure}
\includegraphics[width=0.5\linewidth]{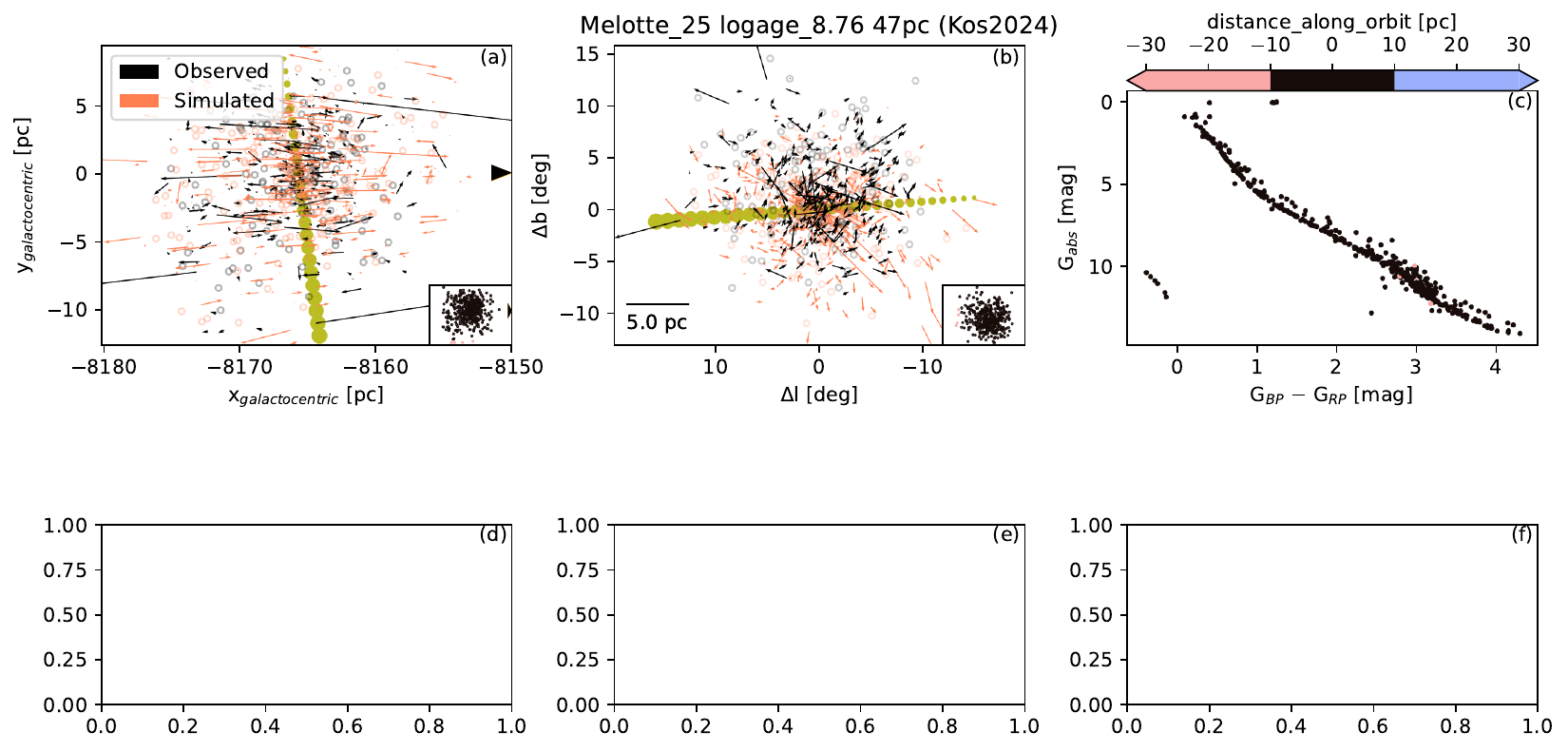}
\includegraphics[width=0.5\linewidth]{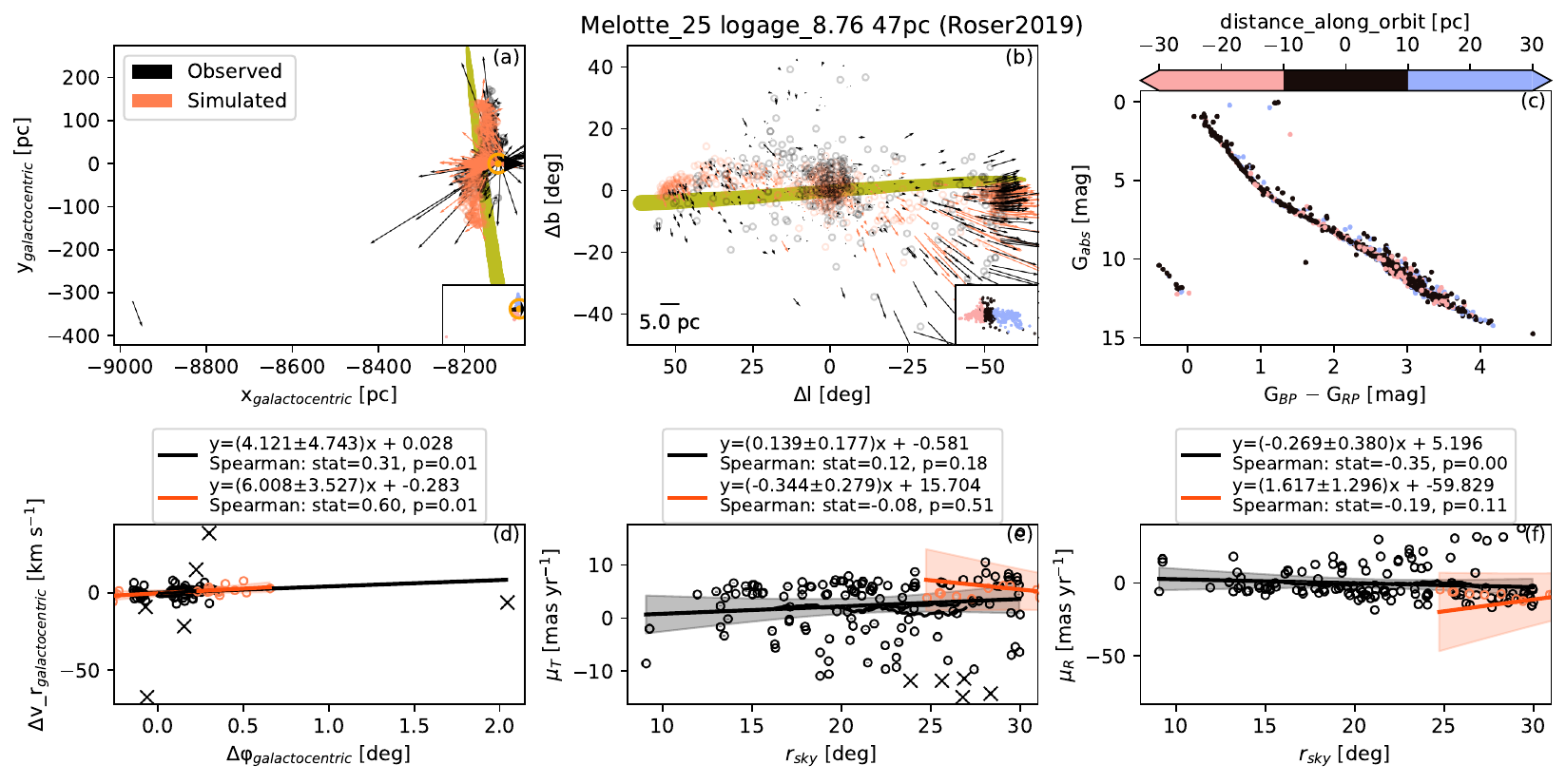}
\includegraphics[width=0.5\linewidth]{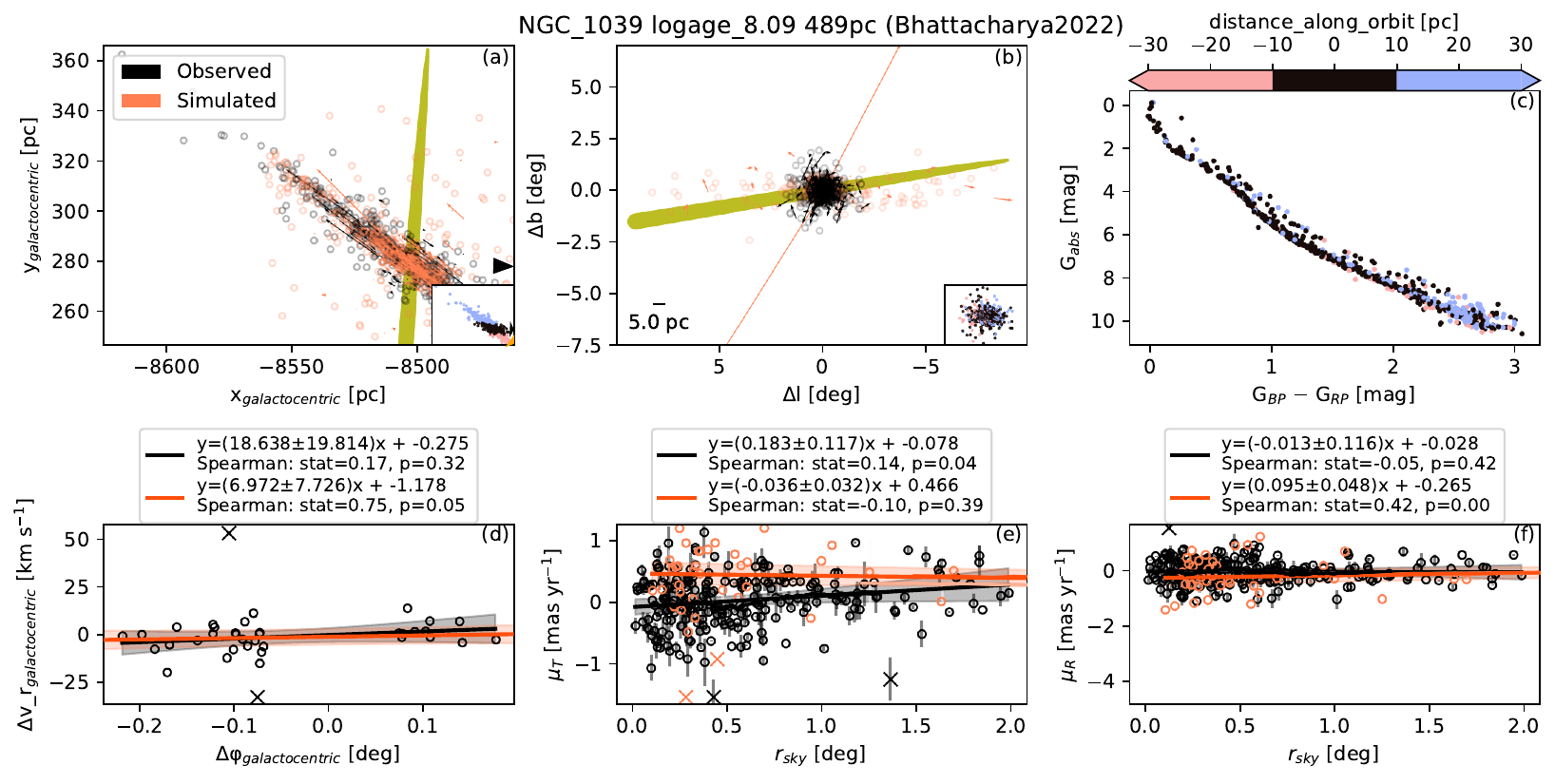}
\includegraphics[width=0.5\linewidth]{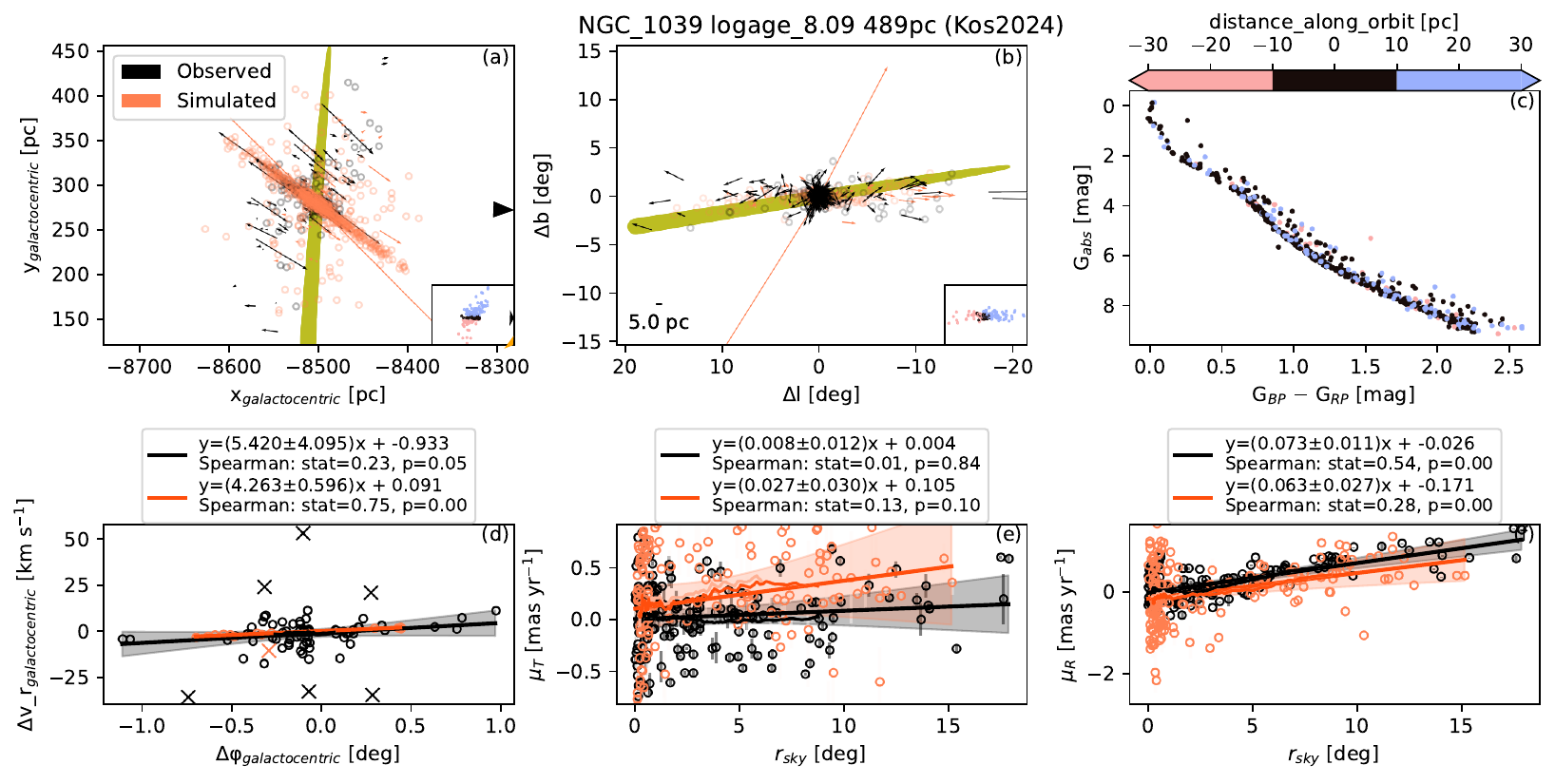}
    \caption{Diagnostic figures for Melotte 25 (Kos2024), Melotte 25 (Roser2019), NGC 1039 (Bhattacharya2022), NGC 1039 (Kos2024).}
        \label{fig:supplementary.NGC_1039.Kos2024}
        \end{figure}
         
        \begin{figure}
\includegraphics[width=0.5\linewidth]{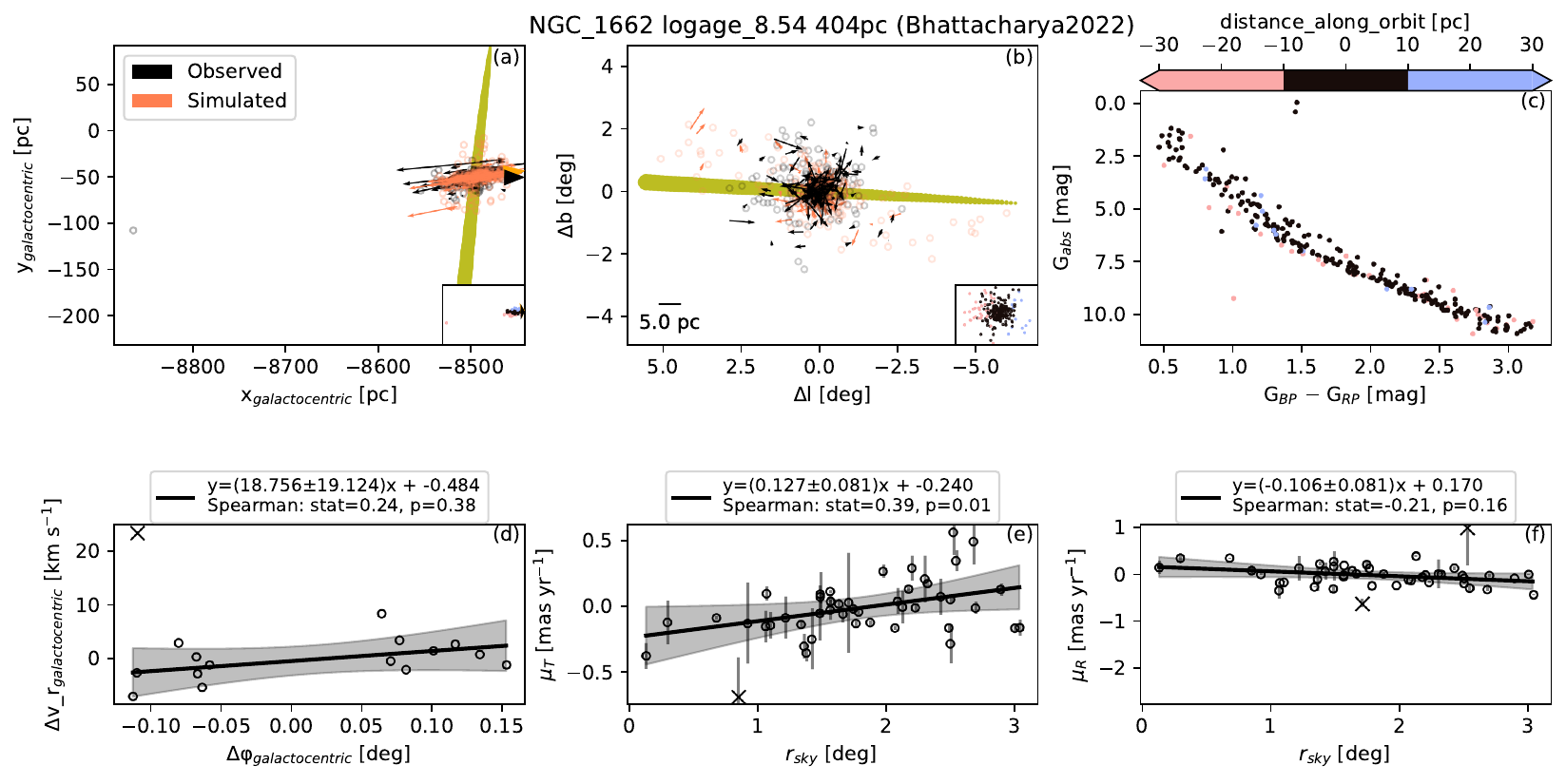}
\includegraphics[width=0.5\linewidth]{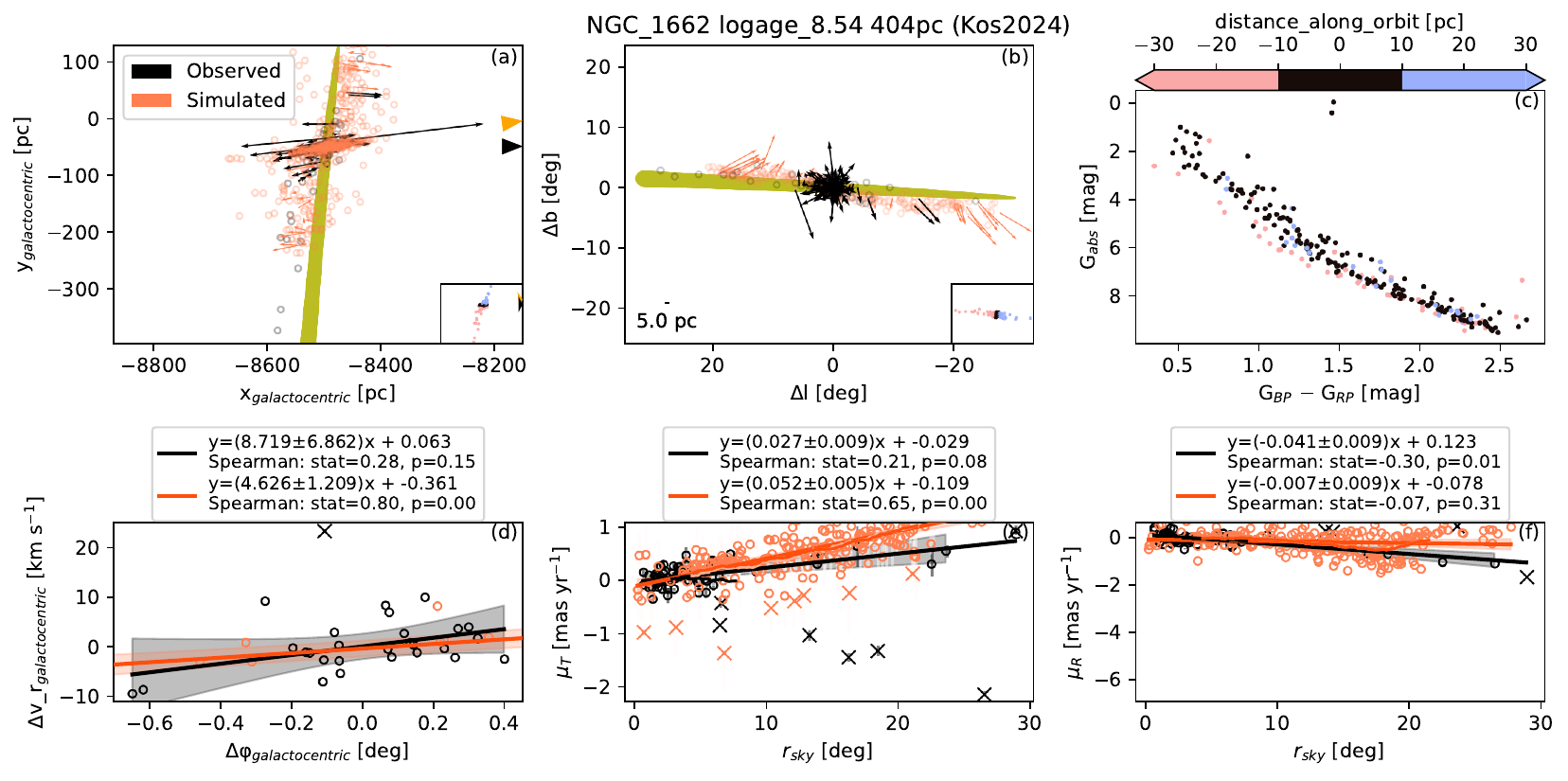}
\includegraphics[width=0.5\linewidth]{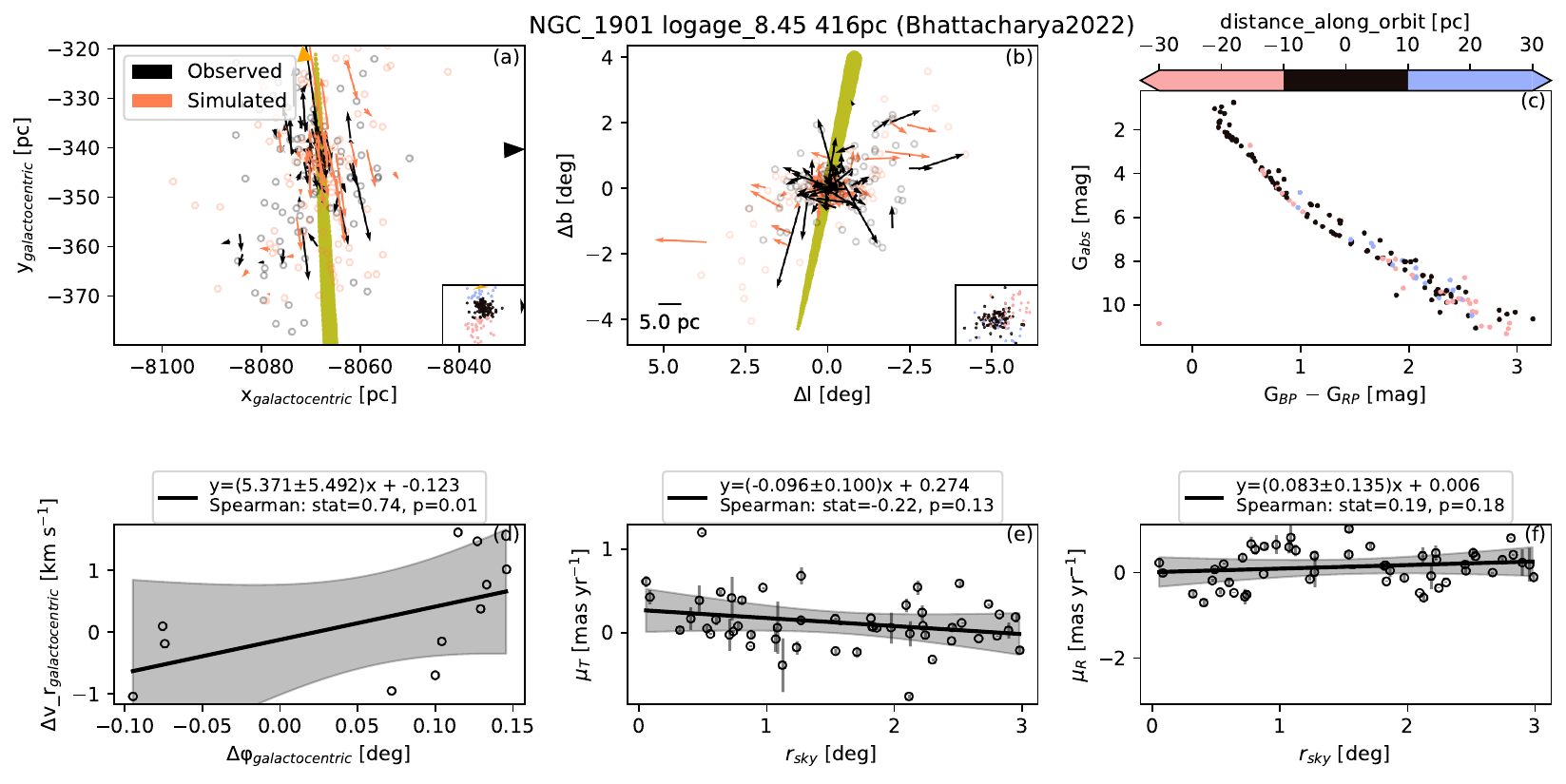}
\includegraphics[width=0.5\linewidth]{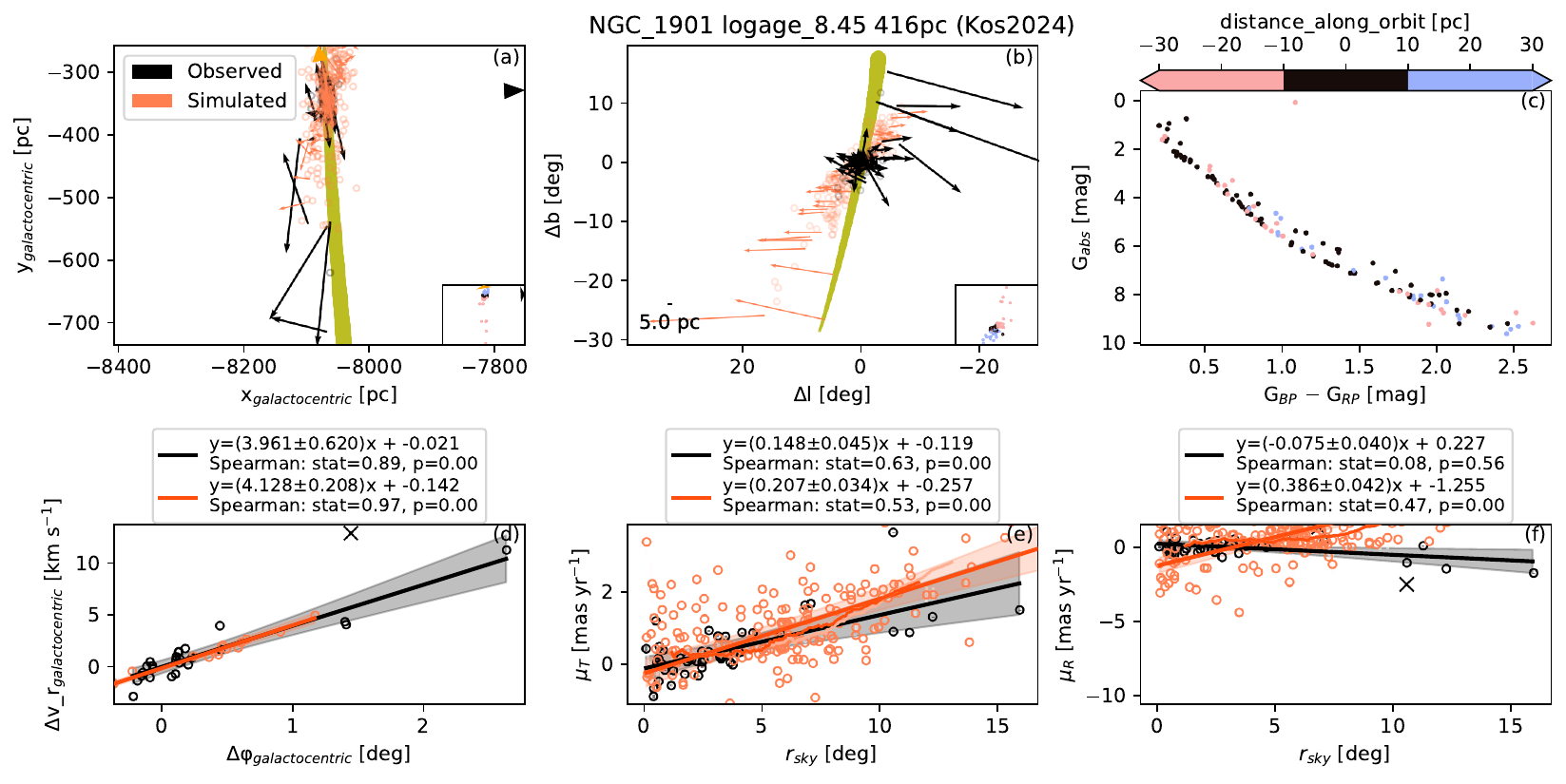}
    \caption{Diagnostic figures for NGC 1662 (Bhattacharya2022), NGC 1662 (Kos2024), NGC 1901 (Bhattacharya2022), NGC 1901 (Kos2024).}
        \label{fig:supplementary.NGC_1901.Kos2024}
        \end{figure}
         
        \begin{figure}
\includegraphics[width=0.5\linewidth]{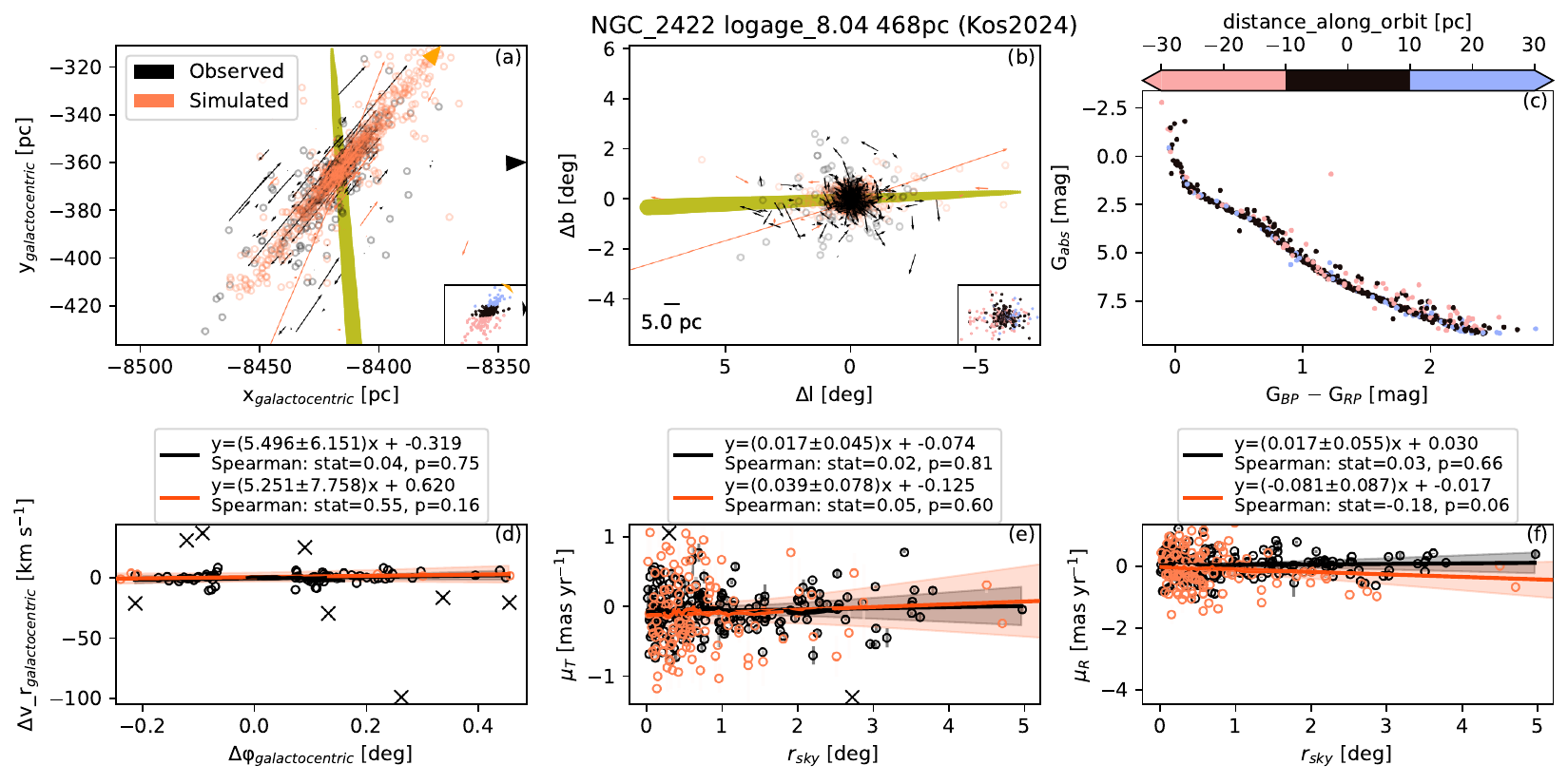}
\includegraphics[width=0.5\linewidth]{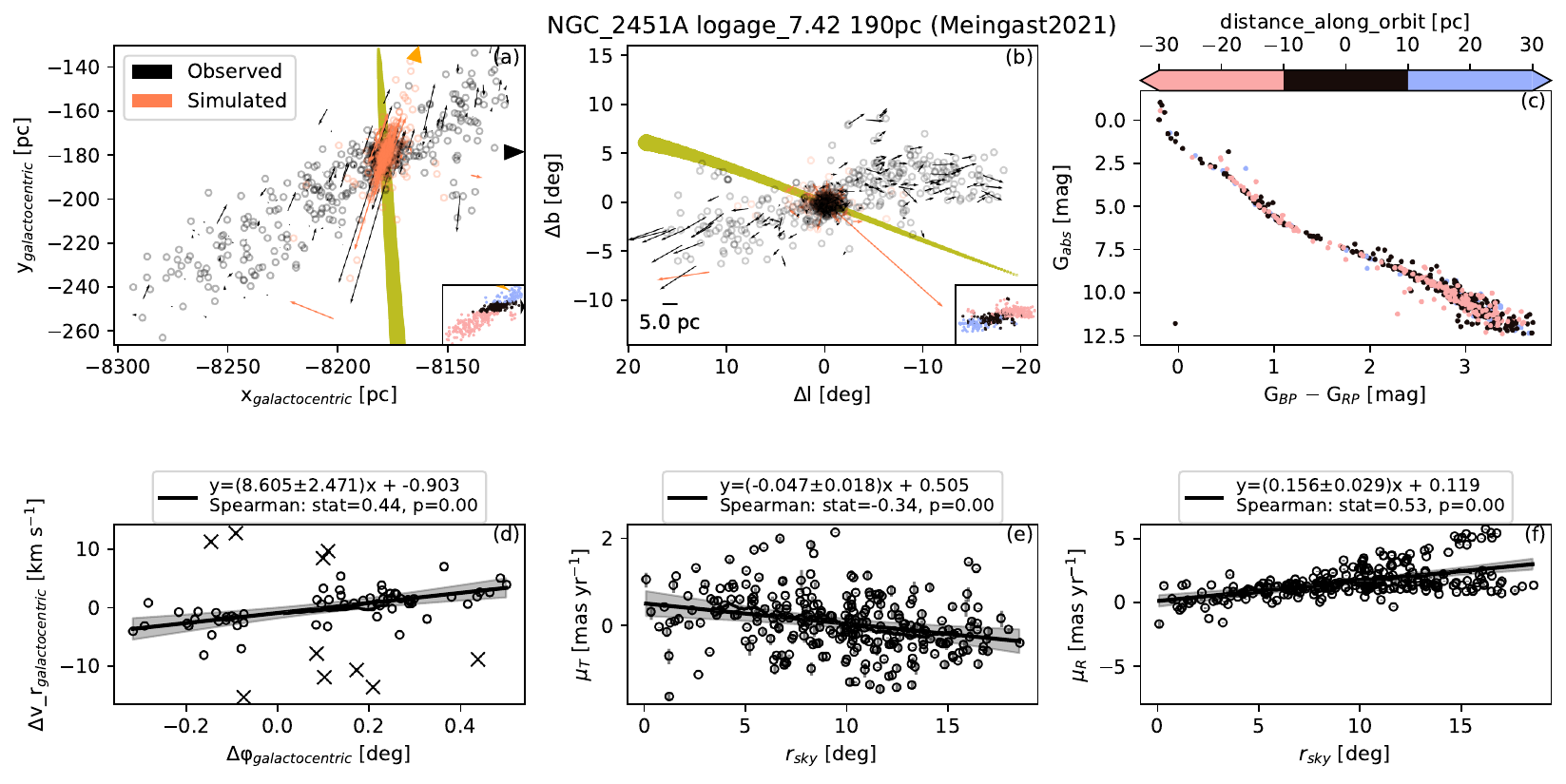}
\includegraphics[width=0.5\linewidth]{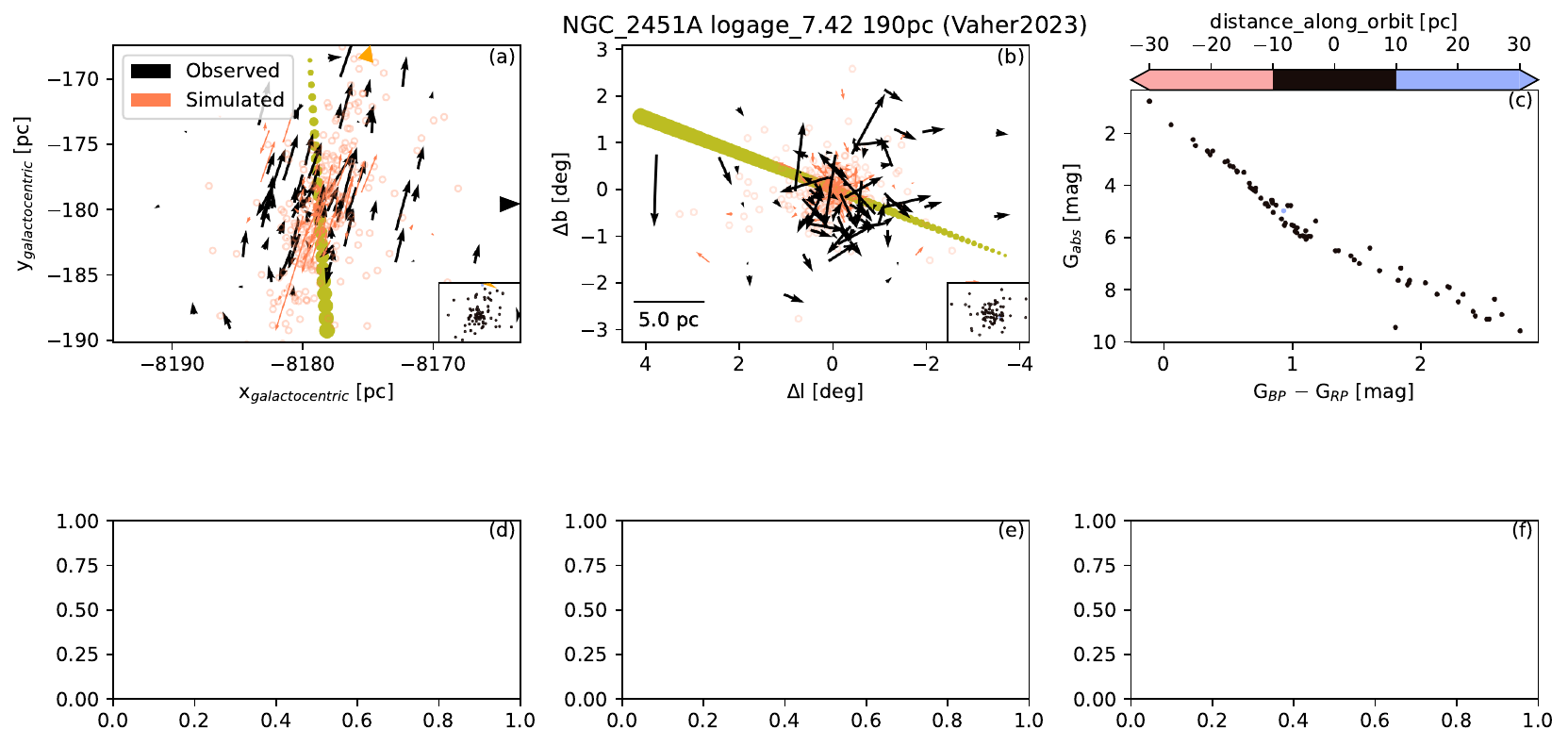}
\includegraphics[width=0.5\linewidth]{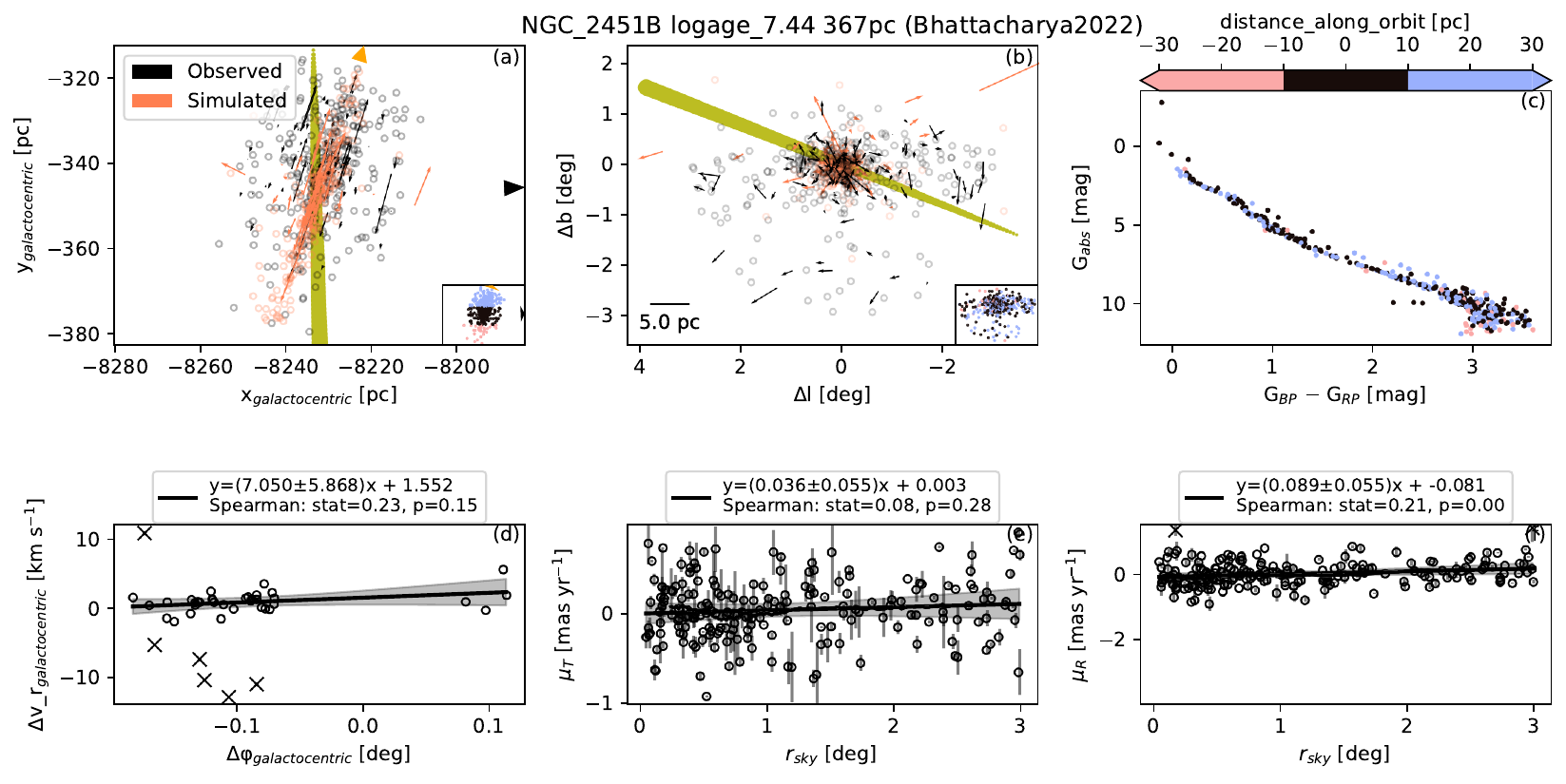}
    \caption{Diagnostic figures for NGC 2422 (Kos2024), NGC 2451A (Meingast2021), NGC 2451A (Vaher2023), NGC 2451B (Bhattacharya2022).}
        \label{fig:supplementary.NGC_2451B.Bhattacharya2022}
        \end{figure}
         
        \begin{figure}
\includegraphics[width=0.5\linewidth]{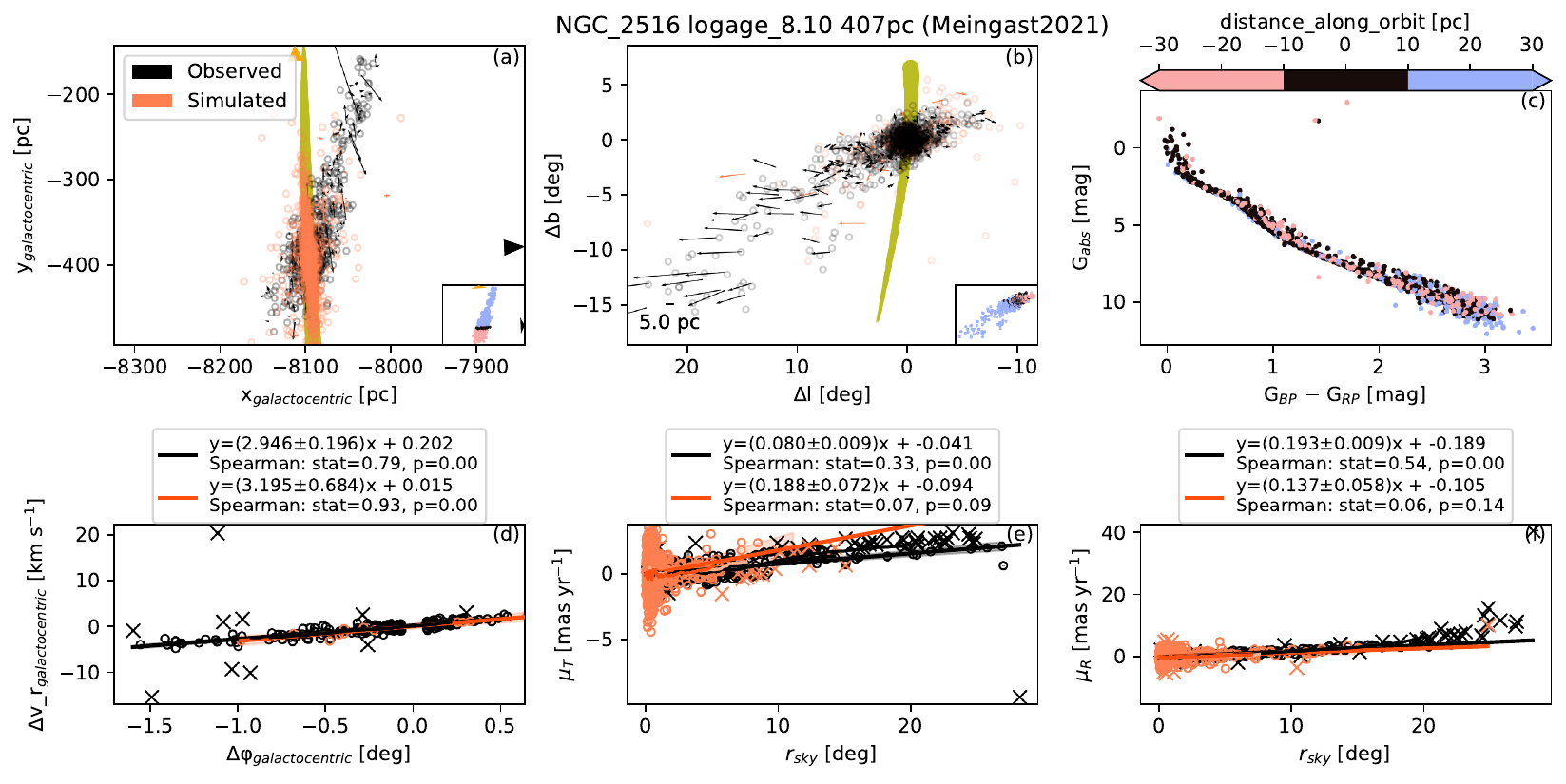}
\includegraphics[width=0.5\linewidth]{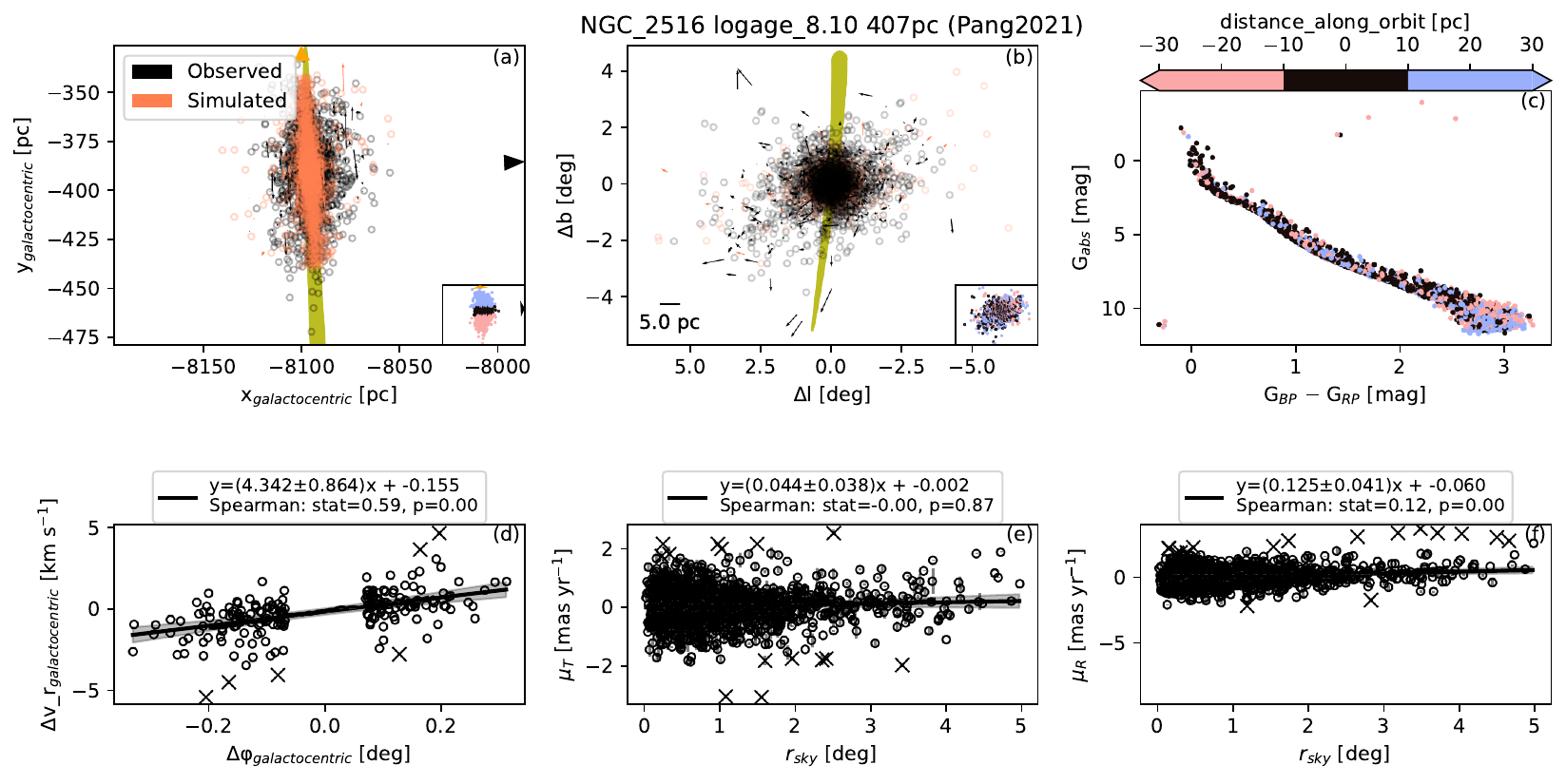}
\includegraphics[width=0.5\linewidth]{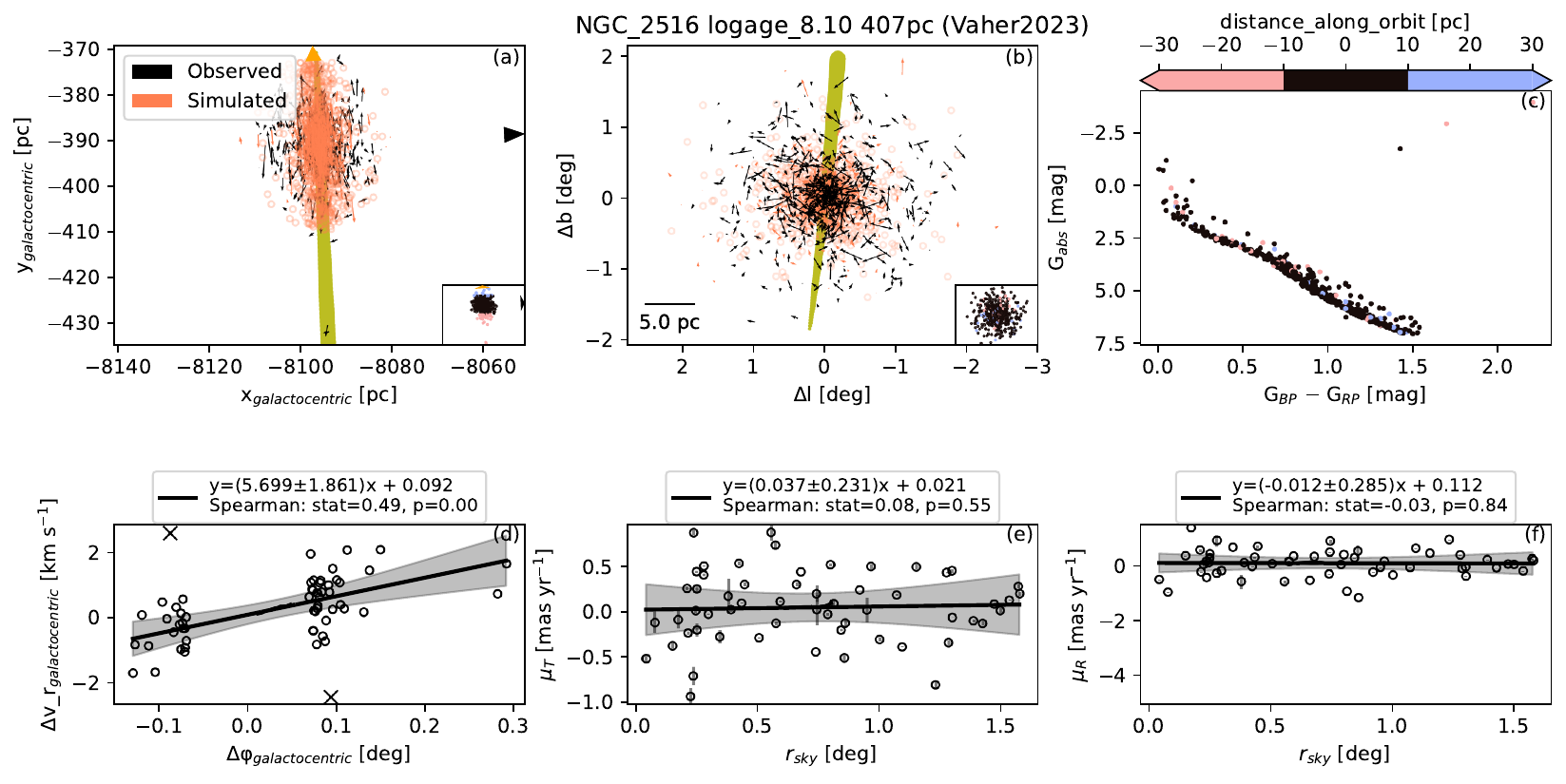}
\includegraphics[width=0.5\linewidth]{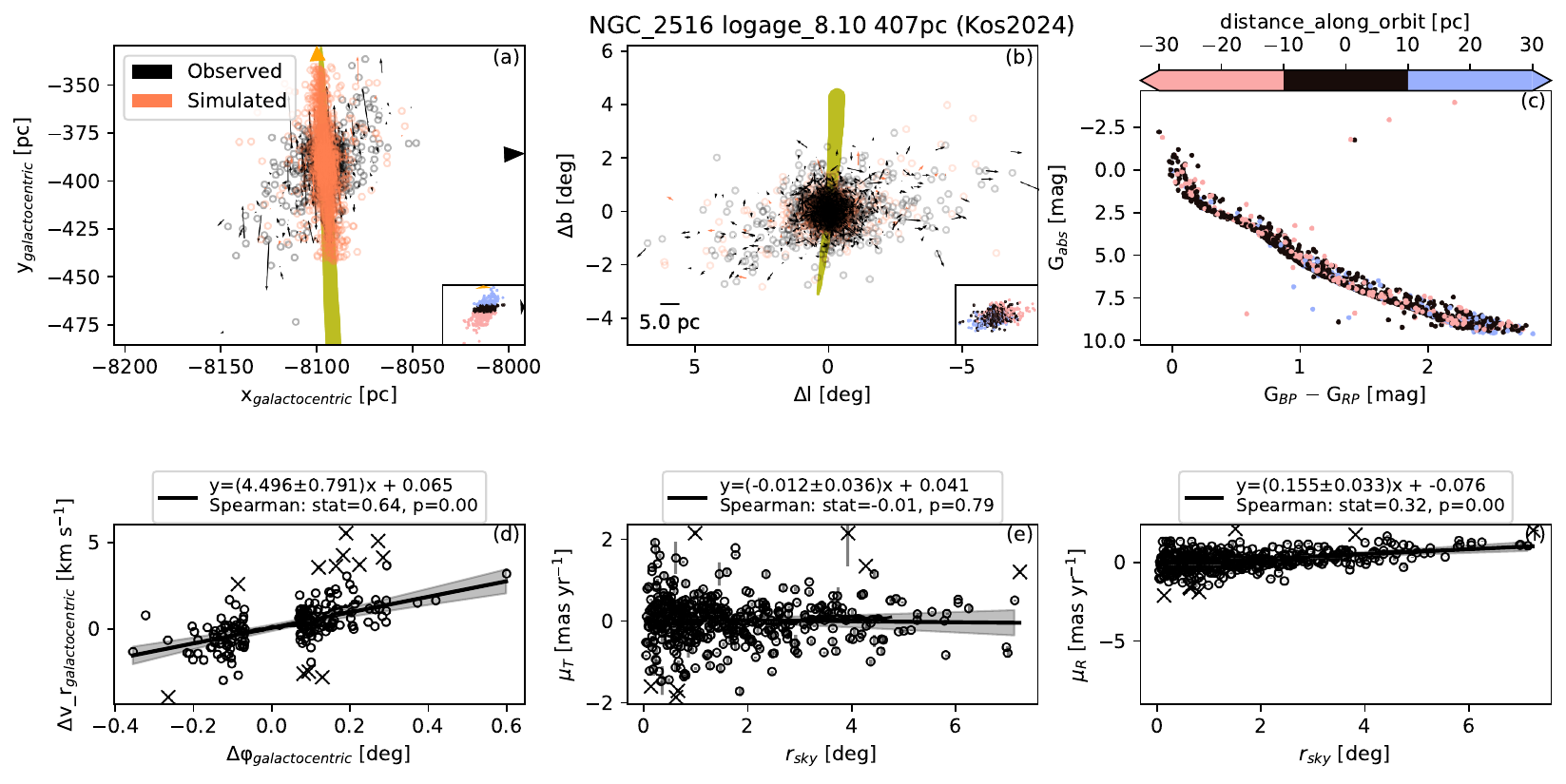}
    \caption{Diagnostic figures for NGC 2516 (Meingast2021), NGC 2516 (Pang2021), NGC 2516 (Vaher2023), NGC 2516 (Kos2024).}
        \label{fig:supplementary.NGC_2516.Kos2024}
        \end{figure}
         
        \begin{figure}
\includegraphics[width=0.5\linewidth]{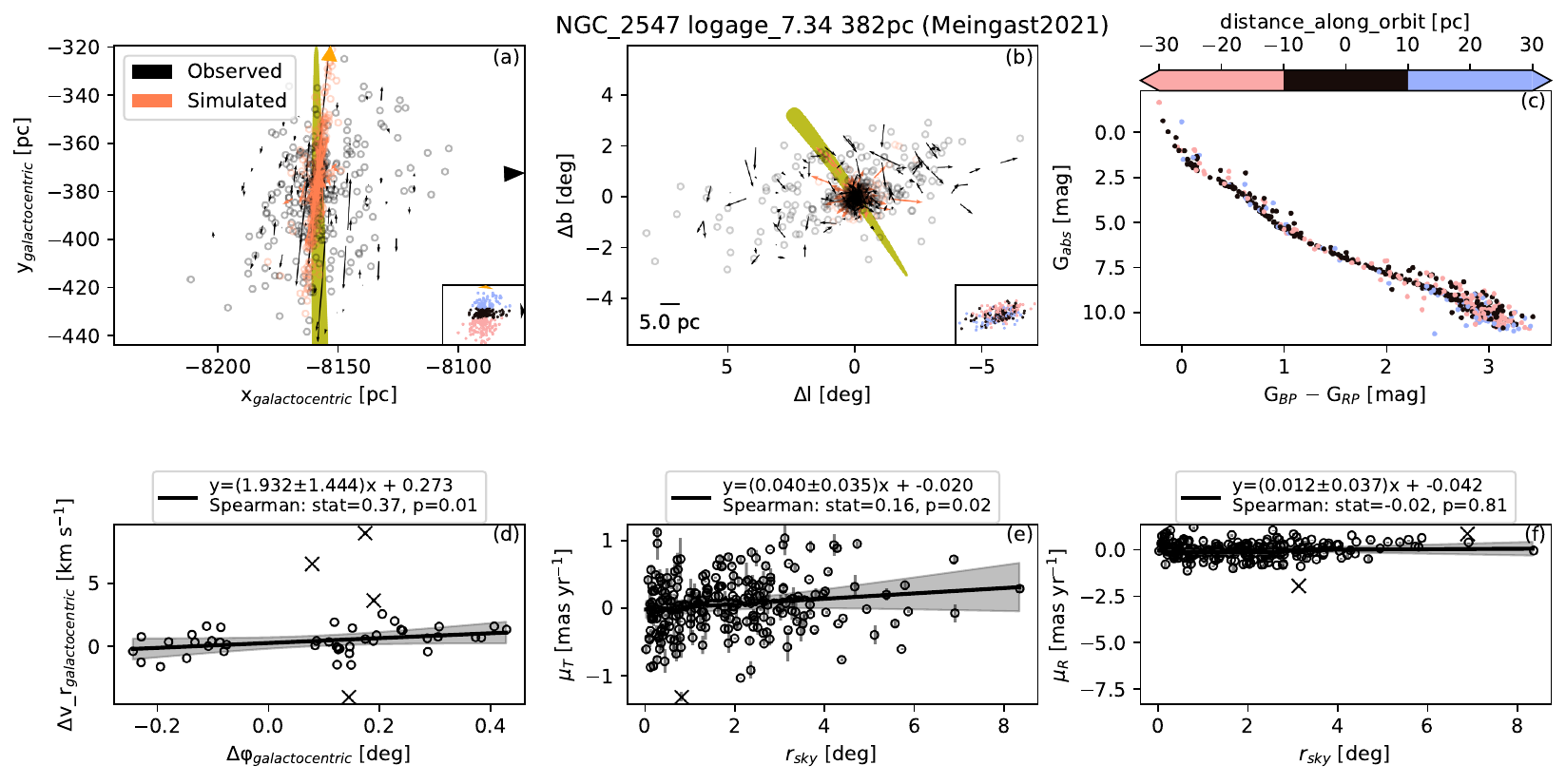}
\includegraphics[width=0.5\linewidth]{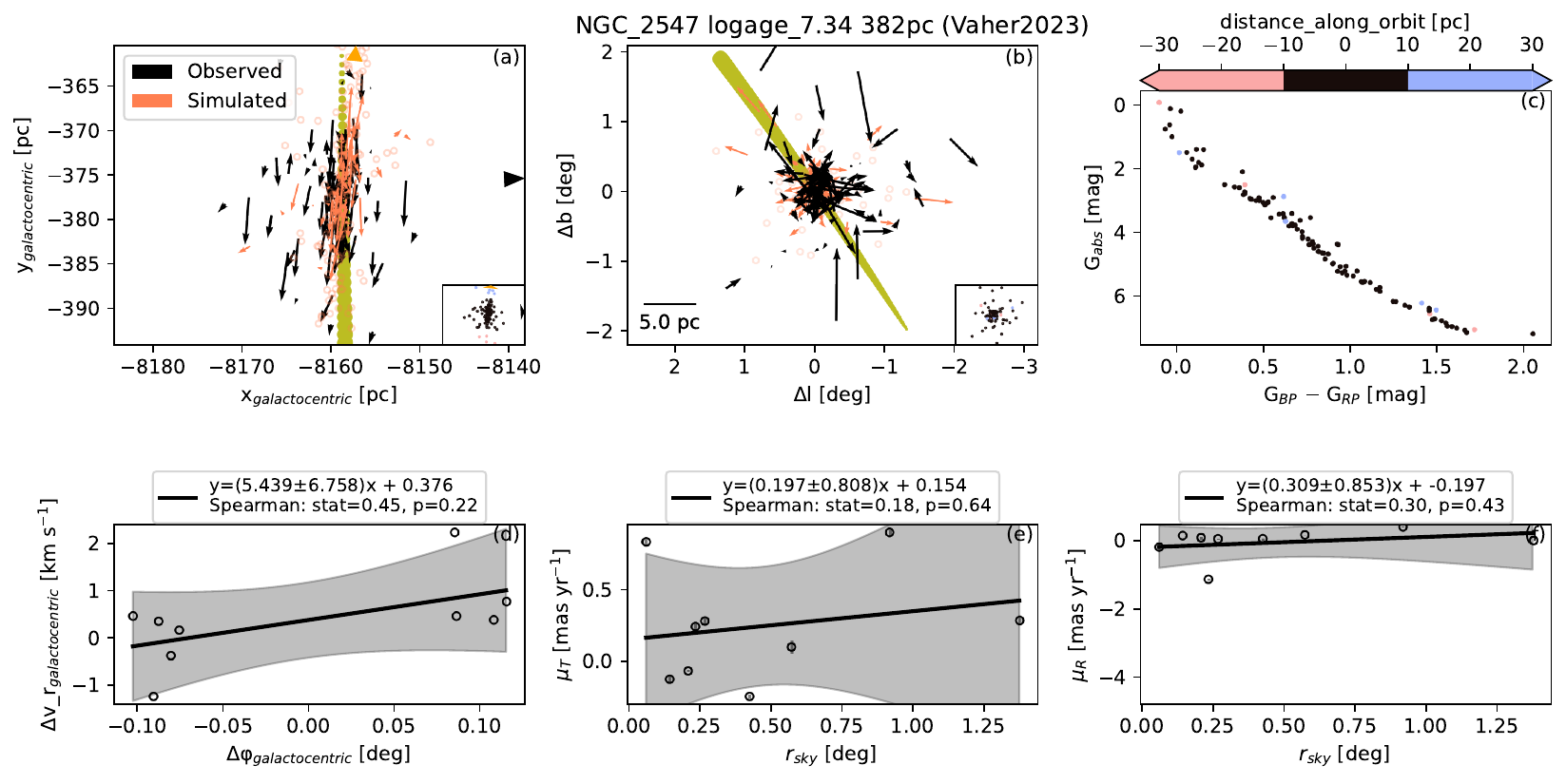}
\includegraphics[width=0.5\linewidth]{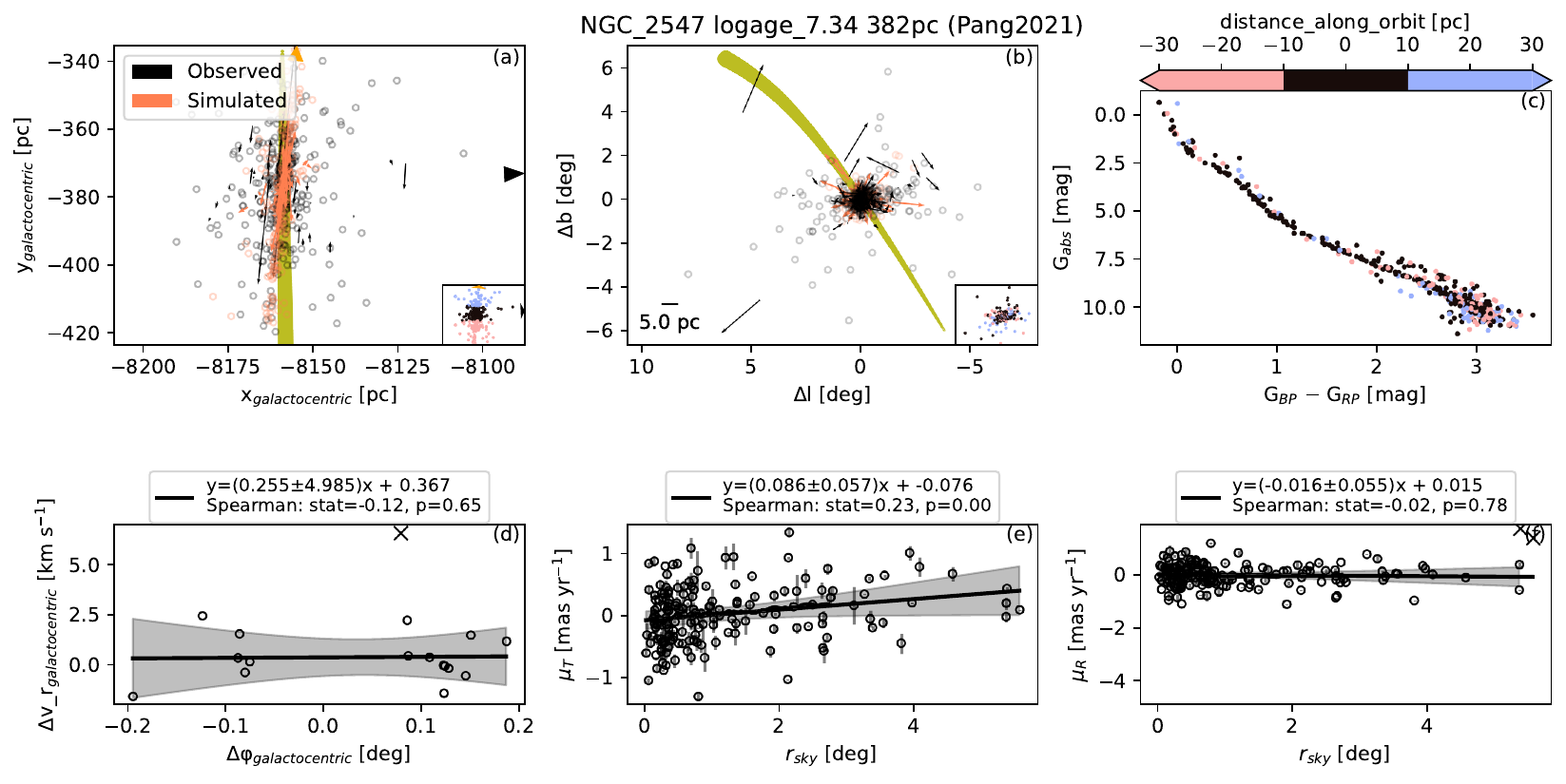}
\includegraphics[width=0.5\linewidth]{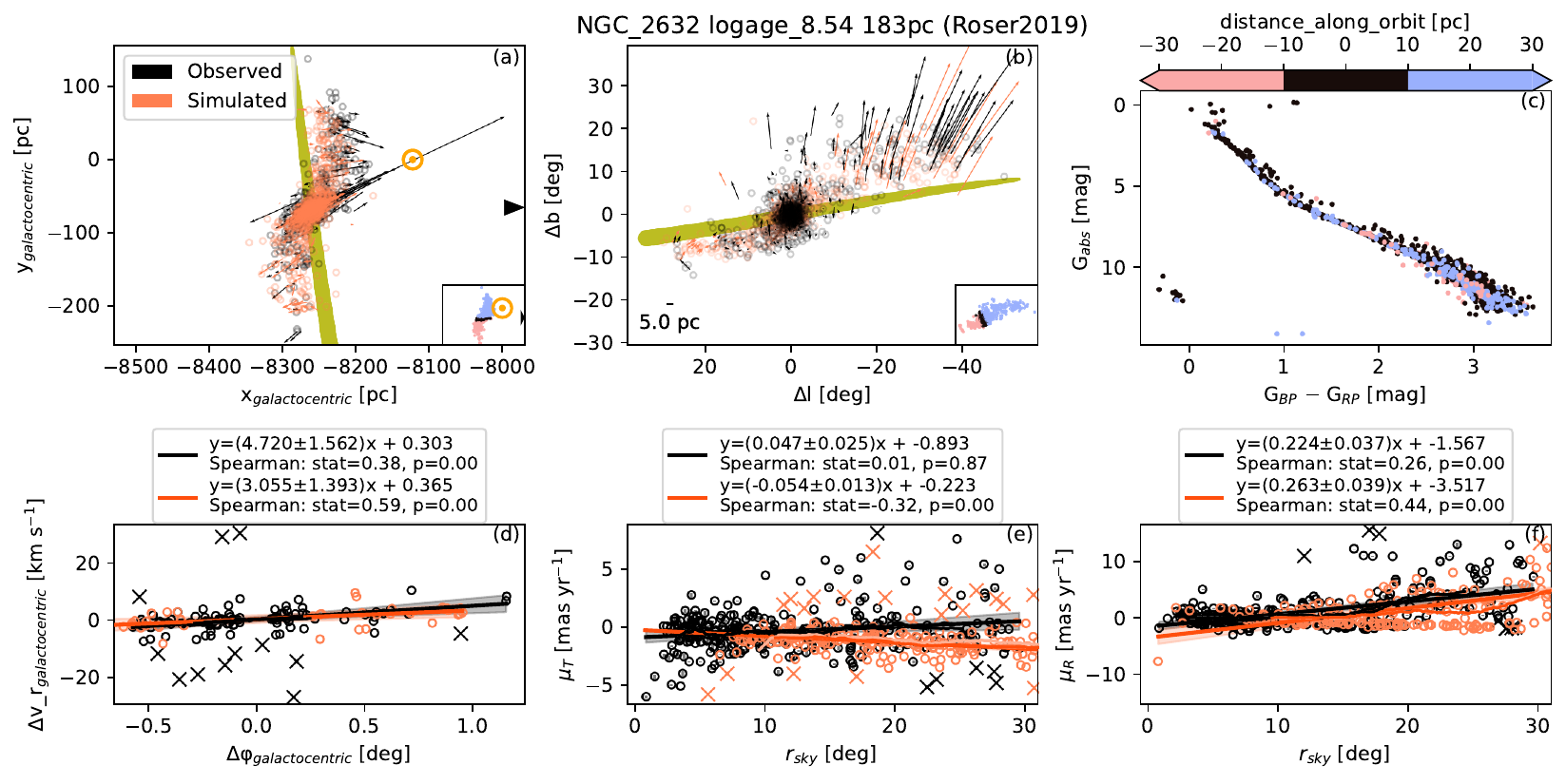}
    \caption{Diagnostic figures for NGC 2547 (Meingast2021), NGC 2547 (Vaher2023), NGC 2547 (Pang2021), NGC 2632 (Roser2019).}
        \label{fig:supplementary.NGC_2632.Roser2019}
        \end{figure}
         
        \begin{figure}
\includegraphics[width=0.5\linewidth]{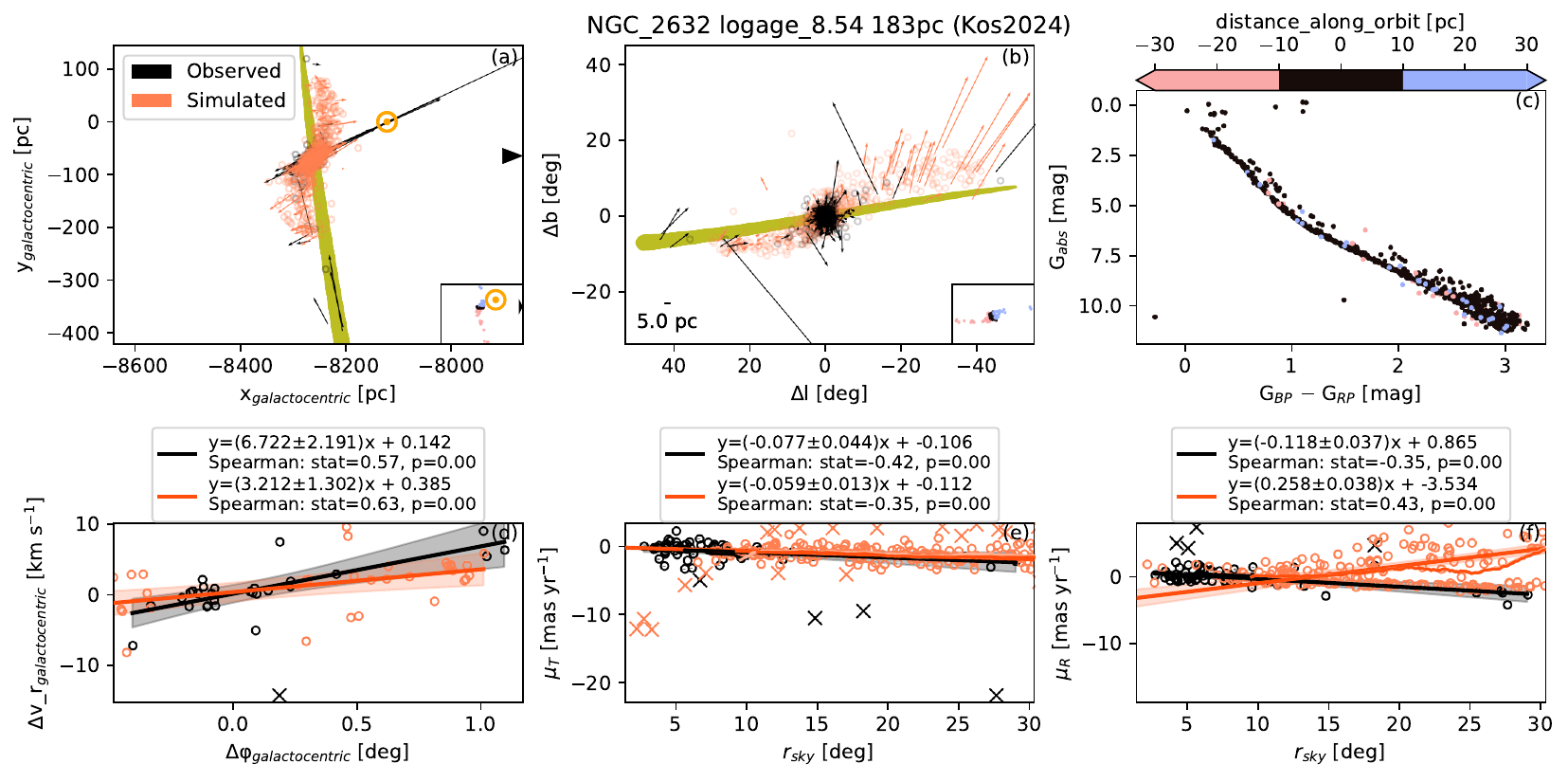}
\includegraphics[width=0.5\linewidth]{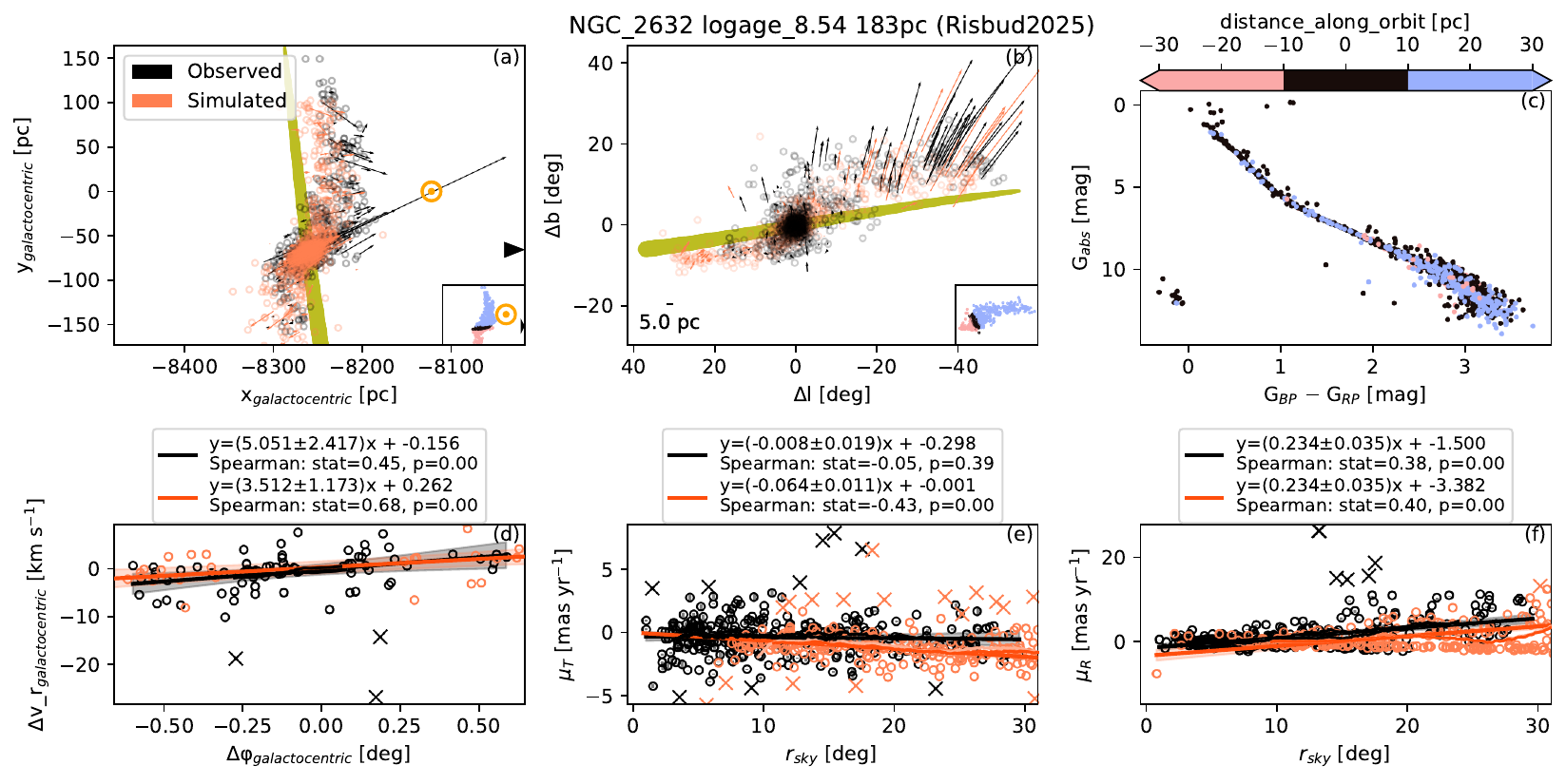}
\includegraphics[width=0.5\linewidth]{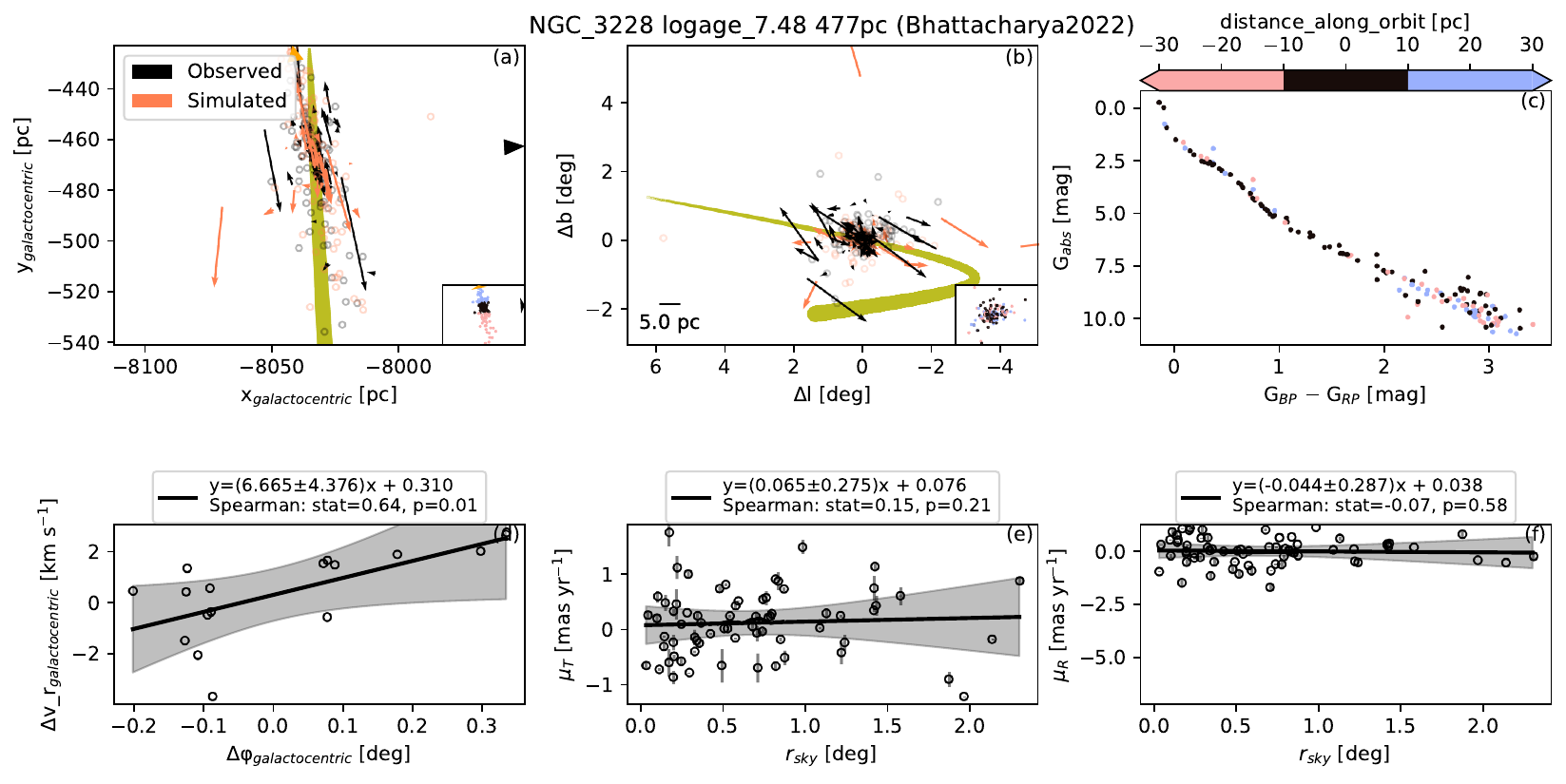}
\includegraphics[width=0.5\linewidth]{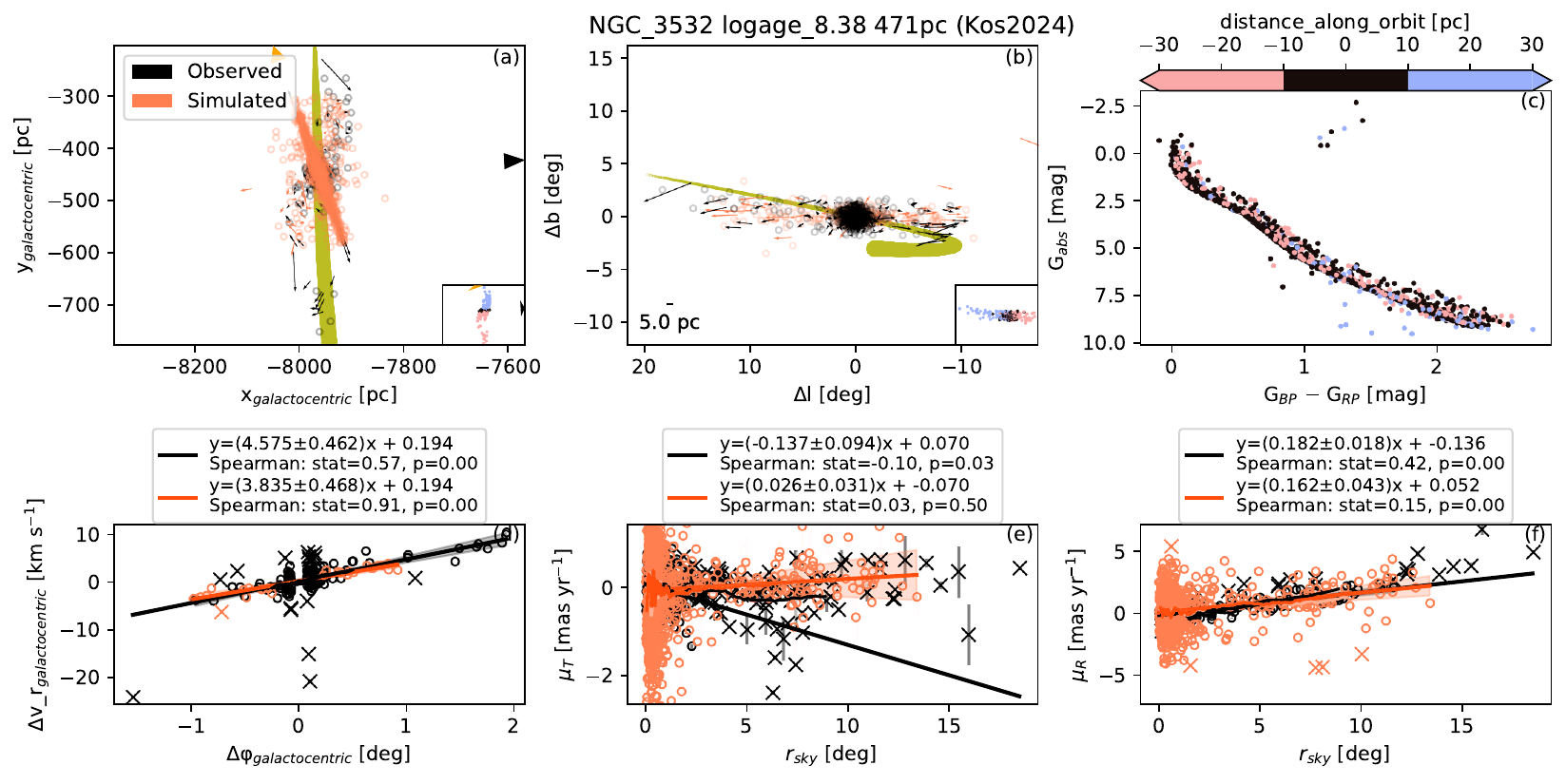}
    \caption{Diagnostic figures for NGC 2632 (Kos2024), NGC 2632 (Risbud2025), NGC 3228 (Bhattacharya2022), NGC 3532 (Kos2024).}
        \label{fig:supplementary.NGC_3532.Kos2024}
        \end{figure}
         
        \begin{figure}
\includegraphics[width=0.5\linewidth]{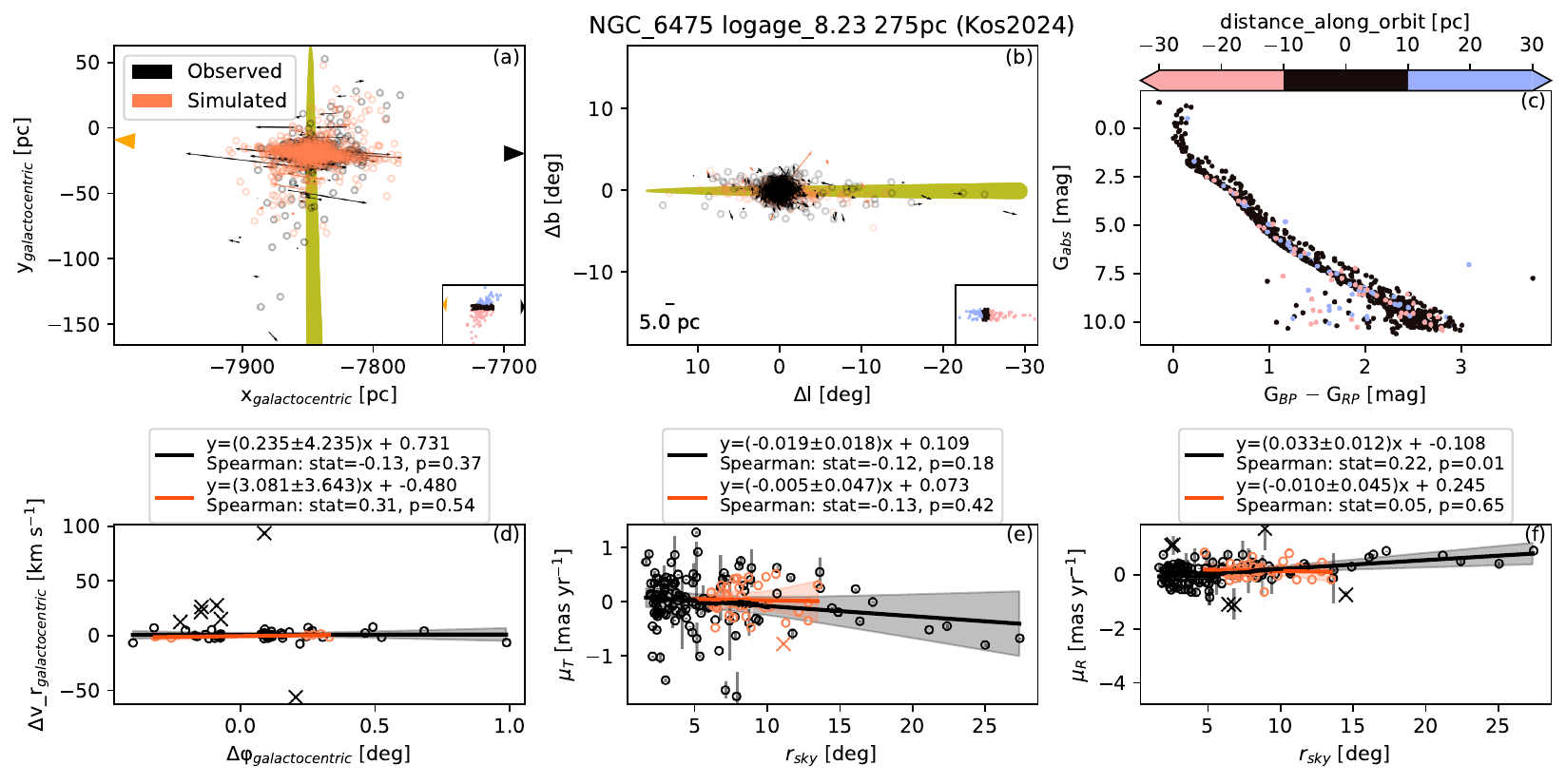}
\includegraphics[width=0.5\linewidth]{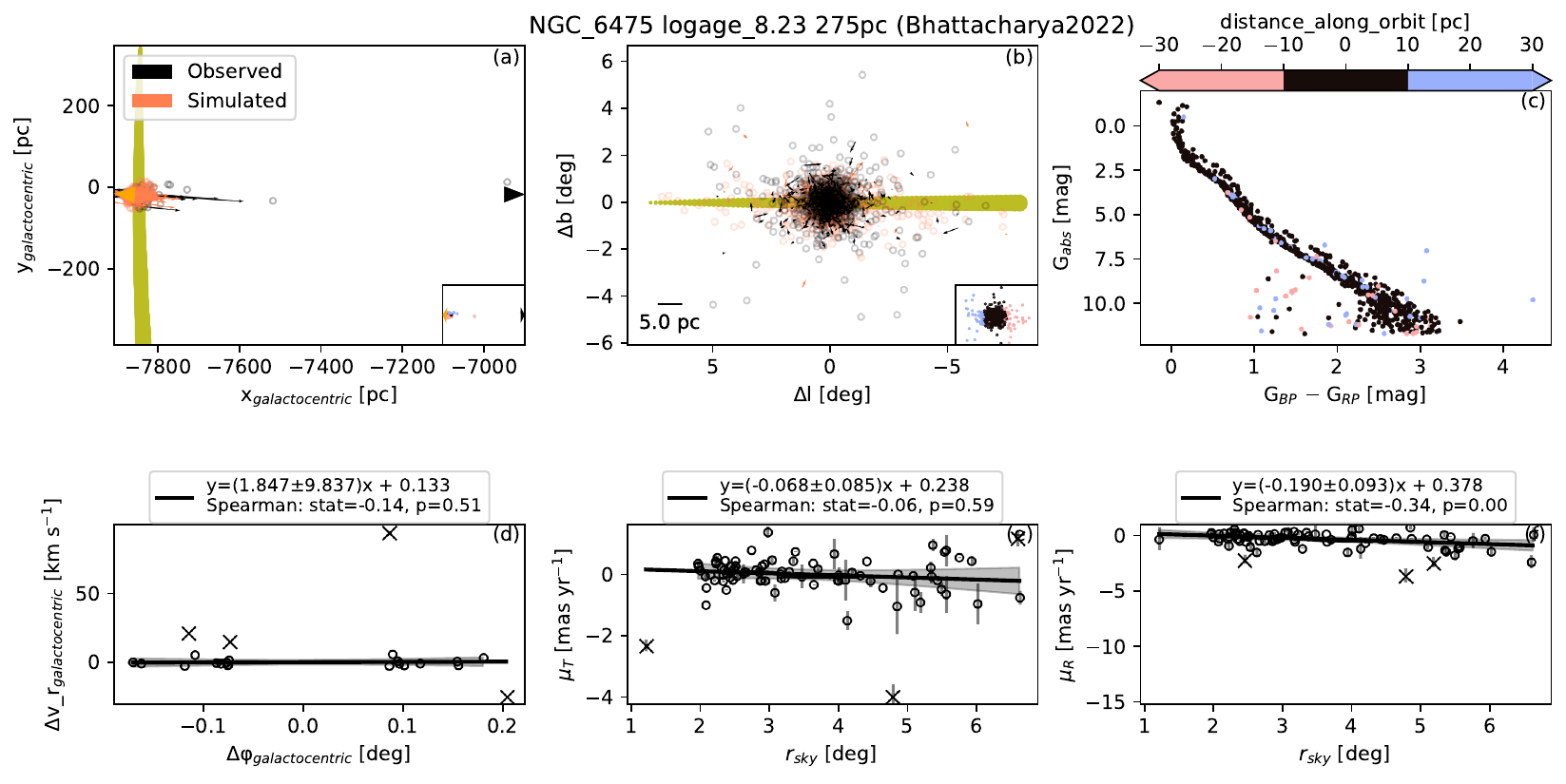}
\includegraphics[width=0.5\linewidth]{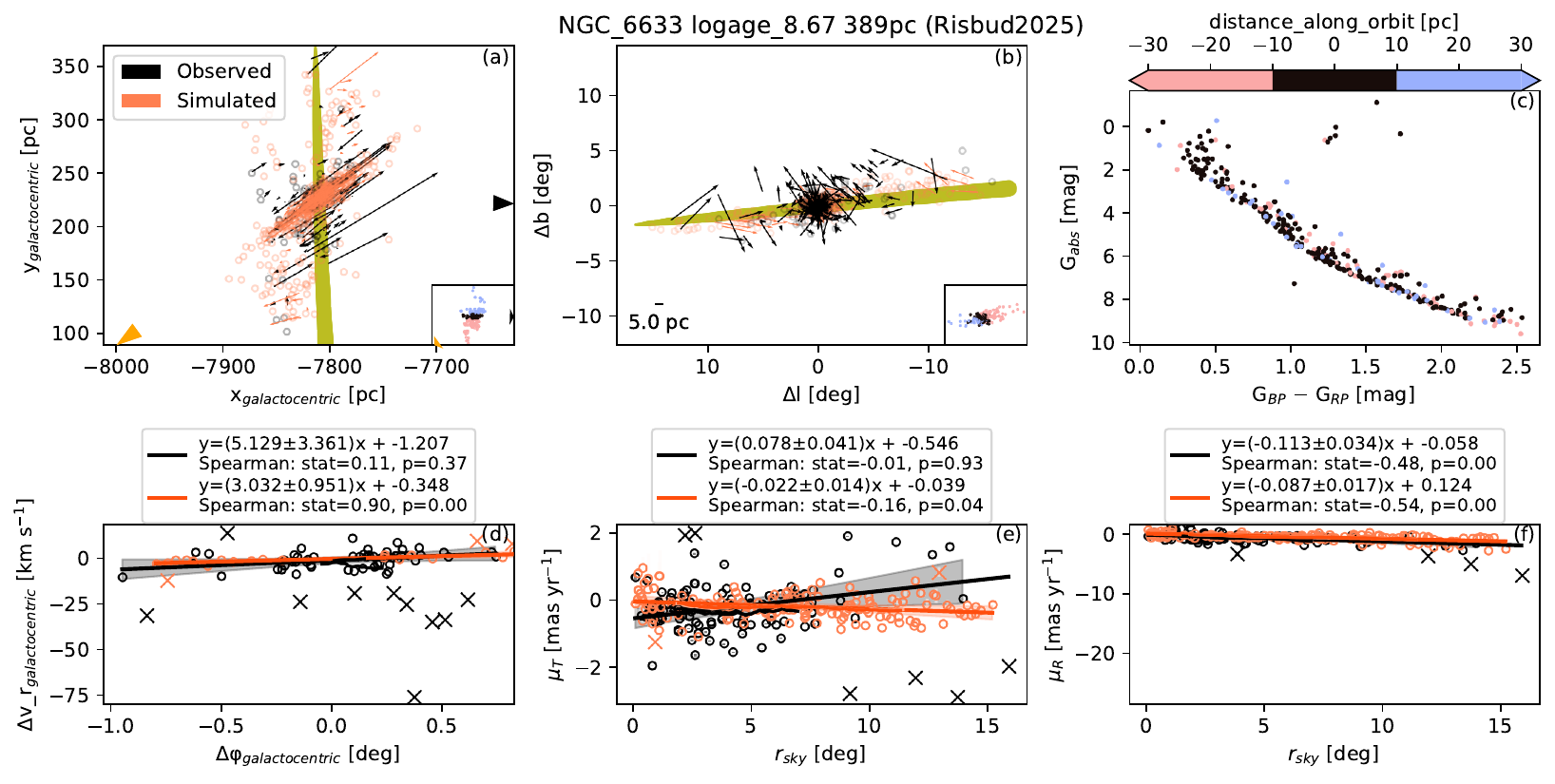}
\includegraphics[width=0.5\linewidth]{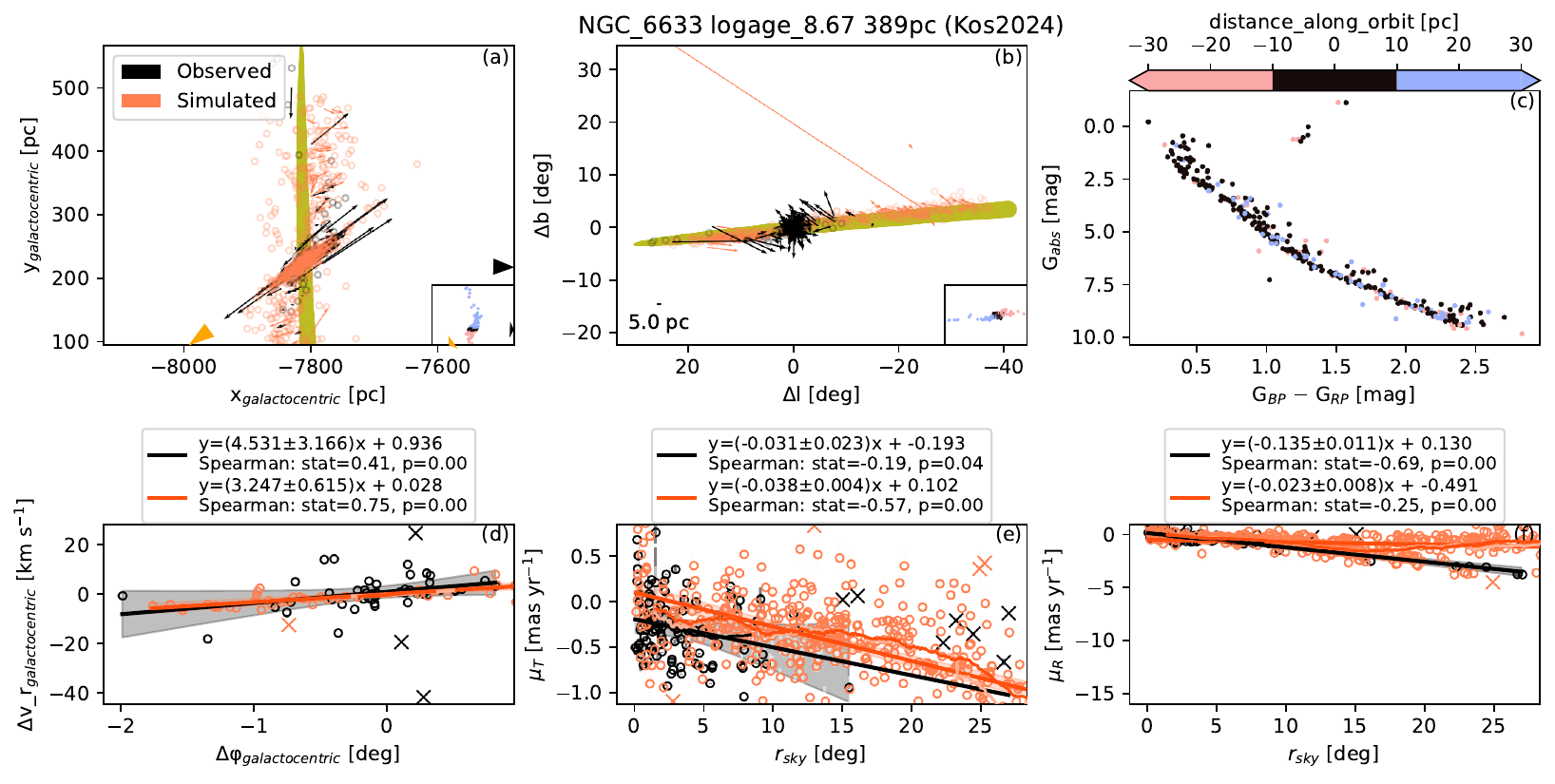}
    \caption{Diagnostic figures for NGC 6475 (Kos2024), NGC 6475 (Bhattacharya2022), NGC 6633 (Risbud2025), NGC 6633 (Kos2024).}
        \label{fig:supplementary.NGC_6633.Kos2024}
        \end{figure}
         
        \begin{figure}
\includegraphics[width=0.5\linewidth]{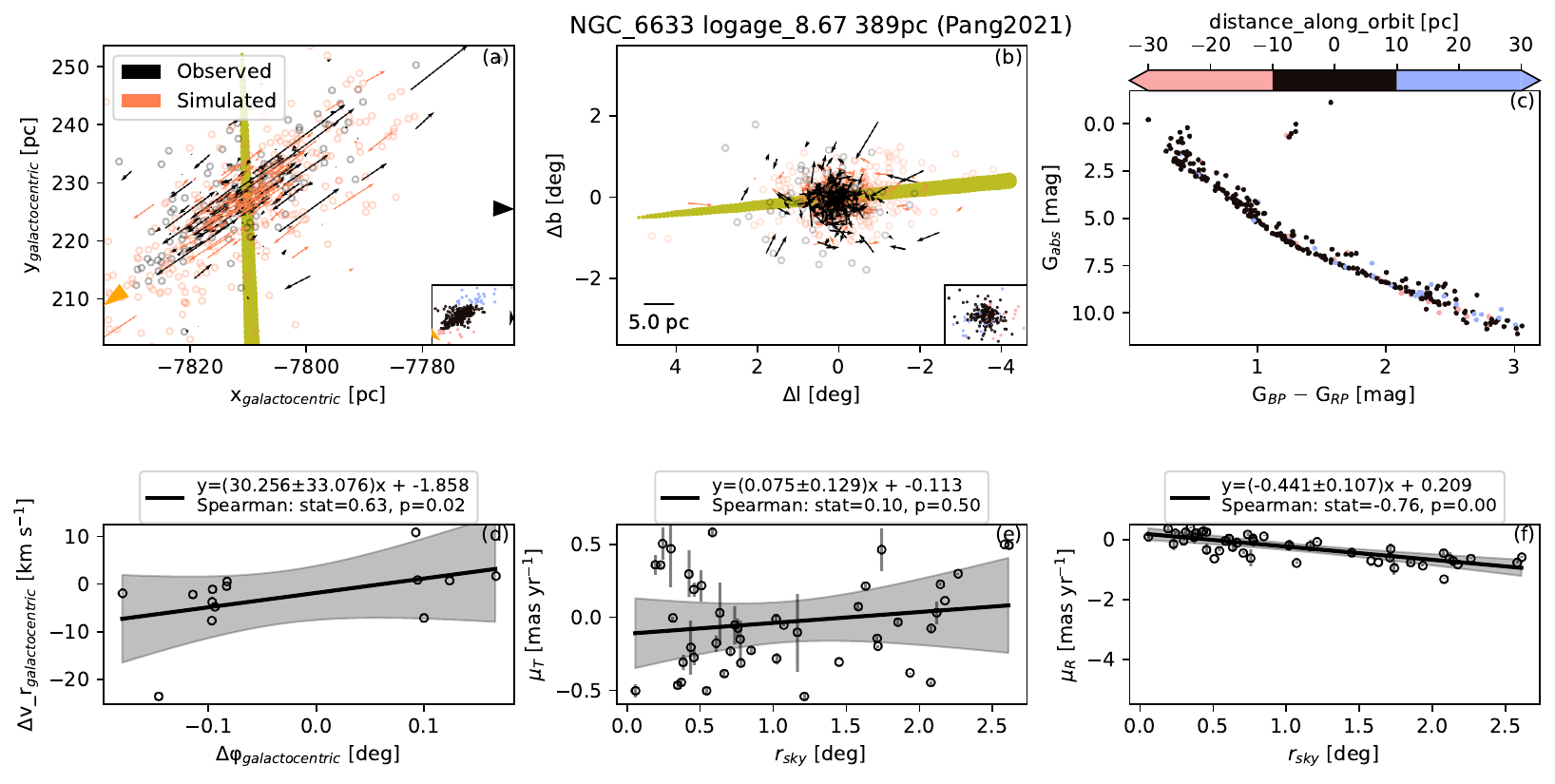}
\includegraphics[width=0.5\linewidth]{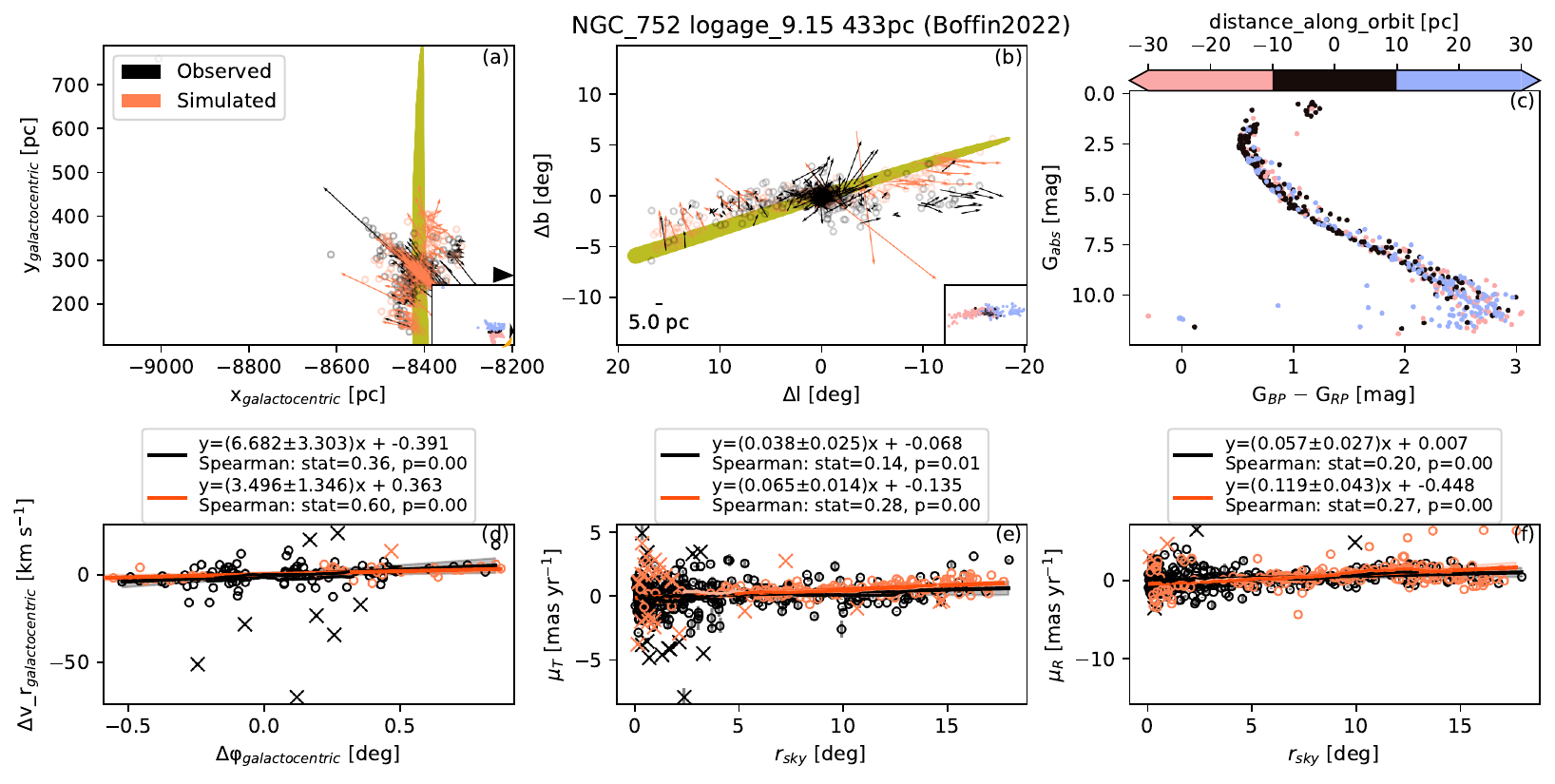}
\includegraphics[width=0.5\linewidth]{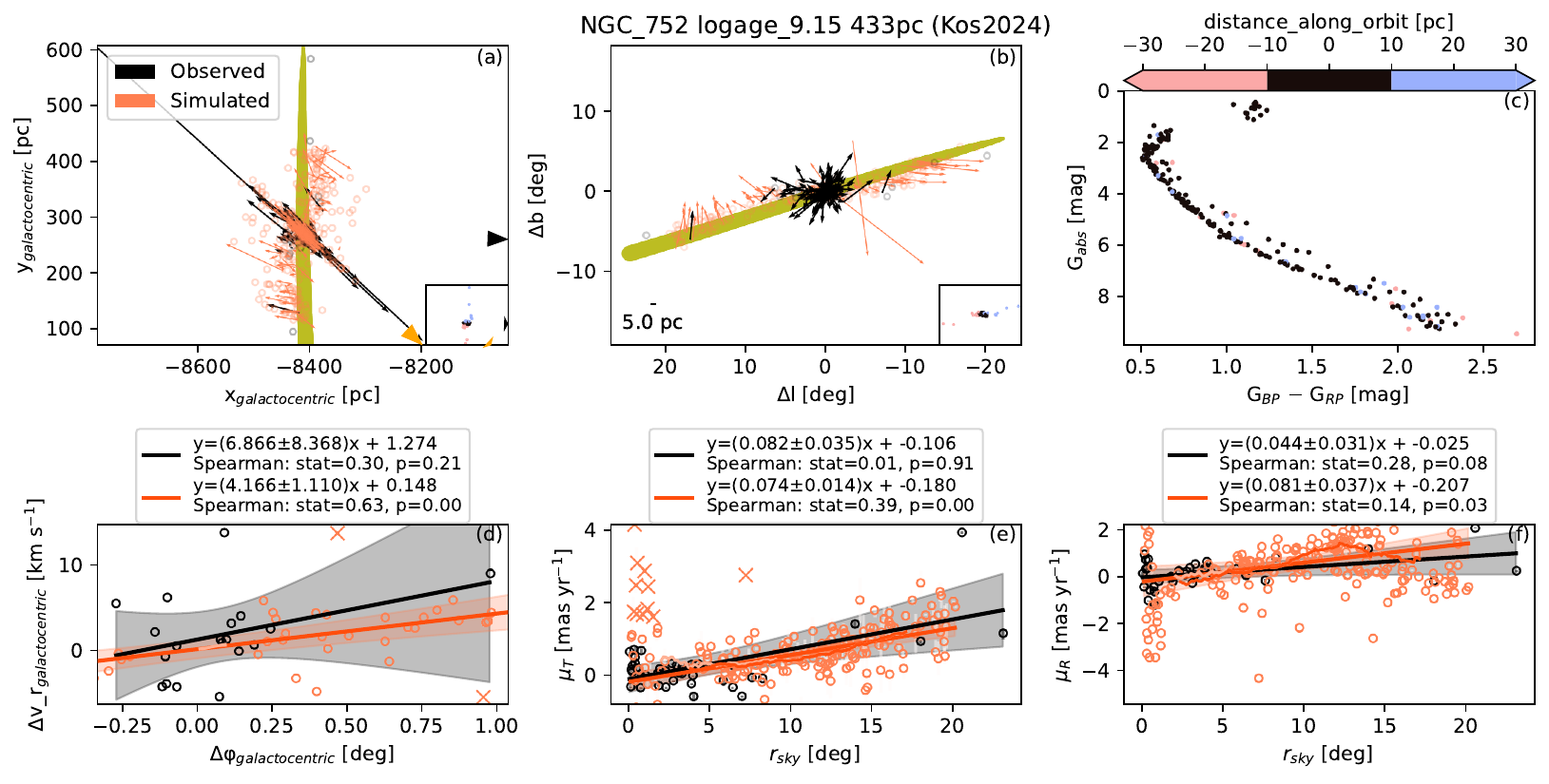}
\includegraphics[width=0.5\linewidth]{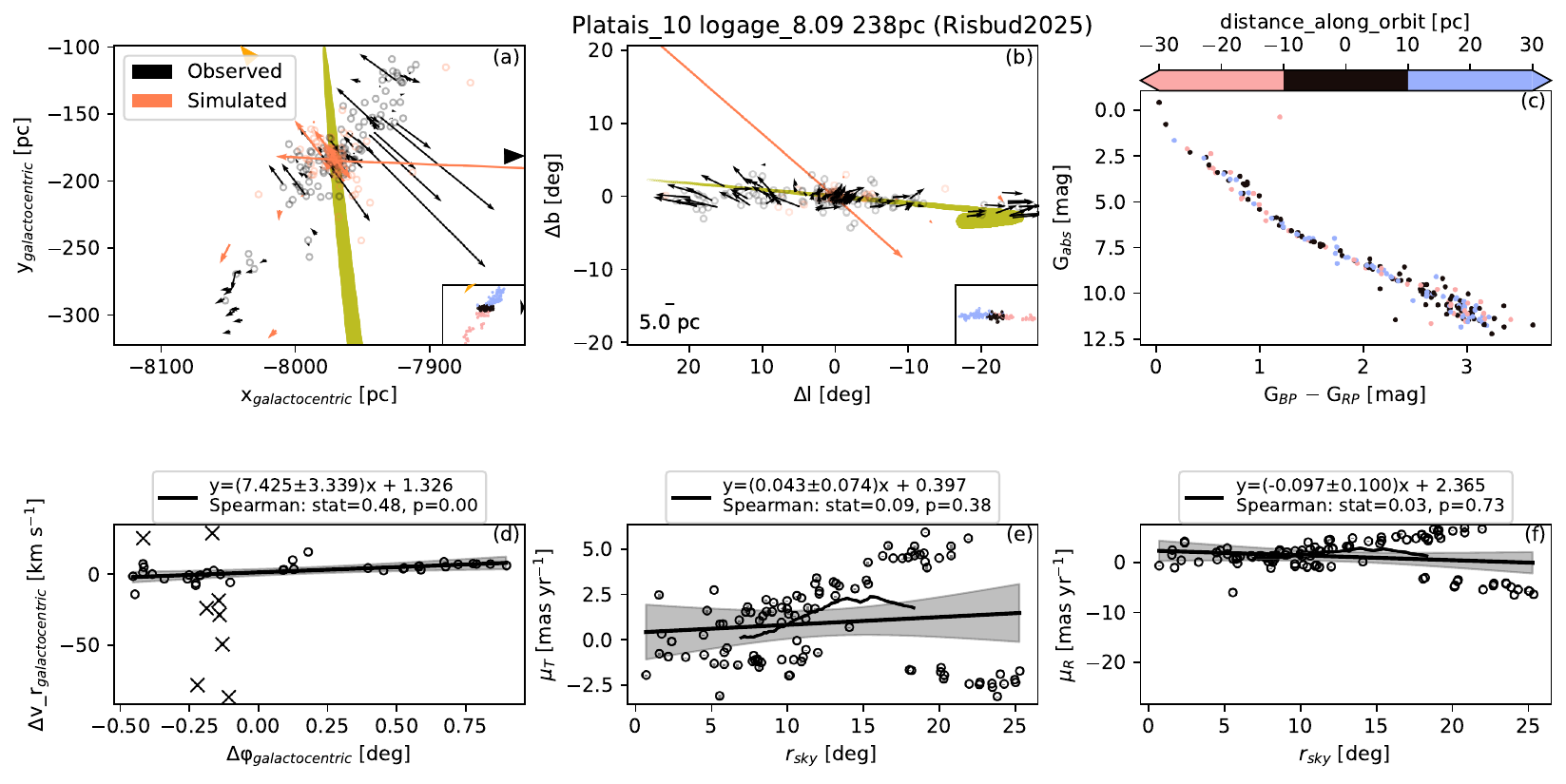}
    \caption{Diagnostic figures for NGC 6633 (Pang2021), NGC 752 (Boffin2022), NGC 752 (Kos2024), Platais 10 (Risbud2025).}
        \label{fig:supplementary.Platais_10.Risbud2025}
        \end{figure}
         
        \begin{figure}
\includegraphics[width=0.5\linewidth]{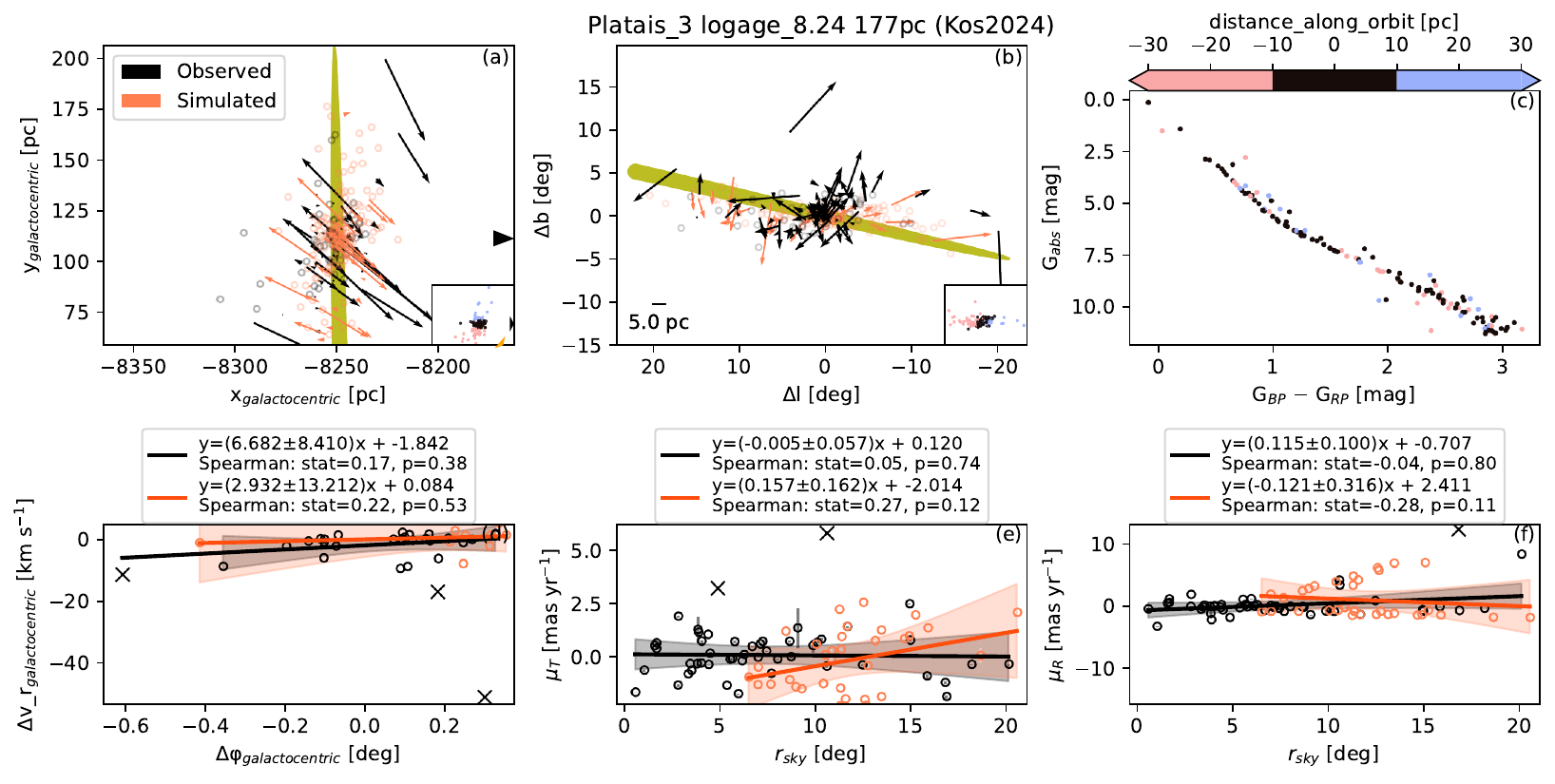}
\includegraphics[width=0.5\linewidth]{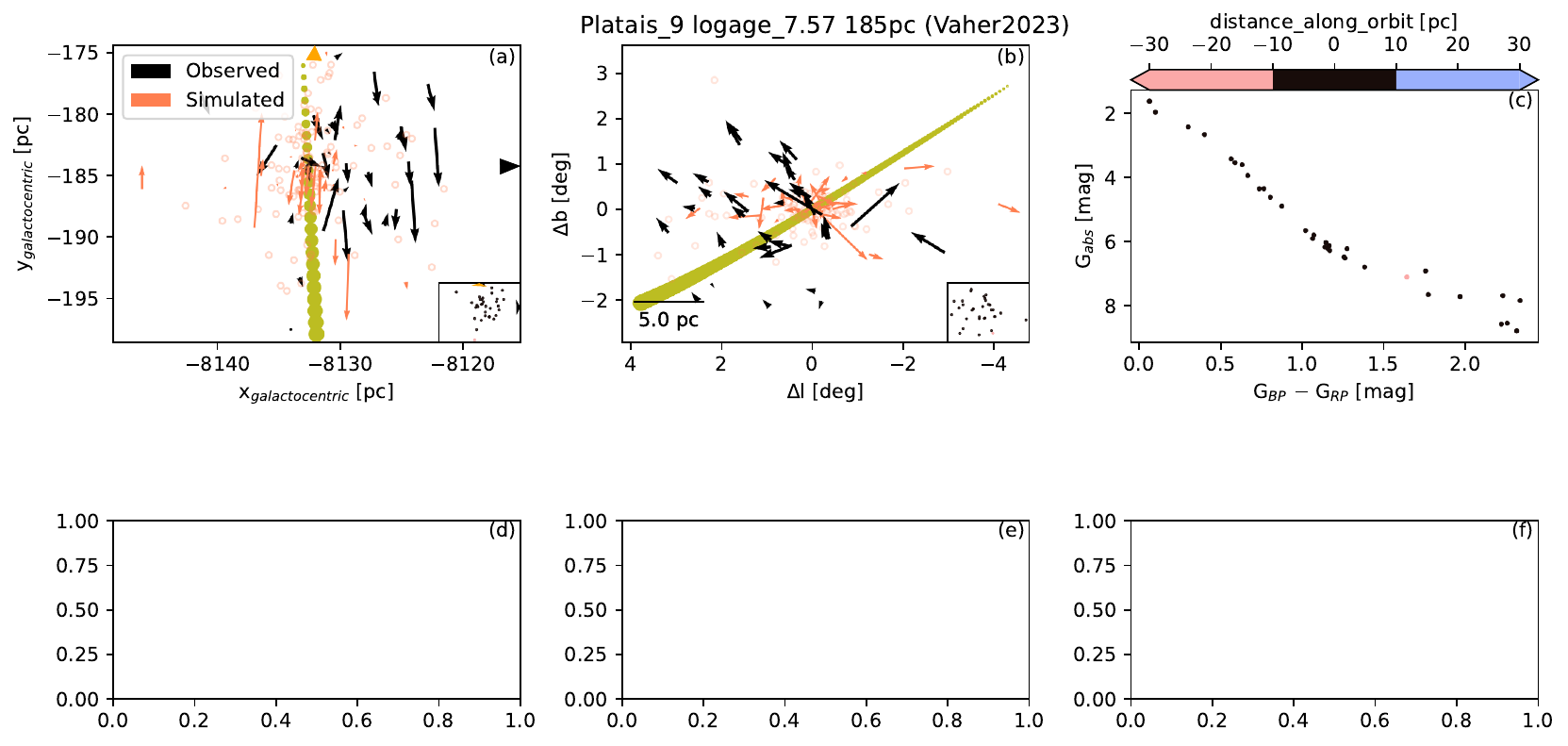}
\includegraphics[width=0.5\linewidth]{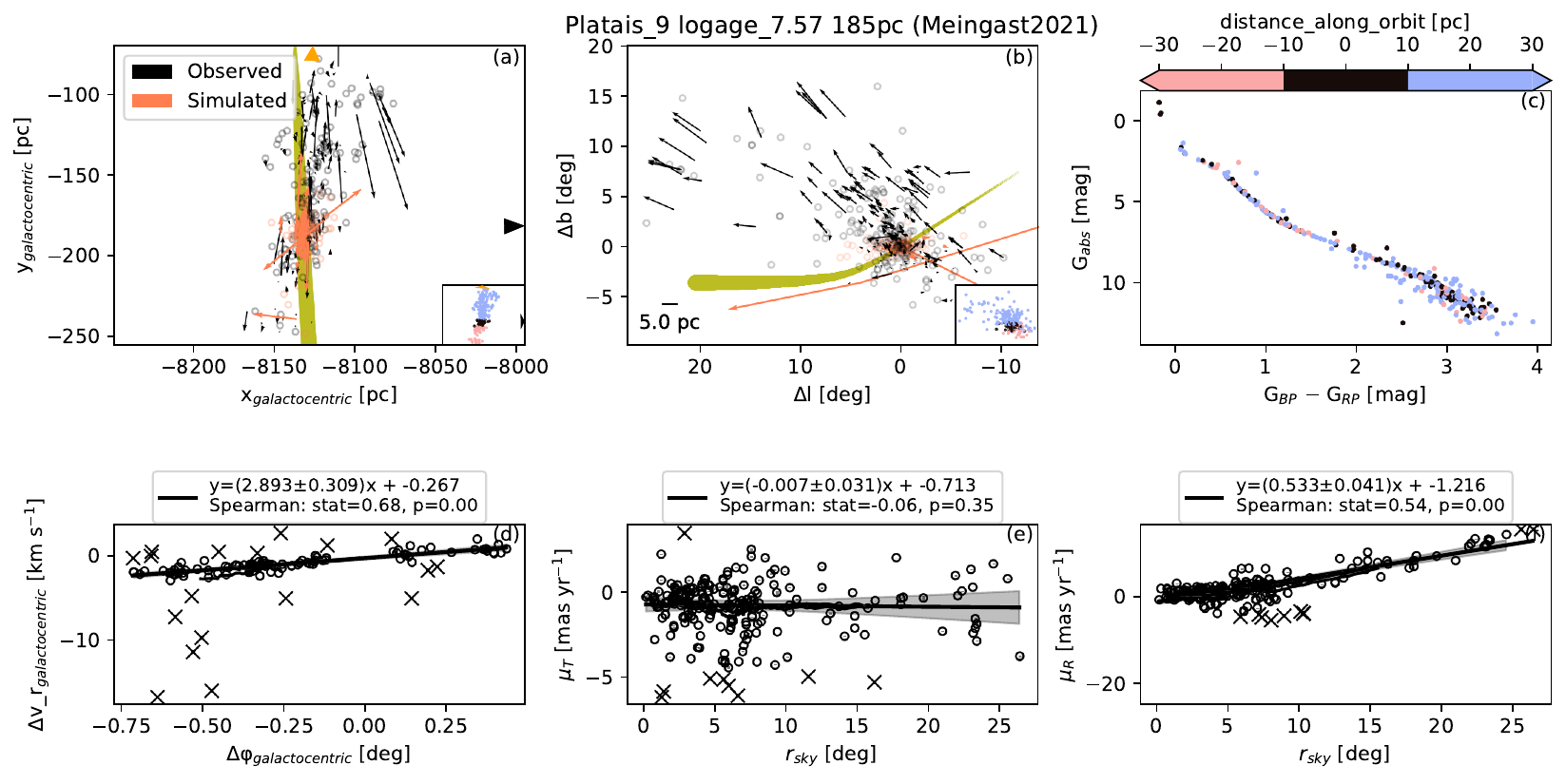}
\includegraphics[width=0.5\linewidth]{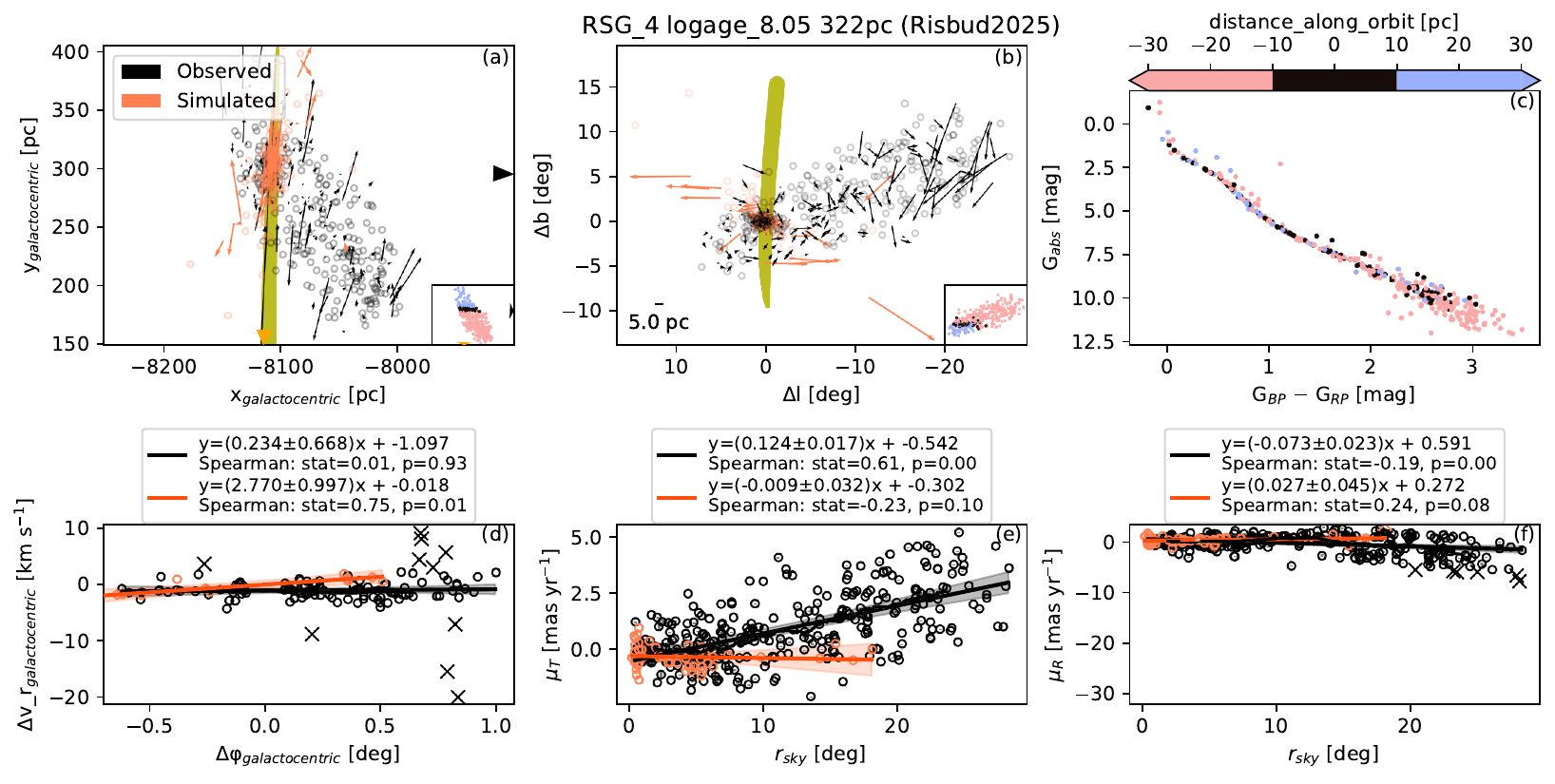}
    \caption{Diagnostic figures for Platais 3 (Kos2024), Platais 9 (Vaher2023), Platais 9 (Meingast2021), RSG 4 (Risbud2025).}
        \label{fig:supplementary.RSG_4.Risbud2025}
        \end{figure}
         
        \begin{figure}
\includegraphics[width=0.5\linewidth]{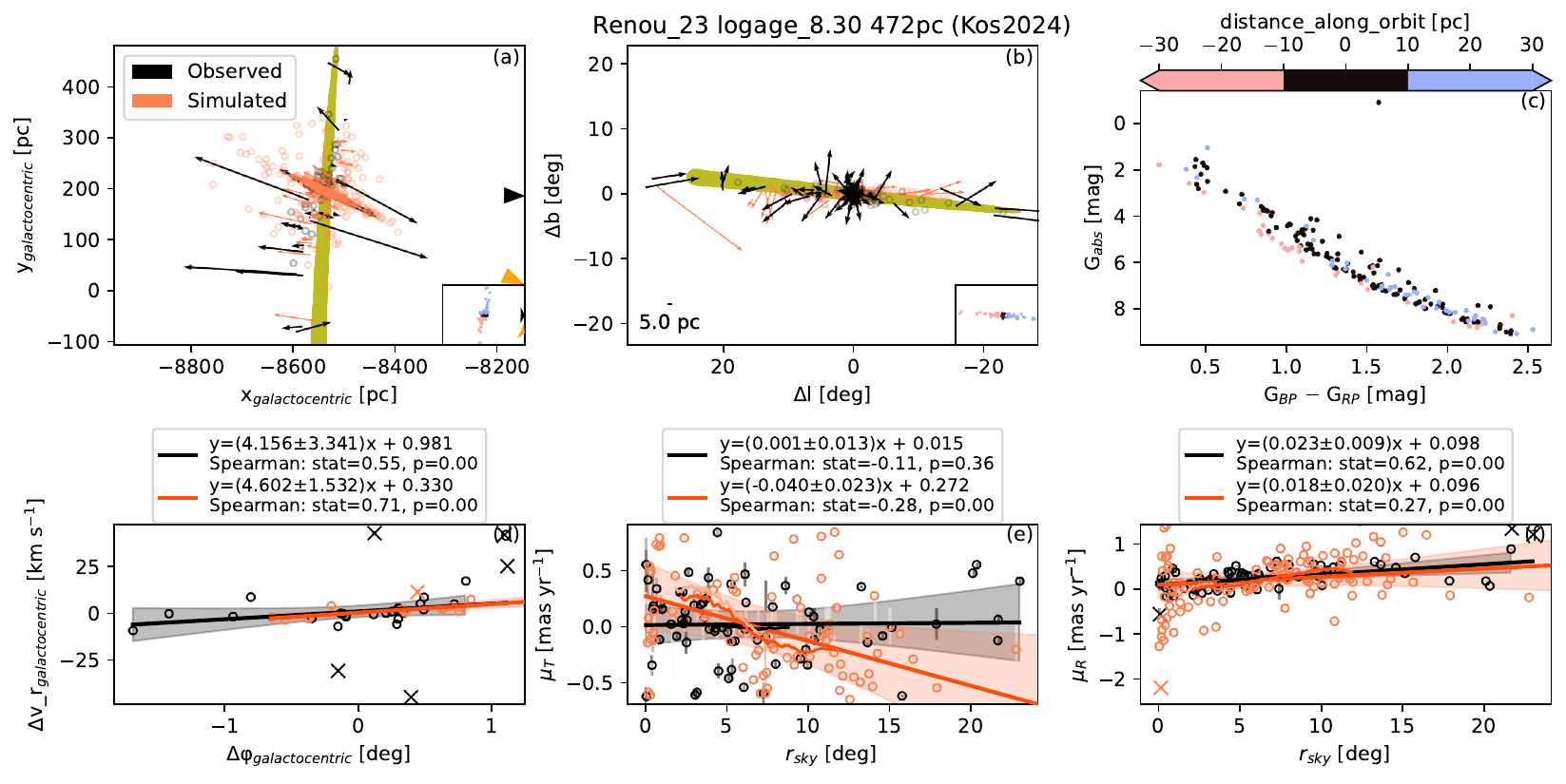}
\includegraphics[width=0.5\linewidth]{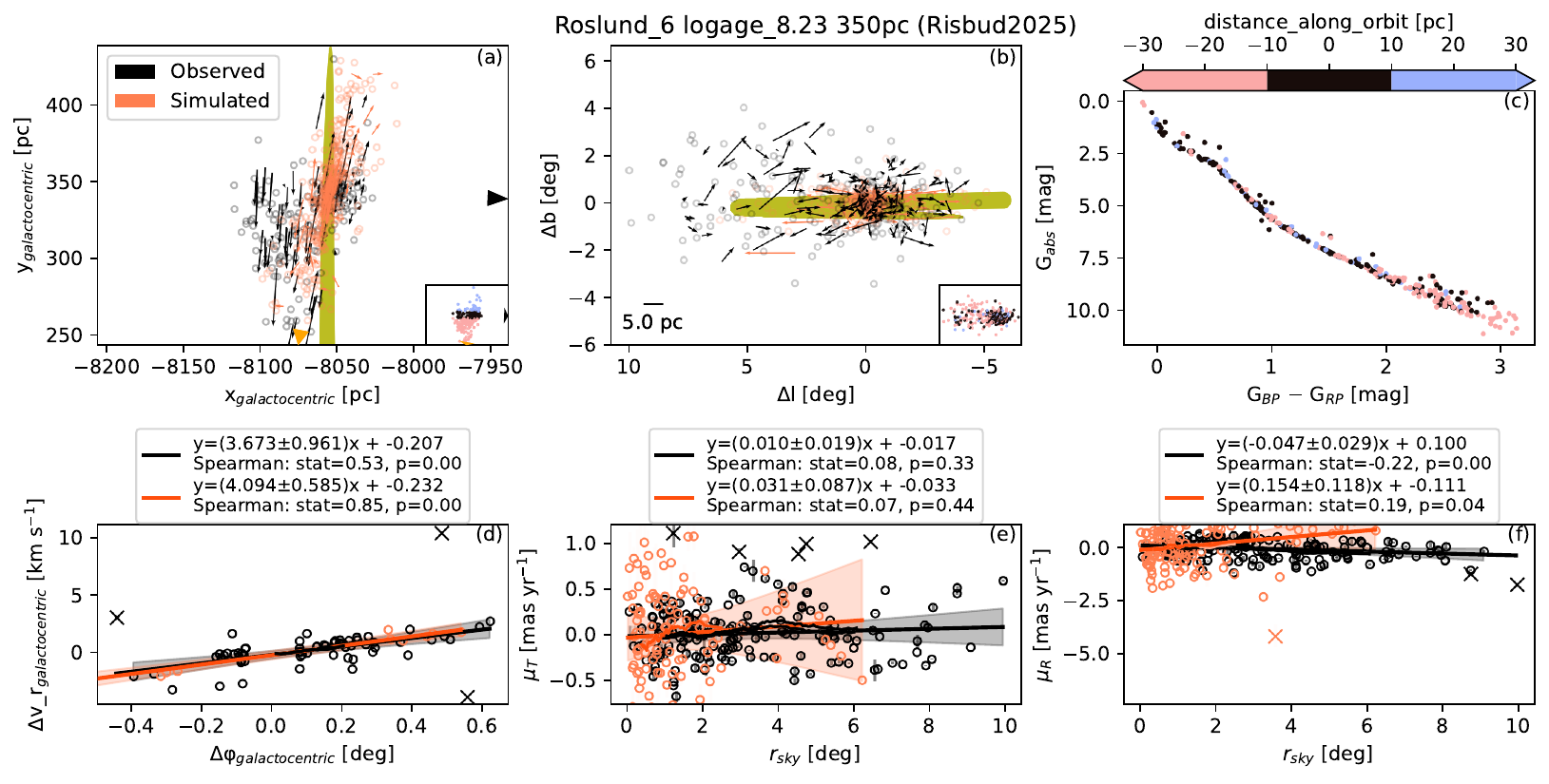}
\includegraphics[width=0.5\linewidth]{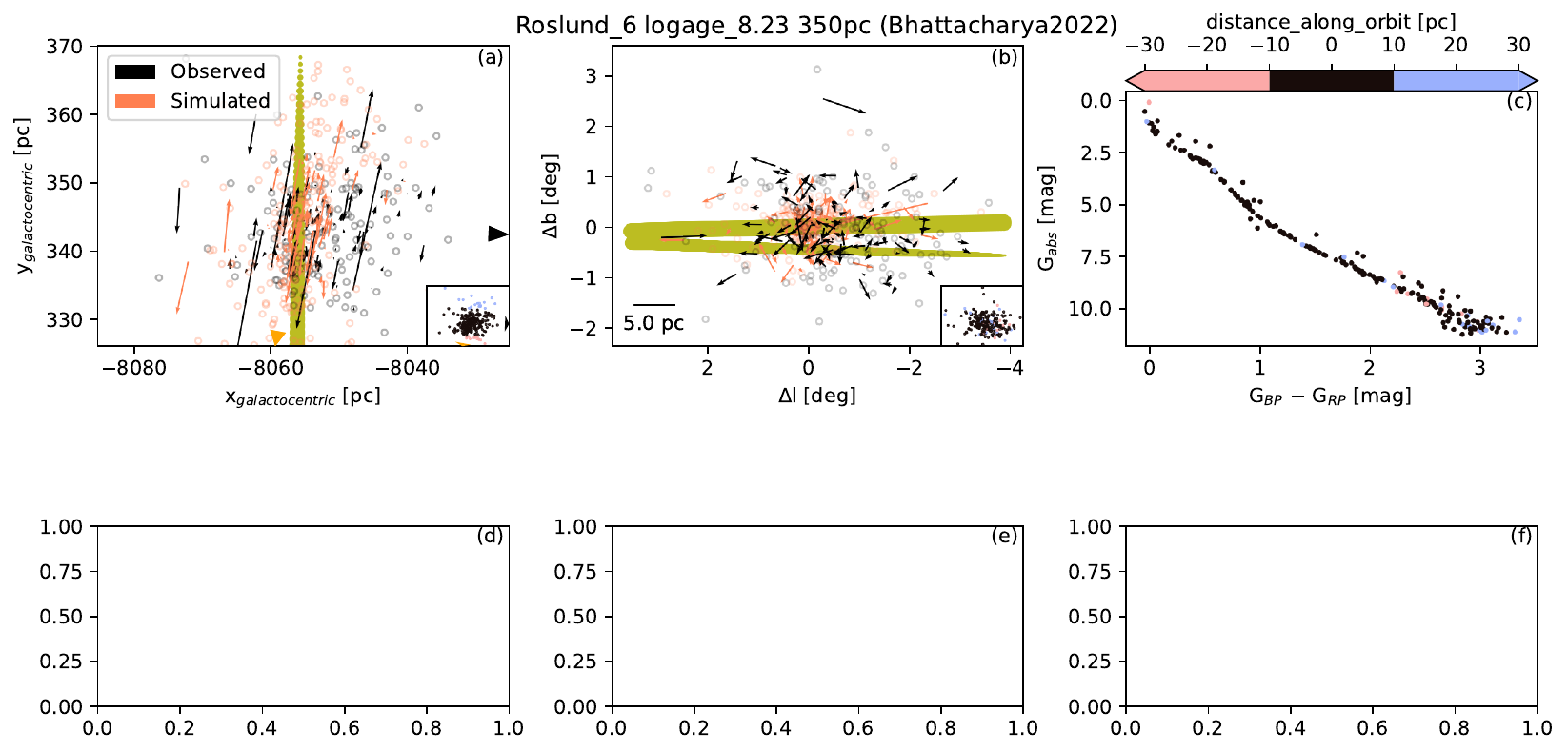}
\includegraphics[width=0.5\linewidth]{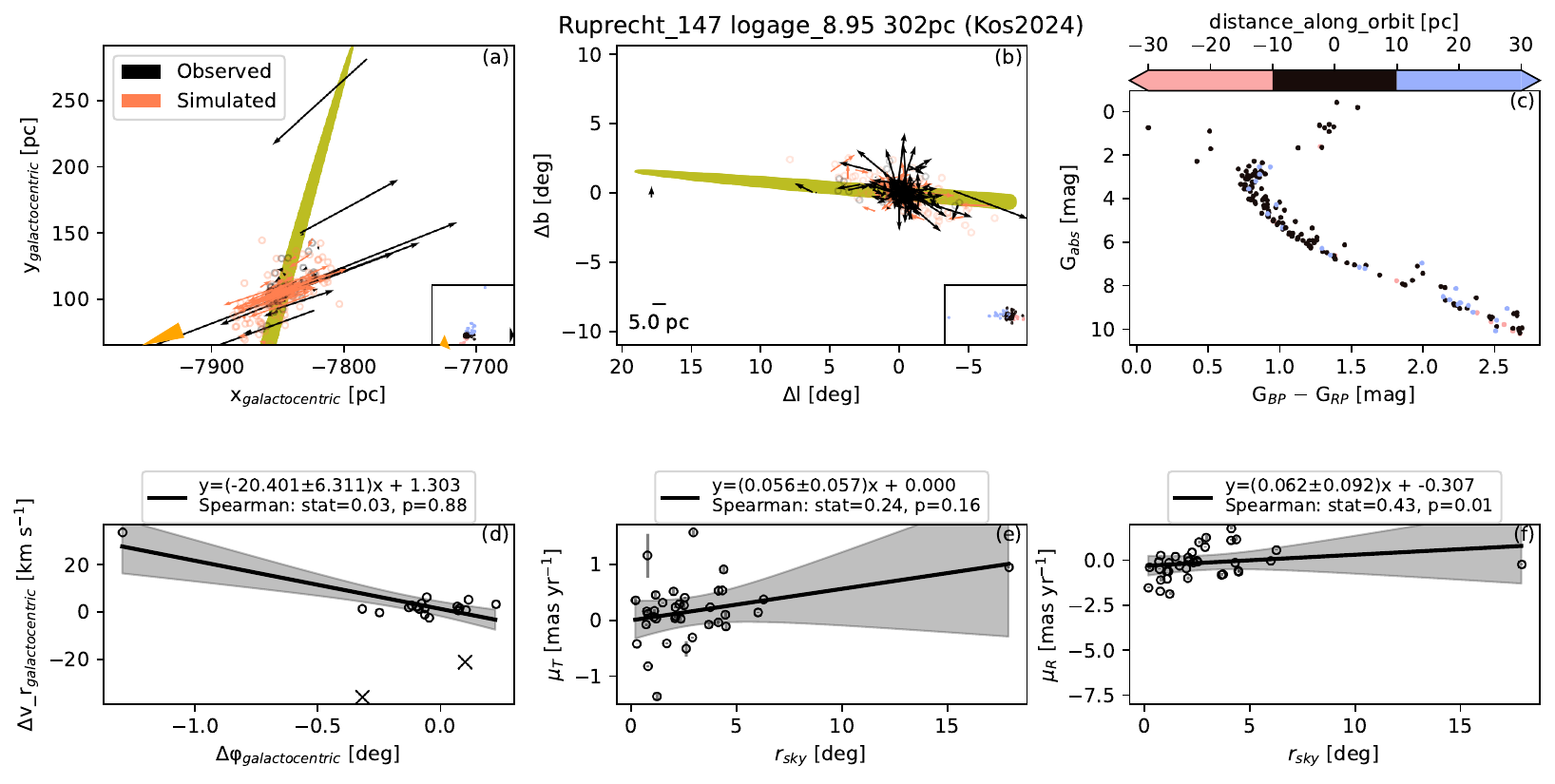}
    \caption{Diagnostic figures for Renou 23 (Kos2024), Roslund 6 (Risbud2025), Roslund 6 (Bhattacharya2022), Ruprecht 147 (Kos2024).}
        \label{fig:supplementary.Ruprecht_147.Kos2024}
        \end{figure}
         
        \begin{figure}
\includegraphics[width=0.5\linewidth]{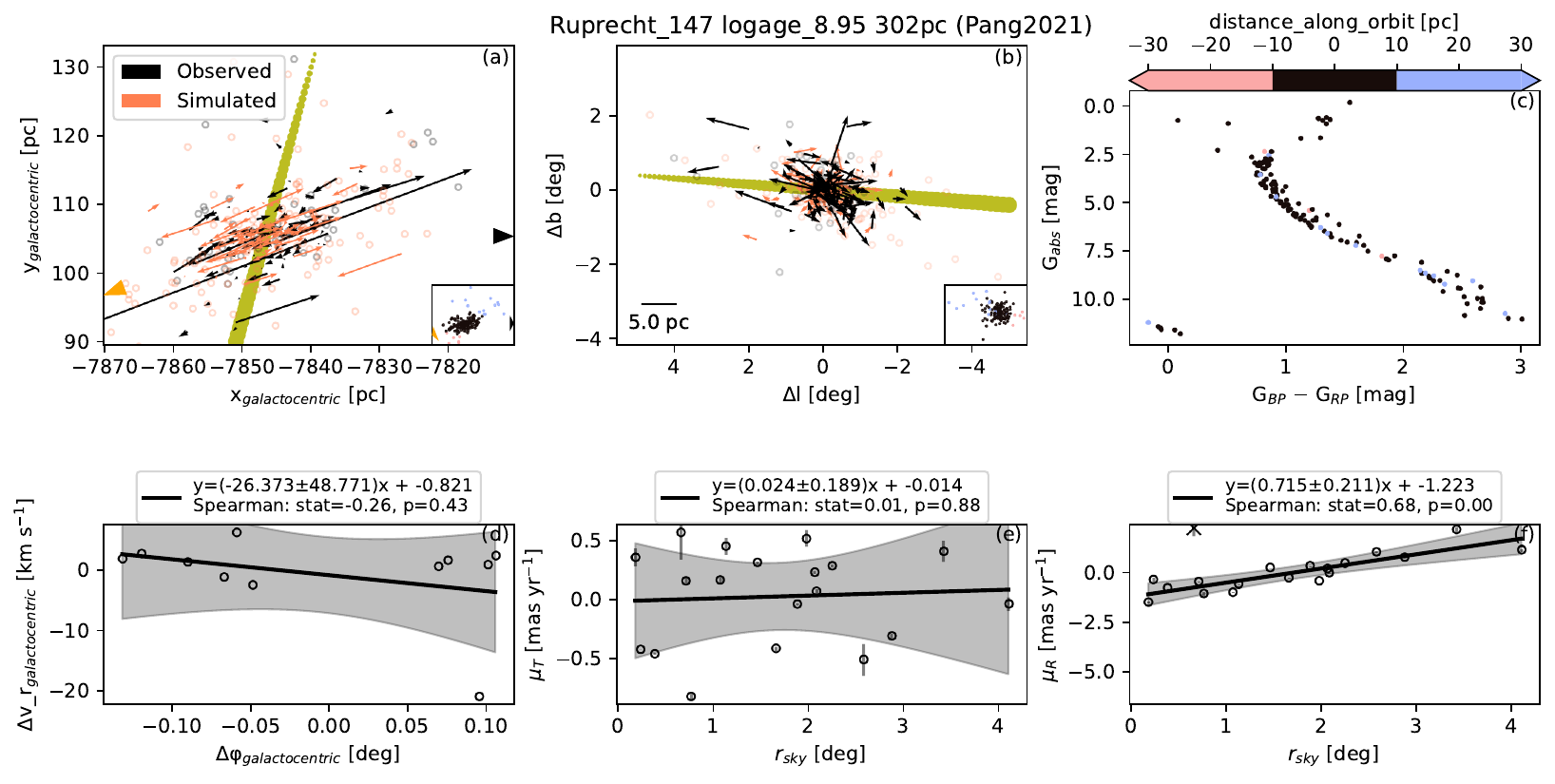}
\includegraphics[width=0.5\linewidth]{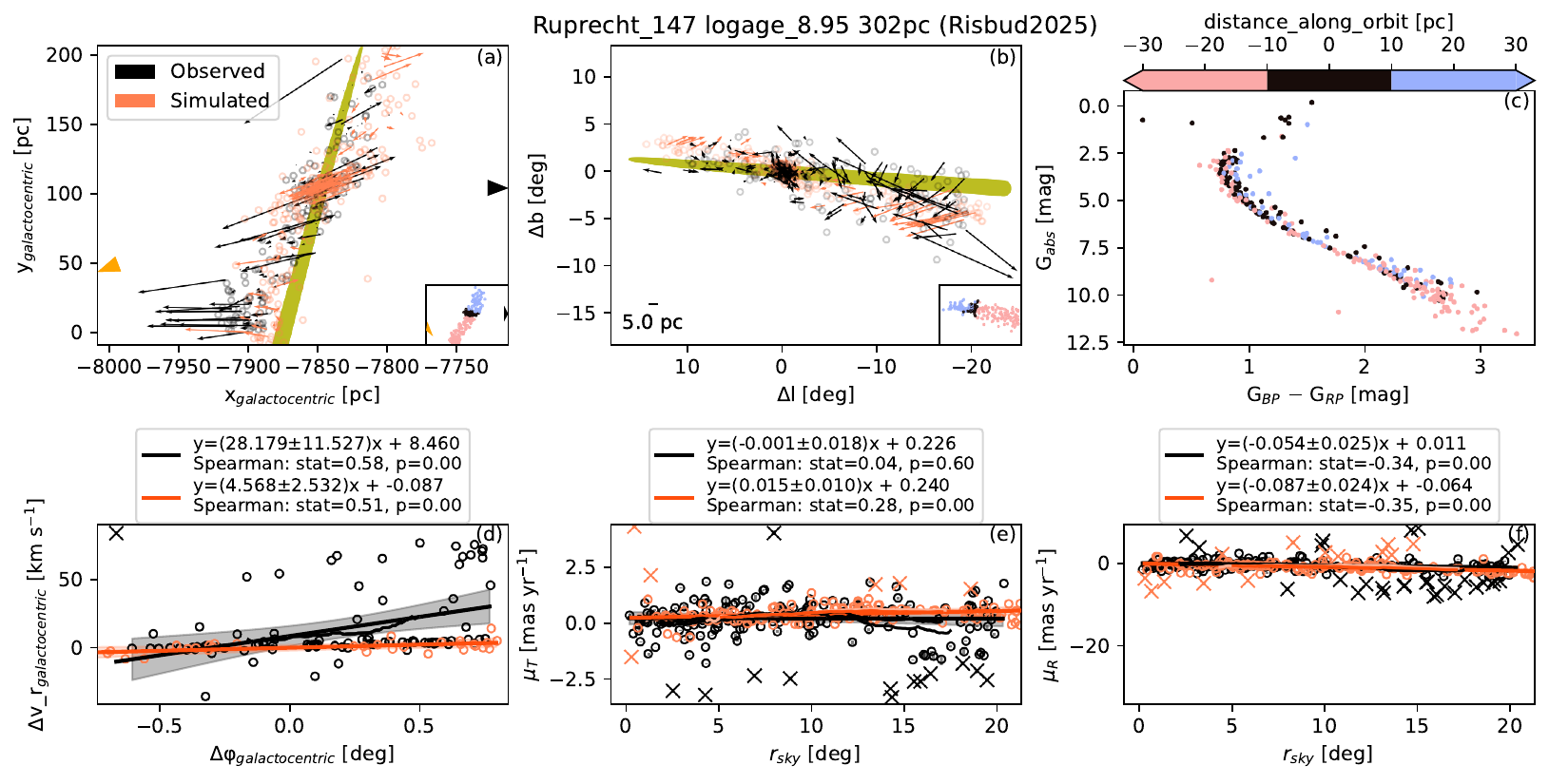}
\includegraphics[width=0.5\linewidth]{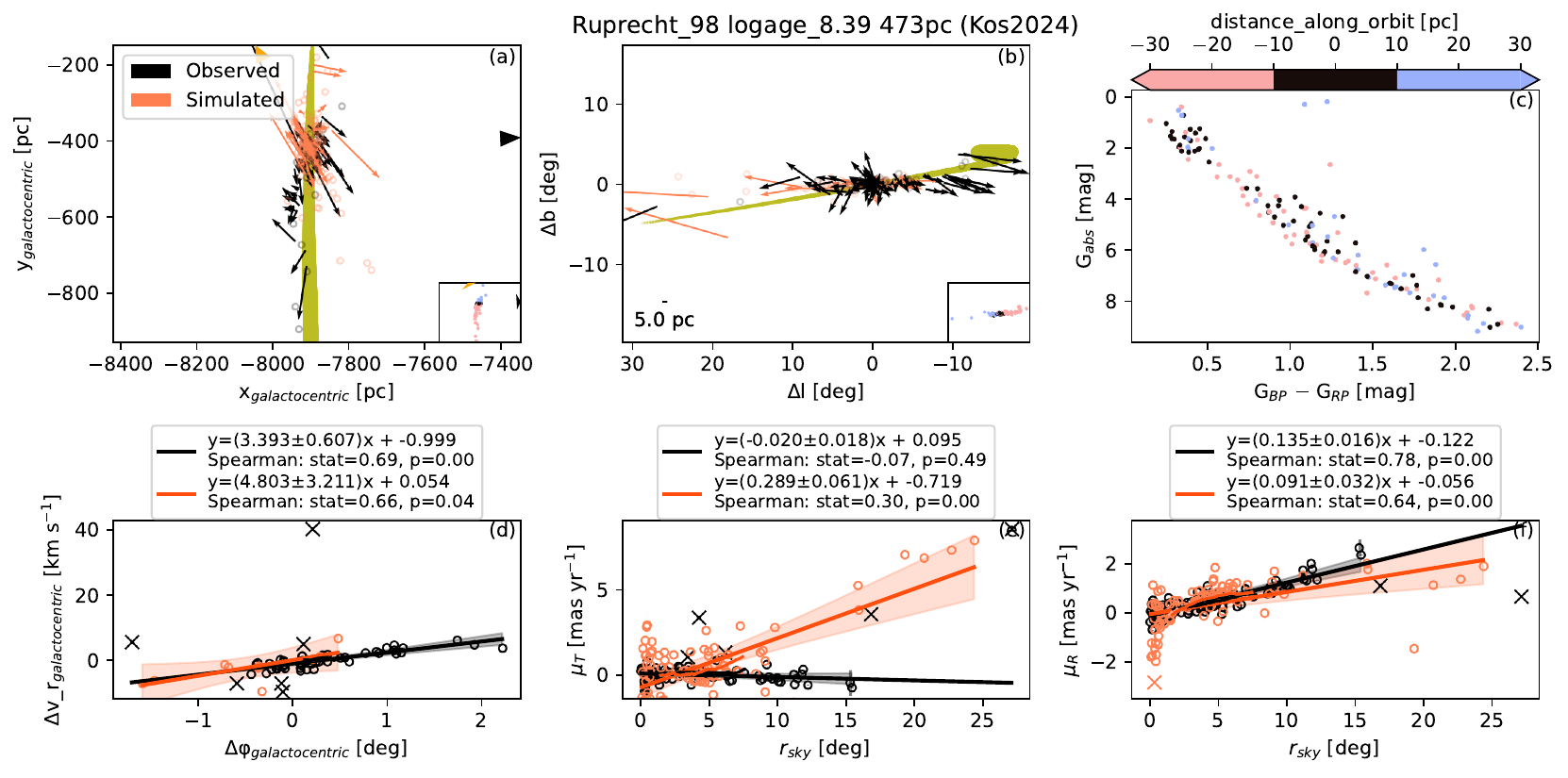}
\includegraphics[width=0.5\linewidth]{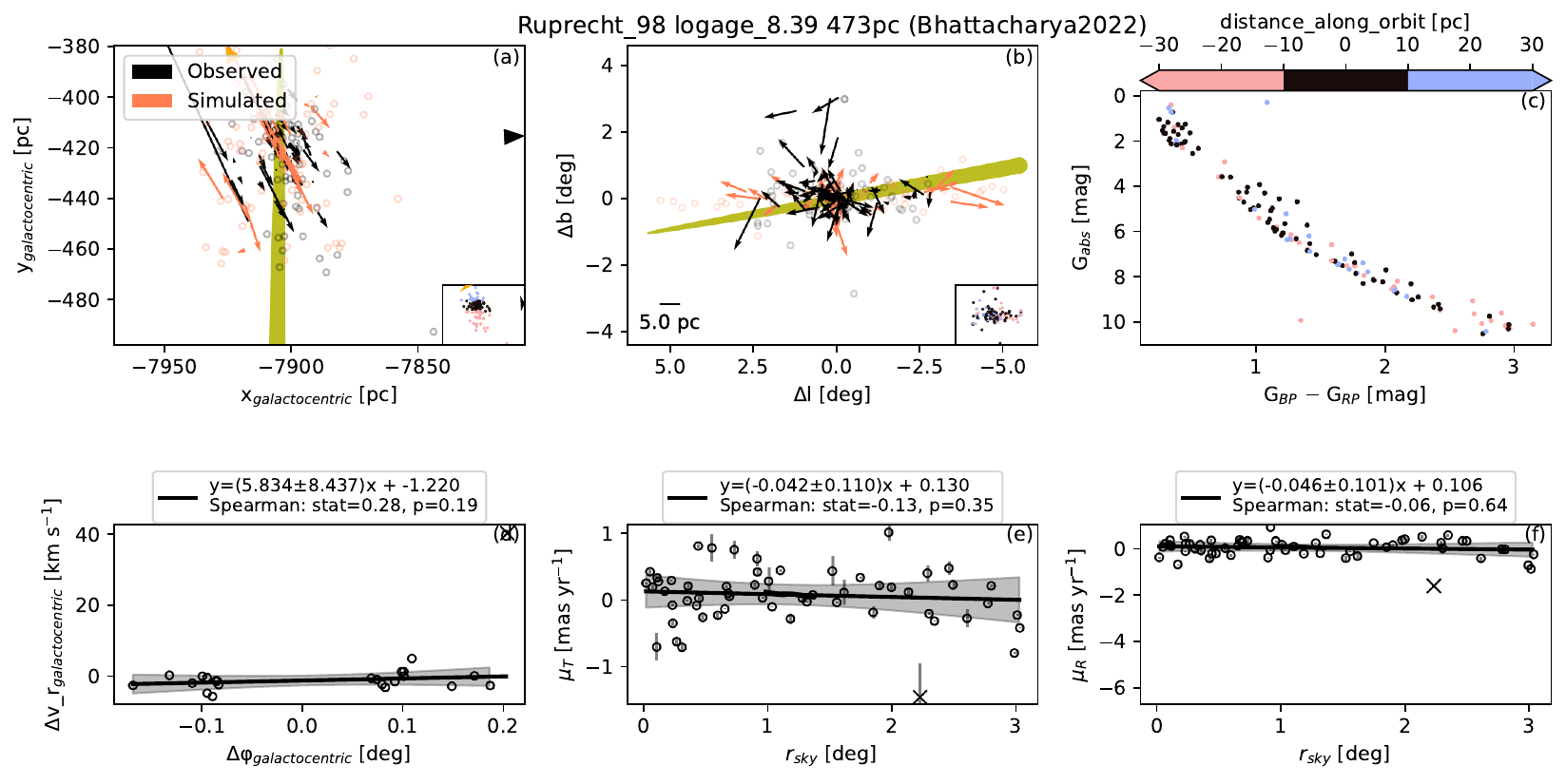}
    \caption{Diagnostic figures for Ruprecht 147 (Pang2021), Ruprecht 147 (Risbud2025), Ruprecht 98 (Kos2024), Ruprecht 98 (Bhattacharya2022).}
        \label{fig:supplementary.Ruprecht_98.Bhattacharya2022}
        \end{figure}
         
        \begin{figure}
\includegraphics[width=0.5\linewidth]{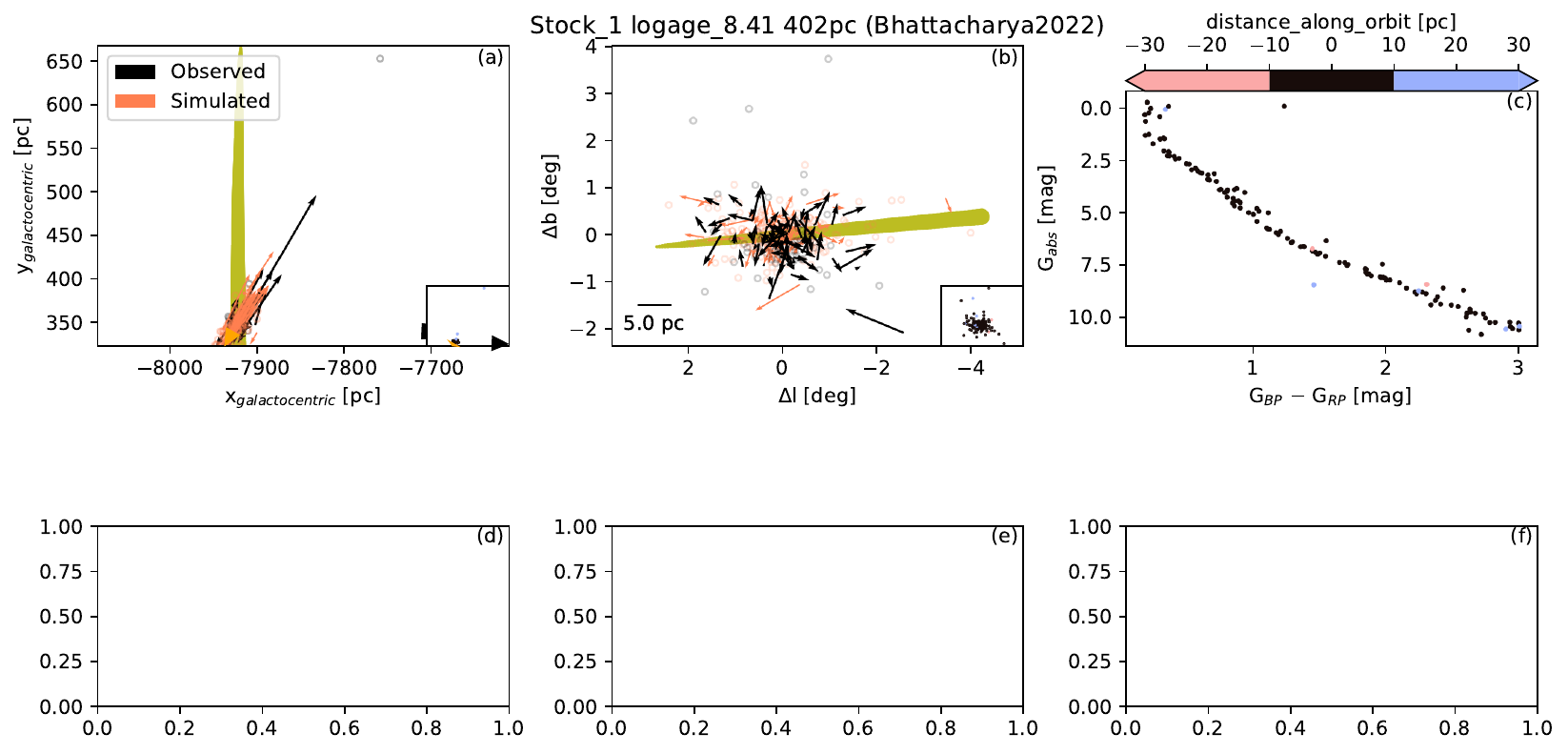}
\includegraphics[width=0.5\linewidth]{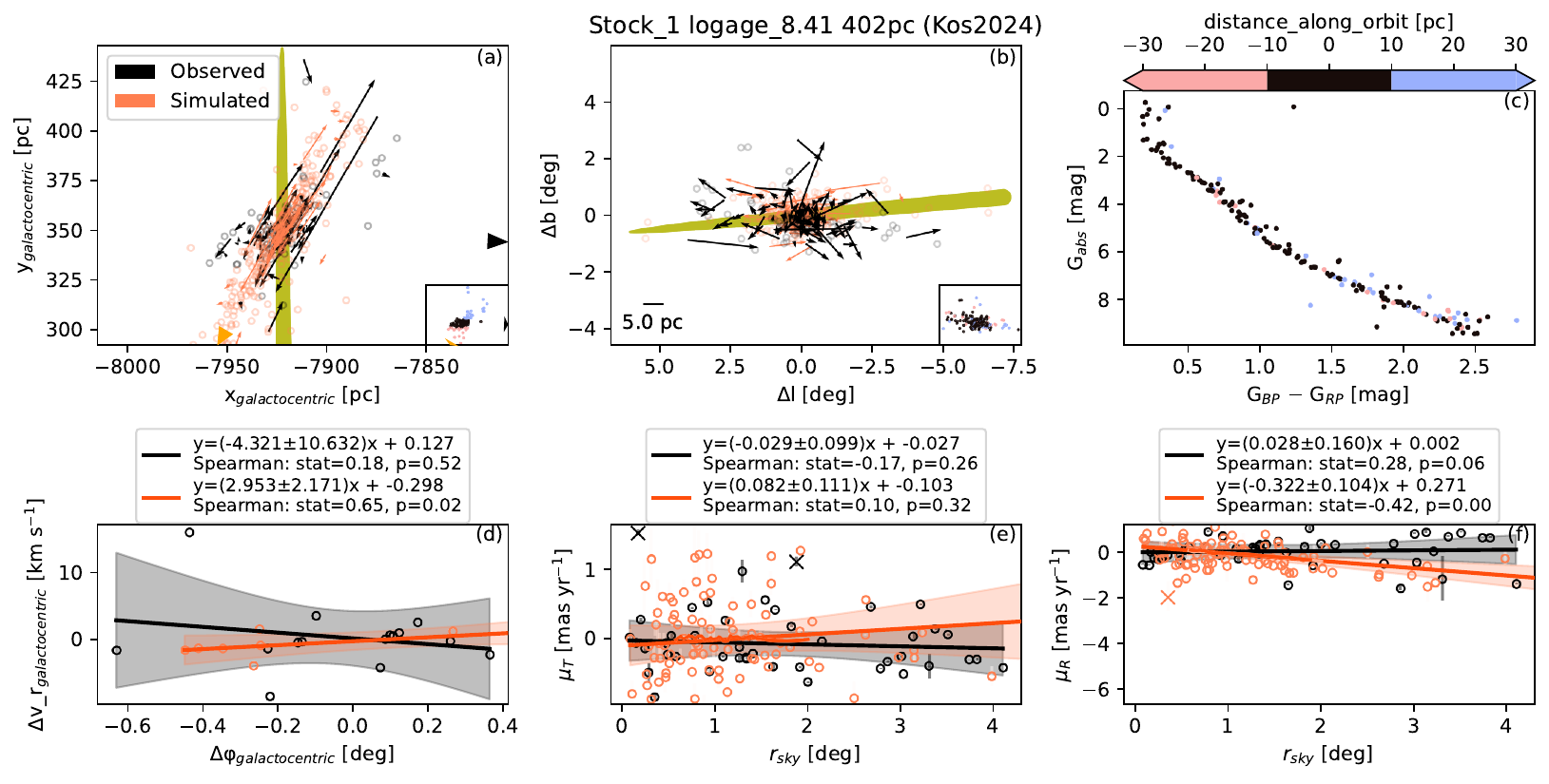}
\includegraphics[width=0.5\linewidth]{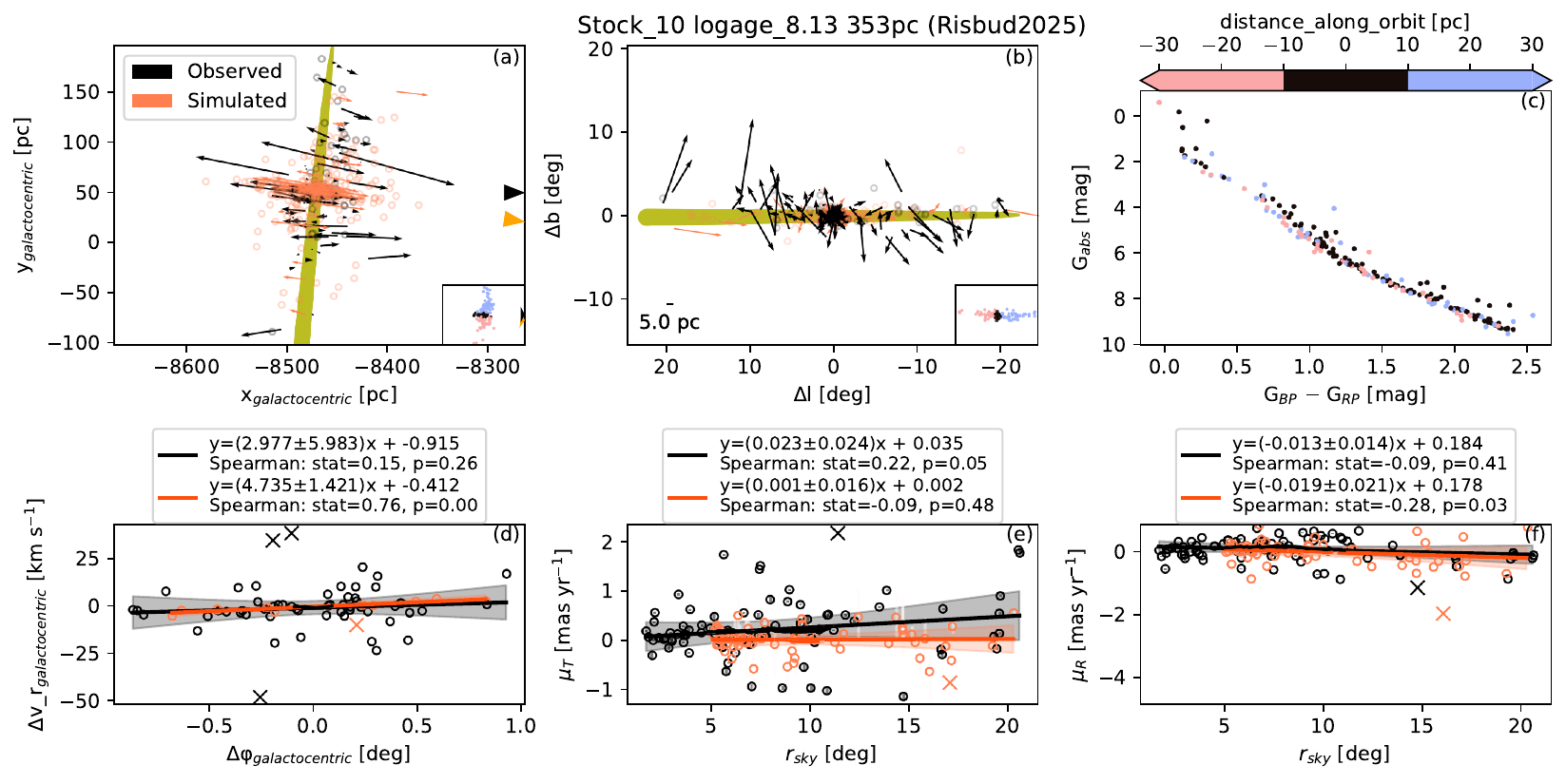}
\includegraphics[width=0.5\linewidth]{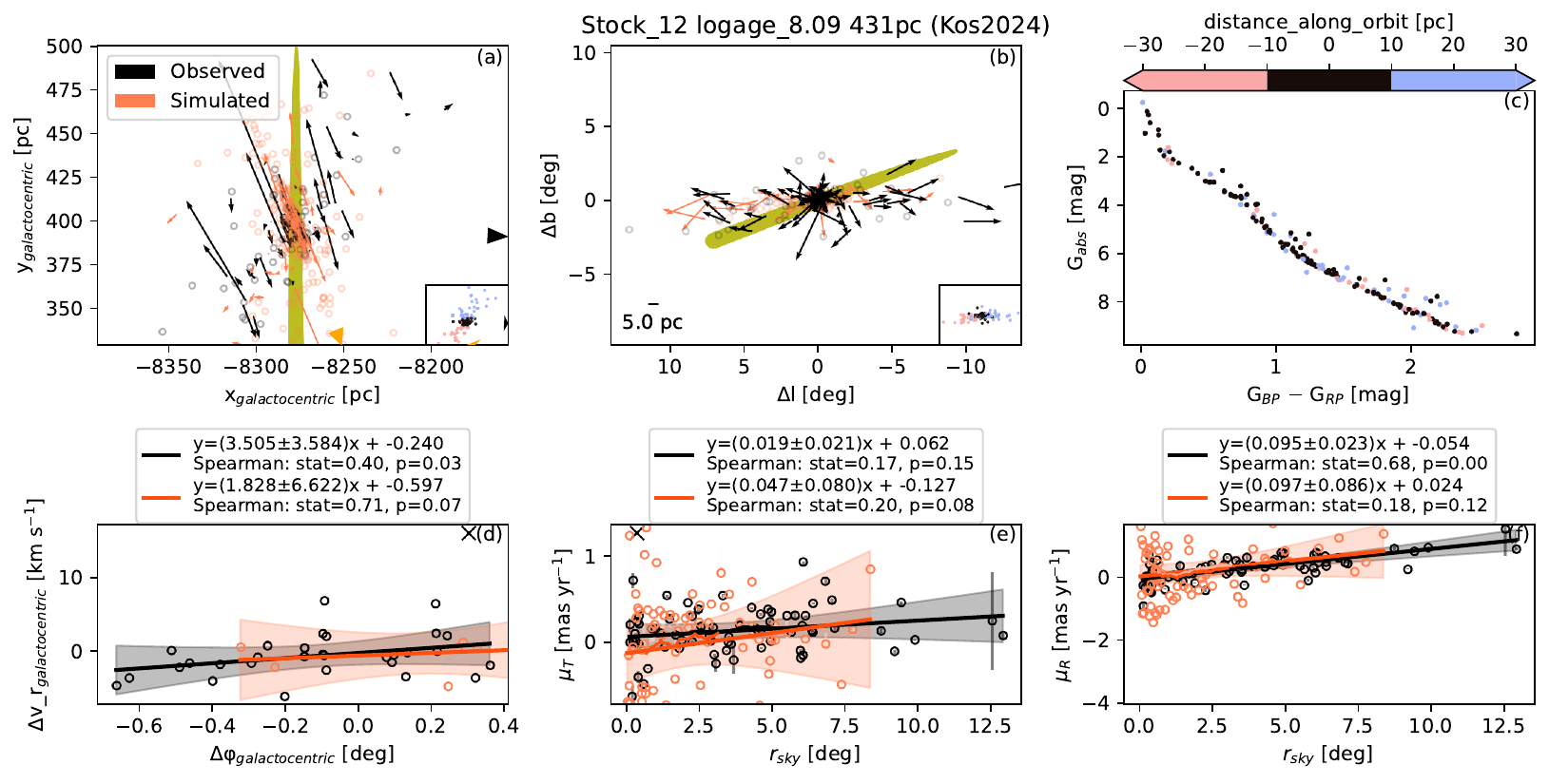}
    \caption{Diagnostic figures for Stock 1 (Bhattacharya2022), Stock 1 (Kos2024), Stock 10 (Risbud2025), Stock 12 (Kos2024).}
        \label{fig:supplementary.Stock_12.Kos2024}
        \end{figure}
         
        \begin{figure}
\includegraphics[width=0.5\linewidth]{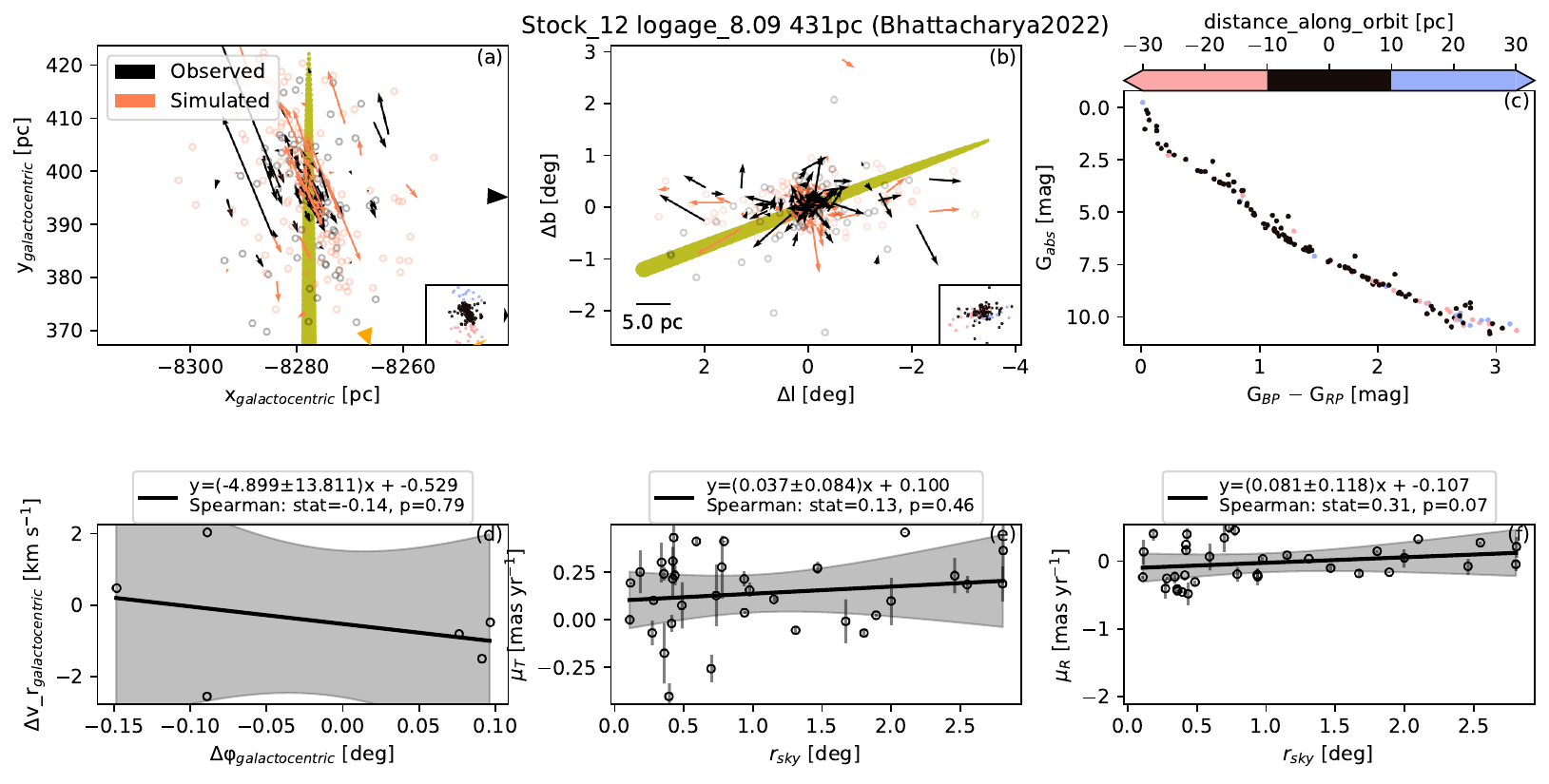}
\includegraphics[width=0.5\linewidth]{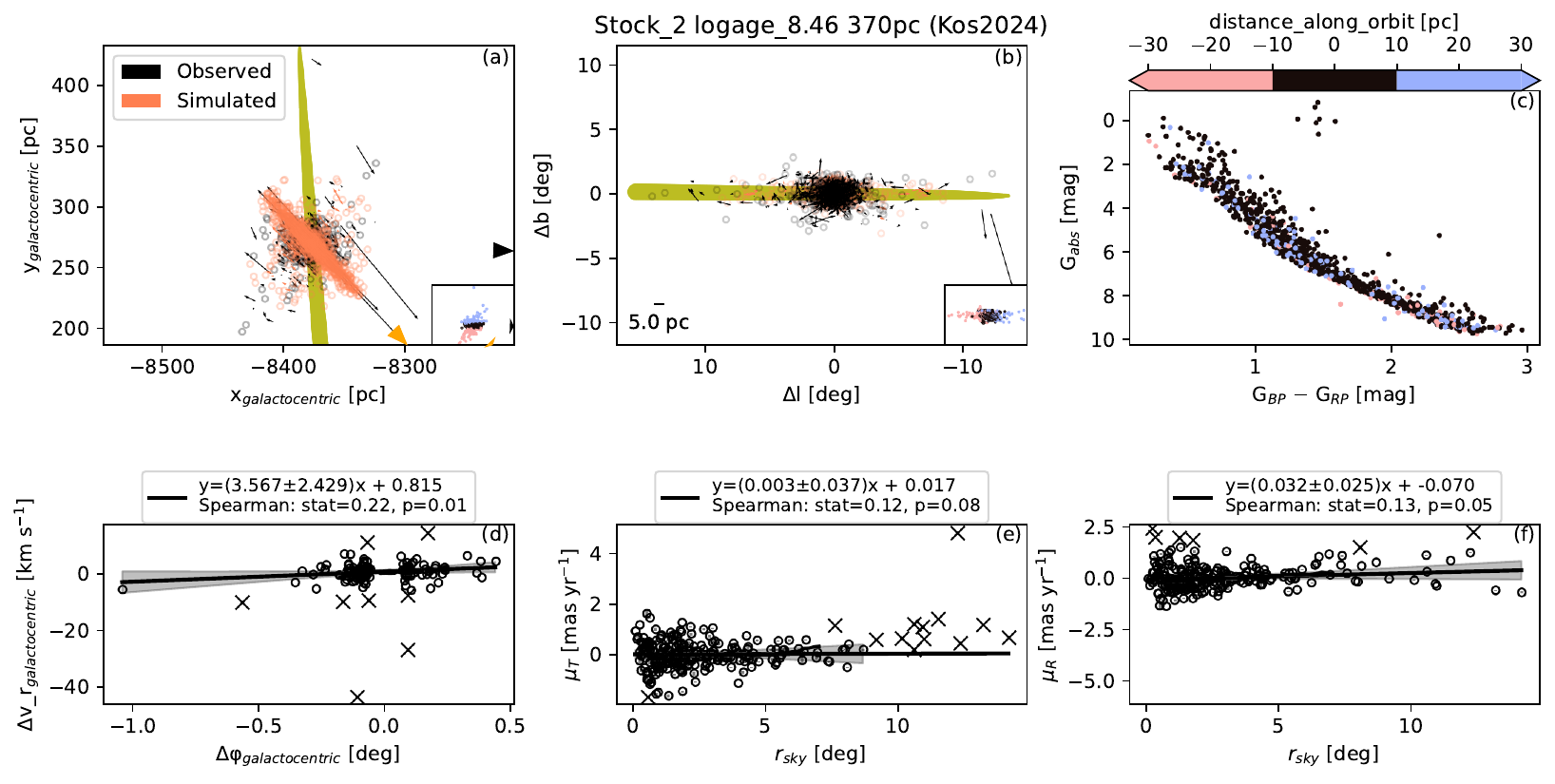}
\includegraphics[width=0.5\linewidth]{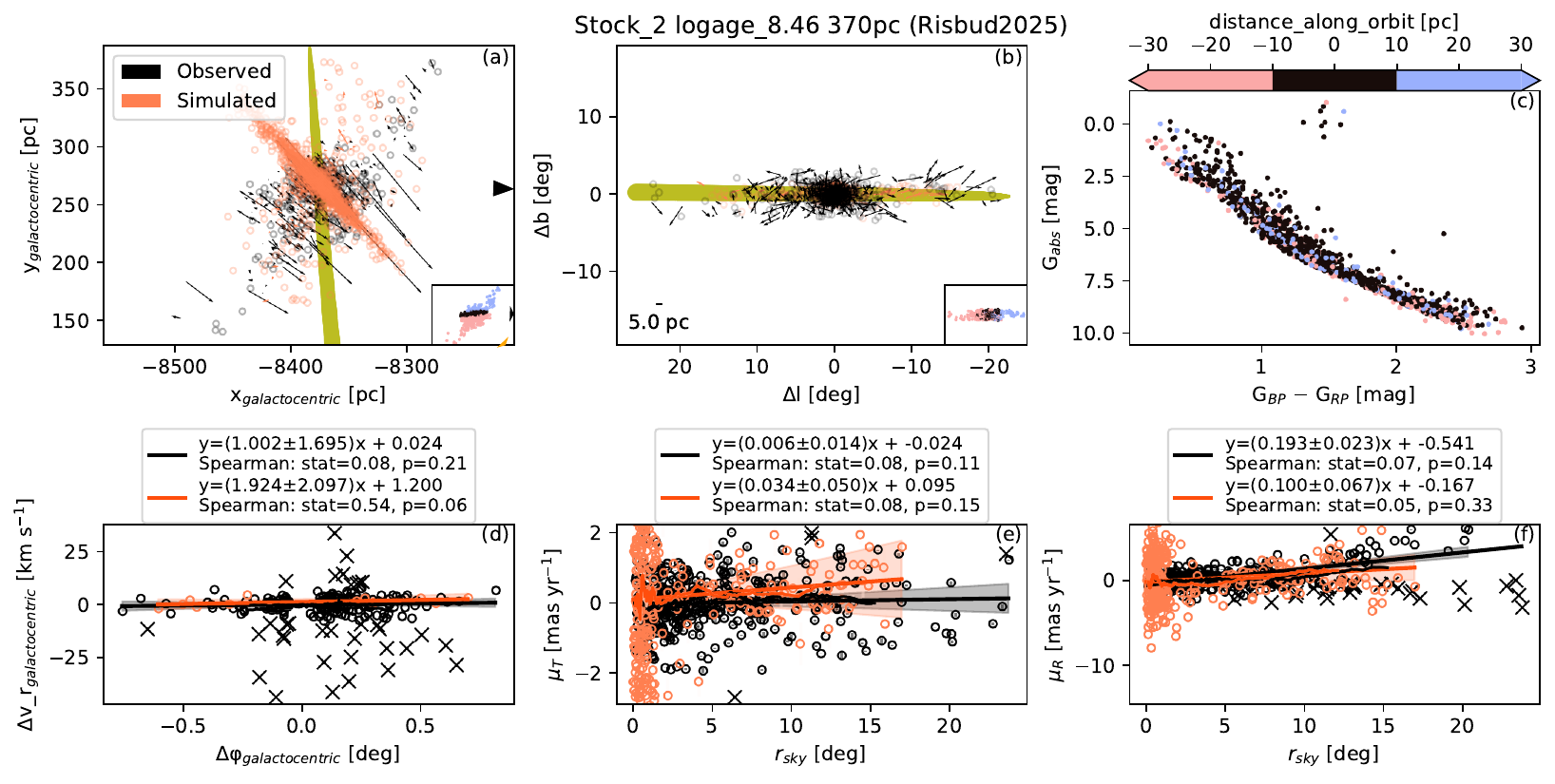}
\includegraphics[width=0.5\linewidth]{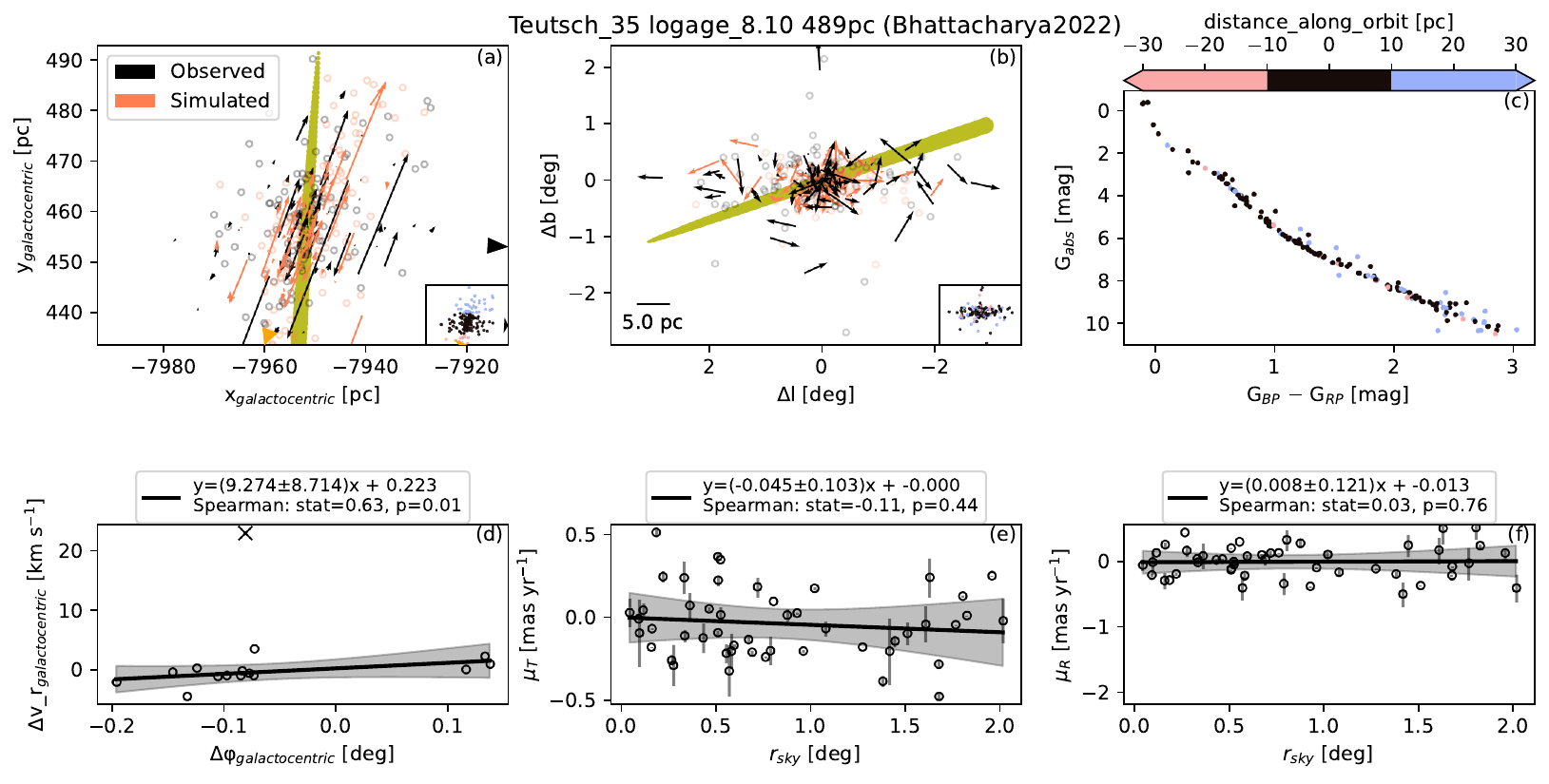}
    \caption{Diagnostic figures for Stock 12 (Bhattacharya2022), Stock 2 (Kos2024), Stock 2 (Risbud2025), Teutsch 35 (Bhattacharya2022).}
        \label{fig:supplementary.Teutsch_35.Bhattacharya2022}
        \end{figure}
         
        \begin{figure}
\includegraphics[width=0.5\linewidth]{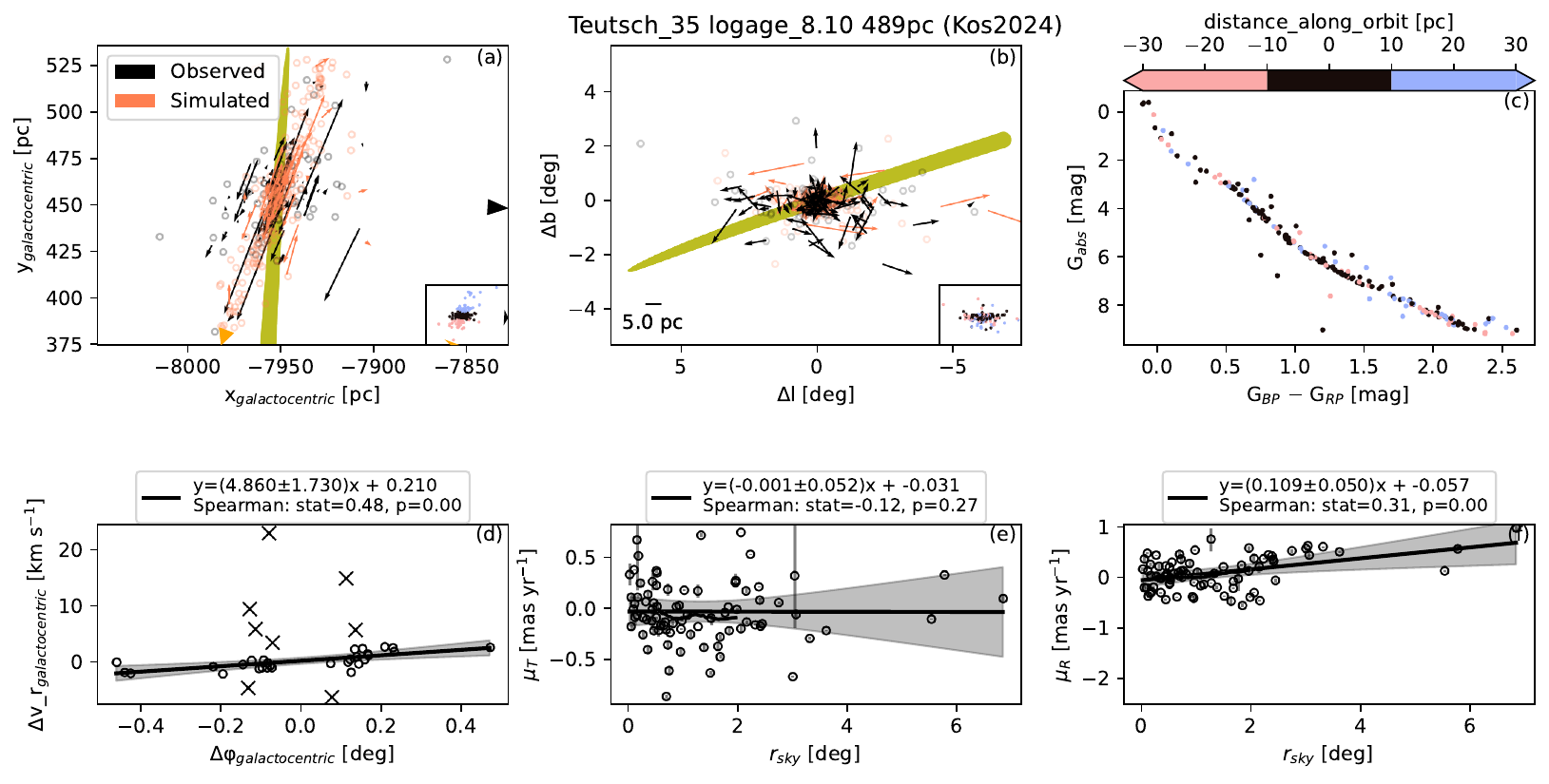}
\includegraphics[width=0.5\linewidth]{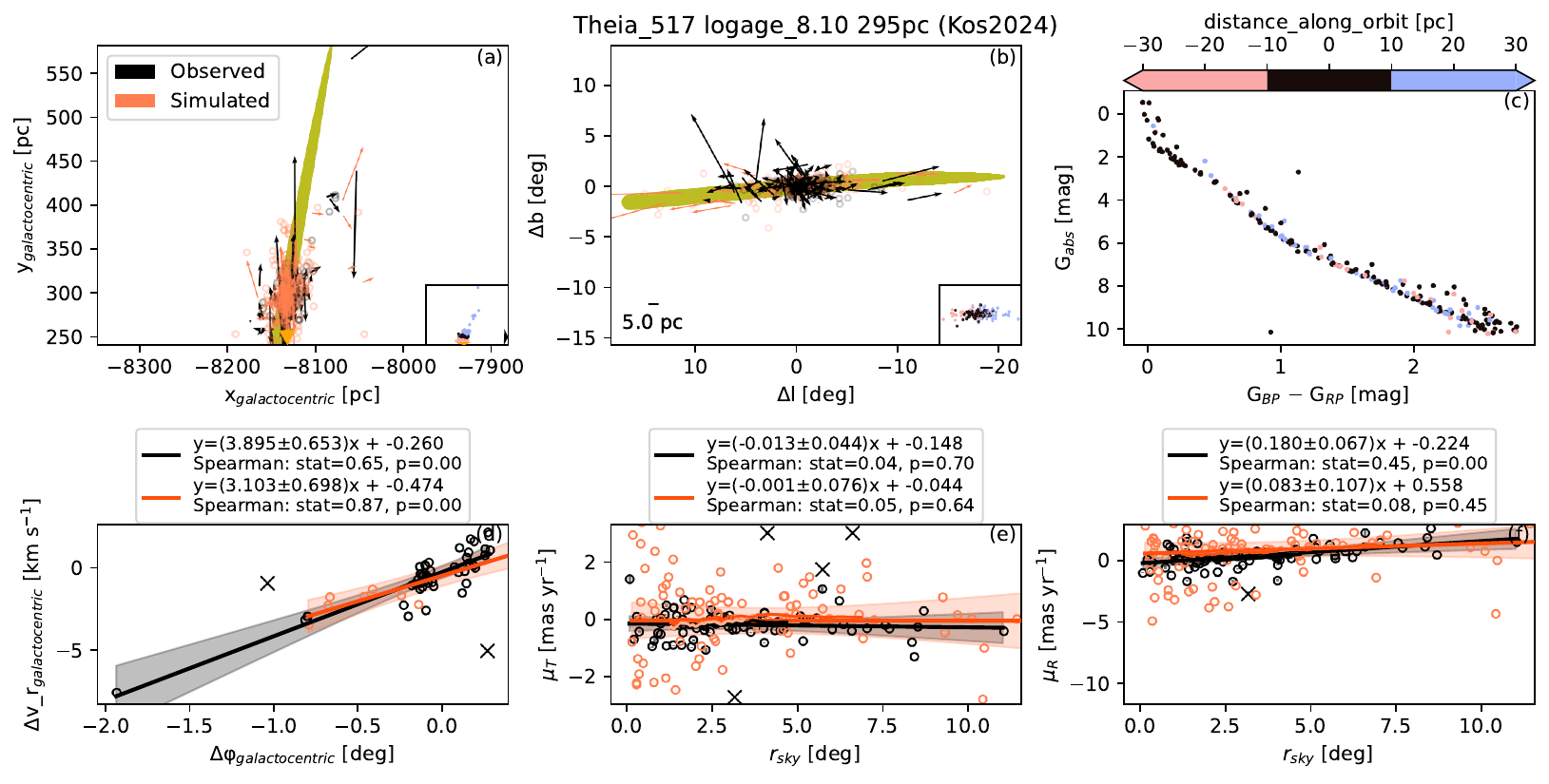}
\includegraphics[width=0.5\linewidth]{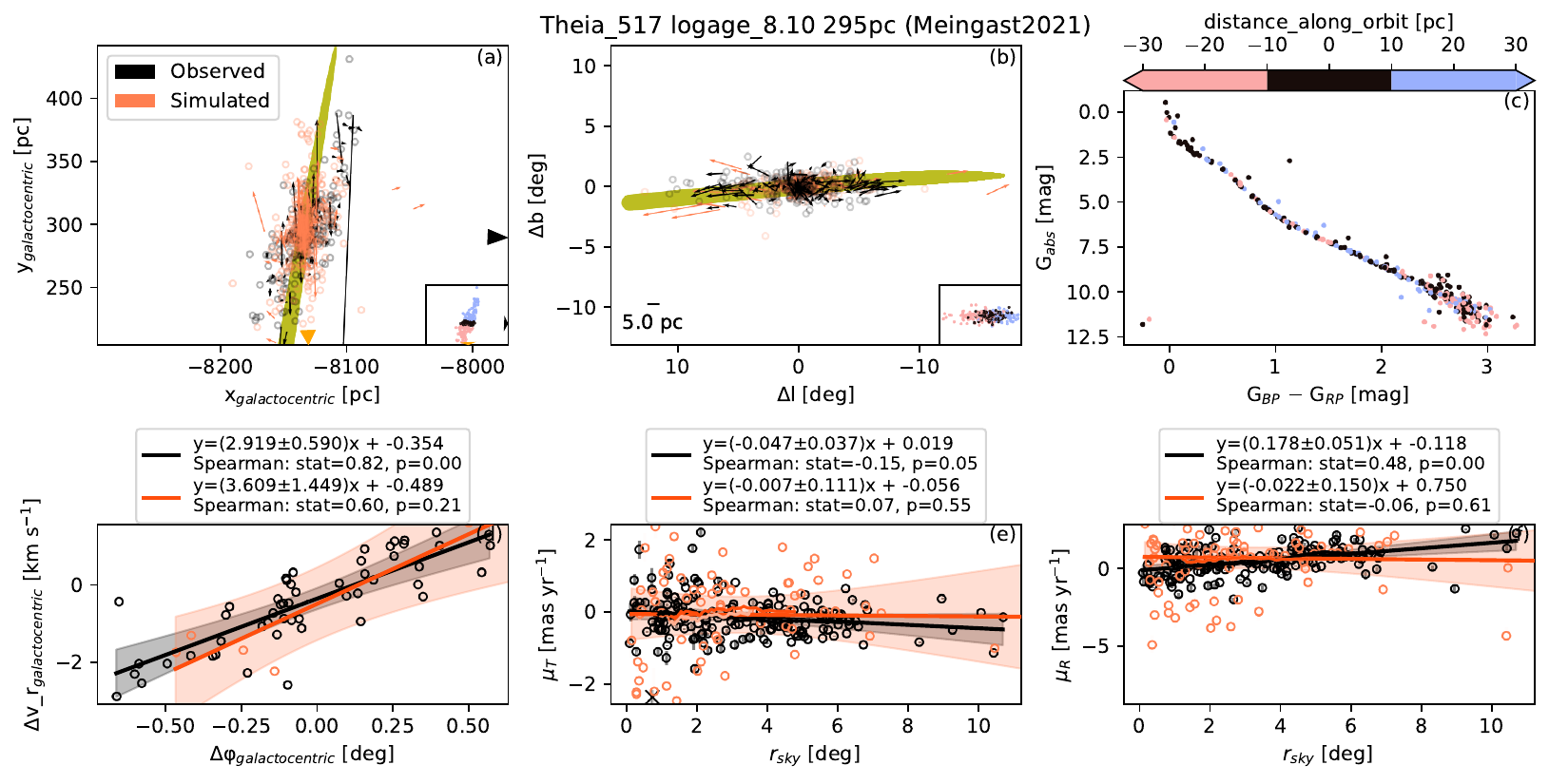}
\includegraphics[width=0.5\linewidth]{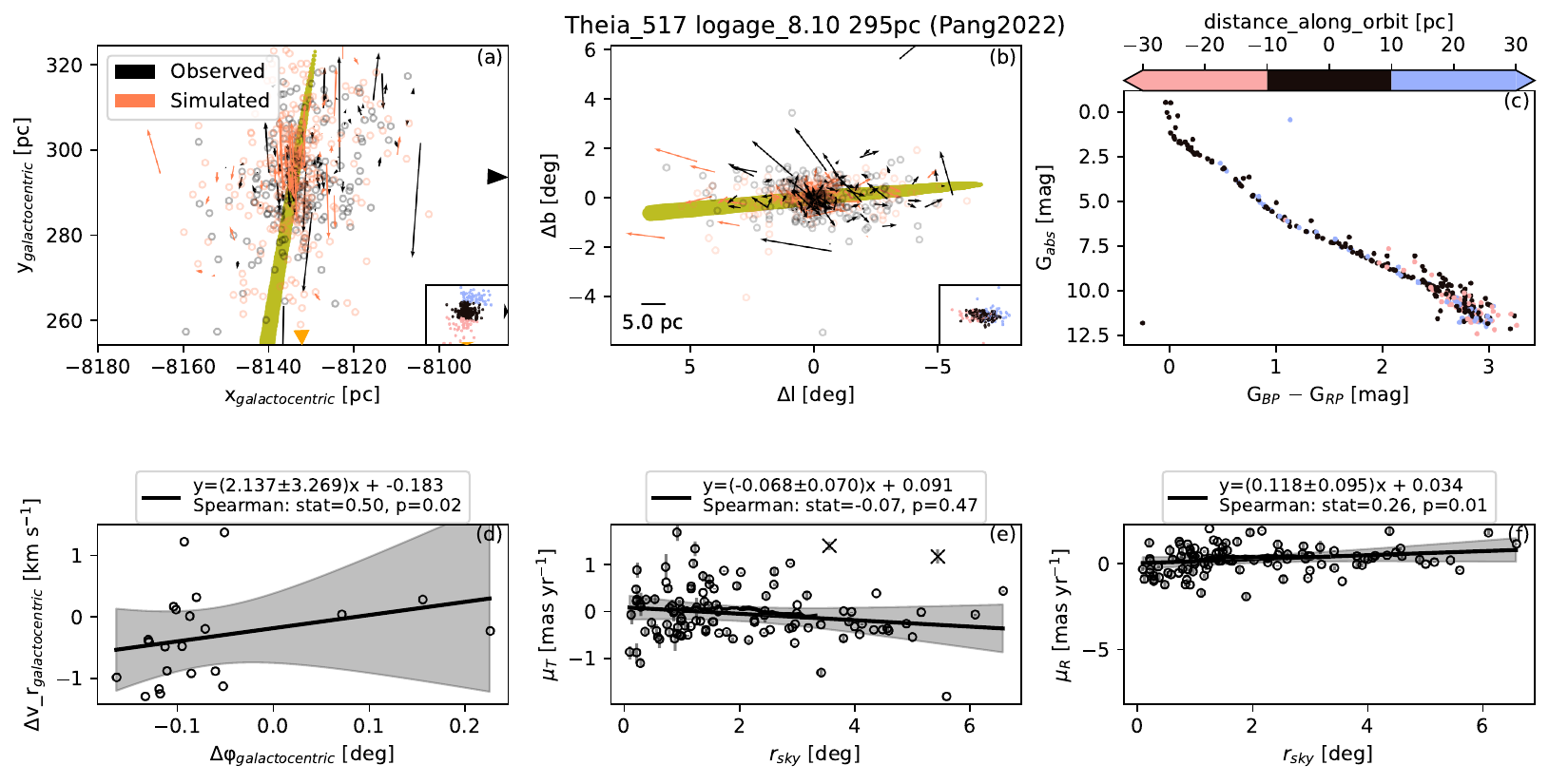}
    \caption{Diagnostic figures for Teutsch 35 (Kos2024), Theia 517 (Kos2024), Theia 517 (Meingast2021), Theia 517 (Pang2022).}
        \label{fig:supplementary.Theia_517.Pang2022}
        \end{figure}
         
        \begin{figure}
\includegraphics[width=0.5\linewidth]{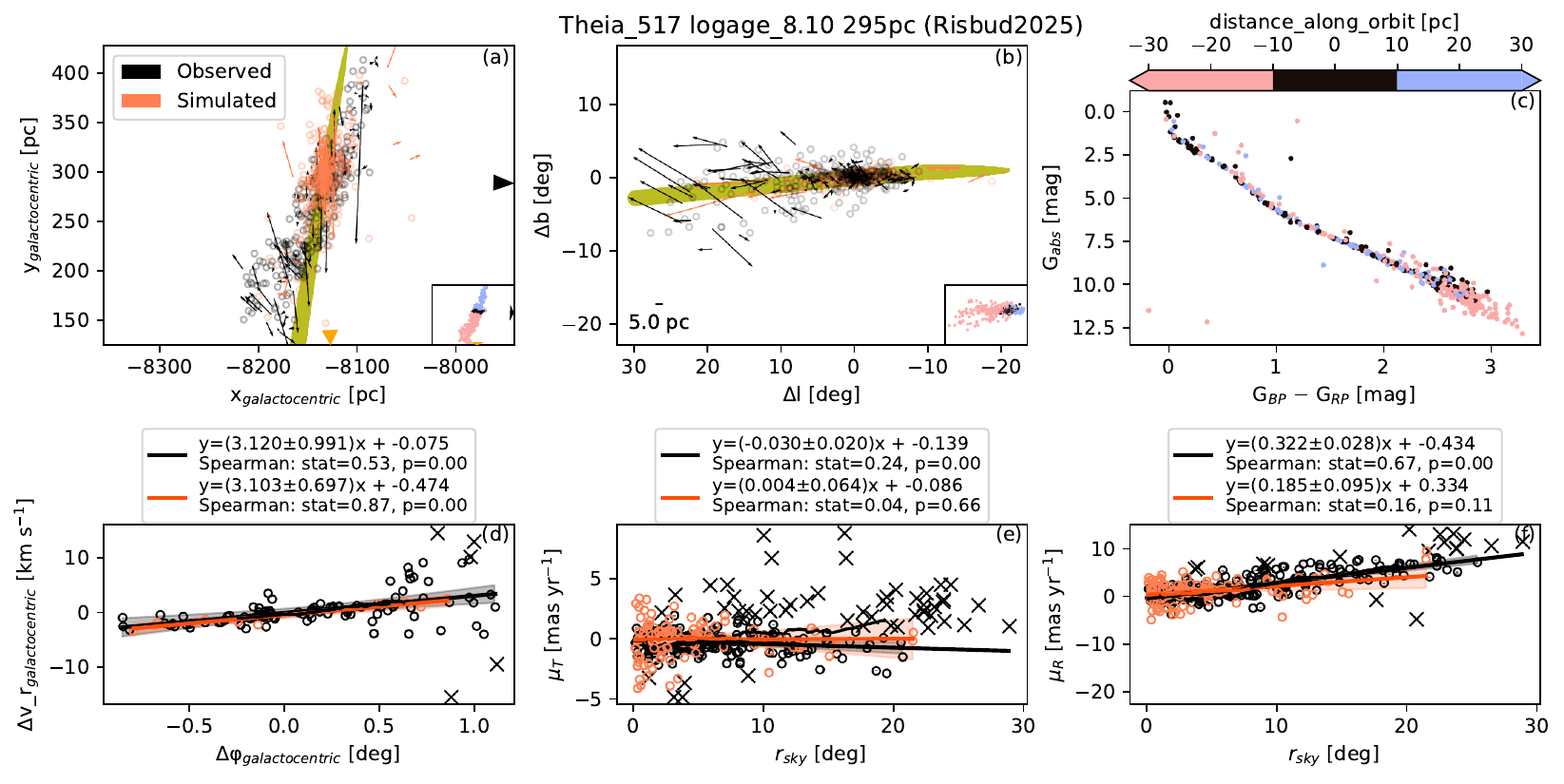}
\includegraphics[width=0.5\linewidth]{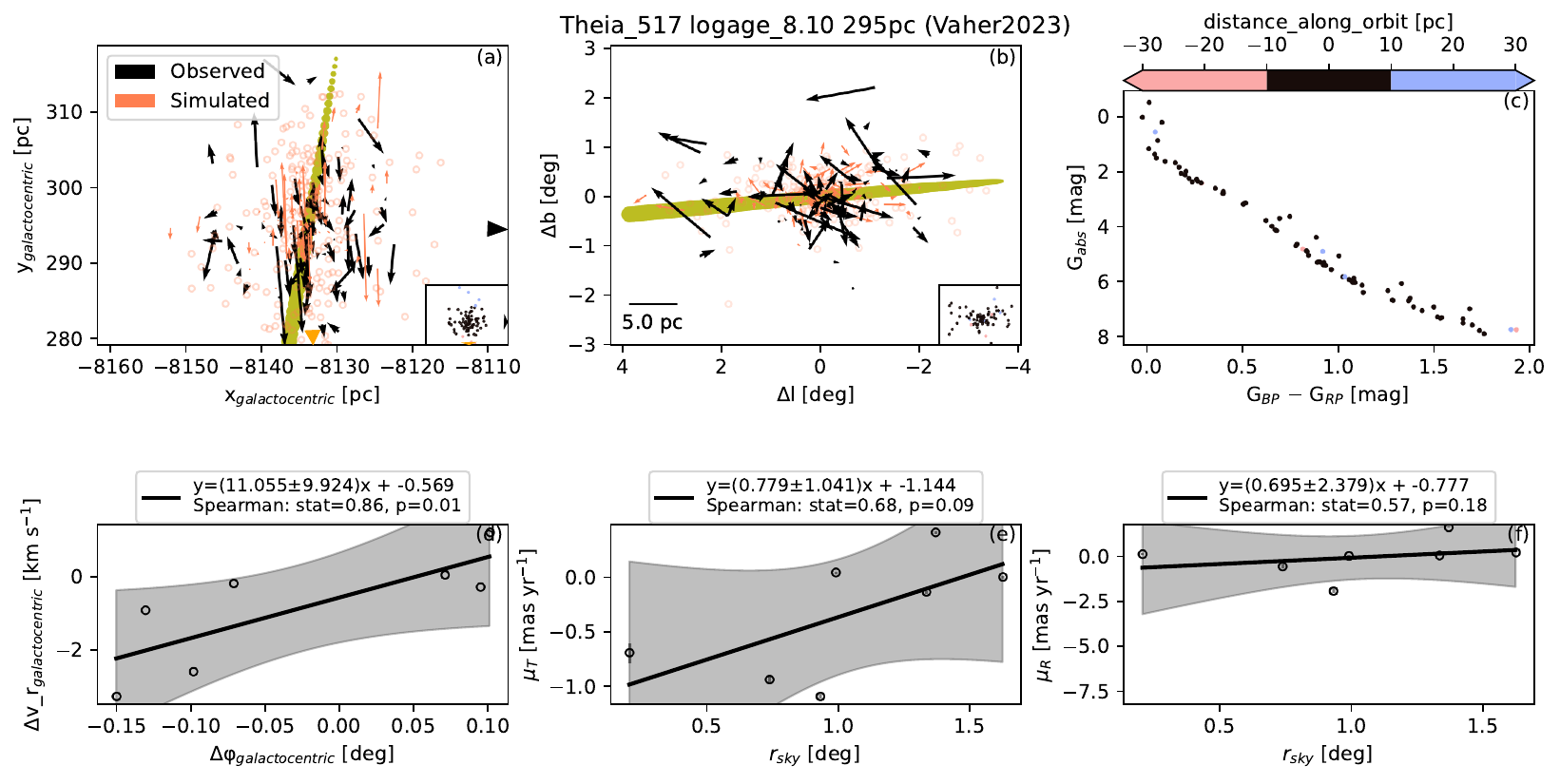}
\includegraphics[width=0.5\linewidth]{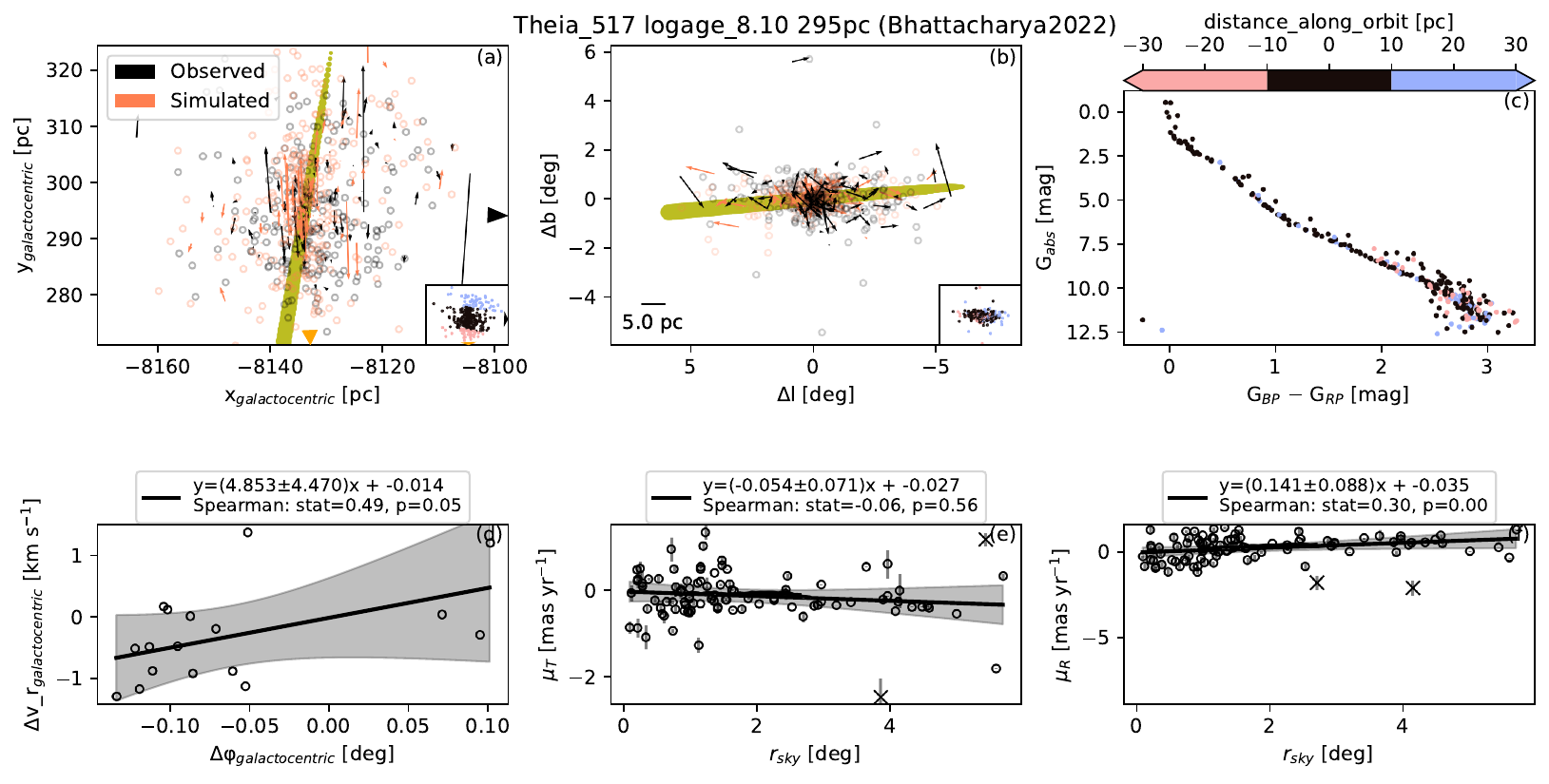}
\includegraphics[width=0.5\linewidth]{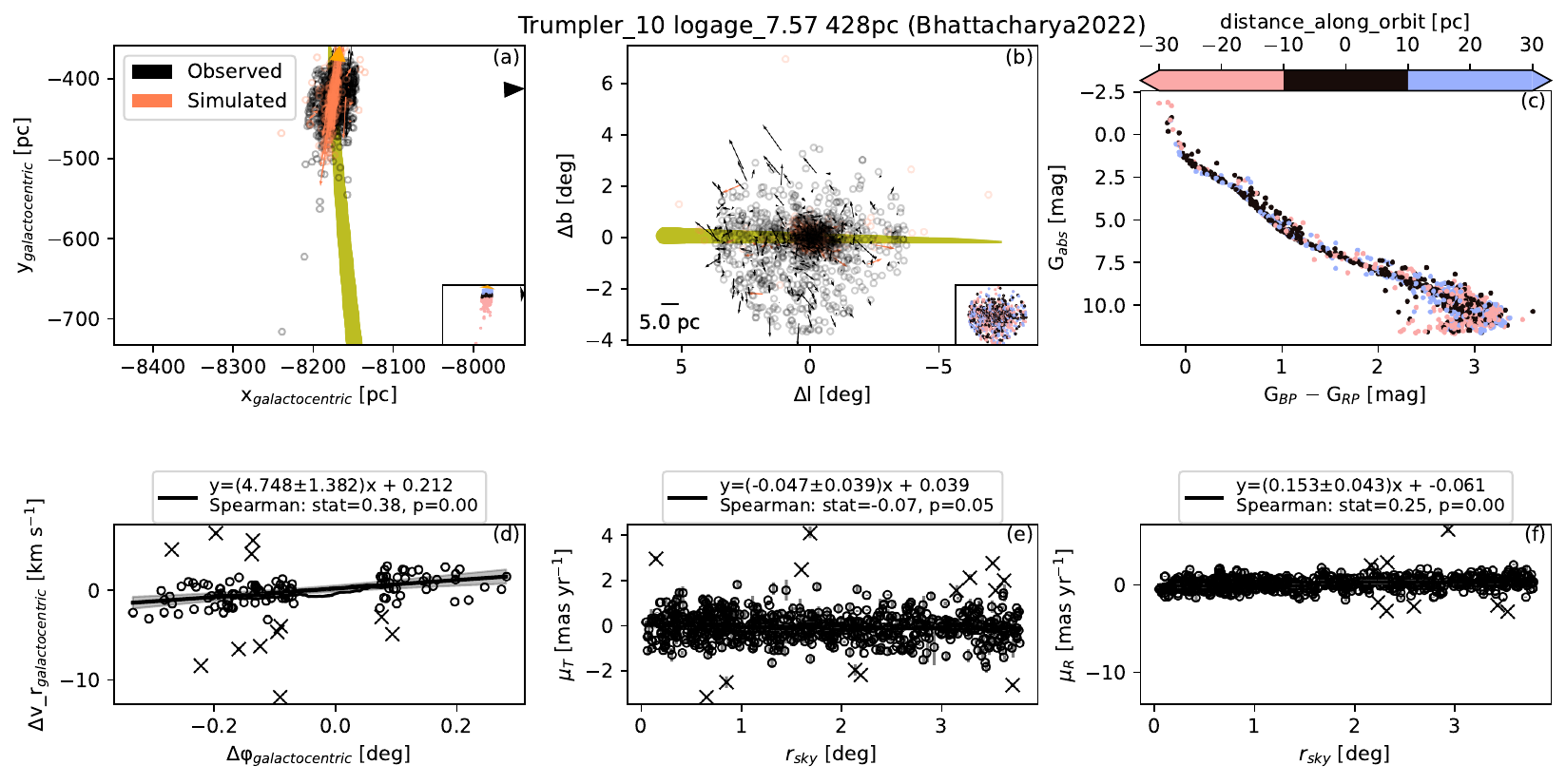}
    \caption{Diagnostic figures for Theia 517 (Risbud2025), Theia 517 (Vaher2023), Theia 517 (Bhattacharya2022), Trumpler 10 (Bhattacharya2022).}
        \label{fig:supplementary.Trumpler_10.Bhattacharya2022}
        \end{figure}
         
        \begin{figure}
\includegraphics[width=0.5\linewidth]{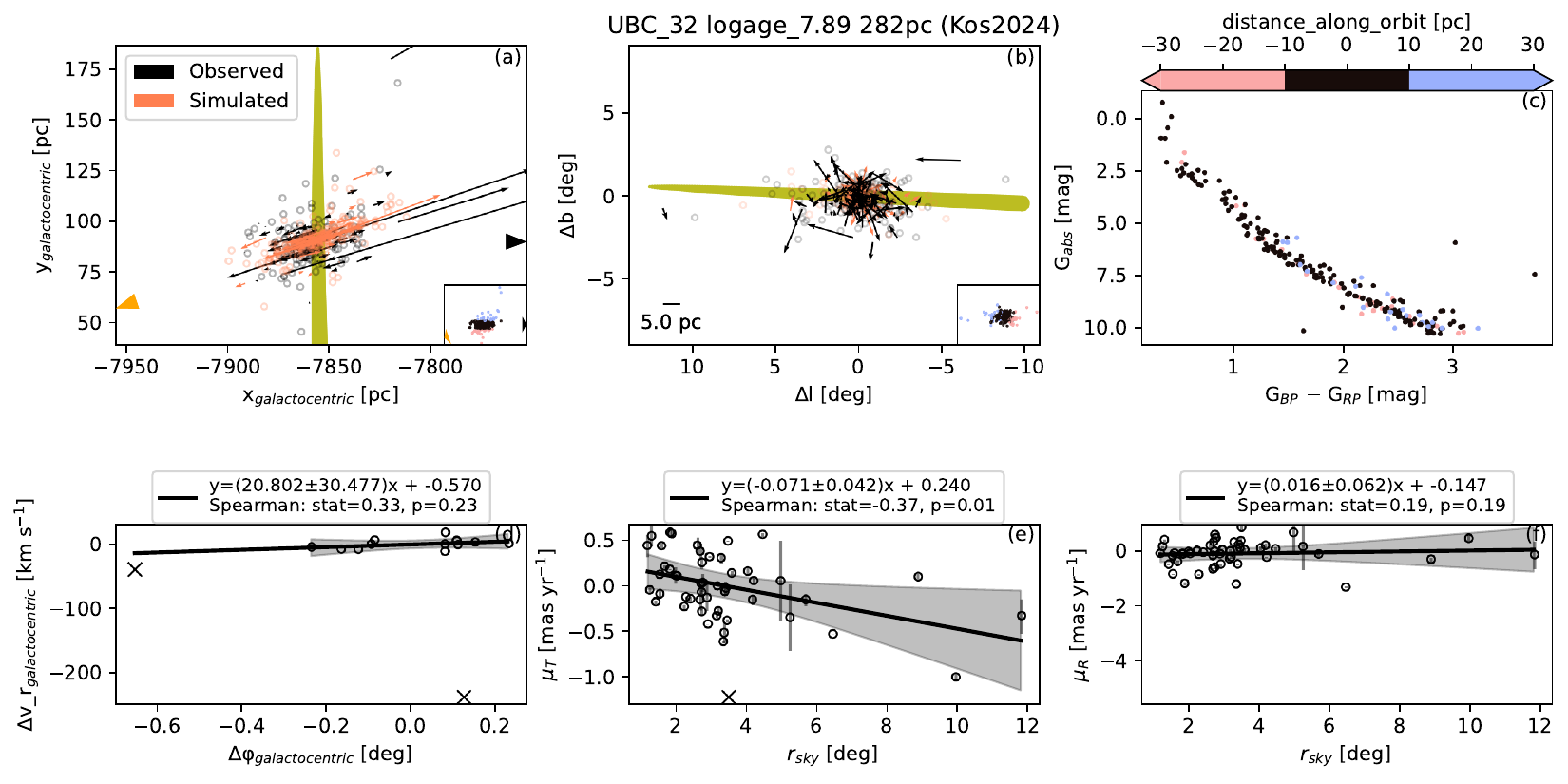}
\includegraphics[width=0.5\linewidth]{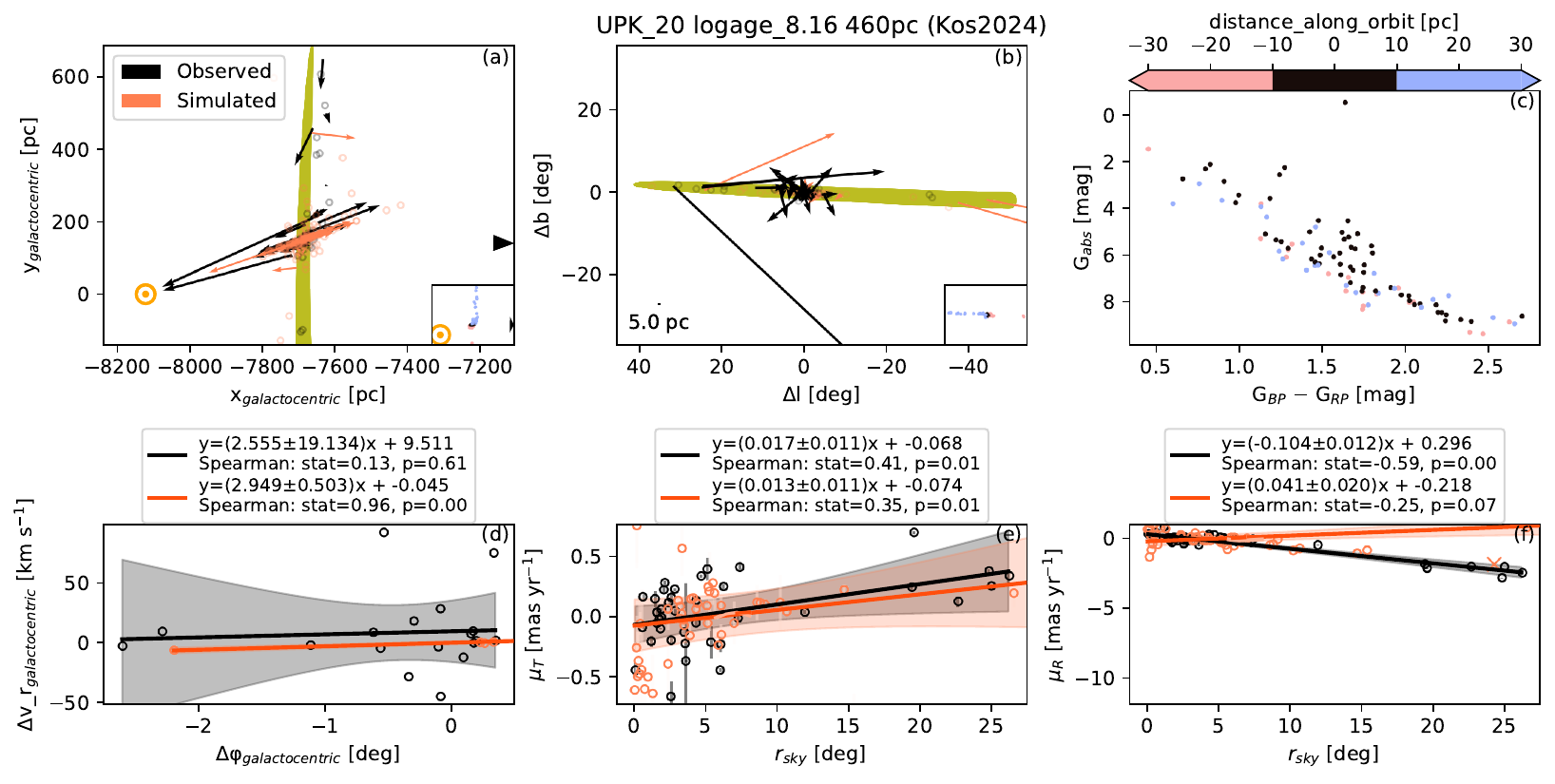}
\includegraphics[width=0.5\linewidth]{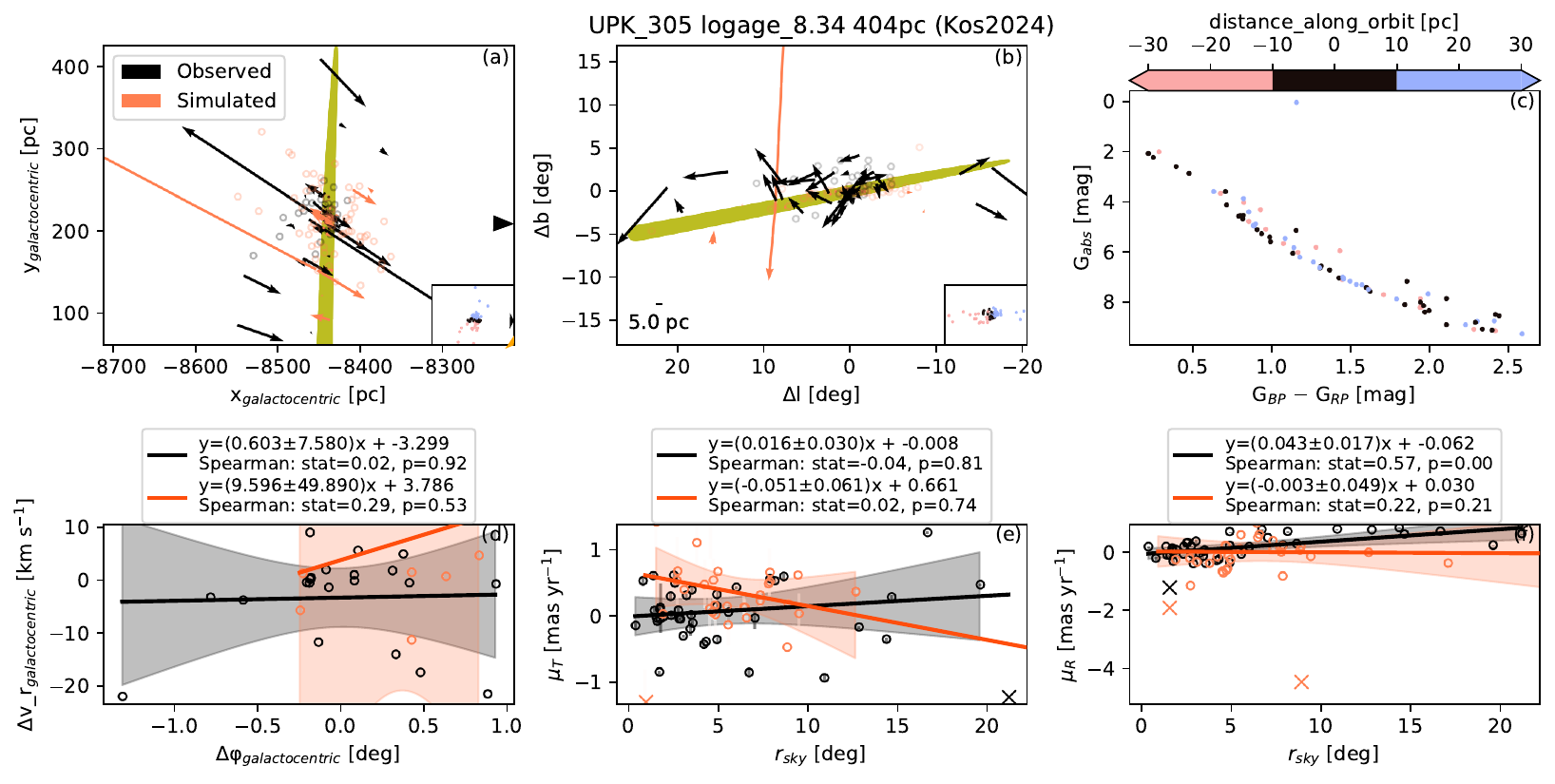}
\includegraphics[width=0.5\linewidth]{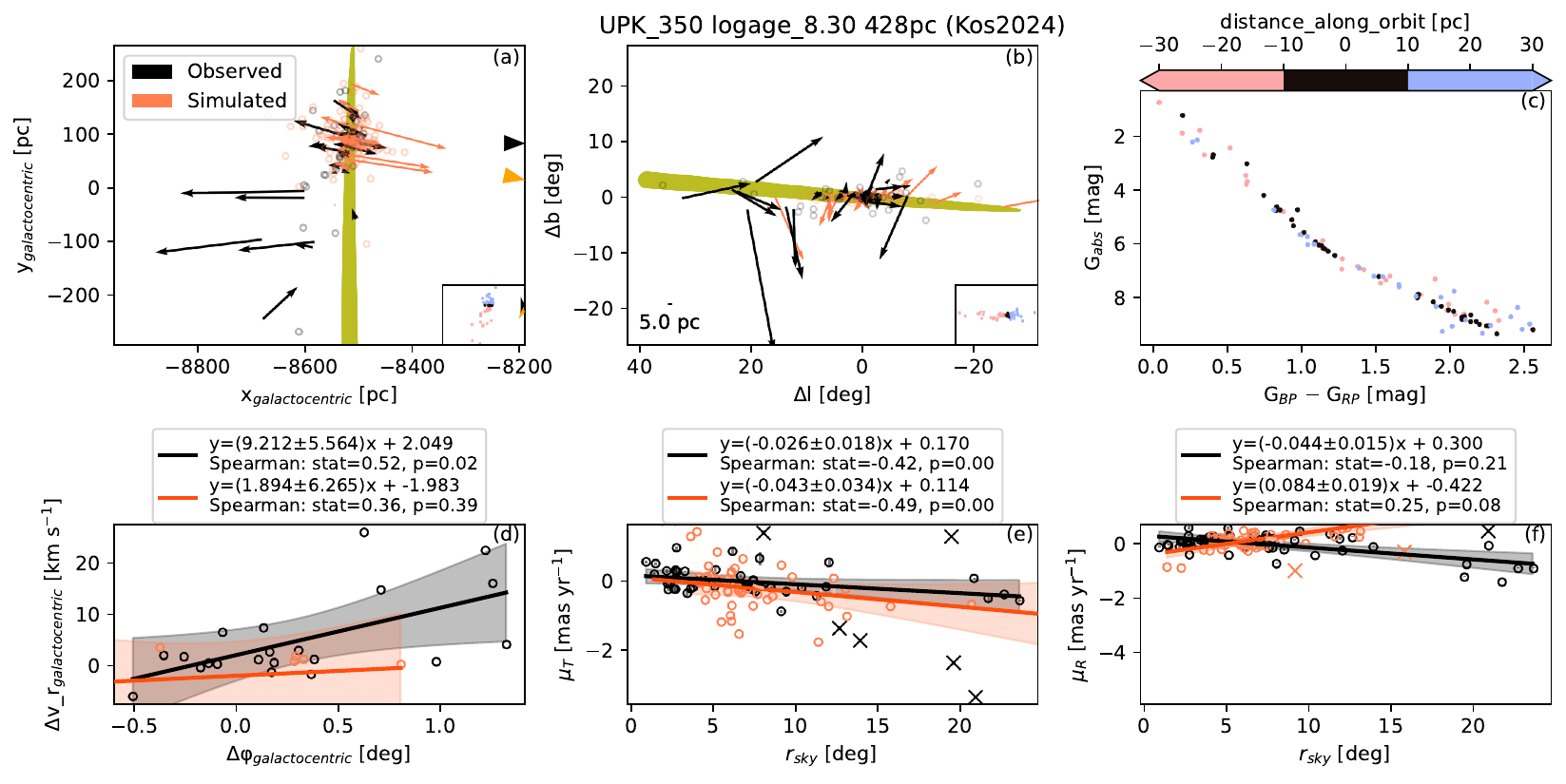}
    \caption{Diagnostic figures for UBC 32 (Kos2024), UPK 20 (Kos2024), UPK 305 (Kos2024), UPK 350 (Kos2024).}
        \label{fig:supplementary.UPK_350.Kos2024}
        \end{figure}
         
        \begin{figure}
\includegraphics[width=0.5\linewidth]{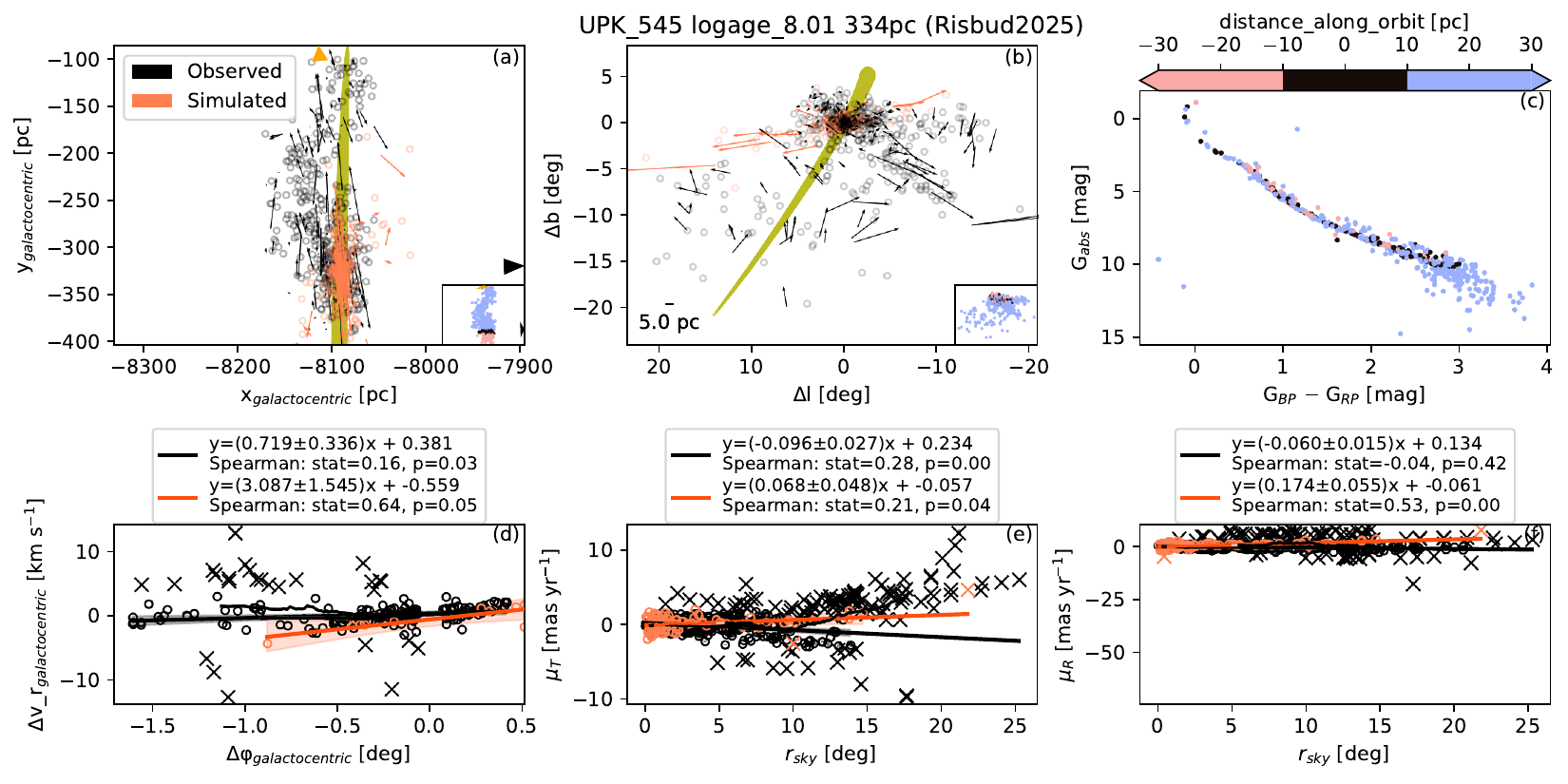}
\includegraphics[width=0.5\linewidth]{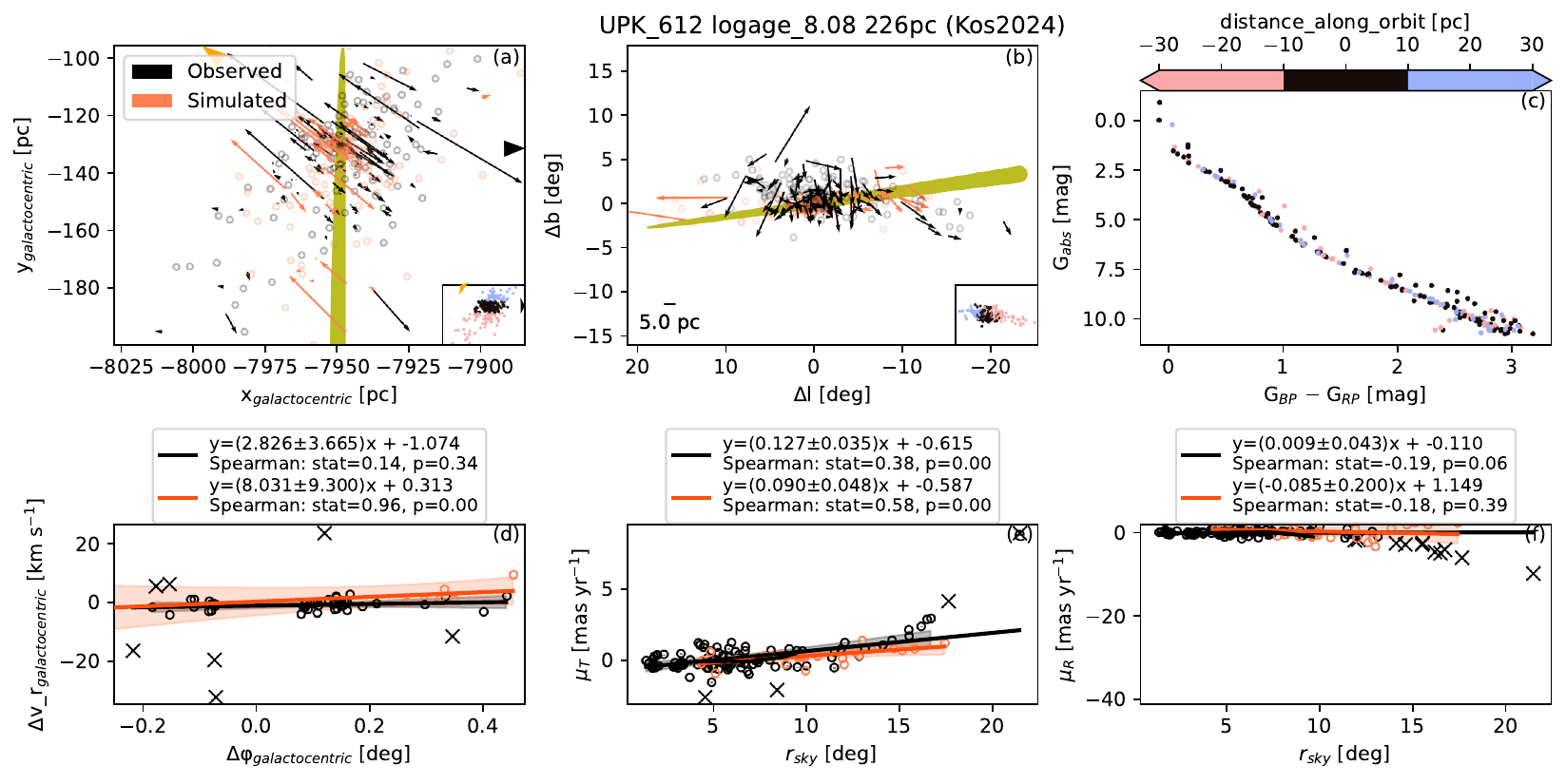}
    \caption{Diagnostic figures for Kos2024 (UPK 545 (Risbud2025), UPK 612 (Kos2024)).}
    \label{fig:supplementary.UPK_612.Kos2024}
    \end{figure}

\end{landscape}

\end{onecolumn}

\end{appendix}

\end{document}